\documentclass[aps,pra,twocolumn,nofootinbib]{revtex4}
\usepackage{amsmath}
\usepackage{amssymb}
\usepackage{epsfig}
\usepackage{graphicx}
\usepackage{bm}
\usepackage{epstopdf}
\pdfoutput=1

\def\be{\begin{equation}}
\def\ee{\end{equation}}
\def\bea{\begin{eqnarray}}
\def\eea{\end{eqnarray}}

\begin{document}

\title{{\Large Vector theory of gravity: }

Universe without black holes and solution of dark energy problem}
\author{Anatoly A. Svidzinsky}
\affiliation{Department of Physics \& Astronomy, Texas A\&M University, College Station,
TX 77843, USA}
\date{\today }

\begin{abstract}
We propose an alternative theory of gravity which assumes that background
geometry of the Universe is fixed four dimensional Euclidean space and
gravity is a vector field $A_{k}$ in this space which breaks the Euclidean
symmetry. Direction of $A_{k}$ gives the time coordinate, while
perpendicular directions are spatial coordinates. Vector gravitational field
is coupled to matter universally and minimally through the equivalent metric 
$f_{ik}$ which is a functional of $A_{k}$. We show that such assumptions
yield a unique theory of gravity, it is free of black holes and, to the best
of our knowledge, passes all available tests. For cosmology our theory
predicts the same evolution of the Universe as general relativity with
cosmological constant and zero spatial curvature. However, the present
theory provides explanation of the dark energy as energy of longitudinal
gravitational field induced by the Universe expansion and yields, with no
free parameters, the value of $\Omega _{\Lambda }=2/3\approx 0.67$ which is
consistent with the recent Planck result $\Omega _{\Lambda }=0.686\pm 0.02$.
Such close agreement with cosmological data indicates that gravity has a
vector, rather than tensor, origin. We demonstrate that gravitational wave
signals measured by LIGO are compatible with vector gravity. They are
produced by orbital inspiral of massive neutron stars which can exist in the
present theory. We also quantize gravitational field and show that quantum
vector gravity is equivalent to QED. Vector gravity can be tested by making
more accurate measurement of the time delay of radar signal traveling near
the Sun; by improving accuracy of the light deflection experiments; or by
measuring propagation direction of gravitational waves relative to laser
interferometer arms. Resolving the supermassive object at the center of our
Galaxy with VLBA could provide another test of gravity and also shed light
on the nature of dark matter.
\end{abstract}

\maketitle
{\small \tableofcontents}

\section{Introduction}

A century ago, Albert Einstein completed the general theory of relativity 
\cite{Eins15}. Einstein's theory then became an accepted theory of gravity.
In general relativity the space-time geometry $g_{ik}$ (metric tensor) is
the gravitational field described by the action%
\begin{equation}
S_{\text{GR}}=-\frac{c^{3}}{16\pi G}\int d^{4}x\sqrt{-g}g^{ik}R_{ik}-\int
\rho \sqrt{g_{ik}\frac{dx^{i}}{dt}\frac{dx^{k}}{dt}}d^{4}x,  \label{aGR}
\end{equation}%
where $G$ is the gravitational constant, $c$ is the speed of light and $%
R_{ik}$ is the Ricci tensor. The second term in Eq. (\ref{aGR}) describes
interaction between gravitational field and matter with the rest mass
density $\rho (t,\mathbf{r})$. Variation of the action (\ref{aGR}) with
respect to $g_{ik}$ yields Einstein equations%
\begin{equation}
R_{ik}=\frac{8\pi G}{c^{4}}\left( T_{ik}-\frac{1}{2}g_{ik}T\right) ,
\label{i1}
\end{equation}%
where $T_{ik}$ is the energy-momentum tensor of matter. Einstein equations (%
\ref{i1}) are a consequence of the postulate that space-time geometry $%
g_{ik} $ is gravitational field.

One should mention that so far general relativity was accurately tested only
at weak gravitational field \cite{Will93,Will06} and thus it is not a theory
fully confirmed experimentally. Observations of binary pulsars yet have not
provided an accurate test of general relativity at strong gravity. Even
though neutron stars in the binary systems are relativistic objects, they
are sufficiently well-separated and all aspects of their orbital behavior
and gravitational wave generation in general relativity are characterized
only by their net masses and angular momentum. As a result, observations of
binary pulsars tested Einstein equations for the weak time-dependent field
and also the equivalence principle which guarantees effacement of the bodies
(relativistic) internal structure. One should note, however, that strong
internal gravitational fields of neutron stars can affect orbital dynamics
and gravitational wave generation in alternative theories of gravity that
violate the strong equivalence principle \cite{Will93,Will06,Will14}. Only
in this sense observations of binary pulsars is a test of strong gravity.

Recent direct detection of gravitational waves from a binary
\textquotedblleft black hole\textquotedblright\ merger by the LIGO team \cite%
{Abbo16,Abbo16a,Abbo17} is also not an accurate test of strong gravity. Such
detection was unable to constrain higher-order post-Newtonian parameters
with a reasonable accuracy \cite{Abbo16b,Abbo16c}. Obtained bounds on
relative deviations in the post-Newtonian parameters are of the order of $%
O(1)$. We show in Sec. \ref{test} of this paper that the LIGO signal can be
interpreted in the framework of vector gravity as being produced by a merger
of massive neutron stars, rather than black holes, which yields radiation
waveform compatible with the LIGO\ data. We also demonstrate that stable
neutron stars with a simple linear \textquotedblleft
causal\textquotedblright\ equation of state can have masses upto about $35$M$%
_{\odot }$ in vector gravity (see Sec. \ref{NSML}). Neutron star mass can be
much larger if \textquotedblleft causality\textquotedblright\ constraint on
the equation of state is not imposed. Moreover, as we show in Sec. \ref{dark}%
, in vector gravity, compact objects composed of dark matter can have masses
exceeding billions solar masses.

In 1998, published observations of Type Ia supernovae by the High-Z
Supernova Search Team \cite{Ries98} followed in 1999 by the Supernova
Cosmology Project \cite{Perl99} suggested that the expansion of the universe
is accelerating. Since then, these observations have been corroborated by
several independent sources. Measurements of the cosmic microwave
background, gravitational lensing, and the large-scale structure of the
cosmos as well as improved measurements of supernovae have been consistent
with the Lambda-CDM model, which is the current standard model of cosmology.
According to this model, a mysterious dark energy gives the main
contribution to the Universe composition. Work done in 2013 based on the
Planck spacecraft observations of the cosmic microwave background gave the
most accurate estimate of about $68$\% of dark energy\ in the Universe \cite%
{Planck14}.

There are questions that general relativity is unable to answer without
extending the model. E.g., general relativity alone can not explain why
Universe is spatially flat; it does not provide a mechanism of matter
generation at the Big Bang; and it can not explain the value of the
cosmological term (mysterious dark energy). It also predicts existence of
singularities such as black holes when a massive star collapses into a point
with zero volume and infinite matter density. One can argue that general
relativity becomes invalid in the vicinity of singularities and a quantum
theory of gravity will remove them. In contrast, the present theory is free
of such singularities at the classical level. Namely, the end point of a
gravitational collapse is not a point singularity but rather a stable star
with a reduced mass. One should mention that black holes have never been
observed directly and \textquotedblleft evidences\textquotedblright\ of
their existence are based on the presumption that general relativity
describes gravity for strong field. Until signatures of the event horizons
are found the existence of black holes will not be proven.

Here we propose an alternative theory of gravity which is a Lagrangian-based
vector field theory in fixed four dimensional Euclidean space. The present
vector theory is a metric theory of gravity \cite{Will93} which means that
space-time is endowed with a symmetric equivalent metric $f_{ik\text{ }}$
formed out of the vector field and Euclidean metric. Matter and
nongravitational fields respond only to the space-time metric $f_{ik\text{ }%
} $. The world lines of test bodies are geodesics of that metric and in
local freely falling frames the nongravitational laws of physics are those
of special relativity. Our theory is prior-geometric for it contains the
fixed background Euclidean geometry (Euclidean metric) and gravity is a
dynamical four-vector field in this geometry which generates the space-time
curvature (equivalent metric $f_{ik\text{ }}$).

Despite the existence of the fixed background geometry, we show that vector
theory of gravity passes all available tests. At strong field our theory
substantially deviates from general relativity and yields no black holes.
For cosmology the present theory gives the same evolution of the Universe as
general relativity with cosmological constant and zero spatial curvature.
However, zero spatial curvature of the Universe is a solution of our
equations, while in general relativity the spatial curvature is a free
parameter. Moreover, the vector theory of gravity yields, with no free
parameters, the value of the cosmological constant $\Omega _{\Lambda
}=2/3\approx 0.67$ which agrees with the recent Planck result $\Omega
_{\Lambda }=0.686\pm 0.02$ \cite{Planck14}. Such precise agreement is a
strong argument in favor of the vector nature of gravity.

Physical explanation of the dark energy (cosmological term) in our theory is
the following. Expansion of the Universe yields change of spatial scale with
time which can be viewed as an increase of the distance between masses. This
generates matter current directed away from an observer. Such current
induces longitudinal vector gravitational field in a similar way as electric
current creates vector potential in classical electrodynamics. Average
energy of the longitudinal gravitational field induced by the Universe
expansion is the mysterious dark energy. Contrary to matter, it has negative
energy density and accelerates expansion of the Universe.

Vector gravity also suggests a mechanism of matter generation at the Big
Bang without involving an additional hypothetical field (inflaton). Namely,
matter was created at the expense of production of negative energy gravitons
and the gravitational field itself caused the stage of inflation and heated
up the Universe. According to the vector gravity, the total energy of the
Universe (which includes the energy of matter and gravitational field) is
equal to zero.

As we show in Sec. \ref{quantization}, quantization of the linearized
equations of vector gravity yields quantum theory which is equivalent to QED.

An interesting feature of our theory is that equations for gravitational
field can be solved analytically for arbitrary static mass distribution (see
Sec. \ref{static}). If point masses are located at $\mathbf{r}_{1}$, $%
\mathbf{r}_{2}$, \ldots\ $\mathbf{r}_{N}$ then exact solution of the field
equations for the equivalent metric is 
\begin{equation}
f_{ik}=\left( 
\begin{array}{cccc}
e^{2\phi } & 0 & 0 & 0 \\ 
0 & -e^{-2\phi } & 0 & 0 \\ 
0 & 0 & -e^{-2\phi } & 0 \\ 
0 & 0 & 0 & -e^{-2\phi }%
\end{array}%
\right) ,  \label{x4}
\end{equation}%
where 
\begin{equation}
\phi (\mathbf{r})=-\frac{m_{1}}{|\mathbf{r}-\mathbf{r}_{1}|}-\ldots -\frac{%
m_{N}}{|\mathbf{r}-\mathbf{r}_{N}|}  \label{x5}
\end{equation}%
and $m_{k}$ ($k=1,\ldots ,N$) are constants determined by the value of
masses. Solution (\ref{x4}) is free of black holes: photons with a radial
velocity component can always escape from gravitationally compact objects.
In recent years, the evidence for the existence of ultra-compact
supermassive objects at centers of galaxies has become very strong. It is
important to note that present solution (\ref{x4}) not only argues that such
objects are not black holes, but also can explain quantitatively their
observed properties and give us a hint about composition of dark matter (see
Sec. \ref{dark} and Ref. \cite{Svid07}).

Before we proceed with building the vector theory of gravity we discuss an
algorithm that we use to construct the theory. Classical electrodynamics is
an example of a successful field theory which is very well tested. It
postulates that electromagnetic field is a four dimensional vector $A_{k}$
in Minkowski space-time. The conserved $4-$current density $j^{k}$ is the
source of the electromagnetic field which is coupled to $A_{k}$ through the
Lorentz invariant term in the action%
\begin{equation}
S_{\text{coupl}}=-\frac{1}{c^{2}}\int d^{4}xA_{k}j^{k}.  \label{se}
\end{equation}

Such postulate allows us to construct classical electrodynamics in a unique
way using symmetries of Eq. (\ref{se}). Namely, conservation of current
yields that $S_{\text{coupl}}$ is invariant under the gauge transformation $%
A_{k}\rightarrow A_{k}+\partial \psi /\partial x^{k}$. Action for the
electromagnetic field $S_{\text{field}}$ must possess the same symmetries as
the coupling term $S_{\text{coupl}}$, namely, it must be Lorentz and gauge
invariant. Such a requirement, together with the condition that $S_{\text{%
field}}$ must be quadratic in field derivatives yields a unique answer for $%
S_{\text{field}}$%
\begin{equation}
S_{\text{field}}=-\frac{1}{16\pi c}\int d^{4}x\left( \frac{\partial A_{k}}{%
\partial x^{i}}-\frac{\partial A_{i}}{\partial x^{k}}\right) \left( \frac{%
\partial A^{k}}{\partial x_{i}}-\frac{\partial A^{i}}{\partial x_{k}}\right)
.
\end{equation}

Variation of the total action of the system $S=S_{\text{field}}+S_{\text{%
coupl}}+S_{\text{matter}}$ with respect to $A_{k}$ yields Maxwell equations
and variation with respect to the particle trajectories gives the Lorentz
force. Thus, the whole classical electrodynamics is uniquely assembled from
the structure of the coupling term (\ref{se}).

In the next sections we will use the same algorithm to construct the
classical vector theory of gravity. Namely, first we postulate how the
gravitational field is coupled to matter. Then, using symmetries of the
coupling term, we uniquely assemble the total action and obtain the
classical field equations from the variational principle. One should note
that symmetries of the coupling term in vector gravity are very different
from the symmetries of Eq. (\ref{se}). As a consequence, the vector theory
of gravity substantially differs from classical electrodynamics.

When we make a transition from the classical to quantum electrodynamics we
must make an additional assumption, namely, that the quantum of the vector
filed $A_{k}$, the photon, is an elementary particle. This postulate yields
that photon is a boson because fermion field does not transform as a vector.
However, a product of two fermion fields can transform as a vector. E.g., a
four-vector field $A_{k}$ can be created from a fermion-antifermion pair as 
\cite{Bjor65} 
\begin{equation*}
A_{k}=\Psi ^{+}\gamma _{0}\gamma _{k}\Psi ,
\end{equation*}%
where $\gamma _{k}$ are $4\times 4$ gamma matrices and $\Psi $ is the
fermion field. In this regard there is an interesting proposal that the
photon is not an elementary particle but rather a composite particle formed
of fermion-antifermion pairs (the so called composite theory of photon). The
idea that the photon is a composite particle dates back to 1932, when Louis
de Broglie \cite{DB32} suggested that the photon is composed of
neutrino-antineutrino pairs. Recall that many composite bosons, such as
Cooper pairs, atoms with total integer spin, deuterons, pions, and kaons,
are not perfect bosons because of their internal fermion structure, however
in the asymptotic limit they are essentially bosons \cite{Perk02}. This
suggests that photon could be a particle composed of spin$-1/2$ fermions as
well. Work on the composite theory of photon continues to be of some
interest \cite{Dvoe01,Perk02,Sen07,Perk13,Perk14}.

The present paper deals with the vector gravity, rather than photons.
However, as we show, the measured value of the energy loss by binary stars
due to emission of gravitational waves can be explained in the vector
gravity only if the quantum of the vector gravitational field (the graviton)
is a composite particle formed of fermion-antifermion pairs. This fact makes
a link between vector gravity and the composite theory of photon. We
quantize gravitational field in Sec. \ref{quantization} assuming that
graviton is a composite particle and obtain a wonderful result that quantum
vector gravity is equivalent to QED.

One of the motivations for the composite photon theory is the lack of a well
defined wave function for a single photon. In particular, the authors of the
classic book on Quantum Optics \cite{Scul97} say: \textquotedblleft There
is, strictly speaking, no such a thing as a photon wave
function\textquotedblright . If this is the case, in order for QED to be
compatible with first quantization the photon can not be an elementary
particle, but rather a particle composed of fermions. Recall that for
fermions, described by the Dirac equation, the wave function is well
defined. Hence, the assumption that photon is composed of fermions seems
natural. For QED such assumption yields essentially no detectable
experimental consequences (see Sec. \ref{quantization} and Ref. \cite{Perk02}%
). However, as we show in Sec. \ref{quantization}, for quantum vector
gravity the consequences are dramatic.

There is also a proposal that a fundamental unified theory of the
gravitational, electroweak and strong interactions can be formulated in
terms of only fermionic degrees of freedom \cite{Hebe03,Wett04}.

\section{Postulates of the vector theory of gravity}

The present vector theory of gravity is based on four postulates:

\textbf{1.} Background geometry of the Universe is a fixed four dimensional
Euclidean space with metric $\delta _{ik}=$diag$(1,1,1,1)$. Such space is
completely isotropic and has no preferred directions.

\textbf{2.} In the four dimensional Euclidean space there is a dynamical $4-$%
vector field $A_{k}$ (the gravitational field) which breaks the symmetry.
Namely, direction of $A_{k}$ is now preferred and this direction becomes the
time coordinate. Directions perpendicular to $A_{k}$ are three spatial
coordinates.

\textbf{3.} Vector gravitational field is coupled to matter and all
nongravitational fields through the equivalent metric $f_{ik}$ which is an
algebraic function of $A_{k}$ and the background Euclidean metric $\delta
_{ik}$. Gravitational field couples universally and minimally to all the
fields of the Standard Model by replacing everywhere the Minkowski metric $%
\eta _{ik}$ with the equivalent metric $f_{ik\text{ }}$ and replacing
partial derivatives with covariant derivatives formed from $f_{ik}$. In
particular, the trajectories of freely falling bodies are geodesics of the
equivalent metric $f_{ik\text{ }}$. Action for a point particle with mass $m$
moving in the gravitational field reads%
\begin{equation}
S_{\text{matter}}=-mc\int \sqrt{f_{ik}dx^{i}dx^{k}},  \label{d1}
\end{equation}%
where $c$ is the speed of light. Action (\ref{d1}) has the same form as in
general relativity, however, the tensor gravitational field $g_{ik}$ of
general relativity is now replaced with the equivalent metric $f_{ik}$. One
should note that the Einstein equivalence principle is a consequence of the
action (\ref{d1}).

\textbf{4. }The quantum of the vector field $A_{k}$ (the graviton) is not an
elementary particle, but rather it is a composite particle formed of
massless fermion-antifermion pairs. Emission and absorption of a graviton
corresponds to creation and annihilation of particle-antiparticle pairs.

For construction of the classical equations for the vector field $A_{k}$ it
does not matter whether the graviton is an elementary particle or a
composite particle. However, processes involving gravitons, e.g. emission of
gravitational waves, might depend on the graviton composition. The present
vector theory of gravity can quantitatively explain the energy loss by
binary stars orbiting each other provided that graviton is a composite
particle.

The postulates 1-3 outlined above allow us to construct the classical vector
theory of gravity in a unique way using symmetries of the action $S_{\text{%
matter}}$ (\ref{d1}). Such symmetries uniquely specify the structure of the
total action. We proceed with assembling the classical theory in the next
sections.

\section{Equivalent metric}

In Appendix \ref{AP1} we show that Einstein equivalence principle yields the
following unique expression for the equivalent metric $f_{ik}$ in terms of
the vector field $A_{k}$ and the background Euclidean metric $\delta _{ik}$%
\begin{equation}
f_{ik}=-\frac{\delta _{ik}}{A}+\left( A+\frac{1}{A}\right) \frac{A_{i}A_{k}}{%
A^{2}},  \label{d2}
\end{equation}%
where%
\begin{equation*}
A=\sqrt{A_{i}A_{k}\delta ^{ik}}.
\end{equation*}

Throughout the paper we use the usual conventions. Namely, unless otherwise
noted, there is summation over repeated indices. Lower case Latin indices ($%
i $, $k$, $l$, ...) label four dimensional coordinates (range $0$, $1$, $2$, 
$3 $), while lower case Greek letters $\alpha $, $\beta $, $\gamma $ denote
spatial coordinates (range $1$, $2$, $3$).

In Cartesian coordinate system if we chose $x^{0}-$axis along the direction
of $A_{k}$ the equivalent metric is diagonal and reads%
\begin{equation*}
f_{ik}=\text{diag}\left( A,-\frac{1}{A},-\frac{1}{A},-\frac{1}{A}\right) .
\end{equation*}

Since $A_{k}$ is a dynamical variable one can make a transformation $%
A_{k}\rightarrow F(A)A_{k}$, where $F$ is an arbitrary function of $A$. Such
a transformation changes the norm of $A_{k}$. It also modifies the
expression for the metric (\ref{d2}) and field equations. However, the
physical answer, e.g., motion of particles in gravitational field is
independent of how we normalize the vector field. Thus, we can choose the
field normalization in any suitable way.

Instead of $A_{k}$, it is convenient to introduce new independent functions,
a scalar $\phi $ and a unit vector $u_{k}$, according to the relations 
\begin{equation*}
A=e^{2\phi },\quad u_{k}=\frac{A_{k}}{A}.
\end{equation*}%
The vector $u_{k}$ has the unit norm 
\begin{equation*}
u_{i}u_{k}\delta ^{ik}=1.
\end{equation*}%
In terms of the unit vector $u_{k}$ and the scalar $\phi $ the equivalent
metric (\ref{d2}) reads%
\begin{equation}
f_{ik}=-e^{-2\phi }\delta _{ik}+2\cosh (2\phi )u_{i}u_{k},  \label{met}
\end{equation}%
while metric $\tilde{f}^{ik}$ inverse to $f_{ik}$, defined as $\tilde{f}%
^{ik}f_{im}=\delta _{m}^{k}$, is%
\begin{equation}
\tilde{f}^{ik}=-e^{2\phi }\delta ^{ik}+2\cosh (2\phi )u^{i}u^{k},
\label{meti}
\end{equation}%
where%
\begin{equation*}
u^{i}=\delta ^{ik}u_{k}.
\end{equation*}
In the present paper raising and lowering of the indices is performed using
Euclidean metric $\delta _{ik}$, unless otherwise stated. We denote
determinant of $f_{ik}$ as $f$. Equation (\ref{met}) yields%
\begin{equation*}
\sqrt{-f}=e^{-2\phi }.
\end{equation*}

One should note that in the literature there were attempts to construct a
vector theory of gravity in background Minkowski (rather than Euclidean)
metric $\eta _{ik}=$diag$(1,-1,-1,-1)$ by Rastall \cite{Rast75,Rast77} and
by the author \cite{Svid09}. The equivalent metric obtained in such theories
has a form similar to our Eq. (\ref{met}) in which Euclidean metric $\delta
_{ik}$ is replaced with $-\eta _{ik}$ and $\cosh (2\phi )$ is replaced with $%
\sinh (2\phi )$. However, the main achievement of the present paper compared
to the previous work on vector gravity is the discovery of how to obtain the
action for the gravitational field $S_{\text{gravity}}$ based on symmetries
of the coupling term $S_{\text{matter}}$. In the previous attempts to
construct the theory $S_{\text{gravity}}$ has not possessed symmetries of $%
S_{\text{matter}}$ which yielded no success.

\section{Action for gravitational field}

Next we construct action of the system in terms of $\phi $ and $u_{k}$. The
total action for the gravitational field and matter is given by 
\begin{equation}
S=S_{\text{gravity}}+S_{\text{matter}},  \label{d3}
\end{equation}%
where $S_{\text{matter}}$ is the action of matter written in curvilinear
coordinates with the metric $f_{ik}$. We obtain the action for the
gravitational field $S_{\text{gravity}}$ using the requirement that
symmetries of $S_{\text{matter}}$ and $S_{\text{gravity}}$ must be the same. 
$S_{\text{matter}}$ possesses the following symmetry: it is invariant under
coordinate transformations that leave background Euclidean metric $\delta
_{ik}$ intact. This symmetry is exact. In addition there are approximate
symmetries. For small deviations of $\phi $ from a constant value $\phi _{0}$
and small deviations of $u_{k}$ from $(1,0,0,0)$ in the rescaled coordinates 
\begin{equation}
x^{0}\rightarrow e^{-\phi _{0}}x^{0},\qquad x^{\alpha }\rightarrow e^{\phi
_{0}}x^{\alpha }  \label{res}
\end{equation}%
the equivalent metric is given by%
\begin{equation}
f_{ik}=\eta _{ik}+\left( 
\begin{array}{cccc}
h_{00} & h_{01} & h_{02} & h_{03} \\ 
h_{01} & h_{00} & 0 & 0 \\ 
h_{02} & 0 & h_{00} & 0 \\ 
h_{03} & 0 & 0 & h_{00}%
\end{array}%
\right) ,  \label{mmm0}
\end{equation}%
where 
\begin{equation}
h_{00}=2(\phi -\phi _{0}),\qquad h_{0\alpha }=2\cosh (2\phi _{0})u_{\alpha },
\label{mmm1}
\end{equation}%
$\alpha =1$, $2$, $3$ and $\eta _{ik}=$diag$(1,-1,-1,-1)$ is the Minkowski
metric. For small deviations of $f_{ik}$ from the Minkowski metric, using%
\begin{equation}
\delta S_{\text{matter}}=-\frac{1}{2c}\int d^{4}x\sqrt{-f}T^{ik}\delta
f_{ik},  \label{s11a}
\end{equation}%
we have 
\begin{equation}
\delta S_{\text{matter}}\approx -\frac{1}{2c}\int d^{4}x\left(
T^{00}h_{00}+2T^{0\alpha }h_{0\alpha }+T^{\alpha \alpha }h_{00}\right) ,
\label{s11}
\end{equation}%
where $T^{ik}$ is the energy-momentum tensor of matter. Recall that the
energy-momentum tensor $T^{ik}$ is obtained as a functional derivative of $%
L_{\text{matter}}$ with respect to the metric tensor $g_{ik}$ \cite{Land95}%
\begin{equation*}
T^{ik}=-\frac{2}{\sqrt{-g}}\frac{\delta \left( \sqrt{-g}L_{\text{matter}%
}\right) }{\delta g_{ik}},
\end{equation*}%
where $L_{\text{matter}}$ is the nongravitational part of the Lagrangian
density of the action written in curvilinear coordinates with metric $g_{ik}$%
\begin{equation*}
S_{\text{matter}}=\frac{1}{c}\int d^{4}x\sqrt{-g}L_{\text{matter}}.
\end{equation*}%
Therefore \cite{Land95} 
\begin{equation*}
\delta S_{\text{matter}}=-\frac{1}{2c}\int d^{4}x\sqrt{-g}T^{ik}\delta g_{ik}
\end{equation*}%
which yields Eq. (\ref{s11a}).

One can see from Eq. (\ref{s11}) that in the rescaled coordinates the action 
$S_{\text{matter}}$ is independent of the background cosmological field $%
\phi _{0}$. This is one of the symmetries of $S_{\text{matter}}$. Another
symmetry can be found by taking into account Eq. (\ref{s11}) and approximate
energy conservation in the Minkowski metric%
\begin{equation*}
\frac{\partial T^{00}}{\partial x^{0}}+\frac{\partial T^{0\alpha }}{\partial
x^{\alpha }}=0
\end{equation*}%
which yields that action $S_{\text{matter}}$ is approximately invariant
under the gauge transformation%
\begin{equation}
h_{00}\rightarrow h_{00}+2\frac{\partial \psi }{\partial x^{0}},\quad
h_{0\alpha }\rightarrow h_{0\alpha }+\frac{\partial \psi }{\partial
x^{\alpha }}  \label{gg1}
\end{equation}%
upto the terms quadratic in the mass velocity $V^{\alpha }=dx^{\alpha }/dt$.
Here $\psi $ is an arbitrary scalar function.

In addition, there is approximate Lorentz invariance. Namely, the line
element that enters the matter action (\ref{d1}) can be approximately
written in the metric (\ref{mmm0}) as%
\begin{equation*}
ds^{2}=f_{ik}dx^{i}dx^{k}\approx
\end{equation*}%
\begin{equation}
\eta _{ik}dx^{i}dx^{k}+h_{00}(dx^{0})^{2}+2h_{0\alpha }dx^{0}dx^{\alpha
}+h_{00}d\mathbf{r}^{2}  \label{gg0}
\end{equation}%
which is invariant (upto the terms of the order of $V^{3}/c^{3}$) under
transformation%
\begin{equation}
x^{0}\rightarrow \left( 1+\frac{V^{2}}{2c^{2}}\right) x^{0}+\frac{1}{c}%
\mathbf{V\cdot r},\quad \mathbf{r}\rightarrow \mathbf{r}+\frac{\mathbf{V}}{c}%
x^{0},  \label{gg2}
\end{equation}%
\begin{equation}
h_{00}\rightarrow h_{00}\left( 1+\frac{2V^{2}}{c^{2}}\right) -2\frac{%
V^{\alpha }}{c}h_{0\alpha },\quad h_{0\alpha }\rightarrow h_{0\alpha }-2%
\frac{V^{\alpha }}{c}h_{00},  \label{gg3}
\end{equation}%
where $\mathbf{V}$ is a constant (velocity) vector and $\alpha =1$, $2$, $3$%
. In Eq. (\ref{gg0}) $h_{0\alpha }$ and $d\mathbf{r}/dx^{0}$ are of the
order of $V/c$.

Requirement that $S_{\text{gravity}}$ must also possess these symmetries,
namely, $S_{\text{gravity}}$ should be invariant under Euclidean
transformations; for small deviations from the background field be
independent of $\phi _{0}$ after scaling transformation (\ref{res}); and
approximately invariant under the gauge (\ref{gg1}) and low-velocity Lorentz
(\ref{gg2}), (\ref{gg3}) transformations, allows us to find $S_{\text{gravity%
}}$ uniquely\footnote{%
As a matter of fact, to fix $S_{\text{gravity}}$ uniquely one should also
use condition of gauge symmetry of the action in a higher post-Newtonian
order. We discuss this in Appendix \ref{AP2}.}. In Appendix \ref{AP2} we
obtain that the gravitational field action in the background four
dimensional Euclidean space is%
\begin{equation*}
S_{\text{gravity}}=\frac{c^{3}}{8\pi G}\int d^{4}x\left[ \frac{\partial \phi 
}{\partial x^{i}}\frac{\partial \phi }{\partial x^{k}}\left( -\delta
^{ik}+\left( 1-3e^{-4\phi }\right) u^{i}u^{k}\right) \right.
\end{equation*}%
\begin{equation*}
+\cosh ^{2}(2\phi )\frac{\partial u_{i}}{\partial x^{k}}\frac{\partial u_{m}%
}{\partial x^{l}}\Big( \delta ^{im}\delta ^{kl}-\delta ^{il}\delta ^{km} -
\end{equation*}%
\begin{equation}
\left. \left( 1+e^{-4\phi }\right) \delta ^{im}u^{k}u^{l}\Big) +2\left(
1+e^{-4\phi }\right) \frac{\partial \phi }{\partial x^{i}}\frac{\partial
u_{m}}{\partial x^{k}}\delta ^{im}u^{k}\right] .  \label{fa2}
\end{equation}%
where $G$ is the gravitational constant.

Action (\ref{fa2}) is written in Euclidean metric, it has no free parameters
and serves as a foundation of the present theory of gravity. Our derivation
of the action (\ref{fa2}) is unique and, hence, the classical vector theory
of gravity is also a unique consequence of the postulates 1-3.

\section{Equations for classical gravitational field}

Taking variation of the total action (\ref{d3}) with respect to $\phi $ and
unit vector $u_{k}$ yields four equations for the classical gravitational
field in the background Euclidean space (see Appendix \ref{AP6} for
derivation details)

\begin{equation*}
\left[ \delta ^{mk}u^{i}-2\delta ^{im}u^{k}+\left( 1+3e^{-4\phi }\right)
u^{m}u^{k}u^{i}\right] \frac{\partial ^{2}\phi }{\partial x^{m}\partial x^{k}%
}
\end{equation*}%
\begin{equation*}
+2\left[ \delta ^{im}+\left( 3e^{-4\phi }-1\right) u^{m}u^{i}\right] \frac{%
\partial \phi }{\partial x^{m}}\frac{\partial \phi }{\partial x^{k}}u^{k}
\end{equation*}%
\begin{equation*}
+2\left[ e^{4\phi }\left( \delta _{l}^{k}\delta ^{im}-\delta _{l}^{i}\delta
^{mk}\right) +\delta _{l}^{i}\delta ^{mk}-\delta _{l}^{m}\delta ^{ik}\right] 
\frac{\partial \phi }{\partial x^{k}}\frac{\partial u^{l}}{\partial x^{m}}
\end{equation*}%
\begin{equation*}
+\left[ 2\left( e^{4\phi }-2e^{-4\phi }-1\right) \delta
_{l}^{i}u^{m}u^{k}-\left( 1+3e^{-4\phi }\right) \delta
_{l}^{m}u^{i}u^{k}\right.
\end{equation*}%
\begin{equation*}
\left. -\left( 2e^{4\phi }+3e^{-4\phi }+1\right) \delta _{l}^{k}u^{m}u^{i} 
\right] \frac{\partial \phi }{\partial x^{k}}\frac{\partial u^{l}}{\partial
x^{m}}
\end{equation*}

\begin{equation*}
+\cosh (2\phi )\left[ e^{2\phi }\frac{\partial }{\partial x^{k}}\left( \frac{%
\partial u^{k}}{\partial x_{i}}-\frac{\partial u^{i}}{\partial x_{k}}\right)
+e^{-2\phi }u_{m}u^{i}\frac{\partial ^{2}u^{m}}{\partial x_{k}\partial x^{k}}%
\right.
\end{equation*}%
\begin{equation*}
+\left. 2\cosh (2\phi )u^{k}u^{l}\frac{\partial ^{2}u^{i}}{\partial
x^{l}\partial x^{k}}-\left( e^{2\phi }+2e^{-2\phi }\right) u^{m}u^{i}\frac{%
\partial ^{2}u^{k}}{\partial x^{k}\partial x^{m}}\right]
\end{equation*}%
\begin{equation*}
+2\cosh ^{2}(2\phi )\left[ \frac{\partial u^{i}}{\partial x^{k}}\frac{%
\partial }{\partial x^{m}}\left( u^{k}u^{m}\right) -\frac{\partial u_{k}}{%
\partial x_{i}}\frac{\partial u^{k}}{\partial x^{l}}u^{l}\right.
\end{equation*}%
\begin{equation*}
\left. -\frac{\partial u^{k}}{\partial x^{m}}\frac{\partial u^{m}}{\partial
x^{k}}u^{i}+\left( 1+2e^{-4\phi }\right) \frac{\partial u^{k}}{\partial x^{m}%
}\frac{\partial u_{k}}{\partial x^{l}}u^{m}u^{l}u^{i}\right]
\end{equation*}%
\begin{equation}
=\frac{8\pi G}{c^{4}}\left( T^{ik}-\frac{T}{2}\tilde{f}^{ik}\right) u_{k},
\label{sss1}
\end{equation}%
where $T^{ik}$ is the energy-momentum tensor of matter, $T=T^{mk}f_{mk}$ is
the trace of the energy-momentum tensor and $\tilde{f}^{ik}$ is the metric
inverse to $f_{ik}$ given by Eq. (\ref{meti}). The right hand side of Eqs. (%
\ref{sss1}) is the source of the gravitational field.

Equations (\ref{sss1}) for the scalar $\phi $ and the unit vector $u_{k}$
are the main equations of the vector theory of gravity. They are written in
Euclidean metric which means that raising and lowering of indexes is carried
out using metric $\delta _{ik}=$diag$(1,1,1,1)$. Equations (\ref{sss1}) play
the same role in vector gravity as Einstein equations in general relativity.

In our theory the motion of particles in gravitational field is described by
the same equation as in general relativity

\begin{equation}
\frac{d^{2}x^{b}}{ds^{2}}=\frac{1}{2}\tilde{f}^{bl}\left[ \frac{\partial
f_{ik}}{\partial x^{l}}-\frac{\partial f_{lk}}{\partial x^{i}}-\frac{%
\partial f_{il}}{\partial x^{k}}\right] \frac{dx^{i}}{ds}\frac{dx^{k}}{ds},
\label{r10aa}
\end{equation}%
where $ds=\sqrt{f_{ik}dx^{i}dx^{k}}$. In Eq. (\ref{r10aa}) the metric $%
g_{ik} $ of general relativity is replaced with the equivalent metric $%
f_{ik} $. We obtain equation of particle motion in Appendix \ref{motion}.

Next we explore solutions of the classical gravitational field equations (%
\ref{sss1}) for various cases.

\section{Static gravitational field}

\label{static}

In this section we consider gravitational field produced by rest matter
distributed with density $\rho (\mathbf{r})$. This situation is idealized
since gravitational interaction will lead to mass motion unless there are
other forces that keep matter at rest. Here we solve gravitational field
equations assuming that masses are static. Self-consistent solution of the
problem requires solving the field equations together with the equation of
motion for masses. Here we replace the latter with the constraint that
position of masses is fixed.

For static field $u_{k}=(1,0,0,0)$, scalar $\phi $ depends only on spatial
coordinates $\mathbf{r}$ and the equivalent metric (\ref{met}) reads 
\begin{equation}
f_{ik}=\left( 
\begin{array}{cccc}
e^{2\phi (\mathbf{r})} & 0 & 0 & 0 \\ 
0 & -e^{-2\phi (\mathbf{r})} & 0 & 0 \\ 
0 & 0 & -e^{-2\phi (\mathbf{r})} & 0 \\ 
0 & 0 & 0 & -e^{-2\phi (\mathbf{r})}%
\end{array}%
\right) ,  \label{s10}
\end{equation}%
$\sqrt{-f}=e^{-2\phi }$, while the inverse metric is 
\begin{equation}
\tilde{f}^{ik}=\left( 
\begin{array}{cccc}
e^{-2\phi } & 0 & 0 & 0 \\ 
0 & -e^{2\phi } & 0 & 0 \\ 
0 & 0 & -e^{2\phi } & 0 \\ 
0 & 0 & 0 & -e^{2\phi }%
\end{array}%
\right) .
\end{equation}

For static masses the energy-momentum tensor of matter has only one nonzero
component $T^{00}$ which depends on spatial coordinates. This component can
be found from the conservation equation 
\begin{equation}
T_{;k}^{ik}=0,  \label{cons}
\end{equation}%
where $";"$ stands for the covariant derivative with metric $f_{ik}$.
Equation (\ref{cons}) yields $T^{00}=\rho c^{2}e^{\phi }$ and $T=\rho
c^{2}e^{3\phi }$, where $\rho $ is the mass density which is independent of $%
\phi $.

For static gravitational field, Eqs. (\ref{sss1}) reduce to a single
equation for $\phi (\mathbf{r})$ 
\begin{equation}
\Delta \phi =\frac{4\pi G}{c^{2}}\rho e^{\phi }.  \label{z10}
\end{equation}

In the Newtonian limit Eq. (\ref{z10}) yields $\Delta \phi =4\pi G\rho
e^{\phi _{0}}/c^{2}$ and, thus, $c^{2}\phi (\mathbf{r})$ has a meaning of
gravitational potential.

Exponential metric solution (\ref{s10}) is free of black holes for any mass
distribution and field strength. For a point mass $M$ located at $r=0$ Eq. (%
\ref{z10}) leads to $c^{2}\Delta \phi =4\pi GM\delta (\mathbf{r})$ and has a
solution $\phi =-GM/c^{2}r$. For $N$ point masses located at $\mathbf{r}%
_{1}, $ \ldots $,$ $\mathbf{r}_{N}$ Eq. (\ref{z10}) yields%
\begin{equation}
\Delta \phi =4\pi \left[ m_{1}\delta (\mathbf{r}_{1})+\ldots +m_{N}\delta (%
\mathbf{r}_{N})\right] ,  \label{r6}
\end{equation}%
where $m_{1}$, \ldots , $m_{N}$ are positive constants. Solution of Eq. (\ref%
{r6}) is 
\begin{equation}
\phi (\mathbf{r})=-\frac{m_{1}}{|\mathbf{r}-\mathbf{r}_{1}|}-\ldots -\frac{%
m_{N}}{|\mathbf{r}-\mathbf{r}_{N}|}.  \label{r7}
\end{equation}

We discuss motion of particles in static gravitational field in Appendix \ref%
{StaticMotion}. For a star of mass $M$ and radius $R$ Eq. (\ref{r7}) reduces
to $\phi (r)=-GM/c^{2}r$ ($r\geq R$) and using Eq. (\ref{u12}) for energy
conservation we obtain that escape velocity for a particle from the stellar
surface is 
\begin{equation}
v=c_{s}\sqrt{1-e^{2\phi (R)}},  \label{x6}
\end{equation}%
where $c_{s}=ce^{2\phi (R)}$ is the speed of light at the stellar surface
(see Eq. (\ref{u21})). Equation (\ref{x6}) shows that escape velocity is
always smaller then $c_{s}$ ($c_{s}\leq c$).

One should note that exponential metric (\ref{s10}) was also obtained for
static field in some alternative theories of gravity \cite%
{Yilm58,Yilm71,Rose71,Chan80,Chan80b,Rast77,Svid09} and based on simple
physical arguments in \cite{Rast75,Lind81,Mart10}. Stability of compact
stars in the exponential metric has been investigated in the literature. It
has been shown that solution (\ref{s10}) predicts that stars do not collapse
into a point singularity but rather form stable compact objects with no
event horizon and finite gravitational redshift \cite{Robe99}.

\section{Post-Newtonian limit}

\label{PPN}

Post-Newtonian limit applies when the gravitational field is weak, and the
motion of the matter is slow. It is sufficiently accurate to encompass all
solar-system tests of gravity performed so far. In the post-Newtonian
formalism the metric is expanded in a small parameter $\epsilon $. The
\textquotedblleft order of smallness\textquotedblright\ is determined
according to the rules that matter velocity is of order $V\sim \epsilon
^{1/2}$ and gravitational constant $G\sim \epsilon $. A consistent
post-Newtonian limit requires determination of $g_{00}$ correction through $%
O(\epsilon ^{2})$, $g_{0\alpha }$ through $O(\epsilon ^{3/2})$, and $%
g_{\alpha \beta }$ through $O(\epsilon )$ \cite{Will93}.

We compare the vector theory of gravity with general relativity in the
post-Newtonian limit in the cosmological reference frame (the mean rest
frame of the Universe in which the Universe appears isotropic) for which
background equivalent metric $f_{ik}$ is diagonal and after rescaling
coordinates reduces to Minkowski metric. As it is shown in Ref. \cite{Will93}%
, comparison of any metric theory of gravity with general relativity can be
done in any suitable reference frame. The rest frame of the Universe is a
convenient choice due to symmetry of the situation.

Let us consider small deviations $h_{ik}$ of the tensor gravitational field $%
g_{ik}$ from the Minkowski metric $\eta _{ik}$%
\begin{equation*}
g_{ik}=\eta _{ik}+h_{ik}.
\end{equation*}%
In the post-Newtonian limit of general relativity in the post-Newtonian
gauge \cite{Will93} 
\begin{equation}
h_{\beta }^{\alpha }=-h_{00}\delta _{\beta }^{\alpha }  \label{pp4}
\end{equation}%
and the tensor gravitational field $g_{ik}$ is described by four independent
functions $h_{0k}$, so that metric is given by 
\begin{equation}
g_{ik}=\eta _{ik}+\left( 
\begin{array}{cccc}
h_{00} & h_{01} & h_{02} & h_{03} \\ 
h_{01} & h_{00} & 0 & 0 \\ 
h_{02} & 0 & h_{00} & 0 \\ 
h_{03} & 0 & 0 & h_{00}%
\end{array}%
\right) .  \label{eee3}
\end{equation}

In the present vector theory of gravity in the post-Newtonian limit after
rescaling of coordinates in the cosmological reference frame the equivalent
metric $f_{ik}$ also has the form of Eq. (\ref{eee3}) (see Eq. (\ref{mmm0}%
)). Therefore, in both theories in the post-Newtonian limit the
gravitational fields are described by an equal number of independent
functions which are coupled with matter in the same way. Since, by
construction of both theories, such coupling uniquely specifies the total
action and, hence, the field equations, the general relativity and the
vector theory of gravity are identical in the post-Newtonian limit.

To convince the reader that this is indeed the case, in Appendix \ref{AP7}
we show directly that in the post-Newtonian limit the general relativity and
the vector theory of gravity give the same equations for $h_{0k}$ in the
cosmological reference frame%
\begin{equation*}
\frac{1}{2}\Delta h_{00}+\frac{3}{2}\frac{\partial ^{2}h_{00}}{\partial
x^{0}\partial x^{0}}-\frac{\partial ^{2}h_{0\beta }}{\partial x^{0}\partial
x^{\beta }}+\frac{1}{2}h_{00}\Delta h_{00}-\frac{1}{2}\left( \nabla
h_{00}\right) ^{2}
\end{equation*}%
\begin{equation}
=\frac{8\pi G}{c^{4}}\left( T_{00}-\frac{1}{2}g_{00}T\right) ,  \label{pn1}
\end{equation}%
\begin{equation}
\frac{1}{2}\Delta h_{0\alpha }-\frac{1}{2}\frac{\partial ^{2}h_{0\beta }}{%
\partial x^{\alpha }\partial x^{\beta }}+\frac{\partial ^{2}h_{00}}{\partial
x^{0}\partial x^{\alpha }}=\frac{8\pi G}{c^{4}}T_{0\alpha }.  \label{pn2}
\end{equation}%
Boundary conditions for $h_{0k}$ are also the same and, therefore, both
theories are equivalent in the post-Newtonian limit.

We made comparison of the theories in the cosmological reference frame
assuming that matter moves relative to this frame with nonrelativistic
speed. Since both theories are equivalent in such a frame in the
post-Newtonian limit they are also equivalent in a frame moving with non
relativistic velocity relative to the mean rest frame of the Universe. This
is the case because in vector gravity the equivalent metric $f_{ik}$, by its
definition, is also a tensor under general coordinate transformations. If
equations of vector gravity and general relativity give the same metric in
one reference frame then in any other frame the two metrics will coincide
because both of them transform in the same way under coordinate
transformation from one frame to another.

This result can be also obtained by noting that Eqs. (\ref{pn1}) and (\ref%
{pn2}) are invariant under the low-velocity Lorentz transformation%
\begin{equation}
\frac{\partial }{\partial x^{0}}\rightarrow \frac{\partial }{\partial x^{0}}-%
\frac{\mathbf{V}}{c}\nabla ,\quad \frac{\partial }{\partial \mathbf{r}}%
\rightarrow \frac{\partial }{\partial \mathbf{r}}-\frac{\mathbf{V}}{c}\frac{%
\partial }{\partial x^{0}}+\frac{\mathbf{V}}{2c^{2}}\left( \mathbf{V}\frac{%
\partial }{\partial \mathbf{r}}\right) ,  \label{lt1}
\end{equation}%
\begin{equation}
h_{00}\rightarrow h_{00}\left( 1+\frac{2V^{2}}{c^{2}}\right) -2\frac{%
V^{\alpha }}{c}h_{0\alpha },\quad h_{0\alpha }\rightarrow h_{0\alpha }-2%
\frac{V^{\alpha }}{c}h_{00},  \label{lt2}
\end{equation}%
\begin{equation}
T^{00}\rightarrow T^{00}+T^{00}\frac{V^{2}}{c^{2}}+2\frac{V^{\alpha }}{c}%
T^{0\alpha },\quad T^{\alpha 0}\rightarrow T^{\alpha 0}+\frac{V^{\alpha }}{c}%
T^{00}  \label{lt3}
\end{equation}%
for which $h_{0k}$ and $T^{0k}$ transform as symmetric tensors (keeping in
mind that in the post-Newtonian limit $h_{\alpha \alpha }=h_{00}$), and $%
\mathbf{V}$ is a constant (velocity) vector. Thus, for any reference frame
moving with nonrelativistic speed relative to the mean rest frame of the
Universe (e.g., frame of the Solar System) equations of vector gravity and
general relativity are equivalent in the post-Newtonian limit and have the
form of Eqs. (\ref{pn1}) and (\ref{pn2}).

As a consequence, the vector theory of gravity, similarly to general
relativity, yields no post-Newtonian preferred frame and preferred location
effects. For vector gravity the ten post-Newtonian parameters introduced to
compare metric theories of gravity with each other \cite{Will93,Will06} have
the same values as in general relativity.

To convince the most sceptical reader, in Appendix \ref{PPNF} we investigate
the post-Newtonian limit of vector gravity in the framework of the
parametrized post-Newtonian formalism following the $9-$step procedure of
Ref. \cite{Will93} and explicitly calculate the ten post-Newtonian
parameters. As expected, they are equal to those in general relativity.

\section{Weak field limit}

\subsection{Linearized gravitational field equations}

In this section we linearize equations for classical gravitational field in
the cosmological reference frame assuming that unit vector $u_{k}$ slightly
deviates from $(1,0,0,0)$. For small deviations of $\phi $ from a constant
value $\phi _{0}$ and $|u_{\alpha }|\ll 1$, keeping linear terms, Eqs. (\ref%
{sss1}) yield%
\begin{equation*}
\Delta \phi +3e^{-4\phi _{0}}\frac{\partial ^{2}\phi }{\partial
x^{0}\partial x^{0}}-2e^{-2\phi _{0}}\cosh (2\phi _{0})\frac{\partial
^{2}u^{\beta }}{\partial x^{0}\partial x^{\beta }}
\end{equation*}%
\begin{equation*}
=\frac{8\pi G}{c^{4}}\left( T^{00}-\frac{T}{2}\tilde{f}^{00}\right) ,
\end{equation*}

\begin{equation*}
\cosh (2\phi _{0})\left( e^{2\phi _{0}}\frac{\partial ^{2}u^{\beta }}{%
\partial x_{\alpha }\partial x^{\beta }}-e^{2\phi _{0}}\Delta u^{\alpha
}+e^{-2\phi _{0}}\frac{\partial ^{2}u^{\alpha }}{\partial x^{0}\partial x^{0}%
}\right)
\end{equation*}%
\begin{equation*}
-2\frac{\partial ^{2}\phi }{\partial x^{\alpha }\partial x^{0}}=\frac{8\pi G%
}{c^{4}}T^{\alpha 0}.
\end{equation*}%
In the rescaled coordinates 
\begin{equation*}
x^{0}\rightarrow e^{-\phi _{0}}x^{0},\qquad x^{\alpha }\rightarrow e^{\phi
_{0}}x^{\alpha }
\end{equation*}%
the equivalent metric is given by%
\begin{equation*}
f_{ik}=\eta _{ik}+\left( 
\begin{array}{cccc}
h_{00} & h_{01} & h_{02} & h_{03} \\ 
h_{01} & h_{00} & 0 & 0 \\ 
h_{02} & 0 & h_{00} & 0 \\ 
h_{03} & 0 & 0 & h_{00}%
\end{array}%
\right) ,
\end{equation*}%
where 
\begin{equation*}
h_{00}=2(\phi -\phi _{0}),\quad h_{0\alpha }=2\cosh (2\phi _{0})u_{\alpha }
\end{equation*}%
and $\alpha =1$, $2$, $3$. In terms of $h_{0k}$ equations for the
gravitational field in the weak field limit read%
\begin{equation}
\Delta h_{00}+3\frac{\partial ^{2}h_{00}}{\partial x^{0}\partial x^{0}}-2%
\frac{\partial ^{2}h_{0\beta }}{\partial x^{0}\partial x^{\beta }}=\frac{%
16\pi G}{c^{4}}\left( T^{00}-\frac{T}{2}\right) ,  \label{wf1}
\end{equation}%
\begin{equation}
\left( \frac{\partial ^{2}}{\partial x^{0}\partial x^{0}}-\Delta \right)
h_{0\alpha }+\frac{\partial ^{2}h_{0\beta }}{\partial x^{\alpha }\partial
x^{\beta }}-2\frac{\partial ^{2}h_{00}}{\partial x^{\alpha }\partial x^{0}}=%
\frac{16\pi G}{c^{4}}T^{\alpha 0}.  \label{wf2}
\end{equation}%
In these equations the energy-momentum tensor of matter $T^{ik}$ is written
in Minkowski metric. Equations (\ref{wf1}) and (\ref{wf2}) are invariant
upto the terms of the order of $V^{2}/c^{2}$ under the low-velocity Lorentz
transformation (\ref{lt1})-(\ref{lt3}). Therefore, they remain the same in
inertial reference frames moving with non relativistic velocity relative to
the rest frame of the Universe.

Equation (\ref{r10aa}) yields that non relativistic motion of particles in
weak gravitational field is described by the following equation%
\begin{equation}
\frac{1}{c^{2}}\frac{dV^{\alpha }}{dt}=\frac{\partial h_{0\alpha }}{\partial
x^{0}}-\frac{1}{2}\frac{\partial h_{00}}{\partial x^{\alpha }}-\left( \frac{%
\partial h_{0\beta }}{\partial x^{\alpha }}-\frac{\partial h_{0\alpha }}{%
\partial x^{\beta }}\right) \frac{V^{\beta }}{c}+\frac{\partial h_{00}}{%
\partial x^{0}}\frac{V^{\alpha }}{c},  \label{em}
\end{equation}%
where $V^{\alpha }=dx^{\alpha }/dt$ is the particle velocity.

In Appendix \ref{AP4} we explore an analogy between weak gravity and
electrodynamics and show that equations for weak classical gravitational
field are analogous to Maxwell's equations in a medium with negative
refractive index.

\subsection{Energy density and energy flux in the classical limit of vector
gravity}

In Appendix \ref{AP5} we derive expression for the energy density and energy
density flux (Poynting vector) for the weak classical gravitational field
interacting with matter that moves with nonrelativistic velocity. We find
the following expression for the energy density%
\begin{equation*}
w=-\frac{c^{4}}{32\pi G}\left[ 3\left( \frac{\partial h_{00}}{\partial x^{0}}%
\right) ^{2}-\left( \nabla h_{00}\right) ^{2}+\left( \frac{\partial \mathbf{h%
}}{\partial x^{0}}\right) ^{2}+\text{curl}^{2}\mathbf{h}\right]
\end{equation*}%
\begin{equation}
+\rho c^{2}+\frac{1}{2}\rho c^{2}h_{00}+\frac{1}{2}\rho V^{2},  \label{ed}
\end{equation}%
where $\rho $ is the mass density, $\mathbf{V}$ is the matter velocity and
three dimensional vector%
\begin{equation*}
\mathbf{h}=h^{0\alpha }.
\end{equation*}

The energy density flux is given by%
\begin{equation}
\mathbf{S=}\frac{c^{5}}{16\pi G}\left[ -\left( 2\frac{\partial \mathbf{h}}{%
\partial x^{0}}+\nabla h_{00}\right) \frac{\partial h_{00}}{\partial x^{0}}+%
\frac{\partial \mathbf{h}}{\partial x^{0}}\times \text{curl }\mathbf{h}%
\right] +\rho c\mathbf{V}.  \label{edf}
\end{equation}

Equation of the energy conservation reads%
\begin{equation*}
\frac{\partial w}{\partial x^{0}}+\text{div}\mathbf{S}=0.
\end{equation*}

\subsection{Gravitational waves}

\label{gwave}

The weak field limit homogeneous equations for the classical gravitational
field 
\begin{equation}
\Delta h_{00}+3\frac{\partial ^{2}h_{00}}{\partial x^{0}\partial x^{0}}-2%
\frac{\partial ^{2}h_{0\beta }}{\partial x^{\beta }\partial x^{0}}=0,
\label{gw1}
\end{equation}%
\begin{equation}
\left( \frac{\partial ^{2}}{\partial x^{0}\partial x^{0}}-\Delta \right)
h_{0\alpha }+\frac{\partial ^{2}h_{0\beta }}{\partial x^{\alpha }\partial
x^{\beta }}-2\frac{\partial ^{2}h_{00}}{\partial x^{\alpha }\partial x^{0}}=0
\label{gw2}
\end{equation}%
have solutions describing waves propagating with the speed of light $c$.
Taking $\partial /\partial x^{0}$ from Eq. (\ref{gw1}) and $(1/2)\partial
/\partial x^{\alpha }$ from Eq. (\ref{gw2}) and adding them together we
obtain%
\begin{equation*}
\frac{\partial ^{2}}{\partial x^{0}\partial x^{0}}\left( \frac{\partial
h_{00}}{\partial x^{0}}-\frac{1}{2}\frac{\partial h_{0\alpha }}{\partial
x^{\alpha }}\right) =0.
\end{equation*}%
Thus, for the time-dependent solutions describing gravitational waves 
\begin{equation*}
\frac{\partial h_{00}}{\partial x^{0}}-\frac{1}{2}\frac{\partial h_{0\alpha }%
}{\partial x^{\alpha }}=0
\end{equation*}%
and Eqs. (\ref{gw1}), (\ref{gw2}) reduce to separate wave equations for $%
h_{00}$ and $h_{0\alpha }$%
\begin{equation*}
\left( \frac{\partial ^{2}}{\partial x^{0}\partial x^{0}}-\Delta \right)
h_{00}=0,
\end{equation*}%
\begin{equation*}
\left( \frac{\partial ^{2}}{\partial x^{0}\partial x^{0}}-\Delta \right)
h_{0\alpha }=0.
\end{equation*}

Field equations have two classes of solutions corresponding to transverse
and longitudinal waves. For transverse waves 
\begin{equation*}
\frac{\partial h_{0\alpha }}{\partial x^{\alpha }}=0,\qquad h_{00}=0
\end{equation*}%
and, e.g., for a transverse wave propagating along the $x-$axis the
equivalent metric reads%
\begin{equation}
f_{ik}^{\text{tr}}=\eta _{ik}+\left( 
\begin{array}{cccc}
0 & 0 & h_{0y}(t,x) & h_{0z}(t,x) \\ 
0 & 0 & 0 & 0 \\ 
h_{0y}(t,x) & 0 & 0 & 0 \\ 
h_{0z}(t,x) & 0 & 0 & 0%
\end{array}%
\right) .  \label{met1}
\end{equation}

According to Eqs. (\ref{ed}) and (\ref{edf}), the energy density and the
energy density flux of the transverse gravitational wave is 
\begin{equation}
w_{\text{tr}}=-\frac{c^{4}}{32\pi G}\left[ \left( \frac{\partial \mathbf{h}}{%
\partial x^{0}}\right) ^{2}+\text{curl}^{2}\mathbf{h}\right] ,  \label{ff4}
\end{equation}%
\begin{equation}
\mathbf{S}_{\text{tr}}\mathbf{=}\frac{c^{5}}{16\pi G}\frac{\partial \mathbf{h%
}}{\partial x^{0}}\times \text{curl}\mathbf{(h}).  \label{ff5}
\end{equation}

Thus, graviton has negative energy in the classical description of the
gravitational field. This result has important implication for cosmology. If
graviton energy is negative and the matter energy is positive this suggests
that at the Big Bang matter was created at the expense of generation of the
negative energy gravitons. We address this issue in Sec. \ref{content}. In
Sec. \ref{quantization} we show that in the quantum limit (present epoch)
the graviton energy is positive. In this limit the energy density and the
energy density flux are given by the same Eqs. (\ref{ff4}) and (\ref{ff5})
but with the opposite sign.

For a plane transverse wave 
\begin{equation*}
\mathbf{h=h}_{0}\cos \left( \omega t-\mathbf{kr}\right)
\end{equation*}%
Eq. (\ref{ff5}) yields%
\begin{equation*}
\mathbf{S}_{\text{tr}}\mathbf{=-}\frac{c^{4}h_{0}^{2}\omega }{16\pi G}%
\mathbf{k}\sin ^{2}\left( \omega t-\mathbf{kr}\right) ,
\end{equation*}%
that is Poynting vector which gives the direction of the energy flow is
opposite to the wave vector $\mathbf{k}$ in the classical limit. This is
analogous to propagation of electromagnetic waves in a medium with negative
refractive index \cite{Vese}. Thus, vacuum for the classical vector
gravitational field is left-handed.

For a longitudinal wave $h_{00}\neq 0$ and $\partial h_{0\alpha }/\partial
x^{\alpha }=2\partial h_{00}/\partial x^{0}$. For such wave propagating
along the $x-$axis the metric oscillates as 
\begin{equation}
f_{ik}^{\text{long}}=\eta _{ik}+\left( 
\begin{array}{cccc}
h_{00}(t,x) & h_{0x}(t,x) & 0 & 0 \\ 
h_{0x}(t,x) & h_{00}(t,x) & 0 & 0 \\ 
0 & 0 & h_{00}(t,x) & 0 \\ 
0 & 0 & 0 & h_{00}(t,x)%
\end{array}%
\right) .  \label{met2}
\end{equation}%
As we show in Sec. \ref{binary}, binary stars orbiting each other do not
emit longitudinal gravitational waves. Quantum mechanical analysis of Sec. %
\ref{FQS} yields the same answer. However, longitudinal gravitational waves
can be generated during star mergers or in early Universe.

One should mention that gravitational waves in the vector gravity
substantially differ from those in general relativity. In general relativity
the metric for weak plane gravitational waves propagating along the $x-$axis
in a properly chosen coordinate system reads \cite{Land95}%
\begin{equation}
g_{ik}=\eta _{ik}+\left( 
\begin{array}{cccc}
0 & 0 & 0 & 0 \\ 
0 & 0 & 0 & 0 \\ 
0 & 0 & h_{yy}(t,x) & h_{yz}(t,x) \\ 
0 & 0 & h_{yz}(t,x) & -h_{yy}(t,x)%
\end{array}%
\right) ,  \label{wa3}
\end{equation}%
where $h_{yy}$ and $h_{yz}$ are small perturbations obeying the wave
equation 
\begin{equation*}
\square h_{yy}=0,\qquad \square h_{yz}=0,
\end{equation*}%
where $\square $ is the d'Alembertian operator. By making proper change of
coordinates one can transform Eqs. (\ref{met1}) and (\ref{met2}) for plane
waves in vector gravity to 
\begin{equation}
f_{ik}^{\text{tr}}=\eta _{ik}+\left( 
\begin{array}{cccc}
0 & 0 & 0 & 0 \\ 
0 & 0 & h_{xy}(t,x) & h_{xz}(t,x) \\ 
0 & h_{xy}(t,x) & 0 & 0 \\ 
0 & h_{xz}(t,x) & 0 & 0%
\end{array}%
\right) ,  \label{met1a}
\end{equation}%
\begin{equation}
f_{ik}^{\text{long}}=\eta _{ik}+\left( 
\begin{array}{cccc}
0 & 0 & 0 & 0 \\ 
0 & -2h(t,x) & 0 & 0 \\ 
0 & 0 & h(t,x) & 0 \\ 
0 & 0 & 0 & h(t,x)%
\end{array}%
\right) ,
\end{equation}%
which have different structure than general relativistic Eq. (\ref{wa3}).
Thus, measuring polarization of gravitational waves with gravitational wave
detectors could provide a test of the vector gravity.

One should note, however, that both in general relativity and vector gravity
the polarization of gravitational waves emitted by orbiting binary stars is
transverse, that is wave produces motion of test particles in the plane
perpendicular to the direction of wave propagation. The reader might get an
impression that Eq. (\ref{met1a}) does not describe a transverse wave. This
illusion appears because metric (\ref{met1a}) is written in coordinate
system in which test particles do not move under the influence of
gravitational wave. The original metric (\ref{met1}) clearly shows that the
wave is transverse. According to Eq. (\ref{em}), a rest particle (or mirrors
of an interferometer) will move under the influence of the gravitational
wave (\ref{met1}) with time-dependent velocity $V^{\alpha }=h_{0\alpha }c$.
Metric (\ref{met1a}) is obtained from (\ref{met1}) by making a coordinate
transformation into the co-moving frame 
\begin{equation*}
x^{\prime \alpha }=x^{\alpha }-\int^{t}V^{\alpha }dt,
\end{equation*}%
which yields the following relation between components of the metrics (\ref%
{met1a}) and (\ref{met1}): $h_{xy}=h_{0y}$, $h_{xz}=h_{0z}$.

\begin{figure}[t]
\centering
\includegraphics[width=8.5cm]{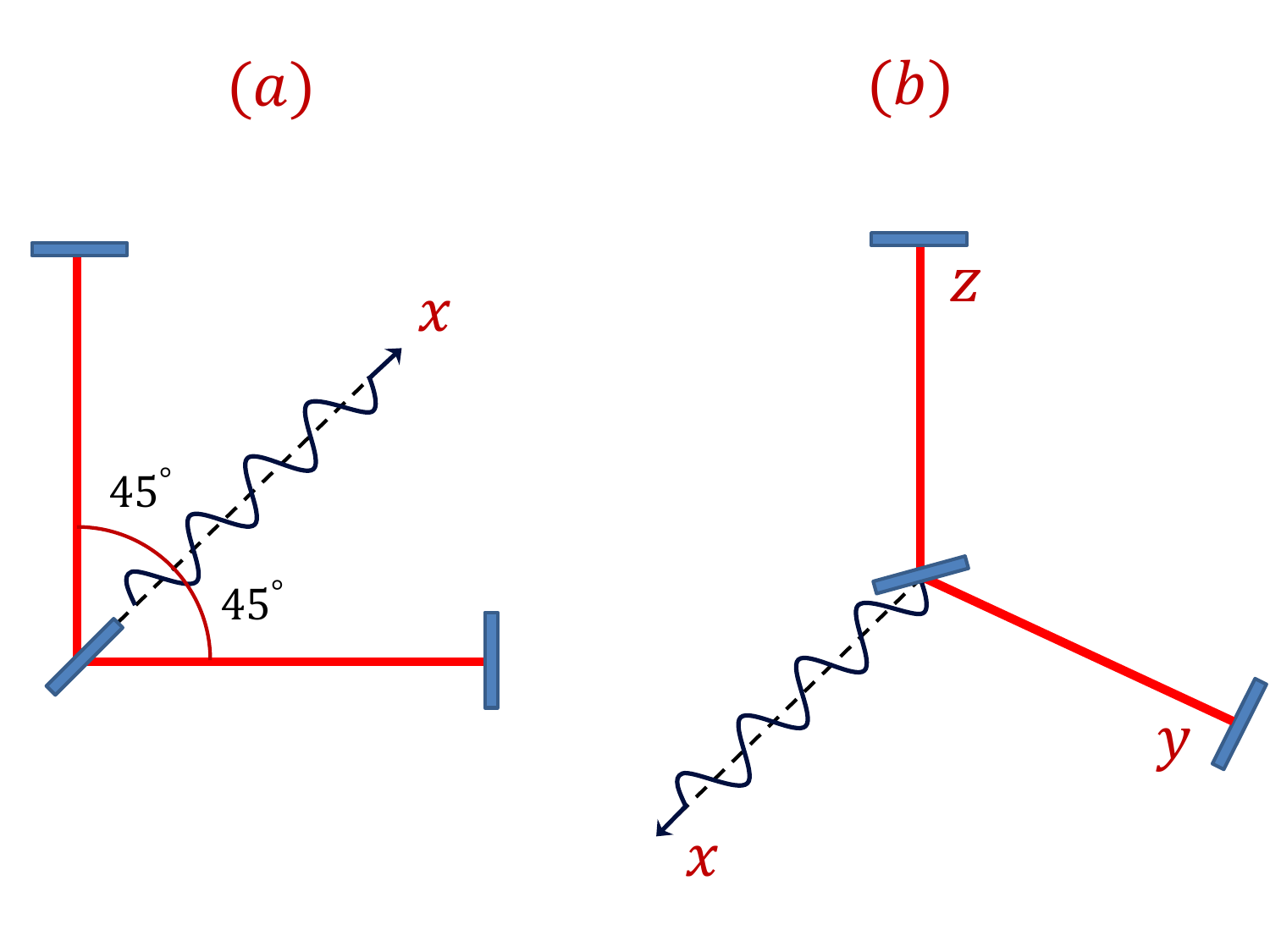}
\vspace{-0.5cm}
\caption{(a) Gravitational wave propagates in the interferometer plane at $%
45^{\circ } $ angle relative to the interferometer arms. Such wave produces
signal in vector gravity but not in general relativity. (b) Gravitational
wave propagates in the direction perpendicular to the interferometer plane.
Such wave yields signal in general relativity but not in vector gravity.}
\label{Arms}
\end{figure}

Signal of the LIGO-like interferometer with arms of length $L$ along the
direction of unit vectors $\hat{a}$ and $\hat{b}$ is proportional to the
relative phase shift $\Delta \varphi $ of electromagnetic waves traveling a
roundtrip distance $2L$ along the two arms (for details see Section \ref%
{GWtest})%
\begin{equation}
\Delta \varphi =\omega \frac{L}{c}h^{\alpha \beta }\left( \hat{a}_{\alpha }%
\hat{a}_{\beta }-\hat{b}_{\alpha }\hat{b}_{\beta }\right) ,  \label{wa4}
\end{equation}%
where $\omega $ is the frequency of electromagnetic wave and $h^{\alpha
\beta }$ is a spatial perturbation of the metric in the reference frame in
which interferometer mirrors do not move (frame of Eqs. (\ref{wa3}) and (\ref%
{met1a})). For the gravitational wave propagating along the $x-$axis Eq. (%
\ref{wa4}) yields for gravitational wave (\ref{wa3}) in general relativity%
\begin{equation}
\Delta \varphi =\omega \frac{L}{c}\left[ h^{yy}\left( \hat{a}_{y}^{2}-\hat{b}%
_{y}^{2}+\hat{b}_{z}^{2}-\hat{a}_{z}^{2}\right) +2h^{yz}\left( \hat{a}_{y}%
\hat{a}_{z}-\hat{b}_{y}\hat{b}_{z}\right) \right] ,  \label{wa5}
\end{equation}%
while for the transverse wave (\ref{met1a}) in vector gravity we obtain%
\begin{equation}
\Delta \varphi =\omega \frac{2L}{c}\left[ h^{xy}\left( \hat{a}_{x}\hat{a}%
_{y}-\hat{b}_{x}\hat{b}_{y}\right) +h^{xz}\left( \hat{a}_{x}\hat{a}_{z}-\hat{%
b}_{x}\hat{b}_{z}\right) \right] .  \label{wa6}
\end{equation}

Equations (\ref{wa5}) and (\ref{wa6}) show that vector gravity and general
relativity predict qualitatively different effect of the gravitational wave
on the interferometer signal. Namely, general relativistic gravitational
wave of any polarization (arbitrary $h^{yy}$ and $h^{yz}$) produces no
signal when gravitational wave propagates parallel to the interferometer
plane at $45^{\circ }$ angle relative to one of its perpendicular arms (see
Fig. \ref{Arms}\textit{a}). For these orientations the gravitational wave in
vector gravity can produce signal.

On the other hand, transverse gravitational wave in vector gravity (for
arbitrary $h^{xy}$ and $h^{xz}$) yields no signal if gravitational wave
propagates in the direction perpendicular to the interferometer plane (see
Fig. \ref{Arms}\textit{b}), or along one of the interferometer arms.

This difference can be used to distinguish between general relativity and
vector gravity in experiments with several gravitational wave
interferometers, e.g., in a joint run of the LIGO-Virgo network. We discuss
details of such experiment in Section \ref{GWtest}.

\section{Cosmology}

\label{cosmology}

In this section we apply our theory to evolution of the Universe.
Cosmological model of the Universe is an effective model which replaces a
complicated infinitely large spatially nonuniform system of moving matter
with those that has uniform density distribution and zero velocity. In such
a model the unit gravitational field vector is $u_{k}=(1,0,0,0)$ and the
scalar $\phi $ depends only on time. However, correct description of the
Universe evolution requires solution of the full system of spatially
nonuniform equations for the gravitational field and then averaging of the
result over the large scales. Symmetry arguments yield that such an
averaging can give an effective gravitational field action with an
additional cosmological term of the form 
\begin{equation}
S_{\text{cosm}}=-c\Lambda \int d^{4}x\sqrt{-f},  \label{cosm}
\end{equation}%
where $\Lambda $ is a constant independent of the gravitational field. The
structure of $S_{\text{cosm}}$ follows from the requirement that after
rescaling of coordinates (\ref{res}) the action should be independent of the
background cosmological field. This is one of the symmetries of the full
action and the effective action must possess such a symmetry. The
cosmological term appears due to replacement of the exact equations with the
equations for the averaged metric which is spatially uniform and isotropic.
Since the term (\ref{cosm}) is an effective, $\Lambda $ depends on the
choice of coordinate system, namely on the reference frame in which we
perform the spatial averaging.

For cosmology, instead of $\phi $, it is convenient to use a scale
(expansion) factor 
\begin{equation*}
a=e^{-\phi }
\end{equation*}%
as a variable. In terms of $a$ the equivalent metric (\ref{met}) reads 
\begin{equation}
f_{ik}=\left( 
\begin{array}{cccc}
\frac{1}{a^{2}} & 0 & 0 & 0 \\ 
0 & -a^{2} & 0 & 0 \\ 
0 & 0 & -a^{2} & 0 \\ 
0 & 0 & 0 & -a^{2}%
\end{array}%
\right) ,  \label{c2}
\end{equation}%
and $\sqrt{-f}=a^{2}$. Components of the energy-momentum tensor of matter
can be obtained from the conservation equation (\ref{cons}). For cold
Universe ($P=0$) there is only one nonzero component which is given by $%
T^{00}=\rho c^{2}/a$ and, hence, $T=\rho c^{2}/a^{3}$, where $\rho $ is a
constant that has a meaning of the matter density for $a=1$.

Spatial part of nonlinear gravitational field equations (\ref{sss1}) ($%
i=\alpha =1,$ $2,$ $3$) gives a simple linear equation for $a(t,\mathbf{r})$%
\begin{equation}
\frac{\partial ^{2}a}{\partial x^{\alpha }\partial t}=0.
\end{equation}%
Solution of this equation with the initial condition $a(t=0,\mathbf{r})=b(%
\mathbf{r})$ is%
\begin{equation}
a(t,\mathbf{r})=a(t)+b(\mathbf{r}),  \label{ab}
\end{equation}%
where $a(t)$ is an arbitrary function of time such that $a(0)=0$. If at the
Big Bang ($t=0$) the Universe was inhomogeneous then subsequent expansion
makes the spatially uniform term $a(t)$ in Eq. (\ref{ab}) much larger than $%
b(\mathbf{r})$ which is time independent. Therefore, we can omit $b(\mathbf{r%
})$ and treat metric as spatially uniform.

Thus, present theory predicts that shortly after Big Bang the Universe
becomes spatially flat and homogeneous on the large scales regardless of the
initial condition. This is not the case for general relativity and known as
the problems of large-scale homogeneity and flatness of the Universe. To
resolve these problems, cosmological models based on general relativity
require stage of inflation and introduction of additional hypothetical
field, the inflaton, responsible for inflation. In contrast, vector gravity
does not need additional fields.

Temporal part of Eqs. (\ref{sss1}) ($i=0$) with the cosmological term yield
the following equation for $a(t)$ 
\begin{equation}
-\frac{d}{dt}(a\dot{a})=\frac{8\pi G}{3}\left( \frac{\rho }{2a^{3}}-\Lambda
\right)  \label{qq1}
\end{equation}%
which shows that matter decelerates expansion of the Universe, while the $%
\Lambda -$term accelerates it (provided $\Lambda >0$). In Eq. (\ref{qq1}) a
dot denotes derivative with respect to time $t$. Integration of Eq. (\ref%
{qq1}) gives%
\begin{equation}
\dot{a}^{2}=\frac{8\pi G}{3}\left( \frac{\rho }{a^{3}}+\Lambda \right) +%
\frac{C}{a^{2}},  \label{w5a}
\end{equation}%
where $C$ is an integration constant which is proportional to the total
energy density of the Universe. Indeed, in the metric (\ref{c2}) the action (%
\ref{d3}) with the additional cosmological term (\ref{cosm}) reads%
\begin{equation*}
S=-\int d^{4}x\left[ \frac{3c}{8\pi G}a^{2}\dot{a}^{2}+c\Lambda a^{2}+\frac{%
\rho c}{a}\right] .
\end{equation*}%
Hence, the Lagrangian density is 
\begin{equation}
L=-\frac{3c^{2}}{8\pi G}a^{2}\dot{a}^{2}-c^{2}\Lambda a^{2}-\frac{\rho c^{2}%
}{a}  \label{Ld}
\end{equation}%
which yields the following conserved Hamiltonian density (the total energy
density) $w$%
\begin{equation}
w=\dot{a}\frac{\partial L}{\partial \dot{a}}-L=-\frac{3c^{2}}{8\pi G}a^{2}%
\dot{a}^{2}+c^{2}\Lambda a^{2}+\frac{\rho c^{2}}{a}.  \label{hd}
\end{equation}%
Therefore, integration constant $C$ in Eq. (\ref{w5a}) is%
\begin{equation}
C=-\frac{8\pi G}{3c^{2}}w.  \label{hd1}
\end{equation}

Observations indicate that $C=0$, or, at least, that in the present epoch
the term $C/a^{2}$ in Eq. (\ref{w5a}) is small compared to the other terms.
Since the term $C/a^{2}$ evolves as $1/a^{2}$, while the matter term is
proportional to $1/a^{3}$ the total energy density in the early Universe
must be equal to zero with very high precision. Namely, for small $a$ the
term $C/a^{2}$ becomes very small as compared to the matter contribution.
Thus, positive energy of matter in the Universe is balanced by the negative
energy of the gravitational field giving zero net energy. This result can be
considered as an observational evidence that matter in the Universe has been
produced at the expense of generation of the gravitational field with
negative energy.

On the other hand, in the metric (\ref{c2}) (for zero spatial curvature),
Einstein equations (\ref{i1}) with the extra cosmological term $-\frac{8\pi G%
}{c^{2}}\Lambda g_{ik}$ in the right hand side are equivalent to the
equations of vector gravity. We will show such equivalence for the
energy-momentum tensor of matter in more general form 
\begin{equation}
T_{k}^{i}=\left( 
\begin{array}{cccc}
\rho (t)c^{2} & 0 & 0 & 0 \\ 
0 & -P(t) & 0 & 0 \\ 
0 & 0 & -P(t) & 0 \\ 
0 & 0 & 0 & -P(t)%
\end{array}%
\right) ,
\end{equation}%
where $P(t)$ is the matter pressure. In the metric (\ref{c2}), Einstein
equation%
\begin{equation*}
R_{00}=\frac{8\pi G}{c^{4}}\left( T_{00}-\frac{1}{2}g_{00}T-\Lambda
c^{2}g_{00}\right)
\end{equation*}%
and the temporal part of Eqs. (\ref{sss1}) ($i=0$) with the cosmological
term yield the same evolution equation%
\begin{equation}
-\frac{d}{dt}(a\dot{a})=\frac{4\pi G}{3}\left( \rho (t)+3\frac{P(t)}{c^{2}}%
-2\Lambda \right)  \label{c7aa}
\end{equation}%
which has two unknown functions $a(t)$ and $\rho (t)$ (equation of state $%
P=P(\rho )$ fixes the relation between $P$ and $\rho $, and, therefore, $P$
is not independent). Additional equation can be taken, e.g., in the form of
the energy-momentum relation $T_{k;i}^{i}=0$ which is the same in both
theories. In the metric (\ref{c2}) this relation yields%
\begin{equation}
\dot{\rho}(t)=-3\left( \rho (t)+\frac{P(t)}{c^{2}}\right) \frac{\dot{a}}{a}.
\label{c7a}
\end{equation}%
For cold Universe $P(t)=0$ and integration of Eq. (\ref{c7a}) yields $\rho
(t)=\rho /a(t)^{3}$, where $\rho $ is independent of $t$. Thus, for cold
Universe, Eq. (\ref{c7aa}) gives the previous Eq. (\ref{qq1}).

Since Eqs. (\ref{c7aa}) and (\ref{c7a}) are identical in both theories,
general relativity for spatially flat metric and vector gravity predict the
same evolution of the Universe which agrees with available cosmological data
for a certain value of $\Lambda $. General relativity, however, does not
predict the value of $\Lambda $.

To find the value of the cosmological constant $\Lambda $ in vector gravity
we must start from the full system of equations (\ref{sss1}) for nonuniform
matter distribution without the cosmological term and average them over the
large scales. We perform this procedure for the linearized equations for
which the answer can be found exactly. Then we will match the nonlinear
equation (\ref{qq1}) with the exact linearized equations (which do not have
the $\Lambda -$term) and obtain the value of $\Lambda $ in the effective
cosmological model.

We calculate $\Lambda $ in our coordinate system, that is at the present
time. To do so we linearize Eq. (\ref{qq1}) near $a=a_{\text{now}}$ and
rescale time and coordinates as $t\rightarrow a_{\text{now}}t$ and $%
x_{\alpha }\rightarrow x_{\alpha }/a_{\text{now}}$. In the vicinity of the
present time the equivalent metric is%
\begin{equation*}
f_{ik}=\left( 
\begin{array}{cccc}
1+h_{00} & 0 & 0 & 0 \\ 
0 & -1+h_{00} & 0 & 0 \\ 
0 & 0 & -1+h_{00} & 0 \\ 
0 & 0 & 0 & -1+h_{00}%
\end{array}%
\right) ,
\end{equation*}%
where $h_{00}=-2(a-a_{\text{now}})/a_{\text{now}}$ and linearization of Eq. (%
\ref{qq1}) yields the following equation for $h_{00}$%
\begin{equation}
3\ddot{h}_{00}+16\pi G\Lambda =8\pi G\frac{\rho }{a_{\text{now}}^{3}}.
\label{cos4}
\end{equation}%
On the other hand, the full system of linearized nonuniform equations
without cosmological term is (see Eqs. (\ref{wf1}) and (\ref{wf2})) 
\begin{equation}
\Delta h_{00}+3\frac{\partial ^{2}h_{00}}{\partial x^{0}\partial x^{0}}-2%
\frac{\partial ^{2}h_{0\beta }}{\partial x^{0}\partial x^{\beta }}=\frac{%
8\pi G}{c^{4}}T_{\text{now}}^{00},  \label{cos5}
\end{equation}%
\begin{equation}
\left( \frac{\partial ^{2}}{\partial x^{0}\partial x^{0}}-\Delta \right)
h_{0\alpha }+\frac{\partial ^{2}h_{0\beta }}{\partial x^{\alpha }\partial
x^{\beta }}-2\frac{\partial ^{2}h_{00}}{\partial x^{\alpha }\partial x^{0}}=%
\frac{16\pi G}{c^{4}}T_{\text{now}}^{0\alpha },  \label{cos6}
\end{equation}%
where we took into account that for nonrelativistic matter $\tilde{f}%
^{00}T=T^{00}$. Taking $2\partial /\partial x^{0}$ from both sides of Eq. (%
\ref{cos5}) and divergence from both sides of Eq. (\ref{cos6}), adding them
together and using the continuity equation%
\begin{equation*}
\frac{\partial T_{\text{now}}^{00}}{\partial x^{0}}+\frac{\partial T_{\text{%
now}}^{0\alpha }}{\partial x^{\alpha }}=0
\end{equation*}%
we obtain%
\begin{equation}
\frac{\partial ^{2}}{\partial x^{0}\partial x^{0}}\left( 2\frac{\partial
h_{00}}{\partial x^{0}}-\frac{\partial h_{0\beta }}{\partial x^{\beta }}%
\right) =0.  \label{cos1}
\end{equation}%
Integration yields%
\begin{equation}
\frac{\partial }{\partial x^{0}}\left( 2\frac{\partial h_{00}}{\partial x^{0}%
}-\frac{\partial h_{0\beta }}{\partial x^{\beta }}\right) =F(\mathbf{r}),
\label{cos8}
\end{equation}%
where $F(\mathbf{r})$ is a function of spatial coordinates.

Equation (\ref{cos1}) has the following physical meaning. Change in time of
the spatial scale (given by $h_{00}$) can be viewed as motion of masses
relative to each other. This matter current produces longitudinal vector
field $h_{0\alpha }$ which, according to Eq. (\ref{cos1}), has nonzero
divergence. Relation between $\partial h_{00}/\partial x^{0}$ and $\partial
h_{0\beta }/\partial x^{\beta }$ should be independent of what causes $%
h_{00} $ to change (expansion of the Universe or motion of a nearby star).
If time dependence of $h_{00}$ is produced by a moving star then solution is
bound and average of the full derivative in the left hand side of Eq. (\ref%
{cos8}) over time vanishes. Since $F(\mathbf{r})$ is independent of time we
obtain $F(\mathbf{r})=\left\langle F(\mathbf{r})\right\rangle _{t}=0$ and,
therefore, Eq. (\ref{cos8}) reduces to 
\begin{equation}
\frac{\partial ^{2}h_{0\beta }}{\partial x^{0}\partial x^{\beta }}=2\frac{%
\partial ^{2}h_{00}}{\partial x^{0}\partial x^{0}}.  \label{cos9}
\end{equation}

Equation (\ref{cos9}) must be also valid for the evolution of the Universe.
Physically, expansion of the Universe changes spatial scale which can be
treated as an effective matter current directed away from a local observer.
Such current produces longitudinal vector field $h_{0\alpha }$ (according to
Eq. (\ref{cos9}) change of $h_{00}$ with time induces $h_{0\alpha }$). This
vector field makes the third term in (\ref{cos5}) nonzero which acts as the
cosmological term in the evolution equation. Since, according to Eq. (\ref%
{ed}), the energy density of the longitudinal vector field is negative the
cosmological term accelerates expansion of the Universe.

Substituting Eq. (\ref{cos9}) into Eq. (\ref{cos5}) we find%
\begin{equation}
\frac{\partial ^{2}h_{00}}{\partial x^{0}\partial x^{0}}=\Delta h_{00}-\frac{%
8\pi G}{c^{4}}T_{\text{now}}^{00}.  \label{cosx}
\end{equation}

Next we average Eq. (\ref{cosx}) over the large scales so that $T_{\text{now}%
}^{00}$ becomes spatially uniform. In the effective cosmological model the
averaged metric (\ref{c2}) is diagonal and depends only on time (see
discussion after Eq. (\ref{ab})). Therefore, spatial averaging of $\Delta
h_{00}$ must be equal to zero and Eq. (\ref{cosx}) yields 
\begin{equation}
\frac{\partial ^{2}}{\partial x^{0}\partial x^{0}}\left\langle
h_{00}\right\rangle =-\frac{8\pi G}{c^{4}}\left\langle T_{\text{now}%
}^{00}\right\rangle .  \label{cos10}
\end{equation}%
On the other hand, spatial averaging of Eq. (\ref{cos9}) gives that%
\begin{equation}
\left\langle \frac{\partial h_{0\beta }}{\partial x^{\beta }}\right\rangle =2%
\frac{\partial }{\partial x^{0}}\left\langle h_{00}\right\rangle
\label{cos11}
\end{equation}%
is not equal to zero. This is consistent with our assumptions. Indeed, Eq. (%
\ref{cos11}) can be satisfied by choosing $h_{0\beta }$ as an odd function
of coordinates, $h_{0\beta }\propto x_{\beta }$. Thus, spatial averaging of $%
h_{0\beta }$ in the local region yields zero and, hence, the averaged metric 
$\left\langle h_{ik}\right\rangle $ is diagonal and spatially homogeneous.

Comparing Eq. (\ref{cos4}) with Eq. (\ref{cos10}) and taking into account
that%
\begin{equation*}
\left\langle T_{\text{now}}^{00}\right\rangle =\frac{\rho c^{2}}{a_{\text{now%
}}^{3}}
\end{equation*}%
we find the following value for the cosmological constant%
\begin{equation*}
\Lambda =\frac{2\rho }{a_{\text{now}}^{3}}.
\end{equation*}

Thus, at the present time the cosmological term contribution (dark energy)
is twice as much as the energy of matter. Therefore, the ratio between the
energy density due to the cosmological constant and the critical density of
the Universe $\Omega _{\text{critical}}=3\dot{a}_{\text{now}}^{2}/8\pi
G=\rho _{\text{matter}}+\Lambda $ is 
\begin{equation}
\frac{\Lambda }{\Omega _{\text{critical}}}=\frac{2}{3}\approx 0.67.
\label{Lambda}
\end{equation}%
This is also the case in any other reference frame. That is an observer who
lives billion years before or after would find the same answer for $\Lambda
/\Omega _{\text{critical}}$ in his reference frame. Our prediction has no
free parameters and agrees with the recent Planck result which measured $%
\Lambda /\Omega _{\text{critical}}$ and obtained $0.686\pm 0.02$ \cite%
{Planck14}.

In the present theory the $\Lambda -$term appears as a solution of
equations. It comes from the gravitational part (left hand side) of Eqs. (%
\ref{sss1}) as a result of averaging of the term $\frac{\partial
^{2}h_{0\beta }}{\partial x^{0}\partial x^{\beta }}$ over the large scales.
Such average does not vanish in Eq. (\ref{cos5}) if we make the transition
to the infinitely large size of the Universe properly.

The physical meaning of the dark energy becomes clear if we compare the
exact expression for the weak field limit Lagrangian density with those
obtained from the effective cosmological action containing the $\Lambda -$%
term. Keeping only the relevant contributions the weak field limit
Lagrangian density in the rescaled coordinates reads (see Eq. (\ref{p2}))%
\begin{equation*}
L=-\frac{3c^{2}}{32\pi G}\left( \dot{h}_{00}\right) ^{2}-\frac{c^{2}}{32\pi G%
}\mathbf{\dot{h}}^{2}-\rho c^{2},
\end{equation*}%
where $\mathbf{h}=h^{0\alpha }$ depends on time and spatial coordinates. On
the other hand, linearization of the effective cosmological Lagrangian (\ref%
{Ld}) yields%
\begin{equation*}
L=-\frac{3c^{2}}{32\pi G}\left( \dot{h}_{00}\right) ^{2}-c^{2}\Lambda -\rho
c^{2}.
\end{equation*}%
Thus, 
\begin{equation*}
\Lambda =\frac{1}{32\pi G}\left\langle \mathbf{\dot{h}}^{2}\right\rangle
\end{equation*}%
and dark energy is the average energy of the longitudinal part of the
gravitational field. From the perspective of a local observer the change in
the spatial scale caused by the Universe expansion is equivalent to a matter
current directed away from the observer. Such current generates the
time-dependent longitudinal field $\mathbf{h}$ which is analogous to
generation of the vector potential $\mathbf{A}$ by a current in classical
electrodynamics. The value of the current depends on the matter density and
on the expansion rate of the Universe which, in turn, is a function of the
matter density. Thus, the value of the cosmological constant $\Lambda $ is
determined by the averaged matter density in the reference frame of the
observer (matter density at the moment the observer measures $\Lambda $).

One should mention that quantization of the gravitational field involves
only radiative part of the field. Since dark energy originates from the non
radiative part it is a pure classical effect which can be described by the
classical field equations.

\section{Contents of the Universe}

\label{content}

\begin{figure}[t]
\centering
\includegraphics[width=0.5\textwidth]{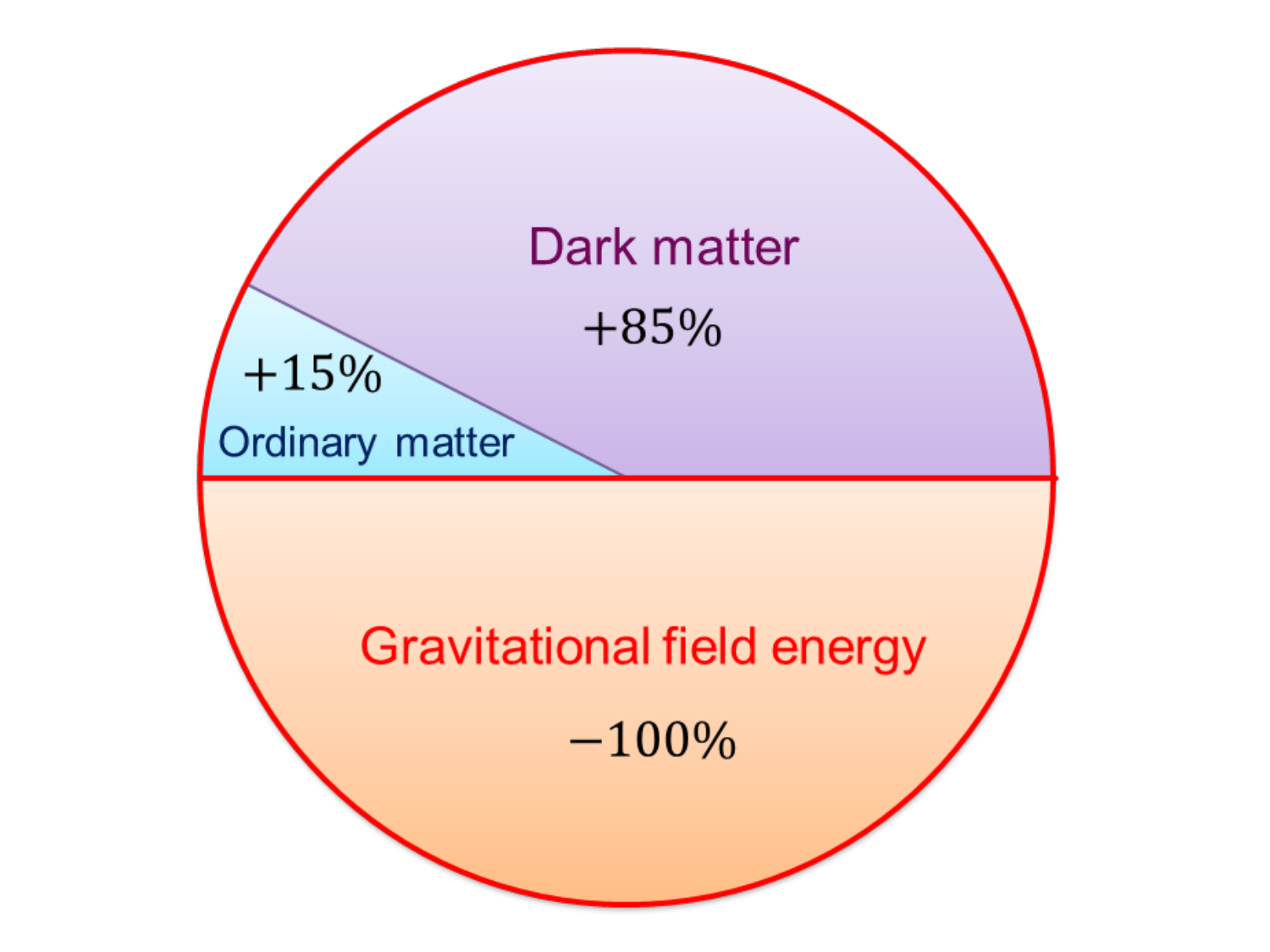}
\caption{Contents of the Universe according to vector gravity. Universe is
made of matter ($85\%$ dark matter and $15\%$ ordinary matter) which has
positive energy and gravitational field which has negative energy. The total
energy of the Universe is equal to zero.}
\label{Ucont}
\end{figure}

Vector theory of gravity explains the nature of dark energy as the energy of
longitudinal gravitational field induced by the Universe expansion. Such
energy is negative and accelerates expansion of the Universe. According to
the present theory, the Universe is made of matter (dark matter and ordinary
matter) and gravitational field (see Fig. \ref{Ucont}). Observations
indicate that the total energy of the Universe is equal to zero (see
discussion after Eq. (\ref{hd1})).

As we show in Sec. \ref{gwave}, classical field equations yield that the
graviton has negative energy. This suggests that at the Big Bang matter was
created at the expense of generation of the negative energy gravitons. Since
the pressure is determined by the wave momentum, rather than energy, the gas
of gravitons has positive pressure $P$. For such a gas the equation of state
reads%
\begin{equation}
P=-\frac{w_{g}}{3},  \label{es}
\end{equation}%
where $w_{g}$ is the graviton energy density. Since the energy momentum
tensor of an isotropic gas is 
\begin{equation*}
T^{ik}=\left( 
\begin{array}{cccc}
w_{g} & 0 & 0 & 0 \\ 
0 & P & 0 & 0 \\ 
0 & 0 & P & 0 \\ 
0 & 0 & 0 & P%
\end{array}%
\right)
\end{equation*}%
we obtain that for the gas of gravitons 
\begin{equation}
T^{00}-\frac{1}{2}\tilde{f}^{00}T=\frac{1}{2}\left( w_{g}+3P\right) =0.
\label{prop}
\end{equation}

According to Eqs. (\ref{sss1}), the combination $T^{00}-\frac{1}{2}\tilde{f}%
^{00}T$ is the source of gravitational field. This combination is also the
source of gravitational field in general relativity. Namely, it determines
gravitational mass of an object (Komar mass) in terms of the energy-momentum
tensor. Since for the primordial gravitational waves $T^{00}-\frac{1}{2}%
\tilde{f}^{00}T=0$ they do not produce gravitational field and, hence, they
do not contribute to the gravitational acceleration measured by an
accelerometer. Equation (\ref{prop}) also yields that primordial gravitons
do not change the cosmological Eq. (\ref{qq1}) and, thus, they have no
effect on expansion rate of the Universe.

According to the postulate $4$ of our theory the graviton is a composite
particle, namely it is composed of fermion-antifermion pairs. Since no more
than one fermion can occupy the same quantum state the matter generation at
the Big Bang has continued until fermion states were filled. The following
Universe expansion practically did not change the fermion occupation number
and states remain filled. These filled states form a vacuum in the present
epoch. As we show in the next section, for the filled vacuum the graviton
energy is positive and, thus, vacuum is stable. For the filled vacuum
creation of a graviton corresponds to creation of fermion-antifermion hole
pairs out of the filled fermion states.

Analogy with the composite photon theory \cite{Perk14} suggests that the
fermion-antifermion pairs that compose the graviton are coupled to matter
with the gravitational constant $G$, while a single fermion interacts with a
much weaker (perhaps zero) coupling constant. As a consequence, the energy
scales associated with the graviton and its constituent fermion are very
different. The Planck energy is the characteristic quantum energy scale for
the graviton and the corresponding Planck frequency is%
\begin{equation*}
\omega _{\text{Pl}}=\sqrt{\frac{c^{5}}{\hbar G}}=1.8\times 10^{43}\text{ s}%
^{-1}\text{.}
\end{equation*}

Perhaps at the Big Bang the gravitational waves have been generated upto the
Planck frequency. However, states of the constituent fermions are filled up
to a much higher frequency which is determined by their coupling constant.
Such frequency can not be predicted in the framework of the present theory.

Constituent fermions have both positive and negative energy states. The net
energy of the graviton gas produced at the Big Bang is negative. Probably
shortly after the Big Bang the negative energy of the graviton gas has been
transferred to the kinetic energy of the Universe expansion (the first term
in the right hand side of Eq. (\ref{hd})) and the negative and positive
energy states became filled symmetrically. This subject, however, requires
detail analysis of cosmological models and is beyond the scope of our paper.

\section{Quantization of Gravitational Field}

\label{quantization}

Here we quantize gravitational field assuming that graviton is composed of
fermion-antifermion pairs. Since quantization procedure is similar to
quantization of electromagnetic field we start from a brief review of the
field quantization in electrodynamics.

\subsection{Classical electrodynamics}

Classical electrodynamics is a vector field theory in four dimensional
Minkowski space-time. Electromagnetic field is a $4-$vector $A^{k}=(A_{0},%
\mathbf{A})$ in this space-time, while electric current density is a $4-$%
vector $j^{k}=(c\rho ,\mathbf{j})$, where $\rho $ and $\mathbf{j}$ are the
electric charge and spatial current densities. The conserved $4-$current
density $j^{k}$ is coupled to $A_{k}$ through the Lorentz and gauge
invariant term in the action%
\begin{equation*}
S_{\text{coupl}}=-\frac{1}{c^{2}}\int d^{4}xA_{k}j^{k}.
\end{equation*}%
The total action of the system is $S=S_{\text{field}}+S_{\text{coupl}}+S_{%
\text{matter}}$, where action for free electromagnetic field is%
\begin{equation*}
S_{\text{field}}=-\frac{1}{16\pi c}\int d^{4}x\left( \frac{\partial A_{k}}{%
\partial x^{i}}-\frac{\partial A_{i}}{\partial x^{k}}\right) \left( \frac{%
\partial A^{k}}{\partial x_{i}}-\frac{\partial A^{i}}{\partial x_{k}}\right)
\end{equation*}%
and for nonrelativistic motion of particles%
\begin{equation*}
S_{\text{matter}}=\int dt\sum_{a}\frac{m_{a}\dot{r}_{a}^{2}}{2},
\end{equation*}%
where the sum is over all particles $a$ having positions $\mathbf{r}_{a}$,
masses $m_{a}$ and electric charges $q_{a}$. Electric current density is
given by%
\begin{equation*}
\mathbf{j}=\sum_{a}q_{a}\mathbf{\dot{r}}_{a}\delta \left( \mathbf{r}-\mathbf{%
r}_{a}(t)\right) .
\end{equation*}

Particle momentum conjugate to $\mathbf{r}_{a}$ is%
\begin{equation*}
\mathbf{p}_{a}=\frac{\partial L}{\partial \mathbf{\dot{r}}_{a}}=m_{a}\mathbf{%
\dot{r}}_{a}+\frac{q_{a}}{c}\mathbf{A(r}_{a}),
\end{equation*}%
while momentum $\pi ^{k}$ conjugate to the field $A_{k}$ reads%
\begin{equation*}
\pi ^{k}=\frac{\partial L}{\partial \dot{A}_{k}}=-\frac{1}{4\pi c}\left\{ 
\begin{array}{c}
0,\quad k=0 \\ 
E^{k},\quad k=1,2,3%
\end{array}%
,\right.
\end{equation*}%
where $L$ is the Lagrangian density and $\mathbf{E}=-\nabla A^{0}-\partial 
\mathbf{A}/\partial x^{0}$ is the electric field.

Electromagnetic field $A_{k}$ has four real components. However, since
momentum $\pi ^{0}$ conjugate to $A_{0}$ vanishes the time component $A_{0}$
is not a dynamical field. This means that $A_{0}$ is not an independent
degree of freedom but rather it is a functional of $\mathbf{A}$ and electric
charge density $\rho $ \cite{Cohe87}. In addition, gauge symmetry implies
that there are only two independent physical degrees of freedom because one
of the degrees of freedom can be eliminated by gauge fixing. These two
independent degrees of freedom of the electromagnetic field corresponding to
radiation are quantized, their quantum is a photon.

Classical Hamiltonian of the system is a functional of $\mathbf{r}_{a}$, $%
\mathbf{p}_{a}$, $A_{k}$ and $\pi _{k}$ 
\begin{equation*}
\mathcal{H}=\sum_{a}\mathbf{p}_{a}\cdot \mathbf{\dot{r}}_{a}+\int
d^{3}x\left( \pi ^{k}\dot{A}_{k}-L\right)
\end{equation*}%
\begin{equation*}
=\int d^{3}x\left[ \frac{1}{8\pi }\left( E^{2}+B^{2}\right) -\frac{1}{4\pi }%
A_{0}\text{div}\mathbf{E}\right]
\end{equation*}%
\begin{equation*}
+\sum_{a}\left[ \frac{1}{2m_{a}}\left( \mathbf{p}_{a}-\frac{q_{a}}{c}\mathbf{%
A}\right) ^{2}+q_{a}A_{0}\right] ,
\end{equation*}%
where $\mathbf{B}=$curl$\mathbf{A}$.

One can decompose $\mathbf{E}$ and $\mathbf{B}$ into longitudinal and
transverse parts%
\begin{equation*}
\mathbf{E}=\mathbf{E}_{\text{lon}}+\mathbf{E}_{\text{tr}},\quad \mathbf{B}=%
\mathbf{B}_{\text{tr}}
\end{equation*}%
such that div$\mathbf{E}_{\text{tr}}=0$ and div$\mathbf{B}_{\text{tr}}=0$.
This decomposition can be done in a straightforward way if we write $\mathbf{%
E}$ as a Fourier series%
\begin{equation*}
\mathbf{E(r)}=\mathbf{\sum_{p}E(p)}e^{i\mathbf{p}\cdot \mathbf{r}}.
\end{equation*}
Then%
\begin{equation*}
\mathbf{E}_{\text{lon}}\mathbf{(r)}=\mathbf{\sum_{p}}\hat{p}(\hat{p}\cdot 
\mathbf{E(p))}e^{i\mathbf{p}\cdot \mathbf{r}},
\end{equation*}
where $\hat{p}$ is a unit vector in the direction of $\mathbf{p}$, and%
\begin{equation*}
\mathbf{E}_{\text{tr}}=\mathbf{E}-\mathbf{E}_{\text{lon}}.
\end{equation*}

Having in mind that photon originates from the transverse electromagnetic
field we decompose 4-vectors $A_{k}$ and $j_{k}$ into transverse spatial
part $\mathbf{A}_{\text{tr}}$, $\mathbf{j}_{\text{tr}}$ and the remaining
piece which contains longitudinal and time-like components, e.g.%
\begin{equation}
A^{k}=A_{\text{tr}}^{k}+A_{_{\text{\textit{l}-t}}}^{k}.  \label{akk}
\end{equation}%
Terms in the right hand side of Eq. (\ref{akk}) do not transform as
4-vectors, only their sum does. In particular, for the spatial part of $%
A^{k} $ we have%
\begin{equation*}
\mathbf{A}=\mathbf{A}_{\text{tr}}+\mathbf{A}_{_{\text{lon}}}.
\end{equation*}

Decomposition of the field into the longitudinal and transverse components
decouples Hamiltonian into two independent pieces $\mathcal{H}=\mathcal{H}_{%
\text{tr}}+\mathcal{H}_{\text{\textit{l}-t}}$, where Hamiltonian of the
transverse field is%
\begin{equation}
\mathcal{H}_{\text{tr}}=\frac{1}{8\pi }\int d^{3}x\left( E_{\text{tr}%
}^{2}+B^{2}\right) +\sum_{a}\frac{1}{2m_{a}}\left( \mathbf{p}_{a}-\frac{q_{a}%
}{c}\mathbf{A}_{\text{tr}}\right) ^{2}  \label{htr}
\end{equation}%
and%
\begin{equation*}
\mathbf{E}_{\text{tr}}=-\frac{\partial \mathbf{A}_{\text{tr}}}{\partial x^{0}%
}.
\end{equation*}%
For the longitudinal and transverse fields Maxwell equations also decouple.
In particular, for transverse field Maxwell equations read%
\begin{equation}
\text{curl}\mathbf{B}=\frac{4\pi }{c}\mathbf{j}_{\text{tr}}+\frac{1}{c}\frac{%
\partial \mathbf{E}_{\text{tr}}}{\partial t},\quad \text{curl}\mathbf{E}_{%
\text{tr}}=-\frac{1}{c}\frac{\partial \mathbf{B}}{\partial t}.  \label{metr}
\end{equation}

\subsection{Quantization of electromagnetic field in elementary photon theory%
}

In conventional quantization of the electromagnetic field the $\mathcal{H}%
_{_{\text{\textit{l}-t}}}$ part of the Hamiltonian remains classical \cite%
{Ginz87}. Part of the transverse field that corresponds to radiation (field
without sources) is quantized and is described by the following Hamiltonian
operator \cite{Cohe87}%
\begin{equation*}
\mathcal{\hat{H}}_{\text{tr}}=\sum_{\mathbf{p,}\mu =1,2}\hbar cp\hat{A}_{%
\mathbf{p,}\mu }^{+}\hat{A}_{\mathbf{p,}\mu }+\sum_{a}\frac{1}{2m_{a}}\left( 
\mathbf{p}_{a}-\frac{q_{a}}{c}\mathbf{\hat{A}}_{\text{tr}}(\mathbf{r}%
_{a})\right) ^{2},
\end{equation*}%
where%
\begin{equation*}
\mathbf{\hat{A}}_{\text{tr}}(\mathbf{r})=\sum_{\mathbf{p,}\mu =1,2}\sqrt{%
\frac{2\pi \hbar c}{pV}}\left( \mathbf{\epsilon }_{\mathbf{p,}\mu }\hat{A}_{%
\mathbf{p,}\mu }e^{i\mathbf{p}\cdot \mathbf{r}}+\text{H.c.}\right) ,
\end{equation*}%
$\mathbf{\epsilon }_{\mathbf{p,}\mu }$ ($\mu =1,2$) are unit
three-dimensional polarization vectors perpendicular to the photon wave
vector $\mathbf{p}$, $V$ is the photon volume and operators $\hat{A}_{%
\mathbf{p,}\mu }$ obey Bose-Einstein commutation relations 
\begin{equation*}
\lbrack \hat{A}_{\mathbf{p},\mu },\hat{A}_{\mathbf{p}^{\prime }\mathbf{,}\mu
^{\prime }}^{+}]=\delta _{\mathbf{p},\mathbf{p}^{\prime }}\delta _{\mu ,\mu
^{\prime }},\quad \mu ,\mu ^{\prime }=1,2
\end{equation*}%
and all other commutators are equal to zero. Operators $\hat{A}_{\mathbf{p}%
,\mu }^{+}$ and $\hat{A}_{\mathbf{p},\mu }$ describe creation and
annihilation of a spin $1$ photon with wave vector $\mathbf{p}$ and
polarization $\mu $. Electric%
\begin{equation*}
\mathbf{\hat{E}}_{\text{tr}}(\mathbf{r})=i\sum_{\mathbf{p,}\mu =1,2}\sqrt{%
\frac{2\pi \hbar cp}{V}}\left( \mathbf{\epsilon }_{\mathbf{p,}\mu }\hat{A}_{%
\mathbf{p,}\mu }e^{i\mathbf{p}\cdot \mathbf{r}}-\text{H.c.}\right) ,
\end{equation*}%
and magnetic%
\begin{equation*}
\mathbf{\hat{B}}(\mathbf{r})=i\sum_{\mathbf{p,}\mu =1,2}\sqrt{\frac{2\pi
\hbar c}{pV}}\left( \mathbf{p}\times \mathbf{\epsilon }_{\mathbf{p,}\mu }%
\hat{A}_{\mathbf{p,}\mu }e^{i\mathbf{p}\cdot \mathbf{r}}-\text{H.c.}\right)
\end{equation*}%
field operators obey commutation relations%
\begin{equation}
\lbrack \hat{A}_{\text{tr}}^{\beta }(\mathbf{r}^{\prime }),\hat{E}_{\text{tr}%
}^{\alpha }(\mathbf{r})]\mathbf{=-}4\pi i\hbar c\delta ^{\beta \alpha
}\delta (\mathbf{r-r}^{\prime }),\quad \alpha ,\beta =1,2,3,  \label{cm1}
\end{equation}%
\begin{equation}
\lbrack \hat{A}_{\text{tr}}^{\beta }(\mathbf{r}^{\prime }),\hat{B}_{\alpha }(%
\mathbf{r})]\mathbf{=}0,  \label{cm2}
\end{equation}%
\begin{equation}
\lbrack \mathcal{\hat{H}}_{0},\mathbf{\hat{E}}_{\text{tr}}]=-ic\hbar \text{%
\thinspace curl}\mathbf{\hat{B},}  \label{cm3}
\end{equation}%
\begin{equation}
\lbrack \mathcal{\hat{H}}_{0},\mathbf{\hat{B}}]=ic\hbar \text{\thinspace curl%
}\mathbf{\hat{E}}_{\text{tr}},  \label{cm4}
\end{equation}%
where 
\begin{equation}
\mathcal{\hat{H}}_{0}=\sum_{\mathbf{p,}\mu =1,2}\hbar cp\hat{A}_{\mathbf{p,}%
\mu }^{+}\hat{A}_{\mathbf{p,}\mu }.  \label{ff1}
\end{equation}

In the Heisenberg picture the Heisenberg equations of motion%
\begin{equation*}
\frac{d\mathbf{\hat{E}}_{\text{tr}}}{dt}=\frac{i}{\hbar }[\mathcal{\hat{H}}_{%
\text{tr}},\mathbf{\hat{E}}_{\text{tr}}],\quad \frac{d\mathbf{\hat{B}}}{dt}=%
\frac{i}{\hbar }[\mathcal{\hat{H}}_{\text{tr}},\mathbf{\hat{B}}]
\end{equation*}%
yield Maxwell's equations for the transverse field (\ref{metr}) in the
operator form%
\begin{equation}
\text{curl}\mathbf{\hat{B}}=\frac{4\pi }{c}\mathbf{j}_{\text{tr}}+\frac{1}{c}%
\frac{\partial \mathbf{\hat{E}}_{\text{tr}}}{\partial t},\quad \text{curl}%
\mathbf{\hat{E}}_{\text{tr}}=-\frac{1}{c}\frac{\partial \mathbf{\hat{B}}}{%
\partial t}.  \label{mqed}
\end{equation}

One should emphasize that only a part of the transverse electromagnetic
field is quantized, namely the part that corresponds to radiation. For
example, magnetic field produced by stationary currents is transverse but is
not quantized and remains classical.

\subsection{Photon as composite particle}

In the composite photon theory the elementary particle is a massless spin
1/2 fermion and photon is composed of the fermion-antifermion pairs. In free
space the Dirac equation for massless spin 1/2 fermion, described by a
four-component spinor $\Psi =\left( 
\begin{array}{c}
\psi _{R} \\ 
\psi _{L}%
\end{array}%
\right) $, reads%
\begin{equation}
\gamma ^{\mu }\partial _{\mu }\Psi =0,  \label{a0}
\end{equation}%
where $\gamma ^{\mu }$ are gamma matrices which can be written in terms of 2$%
\times $2 sub-matrices taken from the Pauli matrices and the 2$\times $2
identity matrix $I$. In the Weyl (chiral) basis the gamma matrices have the
form%
\begin{equation*}
\gamma ^{0}=\left( 
\begin{array}{cc}
0 & -I \\ 
-I & 0%
\end{array}%
\right) ,\quad \gamma ^{1}=\left( 
\begin{array}{cc}
0 & \sigma _{x} \\ 
-\sigma _{x} & 0%
\end{array}%
\right) ,
\end{equation*}%
\begin{equation*}
\gamma ^{2}=\left( 
\begin{array}{cc}
0 & \sigma _{y} \\ 
-\sigma _{y} & 0%
\end{array}%
\right) ,\quad \gamma ^{3}=\left( 
\begin{array}{cc}
0 & \sigma _{z} \\ 
-\sigma _{z} & 0%
\end{array}%
\right)
\end{equation*}%
and Pauli matrices are

\begin{equation*}
\sigma _{x}=\left( 
\begin{array}{cc}
0 & 1 \\ 
1 & 0%
\end{array}%
\right) ,\quad \sigma _{y}=\left( 
\begin{array}{cc}
0 & -i \\ 
i & 0%
\end{array}%
\right) ,\quad \sigma _{z}=\left( 
\begin{array}{cc}
1 & 0 \\ 
0 & -1%
\end{array}%
\right) .
\end{equation*}

Recall that $\gamma ^{\mu }$ are fixed under Lorentz transformations in the
forms given above. Lorentz invariance of the Dirac equation is achieved by
proper transformation of spinor $\Psi $ that counterbalances the
transformation of $\partial _{\mu }$. In the Weyl (or chiral) representation
of the Dirac matrices the Weyl spinors $\psi _{R}$ and $\psi _{L}$ do not
mix under Lorentz transformations.

Solutions of the Dirac equation are arbitrary superposition of four plane
wave spinors%
\begin{equation*}
\Psi _{a}=u_{a}(\mathbf{p})e^{-ipct+i\mathbf{p}\cdot \mathbf{r}},\quad \Psi
_{b}=u_{b}(\mathbf{p})e^{ipct+i\mathbf{p}\cdot \mathbf{r}},
\end{equation*}%
\begin{equation*}
\Psi _{c}=u_{c}(\mathbf{p})e^{ipct+i\mathbf{p}\cdot \mathbf{r}},\quad \Psi
_{d}=u_{d}(\mathbf{p})e^{-ipct+i\mathbf{p}\cdot \mathbf{r}},
\end{equation*}%
where $\mathbf{p}$ is the wave vector, $p=|\mathbf{p}|$ and for $\mathbf{p}$
oriented along the positive direction of the $z-$axis%
\begin{equation*}
u_{a}(\mathbf{p})=\left( 
\begin{array}{c}
1 \\ 
0 \\ 
0 \\ 
0%
\end{array}%
\right) ,\quad u_{b}(\mathbf{p})=\left( 
\begin{array}{c}
0 \\ 
1 \\ 
0 \\ 
0%
\end{array}%
\right) ,
\end{equation*}%
\begin{equation*}
u_{c}(\mathbf{p})=\left( 
\begin{array}{c}
0 \\ 
0 \\ 
1 \\ 
0%
\end{array}%
\right) ,\quad u_{d}(\mathbf{p})=\left( 
\begin{array}{c}
0 \\ 
0 \\ 
0 \\ 
1%
\end{array}%
\right) .
\end{equation*}

Energy of particles described by spinors $\Psi _{a}$ and $\Psi _{d}$ is
positive $\varepsilon =\hbar cp$, while fermions corresponding to spinors $%
\Psi _{b}$ and $\Psi _{c}$ have negative energies $\varepsilon =-\hbar cp$.
The helicity of a particle is right-handed if the direction of its spin is
the same as the direction of its motion. It is left-handed if the directions
of spin and motion are opposite. $u_{a}$ and $u_{c}$ are right-handed
spinors, while $u_{b}$ and $u_{d}$ are left-handed spinors. Helicity of a
massless particle is Lorentz invariant.

General solution of the free Dirac equation can be written as%
\begin{equation*}
\Psi =\sum_{\mathbf{p}}\left[ a_{\mathbf{p}}u_{a}(\mathbf{p})e^{-icpt}+b_{%
\mathbf{p}}u_{b}(\mathbf{p})e^{icpt}\right.
\end{equation*}%
\begin{equation*}
+\left. c_{\mathbf{p}}u_{c}(\mathbf{p})e^{icpt}+d_{\mathbf{p}}u_{d}(\mathbf{p%
})e^{-icpt}\right] e^{i\mathbf{p}\cdot \mathbf{r}},
\end{equation*}%
where $a_{\mathbf{p}}$, $b_{\mathbf{p}}$, $c_{\mathbf{p}}$ and $d_{\mathbf{p}%
}$ are arbitrary constants which transform as scalars under Lorentz
transformation.

Out of $u_{a}(\mathbf{p})$, $u_{b}(\mathbf{p})$, $u_{c}(\mathbf{p})$ and $%
u_{d}(\mathbf{p})$ one can construct four linearly independent $4-$vectors
in $4-$dimensional space-time:%
\begin{equation}
u_{a}^{+}(\mathbf{p})\gamma _{0}\gamma _{\mu }u_{b}(\mathbf{p})=(0,1,-i,0),
\label{pv1}
\end{equation}%
\begin{equation}
u_{d}^{+}(\mathbf{p})\gamma _{0}\gamma _{\mu }u_{c}(\mathbf{p})=(0,-1,-i,0),
\label{pv2}
\end{equation}%
\begin{equation}
u_{b}^{+}(\mathbf{p})\gamma _{0}\gamma _{\mu }u_{b}(\mathbf{p})=u_{c}^{+}(%
\mathbf{p})\gamma _{0}\gamma _{\mu }u_{c}(\mathbf{p})=(1,0,0,-1),
\label{pv3}
\end{equation}%
\begin{equation}
u_{a}^{+}(\mathbf{p})\gamma _{0}\gamma _{\mu }u_{a}(\mathbf{p})=u_{d}^{+}(%
\mathbf{p})\gamma _{0}\gamma _{\mu }u_{d}(\mathbf{p})=(1,0,0,1).  \label{pv4}
\end{equation}%
Other combinations give zero vectors. Vectors (\ref{pv1}) and (\ref{pv2})
are transverse, while (\ref{pv3}) and (\ref{pv4}) are combinations of the
longitudinal and time-like vectors. Recall that wave vector $\mathbf{p}$ is
chosen to be oriented along the positive direction of the $z-$axis.

We map the fermion field into the real transverse field $\mathbf{A}_{\text{tr%
}}$%
\begin{equation}
\mathbf{A}_{\text{tr}}(t,\mathbf{r})=\sum_{\mathbf{p}}\left[ \mathbf{A}_{%
\mathbf{p,}\text{tr}}(t)e^{i\mathbf{p}\cdot \mathbf{r}}+\mathbf{A}_{\mathbf{%
p,}\text{tr}}^{\ast }(t)e^{-i\mathbf{p}\cdot \mathbf{r}}\right]  \label{fer1}
\end{equation}%
such that $\mathbf{A}_{\text{tr}}(t,\mathbf{r})$ obeys Maxwell equations for
free transverse field. In terms of the Fourier components $\mathbf{A}_{%
\mathbf{p,}\text{tr}}(t)$ this can be done in the following way%
\begin{equation}
\mathbf{A}_{\mathbf{p,}\text{tr}}(t)=A_{\mathbf{p},1}e^{-icpt}\mathbf{%
\epsilon }_{\mathbf{p,}1}+A_{\mathbf{p},2}e^{-icpt}\mathbf{\epsilon }_{%
\mathbf{p,}2},  \label{fer2}
\end{equation}%
where $\mathbf{\epsilon }_{\mathbf{p,}1}$ and $\mathbf{\epsilon }_{\mathbf{p,%
}2}$ are the spatial unit polarization vectors of the left and right
circularly polarized photons respectively. Equations (\ref{pv1}) and (\ref%
{pv2}) indicate that $A_{\mathbf{p},1}$ and $A_{\mathbf{p},2}$ should be
chosen as 
\begin{equation*}
A_{\mathbf{p},1}=\sum_{\mathbf{k\parallel p}}F_{1}(\mathbf{p},\mathbf{k})a_{%
\mathbf{p}+\mathbf{k}}b_{-\mathbf{k}},
\end{equation*}%
\begin{equation*}
A_{\mathbf{p},2}=\sum_{\mathbf{k\parallel p}}F_{2}(\mathbf{p},\mathbf{k})d_{%
\mathbf{p}+\mathbf{k}}c_{-\mathbf{k}},
\end{equation*}%
where summation is over all fermion states with wave vectors $\mathbf{k}$
parallel to $\mathbf{p}$.

In order for $\mathbf{A}_{\mathbf{p}\text{,tr}}$ to transform as a
transverse field under Lorentz transformations the spectral functions $F_{1}(%
\mathbf{p},\mathbf{k})$ and $F_{2}(\mathbf{p},\mathbf{k})$ must be scalars.
Therefore they can depend only on the absolute values of the $4-$vectors and
their dot products. $F_{1}(\mathbf{p},\mathbf{k})$ is a factor in front of $%
a_{\mathbf{p}+\mathbf{k}}b_{-\mathbf{k}}$ which is a product of two fermion
states with $4-$wave vectors $(p+k,\mathbf{p}+\mathbf{k})$ and $(-k,-\mathbf{%
k})$. Absolute values of these $4-$vectors and their dot product (with
Minkowski metric) are equal to zero. Therefore, spectral function $F_{1}(%
\mathbf{p},\mathbf{k})$ is independent of $\mathbf{p}$ and $\mathbf{k}$.
Similar arguments yield that $F_{2}(\mathbf{p},\mathbf{k})$ is also a
constant which can be chosen arbitrary. We choose $F_{1}=F_{2}=1/\sqrt{%
N_{\parallel }}$, where $N_{\parallel }$ is the number of fermion states
with wave vectors parallel to $\mathbf{p}$, and assume that $N_{\parallel }$
is independent of $\mathbf{p}$.

\subsubsection{Field quantization in the composite photon theory}

Next we quantize the fermion field by replacing $a_{\mathbf{p}}$, $b_{%
\mathbf{p}}$, $c_{\mathbf{p}}$ and $d_{\mathbf{p}}$ with operators that obey
canonical anticommutation relations%
\begin{equation}
\hat{a}_{\mathbf{p}}\hat{a}_{\mathbf{p}^{\prime }}^{+}+\hat{a}_{\mathbf{p}%
^{\prime }}^{+}\hat{a}_{\mathbf{p}}=\delta _{\mathbf{p},\mathbf{p}^{\prime
}},\quad \hat{b}_{\mathbf{p}}\hat{b}_{\mathbf{p}^{\prime }}^{+}+\hat{b}_{%
\mathbf{p}^{\prime }}^{+}\hat{b}_{\mathbf{p}}=\delta _{\mathbf{p},\mathbf{p}%
^{\prime }},  \label{an1}
\end{equation}%
\begin{equation}
\hat{c}_{\mathbf{p}}\hat{c}_{\mathbf{p}^{\prime }}^{+}+\hat{c}_{\mathbf{p}%
^{\prime }}^{+}\hat{c}_{\mathbf{p}}=\delta _{\mathbf{p},\mathbf{p}^{\prime
}},\quad \hat{d}_{\mathbf{p}}\hat{d}_{\mathbf{p}^{\prime }}^{+}+\hat{d}_{%
\mathbf{p}^{\prime }}^{+}\hat{d}_{\mathbf{p}}=\delta _{\mathbf{p},\mathbf{p}%
^{\prime }}.  \label{an2}
\end{equation}%
All other anticommutators are equal to zero.

The second quantized Hamiltonian for the free fermion field is%
\begin{equation}
\mathcal{\hat{H}}_{0}=\sum_{\mathbf{p}}\left[ \varepsilon _{a}(p)\hat{a}_{%
\mathbf{p}}^{+}\hat{a}_{\mathbf{p}}+\varepsilon _{b}(p)\hat{b}_{\mathbf{p}%
}^{+}\hat{b}_{\mathbf{p}}+\varepsilon _{c}(p)\hat{c}_{\mathbf{p}}^{+}\hat{c}%
_{\mathbf{p}}+\varepsilon _{d}(p)\hat{d}_{\mathbf{p}}^{+}\hat{d}_{\mathbf{p}}%
\right] ,  \label{ffh}
\end{equation}%
where $\varepsilon _{a}(p)=\varepsilon _{d}(p)=\hbar cp$ and $\varepsilon
_{b}(p)=\varepsilon _{c}(p)=-\hbar cp$.

The free transverse field (\ref{fer1}) (field without sources) now becomes
an operator which in the Schr\"{o}dinger picture reads%
\begin{equation}
\mathbf{\hat{A}}_{\text{tr}}(\mathbf{r})=\sum_{\mathbf{p}}\left( \mathbf{%
\hat{A}}_{\mathbf{p,}\text{tr}}e^{i\mathbf{p}\cdot \mathbf{r}}+\mathbf{\hat{A%
}}_{\mathbf{p}\text{,tr}}^{+}e^{-i\mathbf{p}\cdot \mathbf{r}}\right) ,
\label{af}
\end{equation}%
where%
\begin{equation*}
\mathbf{\hat{A}}_{\mathbf{p,}\text{tr}}=\hat{A}_{\mathbf{p},1}\mathbf{%
\epsilon }_{\mathbf{p,}1}+\hat{A}_{\mathbf{p},2}\mathbf{\epsilon }_{\mathbf{%
p,}2}
\end{equation*}%
and%
\begin{equation}
\hat{A}_{\mathbf{p},1}=\frac{1}{\sqrt{N_{\parallel }}}\sum_{\mathbf{%
k\parallel p}}\hat{a}_{\mathbf{p}+\mathbf{k}}\hat{b}_{-\mathbf{k}},
\label{ap1}
\end{equation}%
\begin{equation}
\hat{A}_{\mathbf{p},2}=\frac{1}{\sqrt{N_{\parallel }}}\sum_{\mathbf{%
k\parallel p}}\hat{d}_{\mathbf{p}+\mathbf{k}}\hat{c}_{-\mathbf{k}}.
\label{ap2}
\end{equation}

One should mention that under charge conjugation operation $\hat{a}%
\leftrightarrow \hat{c}$, $\hat{b}\leftrightarrow \hat{d}$. Changing index
of summation in Eqs. (\ref{ap1}) and (\ref{ap2}) we then obtain that under
charge conjugation $\hat{A}_{\mathbf{p},1}\leftrightarrow \hat{A}_{\mathbf{p}%
,2}$. That is in our quantization scheme the photon is identical to its own
antiparticle. This agrees with recent antihydrogen experiments at CERN \cite%
{Ahma17}.

Since fermion $a$ is right-handed the operator $\hat{a}_{\mathbf{p}+\mathbf{k%
}}^{+}$ creates a spin $1/2$ fermion with spin parallel to $\mathbf{p}+%
\mathbf{k}$ and energy $\varepsilon _{a}=\hbar c(p+k)$. On the other hand, $%
\hat{b}_{-\mathbf{k}}^{+}$ creates a spin $1/2$ antifermion with spin
antiparallel to $\mathbf{-k}$ (that is parallel to $\mathbf{k}$) and
negative energy $\varepsilon _{b}=-\hbar ck$. Thus, the combination $\hat{b}%
_{-\mathbf{k}}^{+}\hat{a}_{\mathbf{p}+\mathbf{k}}^{+}$ with $\mathbf{%
k\parallel p}$ creates a fermion-antifermion pair with the total energy $%
\varepsilon =\hbar cp$ and spin $1$ parallel to $\mathbf{p}$. Recall that
for left (right) circularly polarized photon the photon spin is parallel
(antiparallel) to the wave vector $\mathbf{p}$. Therefore, operator $\hat{A}%
_{\mathbf{p},1}^{+}$ ($\hat{A}_{\mathbf{p},2}^{+}$) creates a left (right)
circularly polarized photon with spin $1$. According to Eqs. (\ref{ap1}) and
(\ref{ap2}), emission of a single photon corresponds to creation of $%
N_{\parallel }$ fermion-antifermion pairs with equal probability $%
1/N_{\parallel }$.

Equations (\ref{an1}) and (\ref{an2}) yield the following commutation
relations for operators $\hat{A}_{\mathbf{p},\mu }$ and $\hat{A}_{\mathbf{p}%
,\mu }^{+}$%
\begin{equation*}
\lbrack \hat{A}_{\mathbf{p},1},\hat{A}_{\mathbf{p}^{\prime }\mathbf{,}%
1}^{+}]=\delta _{\mathbf{p},\mathbf{p}^{\prime }}
\end{equation*}%
\begin{equation}
-\frac{1}{N_{\parallel }}\sum_{\mathbf{k\parallel p}}\sum_{\mathbf{k}%
^{\prime }\mathbf{\parallel p}^{\prime }}\left( \delta _{\mathbf{k},\mathbf{k%
}^{\prime }}\hat{a}_{\mathbf{p}^{\prime }+\mathbf{k}}^{+}\hat{a}_{\mathbf{p}+%
\mathbf{k}}+\delta _{\mathbf{p}+\mathbf{k},\mathbf{p}^{\prime }+\mathbf{k}%
^{\prime }}\hat{b}_{-\mathbf{k}^{\prime }}^{+}\hat{b}_{-\mathbf{k}}\right) ,
\label{com1}
\end{equation}%
\begin{equation*}
\lbrack \hat{A}_{\mathbf{p,}2},\hat{A}_{\mathbf{p}^{\prime }\mathbf{,}%
2}^{+}]=\delta _{\mathbf{p},\mathbf{p}^{\prime }}
\end{equation*}%
\begin{equation}
-\frac{1}{N_{\parallel }}\sum_{\mathbf{k\parallel p}}\sum_{\mathbf{k}%
^{\prime }\mathbf{\parallel p}^{\prime }}\left( \delta _{\mathbf{k},\mathbf{k%
}^{\prime }}\hat{d}_{\mathbf{p}^{\prime }+\mathbf{k}}^{+}\hat{d}_{\mathbf{p}+%
\mathbf{k}}+\delta _{\mathbf{p}+\mathbf{k},\mathbf{p}^{\prime }+\mathbf{k}%
^{\prime }}\hat{c}_{-\mathbf{k}^{\prime }}^{+}\hat{c}_{-\mathbf{k}}\right) .
\label{com2}
\end{equation}

Terms under the sum are written in the normal order, that is annihilation
operators are placed to the right of the creation operators. If the total
number of fermion states is very large compared to the number of occupied
states the commutation relations for the vector operator become exactly the
same as in the conventional quantum electrodynamics, namely, in the limit $%
N_{\parallel }\rightarrow \infty $ we obtain%
\begin{equation}
\lbrack \hat{A}_{\mathbf{p},\mu },\hat{A}_{\mathbf{p}^{\prime }\mathbf{,}\mu
^{\prime }}^{+}]=\delta _{\mathbf{p},\mathbf{p}^{\prime }}\delta _{\mu ,\mu
^{\prime }},\quad \mu ,\mu ^{\prime }=1,2  \label{aad}
\end{equation}%
and all other commutators are equal to zero. Roughly, the correction term to
the Bose--Einstein commutation relations is of the order of the ratio of the
number of fermions in the system to the total number of fermion states in
the Universe. Such correction is negligible.

One should mention that the number of fermion states can be much larger than
the number of photon states. For example, for electromagnetic field in a
cavity only certain photon states satisfy boundary conditions. However,
there is no such constraint on the fermion states because fermions do not
interact directly with the cavity walls. The boundary condition constrains
the total sum in Eqs. (\ref{ap1}) and (\ref{ap2}), that is values of the
photon wave vector $\mathbf{p}$. However, there is no constraint on the
values of the summation index $\mathbf{k}$.

In the composite theory of the photon the transverse part of the classical
electromagnetic Hamiltonian (\ref{htr}) describing radiation field
interacting with electric charges is replaced by the operator%
\begin{equation*}
\mathcal{\hat{H}}_{\text{tr}}=\mathcal{\hat{H}}_{0}+\sum_{a}\frac{1}{2m_{a}}%
\left( \mathbf{p}_{a}-\frac{q_{a}}{c}\mathbf{\hat{A}}_{\text{tr}}(\mathbf{r}%
_{a})\right) ^{2},
\end{equation*}
where $\mathcal{\hat{H}}_{0}$ is the Hamiltonian for the free fermion field (%
\ref{ffh}) and $\mathbf{\hat{A}}_{\text{tr}}(\mathbf{r})$ is given by Eqs. (%
\ref{af})-(\ref{ap2}).

In the composite photon theory the electric and magnetic field operators%
\begin{equation}
\mathbf{\hat{E}}_{\text{tr}}(\mathbf{r})=i\sum_{\mathbf{p,}\mu =1,2}\sqrt{%
\frac{2\pi \hbar cp}{V}}\left( \mathbf{\epsilon }_{\mathbf{p,}\mu }\hat{A}_{%
\mathbf{p,}\mu }e^{i\mathbf{p}\cdot \mathbf{r}}-\text{H.c.}\right) ,
\label{elec1}
\end{equation}%
\begin{equation}
\mathbf{\hat{B}}(\mathbf{r})=i\sum_{\mathbf{p,}\mu =1,2}\sqrt{\frac{2\pi
\hbar c}{pV}}\left( \mathbf{p}\times \mathbf{\epsilon }_{\mathbf{p,}\mu }%
\hat{A}_{\mathbf{p,}\mu }e^{i\mathbf{p}\cdot \mathbf{r}}-\text{H.c.}\right)
\label{mag1}
\end{equation}%
obey the same commutation relationships (\ref{cm1})-(\ref{cm4}) as in the
case of the elementary photon theory. Namely, Eqs. (\ref{cm1}) and (\ref{cm2}%
) are a direct consequence of Eq. (\ref{aad}), while relations (\ref{cm3})
and (\ref{cm4}) are satisfied because for the free fermion field Hamiltonian
(\ref{ffh}), as in the case of the free photon Hamiltonian (\ref{ff1}), we
obtain%
\begin{equation*}
\lbrack \mathcal{\hat{H}}_{0},\hat{A}_{\mathbf{p},\mu }]=-\hbar cp\hat{A}_{%
\mathbf{p},\mu },\quad \mu =1,2.
\end{equation*}

Therefore, in the composite photon theory the Heisenberg equation of motion
also yields Maxwell's equations for the transverse field (\ref{mqed}). Thus,
in the limit $N_{\parallel }\rightarrow \infty $ the composite photon theory
yields the same Quantum Electrodynamics as the elementary photon theory and,
in particular, photons obey Bose-Einstein statistics.

In the composite theory the fermions do not bind together to form photons.
It is not the interaction between fermion and antifermion that binds them
together into photons, but rather the manner in which fermions interact with
charged particles that leads to the simplified description of light in terms
of the composite photons \cite{Jord35}.

\subsection{Quantization of gravitational field}

\label{FQS}

In the vector theory of gravity for weak gravitational field and
nonrelativistic motion of masses the Lagrangian of the field interacting
with matter reads 
\begin{equation*}
\mathcal{L}=\frac{c^{4}}{32\pi G}\int d^{3}x\left( -3\frac{\partial h_{00}}{%
\partial x^{0}}\frac{\partial h_{00}}{\partial x^{0}}-\frac{\partial h_{00}}{%
\partial x^{\alpha }}\frac{\partial h_{00}}{\partial x^{\alpha }}-\frac{%
\partial h_{0\alpha }}{\partial x^{0}}\frac{\partial h_{0\alpha }}{\partial
x^{0}}\right.
\end{equation*}%
\begin{equation*}
+\left. \frac{\partial h_{0\alpha }}{\partial x^{\beta }}\frac{\partial
h_{0\alpha }}{\partial x^{\beta }}-\frac{\partial h_{0\alpha }}{\partial
x^{\beta }}\frac{\partial h_{0\beta }}{\partial x^{\alpha }}+2\left[ \frac{%
\partial h_{0\alpha }}{\partial x^{0}}\frac{\partial h_{00}}{\partial
x^{\alpha }}+\frac{\partial h_{0\alpha }}{\partial x^{\alpha }}\frac{%
\partial h_{00}}{\partial x^{0}}\right] \right)
\end{equation*}%
\begin{equation}
+\sum_{a}\left( \frac{m_{a}\dot{r}_{a}^{2}}{2}-\frac{m_{a}}{2}%
c^{2}h_{00}+m_{a}c\mathbf{\dot{r}}_{a}\cdot \mathbf{h}\right) ,  \label{clag}
\end{equation}%
where $h_{0k}$ are components of the equivalent metric, $\mathbf{h}%
=h^{0\alpha }$ and the sum is over all masses $m_{a}$ having positions $%
\mathbf{r}_{a}$ and velocities $\mathbf{\dot{r}}_{a}$. Particle momentum
conjugate to $\mathbf{r}_{a}$ is%
\begin{equation*}
\mathbf{p}_{a}=\frac{\partial \mathcal{L}}{\partial \mathbf{\dot{r}}_{a}}%
=m_{a}\mathbf{\dot{r}}_{a}+m_{a}c\mathbf{h(r}_{a}),
\end{equation*}%
while momentum $\Pi ^{k}$ conjugate to the field $h_{0k}$ reads%
\begin{equation*}
\Pi ^{k}=\frac{\partial L}{\partial \dot{h}_{0k}}=-\frac{c^{3}}{16\pi G}%
\left\{ 
\begin{array}{c}
\pi ^{0},\quad k=0 \\ 
E^{k},\quad k=1,2,3%
\end{array}%
,\right.
\end{equation*}%
where $L$ is the Lagrangian density,%
\begin{equation*}
\pi ^{0}=\text{div}\mathbf{h}+3\frac{\partial h_{00}}{\partial x^{0}},
\end{equation*}%
and%
\begin{equation*}
\mathbf{E}=-\nabla h_{00}-\frac{\partial \mathbf{h}}{\partial x^{0}}.
\end{equation*}%
Classical Hamiltonian of the system is a functional of $\mathbf{r}_{a}$, $%
\mathbf{p}_{a}$, $h_{0k}$ and $\Pi _{k}$ 
\begin{equation*}
\mathcal{H}=\sum_{a}\mathbf{p}_{a}\cdot \mathbf{\dot{r}}_{a}+\int d^{3}x\Pi
^{k}\dot{h}_{0k}-\mathcal{L}
\end{equation*}%
\begin{equation*}
=-\frac{c^{4}}{32\pi G}\int d^{3}x\left( \mathbf{E}^{2}+\mathbf{B}^{2}%
\mathbf{+}\frac{1}{3}\left( \pi ^{0}-\text{div}\mathbf{h}\right) ^{2}-2h_{00}%
\text{div}\mathbf{E}\right)
\end{equation*}%
\begin{equation}
+\sum_{a}\left( \frac{1}{2m_{a}}\left( \mathbf{p}_{a}-m_{a}c\mathbf{h}%
\right) ^{2}+\frac{m_{a}}{2}c^{2}h_{00}\right) ,  \label{gh}
\end{equation}%
where $\mathbf{B}=$curl$\mathbf{h}$. Transverse gravitational field $\mathbf{%
h}_{\text{tr}}$ interacting with matter is described by the Hamiltonian%
\begin{equation}
\mathcal{H}_{\text{tr}}=-\frac{c^{4}}{32\pi G}\int d^{3}x\left( \mathbf{E}_{%
\text{tr}}^{2}+\mathbf{B}^{2}\right) +\sum_{a}\frac{1}{2m_{a}}\left( \mathbf{%
p}_{a}-m_{a}c\mathbf{h}_{\text{tr}}\right) ^{2},  \label{htg}
\end{equation}%
where%
\begin{equation*}
\mathbf{E}_{\text{tr}}=-\frac{\partial \mathbf{h}_{\text{tr}}}{\partial x^{0}%
}.
\end{equation*}%
Please note that the energy of the free classical gravitational field is
negative.

As in the case of electrodynamics, the field quantization procedure replaces
transverse gravitational field corresponding to radiation with the operator%
\begin{equation}
\mathbf{\hat{h}}_{\text{tr}}(\mathbf{r})=\sum_{\mathbf{p,}\mu =1,2}\sqrt{%
\frac{8\pi G\hbar }{pVc^{3}}}\left( \mathbf{\epsilon }_{\mathbf{p,}\mu }\hat{%
h}_{\mathbf{p,}\mu }e^{i\mathbf{p}\cdot \mathbf{r}}+\text{H.c.}\right) ,
\label{htrg}
\end{equation}%
where $\mathbf{\epsilon }_{\mathbf{p,}1}$ and $\mathbf{\epsilon }_{\mathbf{p,%
}2}$ are spatial unit polarization vectors of the right and left circularly
polarized gravitons respectively. The non radiative part of gravitational
field is not quantized.

By analogy with the composite theory of the photon, we assume that graviton
is not an elementary particle but rather it is composed of
fermion-antifermion pairs. In the composite graviton model the part of the
classical Hamiltonian (\ref{htg}) describing radiation field interacting
with matter is replaced with the operator%
\begin{equation*}
\mathcal{\hat{H}}_{\text{tr}}=\mathcal{\hat{H}}_{0}+\sum_{a}\frac{1}{2m_{a}}%
\left( \mathbf{p}_{a}-m_{a}c\mathbf{\hat{h}}_{\text{tr}}\right) ^{2},
\end{equation*}%
where $\mathcal{\hat{H}}_{0}$ is the Hamiltonian of the free fermion field (%
\ref{ffh}) and $\mathbf{\hat{h}}_{\text{tr}}(\mathbf{r})$ is given by Eq. (%
\ref{htrg}) in which 
\begin{equation}
\hat{h}_{\mathbf{p},1}=\frac{1}{\sqrt{N_{\parallel }}}\sum_{\mathbf{%
k\parallel p}}\hat{a}_{-\mathbf{k}}\hat{b}_{\mathbf{p}+\mathbf{k}},
\label{hp1}
\end{equation}%
\begin{equation}
\hat{h}_{\mathbf{p},2}=\frac{1}{\sqrt{N_{\parallel }}}\sum_{\mathbf{%
k\parallel p}}\hat{d}_{-\mathbf{k}}\hat{c}_{\mathbf{p}+\mathbf{k}},
\label{hp2}
\end{equation}%
$N_{\parallel }$ is the number of fermion states with wave vectors $\mathbf{k%
}$ parallel to $\mathbf{p}$ and summation is taken over all such states.

Since fermion $a$ is right-handed the operator $\hat{a}_{-\mathbf{k}}^{+}$
creates a spin $1/2$ fermion with spin parallel to $\mathbf{-k}$ and energy $%
\varepsilon _{a}=\hbar ck$. On the other hand, $\hat{b}_{\mathbf{p}+\mathbf{k%
}}^{+}$ creates a spin $1/2$ antifermion with spin antiparallel to $\mathbf{p%
}+\mathbf{k}$ and negative energy $\varepsilon _{b}=-\hbar c(p+k)$. Thus,
the combination $\hat{b}_{\mathbf{p}+\mathbf{k}}^{+}\hat{a}_{-\mathbf{k}%
}^{+} $ with $\mathbf{k\parallel p}$ creates a fermion-antifermion pair with
the total negative energy $\varepsilon =-\hbar cp$ and spin $1$ antiparallel
to $\mathbf{p}$. The combination $\hat{c}_{\mathbf{p}+\mathbf{k}}^{+}\hat{d}%
_{-\mathbf{k}}^{+}$ with $\mathbf{k\parallel p}$ creates a
fermion-antifermion pair with the same energy $\varepsilon =-\hbar cp$ but
with spin parallel to $\mathbf{p}$. If we adopt the convention that for the
left (right) circularly polarized graviton the graviton spin is parallel
(antiparallel) to the wave vector $\mathbf{p}$ then operator $\hat{h}_{%
\mathbf{p},1}^{+}$ ($\hat{h}_{\mathbf{p},2}^{+}$) creates a right (left)
circularly polarized graviton with spin $1$. Emission of a single graviton
corresponds to creation of $N_{\parallel }$ fermion-antifermion pairs.

Operators $\hat{h}_{\mathbf{p},1}$ and $\hat{h}_{\mathbf{p},2}$ obey the
following commutation relations%
\begin{equation*}
\lbrack \hat{h}_{\mathbf{p},1},\hat{h}_{\mathbf{p}^{\prime }\mathbf{,}%
1}^{+}]=\delta _{\mathbf{p},\mathbf{p}^{\prime }}
\end{equation*}%
\begin{equation}
-\frac{1}{N_{\parallel }}\sum_{\mathbf{k\parallel p}}\sum_{\mathbf{k}%
^{\prime }\mathbf{\parallel p}^{\prime }}\left( \delta _{\mathbf{k},\mathbf{k%
}^{\prime }}\hat{b}_{\mathbf{p}^{\prime }+\mathbf{k}}^{+}\hat{b}_{\mathbf{p}+%
\mathbf{k}}+\delta _{\mathbf{p}+\mathbf{k},\mathbf{p}^{\prime }+\mathbf{k}%
^{\prime }}\hat{a}_{-\mathbf{k}^{\prime }}^{+}\hat{a}_{-\mathbf{k}}\right) ,
\label{c1g}
\end{equation}%
\begin{equation*}
\lbrack \hat{h}_{\mathbf{p,}2},\hat{h}_{\mathbf{p}^{\prime }\mathbf{,}%
2}^{+}]=\delta _{\mathbf{p},\mathbf{p}^{\prime }}
\end{equation*}%
\begin{equation}
-\frac{1}{N_{\parallel }}\sum_{\mathbf{k\parallel p}}\sum_{\mathbf{k}%
^{\prime }\mathbf{\parallel p}^{\prime }}\left( \delta _{\mathbf{k},\mathbf{k%
}^{\prime }}\hat{c}_{\mathbf{p}^{\prime }+\mathbf{k}}^{+}\hat{c}_{\mathbf{p}+%
\mathbf{k}}+\delta _{\mathbf{p}+\mathbf{k},\mathbf{p}^{\prime }+\mathbf{k}%
^{\prime }}\hat{d}_{-\mathbf{k}^{\prime }}^{+}\hat{d}_{-\mathbf{k}}\right) ,
\label{c2g}
\end{equation}%
which also can be written in the form%
\begin{equation*}
\lbrack \hat{h}_{\mathbf{p},1},\hat{h}_{\mathbf{p}^{\prime }\mathbf{,}%
1}^{+}]=-\delta _{\mathbf{p},\mathbf{p}^{\prime }}
\end{equation*}%
\begin{equation}
+\frac{1}{N_{\parallel }}\sum_{\mathbf{k\parallel p}}\sum_{\mathbf{k}%
^{\prime }\mathbf{\parallel p}^{\prime }}\left( \delta _{\mathbf{k},\mathbf{k%
}^{\prime }}\hat{b}_{\mathbf{p}+\mathbf{k}}\hat{b}_{\mathbf{p}^{\prime }+%
\mathbf{k}}^{+}+\delta _{\mathbf{p}+\mathbf{k},\mathbf{p}^{\prime }+\mathbf{k%
}^{\prime }}\hat{a}_{-\mathbf{k}}\hat{a}_{-\mathbf{k}^{\prime }}^{+}\right) ,
\label{c1g1}
\end{equation}%
\begin{equation*}
\lbrack \hat{h}_{\mathbf{p,}2},\hat{h}_{\mathbf{p}^{\prime }\mathbf{,}%
2}^{+}]=-\delta _{\mathbf{p},\mathbf{p}^{\prime }}
\end{equation*}%
\begin{equation}
+\frac{1}{N_{\parallel }}\sum_{\mathbf{k\parallel p}}\sum_{\mathbf{k}%
^{\prime }\mathbf{\parallel p}^{\prime }}\left( \delta _{\mathbf{k},\mathbf{k%
}^{\prime }}\hat{c}_{\mathbf{p}+\mathbf{k}}\hat{c}_{\mathbf{p}^{\prime }+%
\mathbf{k}}^{+}+\delta _{\mathbf{p}+\mathbf{k},\mathbf{p}^{\prime }+\mathbf{k%
}^{\prime }}\hat{d}_{-\mathbf{k}}\hat{d}_{-\mathbf{k}^{\prime }}^{+}\right) .
\label{c2g2}
\end{equation}

In Eqs. (\ref{c1g}) and (\ref{c2g}) the terms under the sum are written in
the normal order, while in Eqs. (\ref{c1g1}) and (\ref{c2g2}) the order is
the opposite. At the moment of Big Bang the fermion states are mostly empty,
that is the total number of fermion states is very large compared to the
number of occupied states. In this case in the limit $N_{\parallel
}\rightarrow \infty $ Eqs. (\ref{c1g}) and (\ref{c2g}) yield Bose--Einstein
commutation relations for operators $\hat{h}_{\mathbf{p},1}$ and $\hat{h}_{%
\mathbf{p},2}$ 
\begin{equation}
\lbrack \hat{h}_{\mathbf{p},\mu },\hat{h}_{\mathbf{p}^{\prime }\mathbf{,}\mu
^{\prime }}^{+}]=\delta _{\mathbf{p},\mathbf{p}^{\prime }}\delta _{\mu ,\mu
^{\prime }},\quad \mu ,\mu ^{\prime }=1,2.  \label{bb1}
\end{equation}%
All other commutators are equal to zero. At this stage of the Universe
evolution the graviton has negative energy $\varepsilon (\mathbf{p})=-\hbar
cp$ which causes cosmological inflation.

Shortly after the Big Bang the fermion states become filled and remain
filled in the present epoch. Such filled states form a new vacuum. If the
total number of fermion states is very large compared to the number of empty
states we must use Eqs. (\ref{c1g1}) and (\ref{c2g2}) that in the limit $%
N_{\parallel }\rightarrow \infty $ give%
\begin{equation}
\lbrack \hat{h}_{\mathbf{p},\mu },\hat{h}_{\mathbf{p}^{\prime }\mathbf{,}\mu
^{\prime }}^{+}]=-\delta _{\mathbf{p},\mathbf{p}^{\prime }}\delta _{\mu ,\mu
^{\prime }},\quad \mu ,\mu ^{\prime }=1,2  \label{bb2}
\end{equation}%
which differs from the Bose--Einstein commutation relations by the minus
sign in the right hand side. However, for operators%
\begin{equation*}
\hat{A}_{\mathbf{p,}\mu }^{+}=\hat{h}_{-\mathbf{p},\mu },\quad \hat{A}_{%
\mathbf{p,}\mu }=\hat{h}_{-\mathbf{p},\mu }^{+}
\end{equation*}%
Eq. (\ref{bb2}) yields Bose--Einstein commutation relations 
\begin{equation}
\lbrack \hat{A}_{\mathbf{p},\mu },\hat{A}_{\mathbf{p}^{\prime }\mathbf{,}\mu
^{\prime }}^{+}]=\delta _{\mathbf{p},\mathbf{p}^{\prime }}\delta _{\mu ,\mu
^{\prime }},\quad \mu ,\mu ^{\prime }=1,2,  \label{bb3}
\end{equation}%
while Eq. (\ref{bb1}) gives relations with the minus sign%
\begin{equation}
\lbrack \hat{A}_{\mathbf{p},\mu },\hat{A}_{\mathbf{p}^{\prime }\mathbf{,}\mu
^{\prime }}^{+}]=-\delta _{\mathbf{p},\mathbf{p}^{\prime }}\delta _{\mu ,\mu
^{\prime }},\quad \mu ,\mu ^{\prime }=1,2.  \label{bb4}
\end{equation}

Operator $\hat{A}_{\mathbf{p},1}^{+}$ ($\hat{A}_{\mathbf{p},2}^{+}$) creates
a graviton with positive energy $\varepsilon (\mathbf{p})=\hbar cp$ and spin 
$1$ parallel (antiparallel) to $\mathbf{p}$. Thus, in the present epoch the
graviton energy is positive. Creation of such graviton corresponds to
annihilation of fermion-antifermion pairs out of the filled vacuum states
(creation of holes). In terms of $\hat{A}_{\mathbf{p},\mu }$ and $\hat{A}_{%
\mathbf{p,}\mu }^{+}$ the field operator (\ref{htrg}) reads%
\begin{equation*}
\mathbf{\hat{h}}_{\text{tr}}(\mathbf{r})=\sum_{\mathbf{p,}\mu =1,2}\sqrt{%
\frac{8\pi G\hbar }{pVc^{3}}}\left( \mathbf{\epsilon }_{\mathbf{p,}\mu }\hat{%
A}_{\mathbf{p,}\mu }e^{i\mathbf{p}\cdot \mathbf{r}}+\text{H.c.}\right) ,
\end{equation*}%
where now $\mathbf{\epsilon }_{\mathbf{p,}1}$ and $\mathbf{\epsilon }_{%
\mathbf{p,}2}$ are spatial unit polarization vectors of the left and right
circularly polarized gravitons respectively.

Commutators of $\hat{A}_{\mathbf{p},\mu }$, gravitoelectric%
\begin{equation*}
\mathbf{\hat{E}}_{\text{tr}}(\mathbf{r})=i\sum_{\mathbf{p,}\mu =1,2}\sqrt{%
\frac{8\pi G\hbar p}{Vc^{3}}}\left( \mathbf{\epsilon }_{\mathbf{p,}\mu }\hat{%
A}_{\mathbf{p,}\mu }e^{i\mathbf{p}\cdot \mathbf{r}}-\text{H.c.}\right) ,
\end{equation*}%
and gravitomagnetic%
\begin{equation*}
\mathbf{\hat{B}}(\mathbf{r})=i\sum_{\mathbf{p,}\mu =1,2}\sqrt{\frac{8\pi
G\hbar }{pVc^{3}}}\left( \mathbf{p}\times \mathbf{\epsilon }_{\mathbf{p,}\mu
}\hat{A}_{\mathbf{p,}\mu }e^{i\mathbf{p}\cdot \mathbf{r}}-\text{H.c.}\right)
\end{equation*}%
field operators with the free field Hamiltonian (\ref{ffh}) are the same as
in electrodynamics, namely%
\begin{equation*}
\lbrack \mathcal{\hat{H}}_{0},\hat{A}_{\mathbf{p},\mu }]=-\hbar cp\hat{A}_{%
\mathbf{p},\mu },\quad \mu =1,2,
\end{equation*}%
\begin{equation*}
\lbrack \mathcal{\hat{H}}_{0},\mathbf{\hat{E}}_{\text{tr}}]=-ic\hbar \,\text{%
curl}\mathbf{\hat{B},}
\end{equation*}%
\begin{equation*}
\lbrack \mathcal{\hat{H}}_{0},\mathbf{\hat{B}}]=ic\hbar \,\text{curl}\mathbf{%
\hat{E}}_{\text{tr}}.
\end{equation*}%
This result is independent of the commutation relation between $\hat{A}_{%
\mathbf{p},\mu }$ and $\hat{A}_{\mathbf{p,}\mu }^{+}$. In addition, the
commutator%
\begin{equation*}
\lbrack \hat{h}_{\text{tr}}^{\beta }(\mathbf{r}^{\prime }),\hat{B}_{\alpha }(%
\mathbf{r})]\mathbf{=}0,\quad \alpha ,\beta =1,2,3
\end{equation*}%
remains the same. However, sign of the commutation relation between $\mathbf{%
\hat{h}}_{\text{tr}}(\mathbf{r})$ and $\mathbf{\hat{E}}_{\text{tr}}(\mathbf{r%
})$ depends on the vacuum state, namely 
\begin{equation}
\lbrack \hat{h}_{\text{tr}}^{\beta }(\mathbf{r}^{\prime }),\hat{E}_{\text{tr}%
}^{\alpha }(\mathbf{r})]\mathbf{=\mp }i\frac{16\pi G\hbar }{c^{3}}\delta
^{\beta \alpha }\delta (\mathbf{r-r}^{\prime }).  \label{hec}
\end{equation}%
The lower sign in Eq. (\ref{hec}) corresponds to Eq. (\ref{bb4}), that is to
the vacuum with empty fermion states. The upper sign follows from Eq. (\ref%
{bb3}) obtained for the filled vacuum and is the same as in quantum
electrodynamics.

In the Heisenberg picture the Heisenberg equations of motion yield
Maxwell-like equations for the transverse gravitational field in the
operator form%
\begin{equation}
\text{curl}\mathbf{\hat{B}}=\pm \frac{16\pi G}{c^{3}}\mathbf{j}_{\text{tr}}+%
\frac{1}{c}\frac{\partial \mathbf{\hat{E}}_{\text{tr}}}{\partial t},\quad 
\text{curl}\mathbf{\hat{E}}_{\text{tr}}=-\frac{1}{c}\frac{\partial \mathbf{%
\hat{B}}}{\partial t},  \label{meg}
\end{equation}%
where $\mathbf{j}_{\text{tr}}$ is the transverse part of the mass current
density%
\begin{equation*}
\mathbf{j}=\sum_{a}m_{a}\mathbf{\dot{r}}_{a}\delta \left( \mathbf{r}-\mathbf{%
r}_{a}(t)\right) .
\end{equation*}%
Hamiltonian (\ref{htg}) gives the following equation of motion of mass $m$
in transverse gravitational field 
\begin{equation}
m\mathbf{\ddot{r}}=c^{2}\left[ m\mathbf{E}_{\text{tr}}+\frac{m}{c}\left( 
\mathbf{\dot{r}}\times \mathbf{B}\right) \right] .  \label{eqmm}
\end{equation}

Comparison of Eqs. (\ref{meg}) and (\ref{eqmm}) with those of quantum
electrodynamics yields that quantum vector gravity, upto irrelevant
numerical factor, is equivalent to QED for the upper sign in Eq. (\ref{meg}%
), that is for the filled vacuum (present epoch).

The lower sign in Eq. (\ref{meg}) corresponds to the vacuum with empty
fermion states. This is the classical limit of the quantum vector gravity
which reproduces the classical weak field equations for the transverse
field. In the present epoch the fermion states are filled and we must take
the upper sign in Eq. (\ref{meg}). Thus, quantum mechanical analysis yields
that evolution equations describing gravitational radiation in the present
epoch are different from those that follow from the classical Lagrangian (%
\ref{clag}). Namely, in classical equations describing radiation, $\mathbf{j}%
_{\text{tr}}$ must be taken with the opposite sign.

One should mention that the difference in equations appears only when we are
dealing with the radiation part of the transverse field which is quantized.
Transverse gravitational field produced by stationary mass currents is not
quantized and is described by the same classical equations. As a
consequence, the Post-Newtonian limit of vector gravity is entirely
classical and Post-Newtonian equations are not modified by the quantum
mechanical analysis. This is also the case for Universe evolution after the
end of the inflation stage. In particular, dark energy comes from the
classical longitudinal part of the gravitational field.

Averaging the operator equations (\ref{meg}) over the state vector yields
Maxwell-like equations for the average fields $\mathbf{E}_{\text{tr}}$ and $%
\mathbf{B}$. The averaged equations lead to the following expression for the
energy of the radiation field interacting with matter 
\begin{equation}
W_{\text{tr}}=\pm \frac{c^{4}}{32\pi G}\int d^{3}x\left( \mathbf{E}_{\text{tr%
}}^{2}+\mathbf{B}^{2}\right) +\sum_{a}\frac{m_{a}\mathbf{\dot{r}}_{a}^{2}}{2}
\label{egrav}
\end{equation}%
and the energy flux density (Poynting vector) of the radiation gravitational
field%
\begin{equation}
\mathbf{S}=\pm \frac{c^{5}}{16\pi G}\mathbf{E}_{\text{tr}}\times \mathbf{B}.
\label{sgrav}
\end{equation}

Energy of the graviton in the classical limit (moment of the Big Bang) is
negative (lower sign in Eqs. (\ref{egrav}) and (\ref{sgrav})). In the
present epoch the energy is positive (upper sign). In this case the analogy
of Eqs. (\ref{egrav}) and (\ref{sgrav}) with the corresponding expressions
in electrodynamics%
\begin{equation}
W_{\text{tr}}=\frac{1}{8\pi }\int d^{3}x\left( \mathbf{E}_{\text{tr}}^{2}+%
\mathbf{B}^{2}\right) +\sum_{a}\frac{m_{a}\mathbf{\dot{r}}_{a}^{2}}{2},\quad 
\mathbf{S}=\frac{c}{4\pi }\mathbf{E}_{\text{tr}}\times \mathbf{B}
\label{wse}
\end{equation}%
is obvious.

Finally we discuss quantization of the longitudinal gravitational waves (\ref%
{met2}) which are described by equations%
\begin{equation}
\frac{\partial h_{00}}{\partial x^{0}}+\frac{1}{2}\text{div}\mathbf{h}_{l}%
\mathbf{=}0,  \label{hl1}
\end{equation}%
\begin{equation}
\left( \frac{\partial ^{2}}{\partial x^{0}\partial x^{0}}-\Delta \right) 
\mathbf{h}_{l}=0.  \label{hl2}
\end{equation}%
Classical analysis yields that such waves are not emitted or absorbed by
orbiting stars (see next Section). Quantum consideration gives the same
answer. Indeed, Eqs. (\ref{pv3}), (\ref{pv4}), (\ref{gh}), (\ref{hl1}) and (%
\ref{hl2}) suggest that longitudinal gravitational waves must be quantized
by replacing $\mathbf{h}_{l}$ and $h_{00}$ with operators%
\begin{equation}
\mathbf{\hat{h}}_{l}(\mathbf{r})=\sum_{\mathbf{p}}\sqrt{\frac{32\pi G\hbar }{%
3pVc^{3}}}\left( \hat{p}\hat{h}_{\mathbf{p}}e^{i\mathbf{p}\cdot \mathbf{r}}+%
\text{H.c.}\right) ,
\end{equation}%
\begin{equation}
\hat{h}_{00}(\mathbf{r})=-\frac{1}{2}\sum_{\mathbf{p}}\sqrt{\frac{32\pi
G\hbar }{3pVc^{3}}}\left( \hat{h}_{\mathbf{p}}e^{i\mathbf{p}\cdot \mathbf{r}%
}+\text{H.c.}\right) ,
\end{equation}%
where $\hat{p}$ is a unit vector in the direction of $\mathbf{p}$ and
operator%
\begin{equation}
\hat{h}_{\mathbf{p}}=\frac{1}{\sqrt{N_{\parallel }}}\sum_{\mathbf{k\parallel
p}}\left( \hat{a}_{\mathbf{k}}\hat{a}_{\mathbf{p}+\mathbf{k}}^{+}+\hat{b}_{%
\mathbf{p}+\mathbf{k}}\hat{b}_{\mathbf{k}}^{+}+\hat{c}_{\mathbf{p}+\mathbf{k}%
}\hat{c}_{\mathbf{k}}^{+}+\hat{d}_{\mathbf{k}}\hat{d}_{\mathbf{p}+\mathbf{k}%
}^{+}\right)
\end{equation}%
describes a composite particle with negative energy $\varepsilon (p)=-\hbar
cp$. Commutator of $\hat{h}_{\mathbf{p}}$ with the free field Hamiltonian (%
\ref{ffh}) is%
\begin{equation*}
\lbrack \mathcal{\hat{H}}_{0},\hat{h}_{\mathbf{p}}]=\hbar cp\hat{h}_{\mathbf{%
p}}.
\end{equation*}%
It is easy to check that in the Heisenberg picture the Heisenberg equations
of motion involving the free field Hamiltonian (\ref{ffh}) yield the free
field Eqs. (\ref{hl1}) and (\ref{hl2}) in the operator form%
\begin{equation}
\frac{\partial \hat{h}_{00}}{\partial x^{0}}+\frac{1}{2}\text{div}\mathbf{%
\hat{h}}_{l}\mathbf{=}0,  \label{hl1a}
\end{equation}%
\begin{equation}
\left( \frac{\partial ^{2}}{\partial x^{0}\partial x^{0}}-\Delta \right) 
\mathbf{\hat{h}}_{l}=0.  \label{hl2a}
\end{equation}

In the limit $N_{\parallel }\rightarrow \infty $ we obtain commutation
relations%
\begin{equation}
\lbrack \hat{h}_{\mathbf{p}},\hat{h}_{\mathbf{p}^{\prime }}^{+}]=[\hat{h}_{%
\mathbf{p}},\hat{h}_{\mathbf{p}^{\prime }}]=0  \label{ch}
\end{equation}%
for both empty and filled vacuum. $\hat{h}_{\mathbf{p}}$ and $\hat{h}_{%
\mathbf{p}}^{+}$ also commute with the transverse field (graviton)
operators. Thus, if longitudinal waves are coupled with matter through the
operators $\hat{h}_{\mathbf{p}}$ and $\hat{h}_{\mathbf{p}}^{+}$ the
commutator of $\hat{h}_{\mathbf{p}}$ with the interaction Hamiltonian will
be equal to zero. As a result, the Heisenberg equations of motion involving
the full Hamiltonian will also give the free field equations (\ref{hl1a})
and (\ref{hl2a}) without sources. This means that evolution of the
longitudinal waves is not affected by matter. Therefore, longitudinal waves
are not produced by matter in the classical (empty vacuum) and quantum
(filled vacuum) limits. However, they can be generated in the early Universe
or at the stage of merger of massive neutron stars.

Commutation relations (\ref{ch}) indicate that longitudinal and time-like
components of the gravitational field behave as classical quantities.

\section{Radiation of gravitational waves by system of masses}

\label{binary}

In this section we consider radiation of gravitational waves by a system of
stars moving with nonrelativistic velocities. Our analysis is also valid for
neutron stars which produce strong gravitational field in their vicinity. We
assume that stars have masses $M_{i}$ ($i=1,2,\ldots $) and move with
velocities $\mathbf{V}_{i}(t)\ll c$. Spacing between starts is large
compared to their dimension, however, the total size of the system is much
smaller that the radiation wavelength.

Strong gravitational field of neutron stars demands to keep nonlinear terms
in the equations for the gravitational field. Such terms are not confined to
a compact region, but extend over all space. However, nonlinear terms decay
as $1/r^{2}$ away from the star and can be large only in the vicinity of a
neutron star. We average the gravitational field equations (\ref{sss1}) over
the volume large compared to the stellar size but much smaller than spacing
between stars. After such averaging the source of the gravitational field
becomes a sum of $\delta -$functions localized at the star positions.
Nonlinear terms in the stellar vicinity are subsumed into the $\delta -$%
function sources. Nonlinear terms far from the star could give a small
correction to the solution in the wave zone of the order of $G^{2}$ which we
neglect.

As a result, after averaging we obtain the following linear equations for
the gravitational field in the cosmological reference frame (cf. Eqs. (\ref%
{wf1}) and (\ref{wf2}))%
\begin{equation}
\Delta h_{00}+3\frac{\partial ^{2}h_{00}}{\partial x^{0}\partial x^{0}}-2%
\frac{\partial ^{2}h_{0\beta }}{\partial x^{0}\partial x^{\beta }}=\frac{%
8\pi G}{c^{2}}\sum_{i}M_{i}\delta \left( \mathbf{r}-\mathbf{r}_{i}(t)\right)
,  \label{w3}
\end{equation}%
\begin{equation*}
\left( \frac{\partial ^{2}}{\partial x^{0}\partial x^{0}}-\Delta \right)
h_{0\alpha }+\frac{\partial ^{2}h_{0\beta }}{\partial x^{\alpha }\partial
x^{\beta }}-2\frac{\partial ^{2}h_{00}}{\partial x^{\alpha }\partial x^{0}}
\end{equation*}%
\begin{equation}
=\frac{16\pi G}{c^{3}}\sum_{i}M_{i}^{\ast }V_{i}^{\alpha }\delta (\mathbf{r}-%
\mathbf{r}_{i}(t)),  \label{w4}
\end{equation}%
where $\mathbf{r}_{i}(t)$ are the radii vectors of the stars. In Eq. (\ref%
{w3}) $M_{i}$ are the stellar masses measured by a distant observer.
However, for neutron stars, $M_{i}^{\ast }$ in Eq. (\ref{w4}) could differ
from $M_{i}$ due to strong-field effects and depend on the stellar equation
of state. Next we show that $M_{i}^{\ast }=M_{i}$ at least upto the second
order in the stellar velocity.

Taking $\partial /\partial x^{0}$ from Eq. (\ref{w3}) and $(1/2)\partial
/\partial x^{\alpha }$ from Eq. (\ref{w4}) and adding these equations
together we obtain%
\begin{equation*}
3\frac{\partial ^{2}}{\partial x^{0}\partial x^{0}}\left( \frac{\partial
h_{00}}{\partial x^{0}}-\frac{1}{2}\frac{\partial h_{0\beta }}{\partial
x^{\beta }}\right) =
\end{equation*}%
\begin{equation}
\frac{8\pi G}{c^{2}}\sum_{i}\left( M_{i}\frac{\partial }{\partial x^{0}}%
\delta (\mathbf{r}-\mathbf{r}_{i}(t))+\frac{1}{c}\frac{\partial }{\partial
x^{\alpha }}M_{i}^{\ast }V_{i}^{\alpha }\delta (\mathbf{r}-\mathbf{r}%
_{i}(t))\right) .  \label{w5}
\end{equation}%
Using%
\begin{equation*}
\frac{\partial }{\partial x^{0}}\delta (\mathbf{r}-\mathbf{r}(t))=-\frac{1}{c%
}\frac{\partial }{\partial x^{\alpha }}\left[ V^{\alpha }\delta (\mathbf{r}-%
\mathbf{r}(t))\right]
\end{equation*}%
one can write Eq. (\ref{w5}) as%
\begin{equation*}
3\frac{\partial ^{2}}{\partial x^{0}\partial x^{0}}\left( \frac{\partial
h_{00}}{\partial x^{0}}-\frac{1}{2}\frac{\partial h_{0\beta }}{\partial
x^{\beta }}\right) =
\end{equation*}%
\begin{equation}
\frac{8\pi G}{c^{2}}\sum_{i}\left( M_{i}-M_{i}^{\ast }\right) \frac{\partial 
}{\partial x^{0}}\delta (\mathbf{r}-\mathbf{r}_{i}(t)).  \label{w6}
\end{equation}%
The left hand side of Eq. (\ref{w6}) is of the order of $(V/c)^{3}$. This
yields that $M_{i}-M_{i}^{\ast }$ in the right hand side are of the order of 
$(V/c)^{2}$.

Thus, with the required accuracy one can take $M_{i}^{\ast }=M_{i}$ in Eqs. (%
\ref{w3}) and (\ref{w4}). Solution of Eqs. (\ref{w3}) and (\ref{w4})
satisfying the proper boundary condition is given by the retarded potentials 
\begin{equation}
h_{00}=-\frac{2G}{c^{2}}\sum_{i}\frac{M_{i}}{\left\vert \mathbf{r}-\mathbf{r}%
_{i}(t_{i})\right\vert },  \label{ss0}
\end{equation}%
\begin{equation}
h_{0\alpha }=\frac{4G}{c^{3}}\sum_{i}\frac{M_{i}V_{i}^{\alpha }(t_{i})}{%
\left\vert \mathbf{r}-\mathbf{r}_{i}(t_{i})\right\vert },  \label{ss1}
\end{equation}%
where $t_{i}$ is solution of the equation $t=t_{i}+|\mathbf{r}-\mathbf{r}%
_{i}(t_{i})|/c$. Retarded potentials describe outgoing waves with phase
velocity directed away from the source. For solution (\ref{ss0})-(\ref{ss1})%
\begin{equation*}
\frac{\partial h_{00}}{\partial x^{0}}-\frac{1}{2}\frac{\partial h_{0\beta }%
}{\partial x^{\beta }}=0.
\end{equation*}

In Appendix \ref{cossup} we provide an alternative derivation of the
gravitational field produced by an orbiting neutron star which is valid in
the $V^{3}/c^{3}$ order in the stellar velocity and arbitrary strength of
the stellar gravitational field $\phi $. The answer for the equivalent
metric is given by Eqs. (\ref{fm1})-(\ref{fm3}). The analytical result
includes both near and far field regions in a single equation which is valid
for arbitrary $\phi $ but omits retardation effects. In the far field the
answer reduces to Eqs. (\ref{ppnf}) and (\ref{ppnh}) of Appendix \ref{cossup}
which match the retarded potentials (\ref{ss0}) and (\ref{ss1}) obtained
here using linearized equations. Such a match justifies omission of the
nonlinear terms in the present Eqs. (\ref{w3}) and (\ref{w4}).

Since $\sum_{i}M_{i}$ is constant, the time dependent part of $h_{00}$ in
Eq. (\ref{ss0}) vanishes at large $r$ as $1/r^{2}$. Thus, $h_{00}$ does not
contribute to the emission of gravitational waves, as in the case of
electromagnetic radiation in classical electrodynamics.

In the present theory, a graviton is not an elementary particle, but rather
it is composed of fermion-antifermion pairs. Solution (\ref{ss1}) is
obtained in the classical limit when vacuum corresponds to empty fermion
states. However, fermion states are filled in the present epoch and, as we
showed in the previous section, for filled vacuum the equations describing
radiation of gravitational waves must be modified by changing the sign of
the mass current to the opposite. As a consequence, solution (\ref{ss1})
must be replaced with 
\begin{equation}
\mathbf{h}=\frac{4G}{c^{3}}\sum_{i}\frac{M_{i}\mathbf{V}_{i}(t_{i})}{%
\left\vert \mathbf{r}-\mathbf{r}_{i}(t_{i})\right\vert },  \label{ss1a}
\end{equation}%
where $\mathbf{h}=h^{0\alpha }=-h_{0\alpha }$. For filled vacuum emission of
a graviton corresponds to absorption of fermion-antifermion pairs out of the
filled vacuum states or creation of fermion-antifermion holes which
propagate away from the source. According to Eq. (\ref{egrav}), for filled
vacuum graviton has positive energy and binary stars orbiting each other are
loosing their energy by emitting gravitational waves. Equation (\ref{ss1a})
has a very similar form to the vector potential produced by moving charges $%
e_{i}$%
\begin{equation}
\mathbf{A}=\frac{1}{c}\sum_{i}\frac{e_{i}\mathbf{V}_{i}(t_{i})}{|\mathbf{r}-%
\mathbf{r}_{i}(t_{i})|},  \label{ss2}
\end{equation}%
and, according to Eqs. (\ref{sgrav}) and (\ref{wse}), expressions for the
Poynting vectors of the radiation field in vector gravity and
electrodynamics are also similar. As a consequence, to obtain the answer for
the power $P$ of emission of gravitational waves by the system of masses we
can apply the formula of classical electrodynamics for the radiation of
electromagnetic waves by a system of charges%
\begin{equation*}
P=\frac{2}{3c^{3}}\mathbf{\ddot{d}}^{2}+\frac{1}{180c^{5}}\dddot{D}_{\alpha
\beta }^{2}+\frac{2}{3c^{3}}\mathbf{\ddot{m}}^{2},
\end{equation*}%
where%
\begin{equation*}
\mathbf{d}=\sum e\mathbf{r}
\end{equation*}%
is the electric dipole moment of the system,%
\begin{equation*}
D_{\alpha \beta }=\sum e\left( 3x_{\alpha }x_{\beta }-r^{2}\delta _{\alpha
\beta }\right)
\end{equation*}%
is the quadrupole moment, 
\begin{equation*}
\mathbf{m}=\frac{1}{2c}\sum e\left( \mathbf{r}\times \mathbf{V}\right)
\end{equation*}%
is the magnetic moment and the sum is over all charges in the system.

Comparison of Eqs. (\ref{ss1a}) and (\ref{ss2}) yields that in order to
obtain power loss due to emission of gravitons one should in the
electrodynamics equations replace the electric charges with $%
e_{i}\rightarrow 4GM_{i}/c^{2}$. In addition, Eq. (\ref{sgrav}) gives that
gravitational energy density flux for the radiation field is%
\begin{equation}
\mathbf{S}=-\frac{c^{5}}{16\pi G}\frac{\partial \mathbf{h}}{\partial x^{0}}%
\times \text{curl}\mathbf{(h}),  \label{pg}
\end{equation}%
which is in a factor $c^{4}/4G$ greater than the corresponding expression
for the electromagnetic waves 
\begin{equation}
\mathbf{S}=\frac{c}{4\pi }\mathbf{E}_{\text{tr}}\times \mathbf{H}=-\frac{c}{%
4\pi }\frac{\partial \mathbf{A}}{\partial x^{0}}\times \text{curl}\mathbf{(A}%
).  \label{pem}
\end{equation}%
Thus, $P$ for gravity must be also multiplied by $c^{4}/4G$.

Combining all factors together we finally obtain the following expression
for the power loss by the system of masses due to emission of gravitational
waves%
\begin{equation}
P=\frac{8G}{3c^{3}}\mathbf{\ddot{d}}^{2}+\frac{G}{45c^{5}}\dddot{D}_{\alpha
\beta }^{2}+\frac{2G}{3c^{5}}\mathbf{\ddot{L}}^{2},  \label{v1}
\end{equation}%
where we introduced the dipole moment of the system%
\begin{equation*}
\mathbf{d}=\sum M\mathbf{r,}
\end{equation*}%
the quadrupole moment of masses%
\begin{equation*}
D_{\alpha \beta }=\sum M\left( 3x_{\alpha }x_{\beta }-r^{2}\delta _{\alpha
\beta }\right)
\end{equation*}%
and the net angular momentum%
\begin{equation*}
\mathbf{L}=\sum M\left( \mathbf{r}\times \mathbf{V}\right) .
\end{equation*}%
In these equations the summation is over all masses.

With the required accuracy, the stellar trajectories can be calculated in
the Newtonian limit. In such limit%
\begin{equation*}
\mathbf{\dot{d}}=\sum_{i}M_{i}\mathbf{V}_{i}
\end{equation*}%
is the total linear momentum of the isolated system which is a conserved
quantity. Therefore, $\mathbf{\ddot{d}}$ vanishes and, hence, there is no
dipole radiation. For an isolated system the total angular momentum $\mathbf{%
L}$ is also conserved and, thus, the last term in Eq. (\ref{v1}) also
vanishes. The fact that $M_{i}^{\ast }=M_{i}$ in Eqs. (\ref{w3}) and (\ref%
{w4}) guarantees that inertial mass that determines the dipole moment is the
same as mass that generates gravitational waves. One should mention that
possible small deviation of $M_{i}^{\ast }$ from $M_{i}$ in our Eq. (\ref{w4}%
) due to strong-field effects (which is of the order of $(V/c)^{2}$) might
yield contribution to the dipole radiation of the order of $\mathbf{\ddot{d}}%
^{2}\propto (V/c)^{8}$, while $\mathbf{\ddot{L}}^{2}\propto (V/c)^{10}$.
These contributions are much smaller than the quadrupole emission which is
proportional to $(V/c)^{6}$.

As a result, the quadrupole radiation\ gives the dominant contribution to
the energy loss. The quadrupole term in Eq. (\ref{v1}) coincides with the
Einstein's formula obtained in general relativity. The rate of loss of
angular momentum from a system of bodies emitting gravitational waves is
also given by the same equation as in general relativity%
\begin{equation}
\frac{dL_{\alpha }}{dt}=-\frac{2G}{45c^{5}}e_{\alpha \beta \gamma }\dddot{D}%
_{\beta \delta }\ddot{D}_{\delta \gamma },  \label{mom1}
\end{equation}%
where $e_{\alpha \beta \gamma }$ is the Levi-Civita symbol. Equation (\ref%
{mom1}) is obtained in \cite{Land95} directly from the rate of energy loss
by the system which is given by the same quadrupole formula in both
theories. As a consequence, for nonrelativistic motion the present theory
yields the same orbital decay of binary stars from gravitational radiation
as general relativity.

The energy flux at infinity carried out by gravitational radiation is
balanced by an equal loss of mechanical or orbital energy by the system $W$.
This loss of energy results in a decrease in the orbital period $T$ given by
Kepler's third law \cite{Will93}%
\begin{equation*}
\frac{\dot{T}}{T}=-\frac{3}{2}\frac{\dot{W}}{W}.
\end{equation*}%
Such decrease in the orbital period has been measured for several binary
systems and agreed with the predictions of general relativity. Thus, it also
agrees with the vector theory of gravity.

In other alternative theories of gravity, while the inertial dipole moment
may remain uniform, the \textquotedblleft gravity wave\textquotedblright\
dipole moment need not, because the mass that generates gravitational waves
depends differently on the internal gravitational binding energy of each
body than does the inertial mass \cite{Will93,Will06}. In such theories, the
additional form of gravitational radiation damping (dipole radiation) could
be significantly stronger for neutron stars than the usual quadrupole
damping. Our vector theory of gravity predicts no dipole gravitational
radiation because it satisfies the strong equivalence principle at least to
the post-Newtonian order.

\section{Neutron star mass limit in vector gravity}

\label{NSML}

According to general relativity an object of nuclear density and more than
about $3M_{\odot }$ would be a black hole \cite{Frie87,Kalo96}. Here we
examine the neutron star upper mass limit in the vector theory of gravity.
To calculate the maximum mass of a neutron star, one must have an equation
of state for matter at high density which is very uncertain at present.

Taking $T_{k}^{i}$ as the energy momentum tensor of a perfect fluid, namely%
\begin{equation*}
T_{0}^{0}=\varepsilon ,\quad T_{\alpha }^{\alpha }=-P\text{,}
\end{equation*}%
all others are $0$, with energy density $\varepsilon =\varepsilon (r)$ and
pressure $P=P(r)$ the field equations for static gravitational field
described by the equivalent metric%
\begin{equation*}
f_{ik}=\left( 
\begin{array}{cccc}
e^{2\phi } & 0 & 0 & 0 \\ 
0 & -e^{-2\phi } & 0 & 0 \\ 
0 & 0 & -e^{-2\phi } & 0 \\ 
0 & 0 & 0 & -e^{-2\phi }%
\end{array}%
\right) ,
\end{equation*}%
become%
\begin{equation*}
\Delta \phi =\frac{4\pi G}{c^{4}}\left( \varepsilon +3P\right) e^{-2\phi }.
\end{equation*}%
The energy-momentum relation 
\begin{equation}
T_{i;k}^{k}=0,  \label{m02}
\end{equation}%
where covariant derivative is taken using equivalent metric $f_{ik}$ 
\begin{equation*}
T_{i;k}^{k}=\frac{1}{\sqrt{-f}}\frac{\partial }{\partial x^{k}}\left( \sqrt{%
-f}T_{i}^{k}\right) -\frac{1}{2}\frac{\partial f_{kl}}{\partial x^{i}}T^{kl},
\end{equation*}%
yields equation of hydrostatic equilibrium%
\begin{equation*}
\frac{\partial P}{\partial r}=-(P+\varepsilon )\frac{\partial \phi }{%
\partial r},
\end{equation*}%
where $P$ is related to $\varepsilon $ by an equation of state%
\begin{equation*}
P=P(\varepsilon ).
\end{equation*}

The boundary conditions at the stellar center $r=0$ and the surface $r=R$
read%
\begin{equation*}
\phi ^{\prime }(0)=0,\quad P(R)=0.
\end{equation*}
Taking into account that outside the star $\phi (r)=-GM/rc^{2}$, continuity
of $\phi $ and $\phi ^{\prime }$ yields additional boundary condition at the
star surface%
\begin{equation*}
R\phi ^{\prime }(R)=-\phi (R).
\end{equation*}
Matching solution inside and outside the star yields expression for the
stellar mass $M$ in terms of $\phi (R)$%
\begin{equation*}
M=-\frac{Rc^{2}}{G}\phi (R).
\end{equation*}

It is now necessary to choose an equation of state. We take it in the form
which have been studied previously in general relativity \cite{Brec76}%
\begin{equation}
P=b\left[ \varepsilon -\varepsilon _{0}(\phi )\right] ,  \label{m07}
\end{equation}%
where $b$ is a constant and $\varepsilon _{0}(\phi )$ is the energy density
above which the equation of state becomes \textquotedblleft
stiff.\textquotedblright

Very often in literature the so called \textquotedblleft
causality\textquotedblright\ condition is imposed to the equation of state: $%
dP/d\varepsilon \leq 1$, that is $b\leq 1$ in Eq. (\ref{m07}). The condition
requires that the speed of sound in the stellar matter $c_{s}$ can not
exceed the speed of light $c$. However, though the last statement is true it
does not mean that we must impose the restriction $dP/d\varepsilon \leq 1$
to the possible equation of state. The point is that equations of
hydrodynamics, which result in $c_{s}=c\sqrt{dP/d\varepsilon }$, are derived
under the assumption of instantaneous interactions between particles (local
approximation). However, real interactions propagate with the speed of
light. As a result, if $dP/d\varepsilon >1$ the equations become
substantially nonlocal and static compressibility $dP/d\varepsilon $ no
longer describes the speed of sound. In this regime the speed of interaction
propagation imposes the restriction on the speed of compression waves in
matter. As a consequence, the speed of sound never exceeds the speed of
light no matter what is the value of the static compressibility $%
dP/d\varepsilon $.

One should mention that some authors also express caution about the
\textquotedblleft causality\textquotedblright\ condition constraint on the
equation of state. E.g., Kalogera and Baym say \cite{Kalo96}:
\textquotedblleft The connection between the zero frequency sound velocity
being greater than the speed of light and violation of causality, while
physically plausible, is a tricky question ... We are not aware of a general
proof yet that the ground state of matter must obey $dP/d\varepsilon \leq 1.$%
\textquotedblright

In Eq. (\ref{m07}) $\varepsilon _{0}(\phi )$ depends on the gravitational
potential $\phi $. To find this dependence we consider spatially uniform
fluid placed in a spatially uniform gravitational field $\phi $ that depends
on time. Such consideration is similar to the cosmological model of the
Universe. Since $\varepsilon =\varepsilon (\phi )$ and $P=P(\phi )$ the
change of $\phi $ causes change of $\varepsilon $ and $P$. Then the
energy-momentum relation (\ref{m02}) yields the equation

\begin{equation*}
\dot{\varepsilon}=3(P+\varepsilon )\dot{\phi}.
\end{equation*}%
Assuming that $\varepsilon (\phi )$, $\varepsilon _{0}(\phi )$ and $P(\phi )$
are proportional to the same function of $\phi $ we obtain

\begin{equation*}
\varepsilon (\phi ),P(\phi )\propto e^{\alpha \phi },
\end{equation*}%
where%
\begin{equation*}
\alpha =3\left( 1+\frac{P}{\varepsilon }\right)
\end{equation*}%
is independent of $\phi $. For the equation of state (\ref{m07}) we find%
\begin{equation}
\varepsilon _{0}(\phi )=\varepsilon _{0}\exp \left[ 3\left( 1+b-\frac{%
b^{2}\varepsilon _{0}(\phi )}{P(\phi )+b\varepsilon _{0}(\phi )}\right) \phi %
\right]  \label{m08}
\end{equation}%
which is an algebraic equation for $\varepsilon _{0}(\phi )$. In Eq. (\ref%
{m08}) $\varepsilon _{0}$ is the value of $\varepsilon _{0}(\phi )$ at $\phi
=0$. It is convenient to introduce dimensionless coordinate, pressure and
energy density as 
\begin{equation*}
r\rightarrow r_{0}r,\quad P\rightarrow \varepsilon _{0}P,\quad \varepsilon
\rightarrow \varepsilon _{0}\varepsilon
\end{equation*}%
where%
\begin{equation*}
r_{0}=\frac{c^{2}}{\sqrt{4\pi G\varepsilon _{0}}}.
\end{equation*}%
For the dimensionless functions the field equation and equation of
hydrostatic equilibrium read%
\begin{equation*}
\Delta \phi =\left( \varepsilon +3P\right) e^{-2\phi },
\end{equation*}%
\begin{equation*}
\frac{\partial P}{\partial r}=-(P+\varepsilon )\frac{\partial \phi }{%
\partial r},
\end{equation*}%
while the boundary conditions are%
\begin{equation*}
\phi ^{\prime }(0)=0,\quad P(R)=0,\quad R\phi ^{\prime }(R)=-\phi (R).
\end{equation*}%
Dimensionless equation of state (\ref{m07}) is 
\begin{equation}
P=b(\varepsilon -\xi ),  \label{des}
\end{equation}%
where $\xi $ is given by a solution of the dimensionless algebraic equation%
\begin{equation*}
\xi =\exp \left[ 3\left( 1+b-\frac{b^{2}\xi }{P+b\xi }\right) \phi \right] .
\end{equation*}%
Stellar mass $M$ is obtained from the formula%
\begin{equation*}
\frac{M}{M_{0}}=-R\phi (R),
\end{equation*}%
where%
\begin{equation*}
M_{0}=\frac{c^{2}r_{0}}{G}=\frac{c^{4}}{\sqrt{4\pi G^{3}\varepsilon _{0}}}.
\end{equation*}%
For $\varepsilon _{0}/c^{2}=10^{14}$ g/cm$^{3}$ we have%
\begin{equation*}
r_{0}=32.8\text{ km},\quad M_{0}=22.3M_{\odot }.
\end{equation*}

\begin{figure}[t]
\centering
\includegraphics[width=0.6\textwidth]{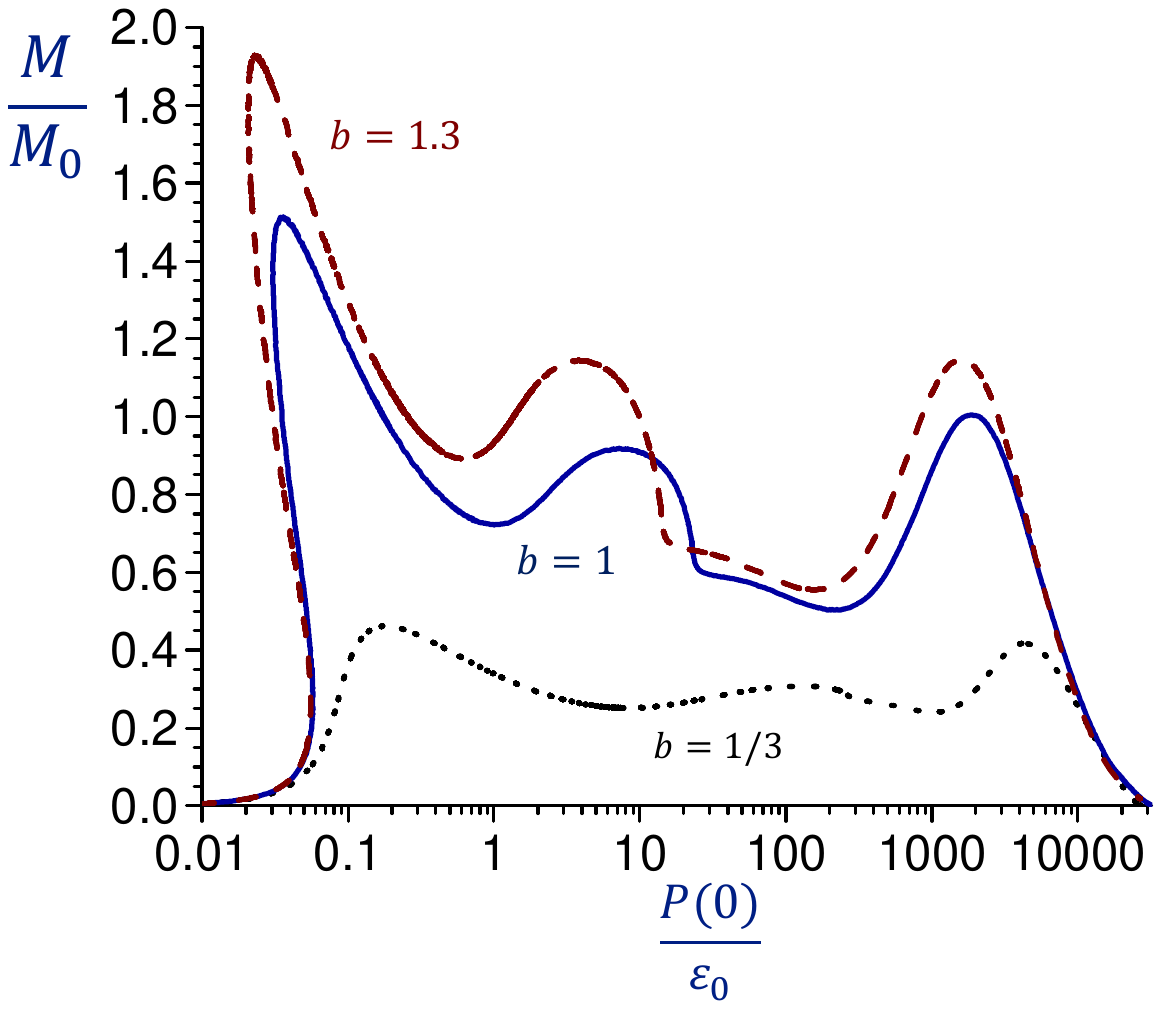}
\vspace{-0.5cm} \vspace{-0.3cm}
\caption{Mass of a neutron star as a function of central pressure $P(0)$ in
vector theory of gravity for equation of state (\protect\ref{des}) with $%
b=1/3$, $1$ and $1.3$. Unit of mass is $M_{0}=c^{4}/\protect\sqrt{4\protect%
\pi G^{3}\protect\varepsilon _{0}}$ which for $\protect\varepsilon %
_{0}/c^{2}=10^{14}$ g/cm$^{3}$ yields $M_{0}=22.3M_{\odot }$.}
\label{Mass}
\end{figure}

\begin{figure}[t]
\centering
\vspace{-0.8cm}
\includegraphics[width=0.6\textwidth]{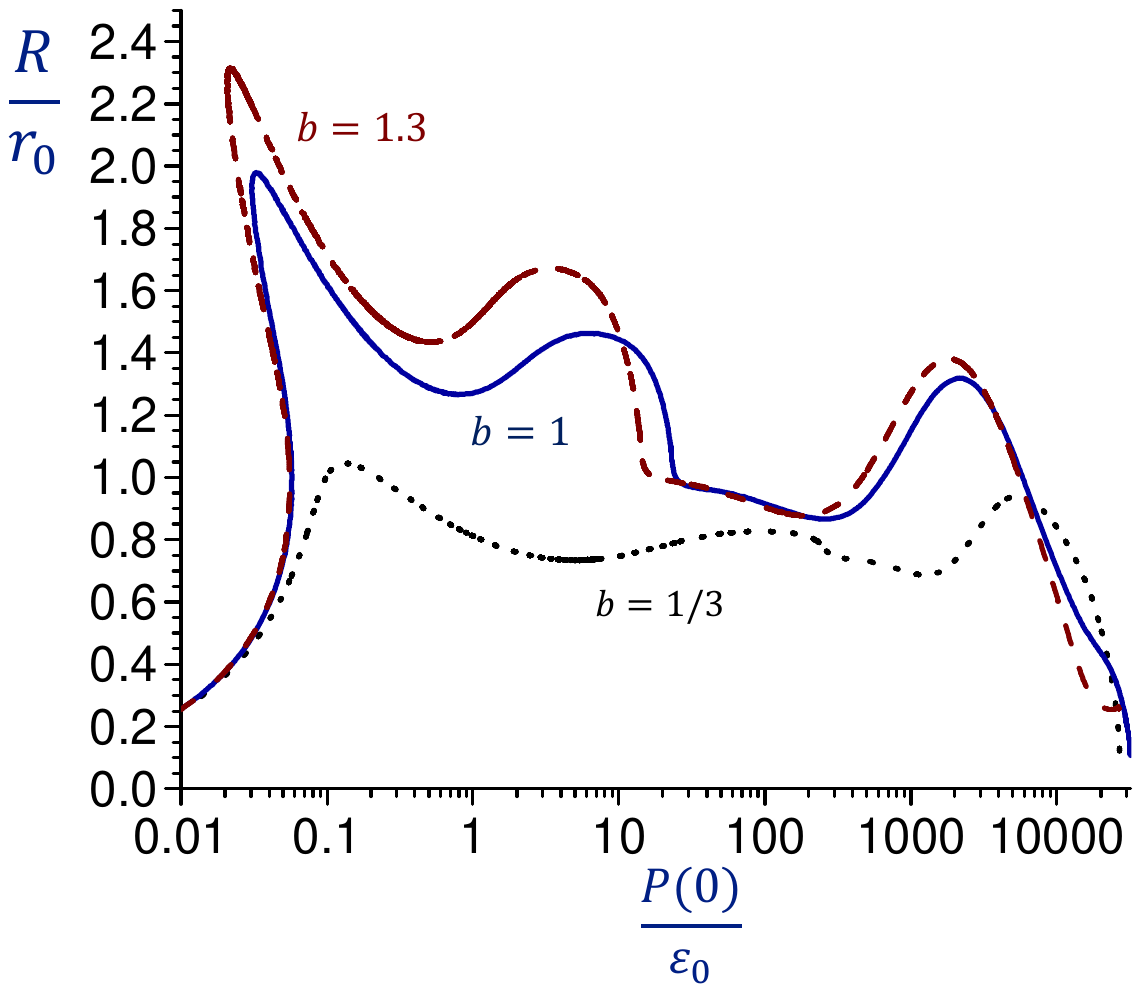}
\vspace{-1cm}
\caption{Radius of a neutron star as a function of central pressure $P(0)$
in vector theory of gravity for equation of state (\protect\ref{des}) with $%
b=1/3$, $1$ and $1.3$. Unit of radius is $r_{0}=c^{2}/\protect\sqrt{4\protect%
\pi G\protect\varepsilon _{0}}$ which for $\protect\varepsilon %
_{0}/c^{2}=10^{14}$ g/cm$^{3}$ yields $r_{0}=32.8$ km.}
\label{Radius}
\end{figure}

\begin{figure}[t]
\centering
\includegraphics[width=0.56\textwidth]{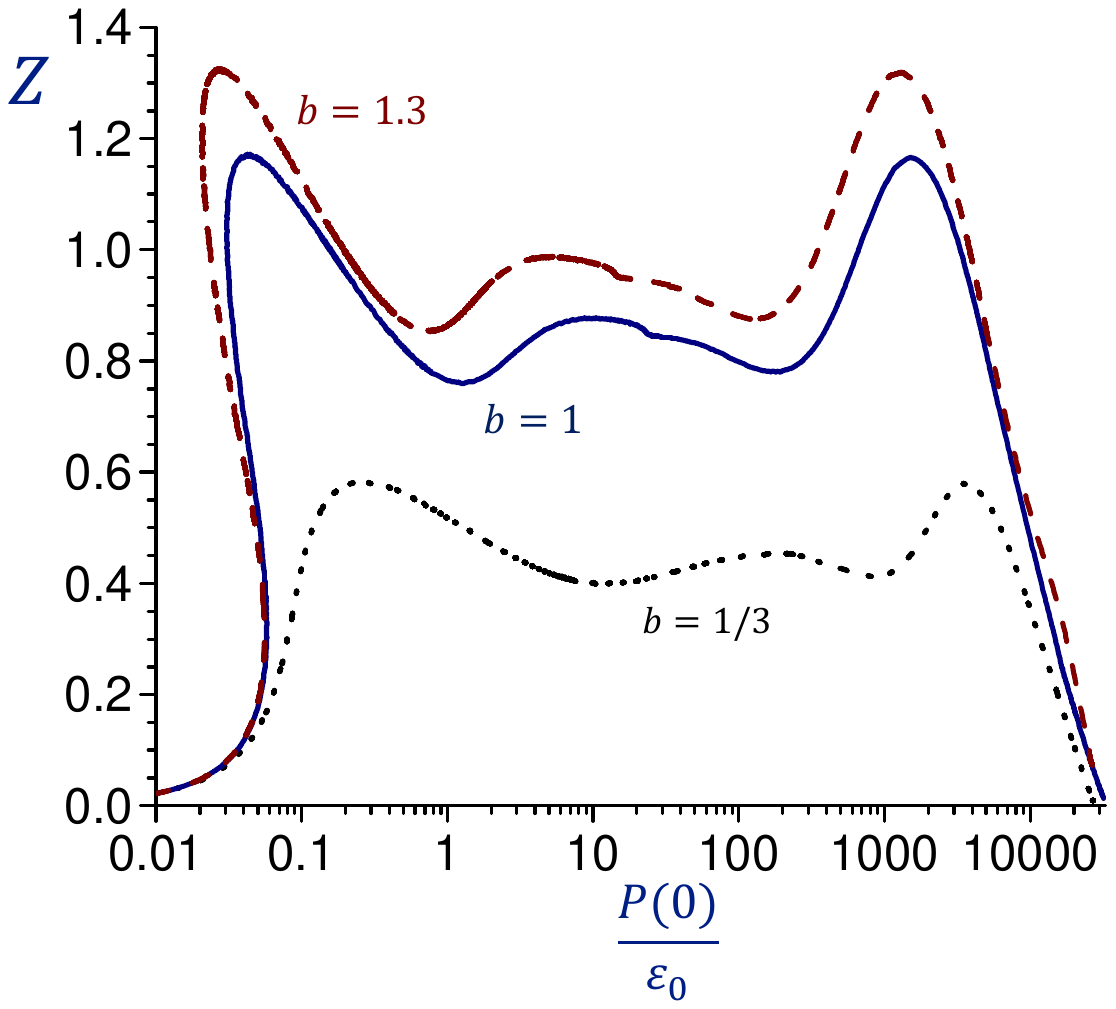}
\vspace{-1cm}
\caption{Gravitational redshift $z$ at the surface of a neutron star as a
function of central pressure $P(0)$ in vector theory of gravity for equation
of state (\protect\ref{des}) with $b=1/3$, $1$ and $1.3$.}
\label{Redshift}
\end{figure}

In Figs. \ref{Mass}, \ref{Radius} and \ref{Redshift} we plot mass of a
neutron star, its radius and surface gravitational redshift as a function of
the central pressure $P(0)\equiv P(r=0)$ in the vector theory of gravity for
the equation of state (\ref{des}) for various values of the stiffness
parameter $b=1/3$, $1$ and $1.3$. Figure \ref{Mass} shows that the stiffer
the equation of state, the higher the star mass limit. For example, for $b=1$
the maximum gravitational mass of a stable neutron star is $M_{\text{max}%
}=1.52M_{0}=34M_{\odot }$ for $\varepsilon _{0}/c^{2}=10^{14}$ g/cm$^{3}$.
This mass is obtained for $dP/d\varepsilon \leq 1$. Radius of star with
maximum mass is $R=1.98r_{0}=65$ km. The effects of stellar rotation may
increase the maximum mass by about $25\%$ \cite{Frie87}.

Star is stable provided $\partial M/\partial P(0)>0$ \cite{Glen97}. For $b=1$
there are three intervals of stability for nonrotating stars, namely $%
M<0.31M_{0}$, $0.5M_{0}<M<1.00M_{0}$ and $1.37M_{0}<M<1.52M_{0}$ which for $%
\varepsilon _{0}/c^{2}=10^{14}$ g/cm$^{3}$ give $M<7M_{\odot }$, $11M_{\odot
}<M<22M_{\odot }$ and $30M_{\odot }<M<34M_{\odot }$ respectively. It is
interesting to note that masses of compact objects in the merging binary
systems obtained by the LIGO team based on the gravitational wave detection
reported so far ($29_{-4}^{+4}M_{\odot }$ and $36_{-4}^{+5}M_{\odot }$ \cite%
{Abbo16}; $14.2_{-3.7}^{+8.3}M_{\odot }$ and $7.5_{-2.3}^{+2.3}M_{\odot }$ 
\cite{Abbo16a}; $31.2_{-6.0}^{+8.4}M_{\odot }$ and $19.4_{-5.9}^{+5.3}M_{%
\odot }$ \cite{Abbo17}), within the error bar fit in the mass intervals for
which neutron stars are stable.

However, stability intervals are sensitive to the choice of the equation of
state which is very uncertain at high matter density. Our choice of the
equation of state (\ref{des}) is only an example. In addition, inclusion of
stellar rotation can considerably widen stability regions. Nevertheless,
vector gravity predicts existence of gaps in the neutron star mass
distribution, although position of the gaps depends on the uncertain
equation of state. One should note that stellar-mass compact objects with
masses up to about $16M_{\odot }$ have been discovered in X-ray binaries 
\cite{Casa14}. Thus, there is a wide gap between $19M_{\odot }$ and $%
29M_{\odot }$ measured by LIGO. Future observations will fill this interval
with more data and test the prediction of vector gravity about existence of
gaps. It is interesting to note that a $3-5M_{\odot }$ gap has been found in
the low-mass part of the measured compact object mass distribution in the
Galaxy \cite{Ozel10,Farr11}. If position of the mass gaps is obtained from
observations this information can be used to determine the equation of state
of matter in the dense stellar cores by matching stability regions with the
astronomical data.

Supermassive compact objects with masses of $>10^{5}M_{\odot }$ have been
discovered in galactic centers \cite{Gree07}. In Sec. \ref{dark} we argue
that such supermassive compact objects are made of dark matter and,
according to vector gravity, can have masses in a range $\sim
10^{5}-10^{10}M_{\odot }$. It is interesting to note that compact objects
with masses in the interval $\sim 10^{2}-10^{5}M_{\odot }$ have not yet been
detected beyond doubt.

We found that, as in general relativity, the neutron star mass limit varies
roughly as $1/\sqrt{\varepsilon _{0}}$, where $\varepsilon _{0}$ is the
energy density above which the equation of state becomes \textquotedblleft
stiff.\textquotedblright\ Unlike general relativity, the stiffer the
equation of state, the higher the mass limit. Our numerical simulations show
that $M_{\text{max}}$ increases with increasing $b$. E.g., for $b=3$ we find 
$M_{\text{max}}=13.4M_{0}\approx 300M_{\odot }$. This result is somewhat
similar to those in the bimetric theory of gravitation which for $%
dP/d\varepsilon >1$ yields that the upper mass limit $M_{\text{max}}$,
unlike the general-relativistic case \cite{Brec76}, can be arbitrary large 
\cite{Capo77}.

In vector gravity, as in general relativity, if the stellar mass exceeds a
certain value $M_{\text{max}}$, there does not exist any static solution of
the field equations, and therefore the star must undergo collapse. However,
there is a great difference between the predictions of the two theories as
to what will happen during the process of collapse. In the case of
gravitational collapse in the framework of general relativity, once the
surface of the star has entered the Schwarzschild sphere, one has a black
hole. It is believed that the matter of the star and its radiation are then
permanently trapped in the black hole \cite{Misn73}.

In vector gravity, as well as in other alternative theories of gravity with
no event horizons, since there does not appear to be anything corresponding
to the Schwarzschild sphere, the inner part of an unstable star will first
contract and then expand \cite{Rose75}. The contracting outer part could
collide with the expanding inner part. The collision could result in the
ejection of the outer envelope and hence in a loss of mass which makes the
star stable again \cite{Rose75}. Another possible scenario is that a neutron
star with a limiting mass swallowing baryons will radiate their mass
equivalents for stability yielding substantial additional radiation of
internal origin. Such radiation could appear as copious photon or neutrino
emissions or gravitational radiation \cite{Robe99}.

The end point of a gravitational collapse in vector gravity is not a point
singularity but rather a stable object with a reduced mass. Merger of two
neutron starts with masses close to the upper limit leads to formation of a
stable star with a higher baryon number but not the mass. The net mass of
the two merging starts is reduced due to greater gravitational binding
energy of the merger. The excess energy is radiated away, e.g., by neutrino
emission or by other mechanisms.

In vector gravity there is no gravitational collapse of a star into a point
singularity because such an object would have zero mass. Indeed, let us
consider two static point masses separated by a distance $R$ and assume that 
$m_{1}$ and $m_{2}$ are the values of masses at infinite separation. In
vector gravity, static gravitational field is described by Eq. (\ref{z10})
which for the case of two masses reduces to%
\begin{equation*}
\Delta \phi =\frac{4\pi G}{c^{2}}\left[ m_{1}\delta (\mathbf{r}-\mathbf{r}%
_{1})+m_{2}\delta (\mathbf{r}-\mathbf{r}_{2})\right] e^{\phi }
\end{equation*}%
and has the following solution%
\begin{equation}
\phi (\mathbf{r})=-\frac{G}{c^{2}}\left[ \frac{m_{1}e^{\phi _{2}(R)}}{|%
\mathbf{r}-\mathbf{r}_{1}|}+\frac{m_{2}e^{\phi _{1}(R)}}{|\mathbf{r}-\mathbf{%
r}_{2}|}\right] ,  \label{z10a}
\end{equation}%
where 
\begin{equation*}
\phi _{1,2}(R)=-\frac{Gm_{1,2}}{c^{2}R}.
\end{equation*}%
The net mass of the system $m$ is determined by the asymptotic of Eq. (\ref%
{z10a}) at large $r$ which gives 
\begin{equation}
m=m_{1}\exp \left( -\frac{Gm_{2}}{c^{2}R}\right) +m_{2}\exp \left( -\frac{%
Gm_{1}}{c^{2}R}\right) .
\end{equation}%
For $R\rightarrow 0$ the net mass vanishes. If we gradually move masses
closer to each other the net mass of the system decreases to zero due to
negative contribution of the gravitational potential energy. A star
collapsed to a point would also have zero kinetic energy (in the classical
consideration) and, hence, the total energy of such collapsed object would
be equal to zero. As a consequence, spatial point singularities do not exist
in vector gravity. However, for simplicity, in many problems masses can be
approximated as point masses similar to the concept of point charges in
electrodynamics.

\section{Tests of the theory of gravity}

\label{test}

In this section we compare predictions of the vector theory of gravity with
observations. Refs. \cite{Will93,Will06,Will14} provide a detail procedure
of how to compare metric theories of gravity with experimental tests and
show viability of the theory. Here we follow this procedure step by step and
show that vector gravity passes all available tests.

\textit{\textbf{Post-Newtonian limit}.} In vector gravity there is a
preferred cosmological reference frame in which background vector
gravitational field has only time component. This is a reference frame in
which the large-scale distribution of matter is isotropic (presumably the
rest frame of the cosmic background radiation). As we show in Sec. \ref{PPN}%
, in such cosmological frame in the post-Newtonian limit, equations of the
vector gravity as well as the boundary conditions are equivalent to those in
general relativity. Thus, in any frame moving with a non relativistic speed
with respect to the cosmological reference frame (such as our solar system)
both theories remain equivalent provided we are not going beyond the
post-Newtonian approximation. As a consequence, vector gravity yields the
same values for the ten PPN parameters as general relativity \cite%
{Will93,Will06,Will14}. In Appendix \ref{PPNF} we investigate the
post-Newtonian limit of vector gravity in the framework of the parametrized
post-Newtonian formalism and explicitly calculate the ten post-Newtonian
parameters. As expected, they are equal to those in general relativity.

The post-Newtonian limit is sufficient to describe the gravitational physics
of the solar system and the experimental tests one can perform there \cite%
{Will93,Will06,Will14}. To some degree, it can also describe the gravity of
binary-pulsar systems. Since vector gravity and general relativity are
equivalent in the post-Newtonian limit, they both pass every high-precision
test in the solar system, where gravitational fields are relatively weak. In
those familiar precincts, they correctly predict redshifting, light
deflection by a massive body, Shapiro time delay, precession of planetary
orbits, the strict equivalence of gravitational and inertial mass, lack of
the preferred-frame and preferred-location effects, etc.

Vector gravity, as well as general relativity, is built on the Einstein
equivalence principle which states that matter is coupled in a universal
manner to a single tensorial field, the metric. Extension of the Einstein
equivalence principle to gravitational experiments is known as the strong
equivalence principle which states that local gravitational physics is
independent of the position and velocity of the local reference frame.
Alternative theories of gravity involving additional fields or fixed
background geometry tend to violate the strong equivalence principle \cite%
{Will93,Will06,Will14}. However, since vector gravity is equivalent to
general relativity in the post-Newtonian limit the vector gravity obeys the
strong equivalence principle in this limit.

\textit{\textbf{Gravitational radiation by binary pulsars}.} There are
several tests of gravity beyond the solar system. Gravitational radiation by
binary pulsars provides a tool for testing relativistic gravity. In general
relativity, the gravitational waves emitted by a slowly-moving system are
dominantly quadrupole and there is no monopole or dipole radiation. This is
because the field equations of general relativity insist that monopole and
dipole moments are the total mass and the total momentum of the system which
are constants if the system is isolated \cite{Will93,Will06,Will14}. There
is no reason to expect that a generic alternative theory will predict the
suppression of monopole and dipole emission. However, as we show in Sec. \ref%
{binary}, for vector gravity there is no monopole and dipole radiation due
to the same reason as for general relativity. Moreover, we show that in our
theory the gravitational energy loss by binary stars is described by the
same formula as in general relativity. Our analysis of Sec. \ref{binary} is
also valid for neutron stars which are relativistic objects. Energy loss by
binary pulsars due to emission of gravitational radiation was measured for
several systems and served as a quantitative test of Einstein equations for
weak time-dependent field. The present theory also passes this test.

One should note that studies carried out in the wake of the discovery of the
Hulse-Taylor binary pulsar in 1974 \cite{Huls75} revealed that a number of
otherwise respectable alternative theories of gravity predicted the emission
of the negative energy \cite{Will93}. Once the orbital period of the binary
pulsar was shown to decrease in response to the emission of gravitational
waves (that is total energy of the binary system decreases with time), these
theories were ruled out. Since in the present theory the graviton energy is
negative in the classical limit, the vector theory of gravity would also
predict emission of the negative energy by binary pulsars if we would not
postulate that graviton is composed of fermion-antifermion pairs by analogy
with the composite theory of photon.

One should mention that theories with negative graviton energy are somewhat
appealing because they provide a natural mechanism of matter generation at
the Big Bang without involving additional cosmological fields. In the
present theory, as soon as fermion states are filled shortly after the Big
Bang the matter generation stops and Universe subsequently evolves according
to the usual hot Universe scenario. For filled vacuum the graviton energy
becomes positive (see Sec. \ref{quantization}) and emission of a graviton
corresponds to creation of fermion-antifermion hole pairs out of the filled
fermion states. Since fermion states are filled in the present epoch the
binary pulsars orbiting each other emit positive energy gravitons which, as
we show in Sec. \ref{binary}, yields exactly the same energy loss by the
binary systems as predicted by general relativity.

\textit{\textbf{Structure and motion of compact objects}.} Precise orbital
data obtained for binary pulsars permits the direct measurement of the mass
of a neutron star and the study of relativistic orbital effects (such as
periastron shifts) in systems containing compact objects possessing strong
gravitational field. In alternative theories of gravity, strong
gravitational field involved in the neutron star can make significant
differences in relativistic orbital effects. When dealing with a system such
as the binary pulsar one must employ a method for deriving equations of
motion for compact objects that involves solving the full relativistic
equations for the regions inside and near each body, solving the
post-Newtonian equations in the interbody region and matching these
solutions \cite{Will93}.

Most alternative theories of gravity possess additional gravitational fields
(dynamical or fixed), whose values in the matching region can influence the
structure of each body, and, as a consequence, affect its motion. Namely,
mass of the compact object may depend on the boundary values of the
auxiliary fields leading to modification of the body's motion. Thus, the
location and velocity of the body relative to the external gravitational
environment can affect its structure and motion. This is known as the
preferred location and preferred frame effects. Orbital data obtained for
binary pulsars show lack of such effects for compact objects at least in the 
$V^{2}/c^{2}$ order, where $V$ is the neutron star velocity \cite{Will93}.
Namely, observations show that binary pulsars move the same way as if they
were weak-field post-Newtonian bodies.

Present vector theory of gravity contains an auxiliary nondynamical field,
the flat Euclidean metric $\delta _{ik}$, which yields a possibility of the
preferred frame and preferred location effects. In Appendix \ref{cossup} we
investigate this question and demonstrate lack of such effects for binary
pulsars in the $V^{2}/c^{2}$ order (this might also be valid in higher
orders). Namely, we show that equivalent metric produced by a moving neutron
star is independent of the external gravitational background and of the star
velocity relative to the background. The metric is characterized only by the
object's Kepler-measured mass $M$, and is independent of its internal
structure. In the region far from the neutron star in the post-Newtonian
limit the metric in our theory is given by the same formula as in general
relativity. Thus, the matching procedure described above must yield the same
result, whether the body is a neutron star of mass $M$ or a post-Newtonian
body of mass $M$. Therefore, in vector gravity motion of compact objects is
described by the same equations as motion of weak-field stars and coincides
with predictions of general relativity in the $V^{2}/c^{2}$ order. Hence,
vector gravity passes the binary pulsar test.

\textit{\textbf{Cosmological test}.} Cosmology provides another important
test of gravitational theories. As we show in Sec. \ref{cosmology}, for
cosmology with a general equation of state of matter the present theory
gives the same evolution of the Universe as general relativity with
cosmological constant and zero spatial curvature. Thus, vector theory of
gravity passes the cosmological test and, in particular, provides the same
explanation for the cosmic microwave background radiation and the helium
abundance as general relativity. Moreover, vector gravity yields, with no
free parameters, the value of the cosmological constant $\Omega _{\Lambda
}=2/3\approx 0.67$ which agrees with the recent Planck result $\Omega
_{\Lambda }=0.686\pm 0.02$ \cite{Planck14}. Thus, vector gravity also passes
the \textquotedblleft dark energy\textquotedblright\ test.

\begin{figure}[h]
\centering
\includegraphics[width=0.47\textwidth]{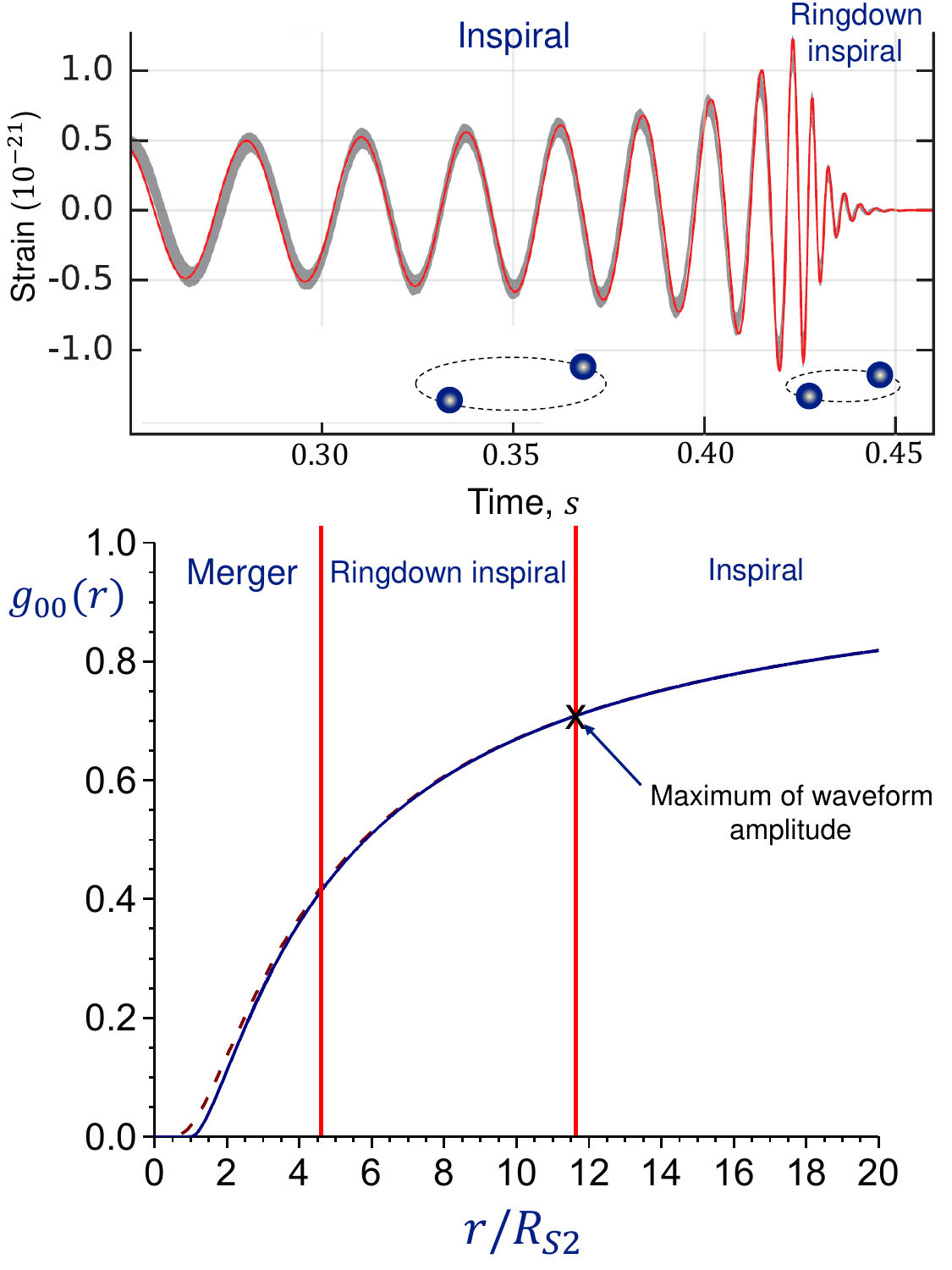}
\caption{\textit{Top:} Estimated gravitational-wave strain amplitude from
GW150914 event as a function of time obtained by numerical relativity
(Adopted from \protect\cite{Abbo16}). \textit{Bottom:} Metric component $%
g_{00}$ as a function of interstellar separation $r$ in general relativity
(solid line) and vector gravity (dashed line). The unit of distance is
Schwarzschild radius in isotropic coordinates $R_{S2}$. The cross denotes
the location of the maximum of the radiation waveform amplitude. Vertical
lines separate regions of orbital inspiral, ringdown inspiral and merger of
two neutron stars of $60$ km radii and masses $35M_{\odot }$.}
\label{vgmerge}
\end{figure}

\textit{\textbf{Direct detection of gravitational waves by laser
interferometers}.} Possibility of gravitational wave detection by laser
interferometers was first suggested in 1962 \cite{Gert63}, shortly after
invention of a laser. Recently LIGO team reported first observation of a
transient gravitational-wave signal from orbital inspiral and merger of two
compact objects loosing their energy due to gravitational-wave emission \cite%
{Abbo16}. Over $0.2$ s, the signal increased in frequency and amplitude in
about $8$ cycles from $35$ to $150$ Hz, where the amplitude reached a
maximum. Then waveform decayed undergoing damped oscillations (see top part
of Fig. \ref{vgmerge}). Interpretation of the signal in Ref. \cite{Abbo16}
is based on general relativity which yields that the merging objects are two
black holes with masses $29_{-4}^{+4}M_{\odot }$ and $36_{-4}^{+5}M_{\odot }$%
. Numerical relativity gives that at the maximum of the waveform amplitude
the objects are separated by a distance $r_{\max }\approx 350$ km \cite%
{Abbo16}.

Here we discuss interpretation of the LIGO signal GW150914 in the framework
of vector gravity and calculate the gravitational radiation waveform. For
simplicity we suppose that the two orbiting bodies have equal masses $m$ and
move with velocity $V$ along circular trajectories of diameter $r$ around
their common centre of mass. In Schwarzschild coordinates, the line element
for the Schwarzschild metric of a point mass $m$ has the form

\begin{equation}
ds^{2}=\left( 1-\frac{2Gm}{c^{2}r}\right) c^{2}dt^{2}-\left( 1-\frac{2Gm}{%
c^{2}r}\right) ^{-1}dr^{2}-r^{2}d\Omega ^{2}.  \label{sm1}
\end{equation}%
In these coordinates Eq. (\ref{sm1}) yields that for $m=35M_{\odot }$ the
Schwarzschild radius is 
\begin{equation*}
R_{S1}=\frac{2Gm}{c^{2}}=103\text{ km}
\end{equation*}%
which is comparable with the $350$ km separation between objects at the peak
of the waveform amplitude. Based on this observation, the decay of the
radiation waveform in the LIGO signal is commonly interpreted as damped
oscillations of two merging black holes relaxing to a final stationary Kerr
configuration \cite{Abbo16}.

This interpretation, however, should be taken with caution due to general
covariance of Einstein's theory. Recall that suitable nonlinear change of
coordinates can make the value of the gravitational radius much smaller then
spacing between two objects. For comparison with vector gravity that yields
a spatially isotropic line element%
\begin{equation}
ds^{2}=\exp \left( -\frac{2Gm}{c^{2}r}\right) c^{2}dt^{2}-\exp \left( \frac{%
2Gm}{c^{2}r}\right) (dx^{2}+dy^{2}+dz^{2})  \label{vgm1}
\end{equation}%
we must write Schwarzschild metric (\ref{sm1}) in isotropic coordinates by
making nonlinear coordinate transformation $r\rightarrow (1+Gm/2c^{2}r)^{2}r$%
. This transformation reduces the value of the Schwarzschild radius $4$
times but changes $r_{\max }$ only a little from $350$ km to $\approx 300$
km. In isotropic coordinates the Schwarzschild line element reads%
\begin{equation}
ds^{2}=\frac{\left( 1-\frac{Gm}{2c^{2}r}\right) ^{2}}{\left( 1+\frac{Gm}{%
2c^{2}r}\right) ^{2}}c^{2}dt^{2}-\left( 1+\frac{Gm}{2c^{2}r}\right)
^{4}(dx^{2}+dy^{2}+dz^{2}).  \label{sm2}
\end{equation}%
Eq. (\ref{sm2}) gives that for $m=35M_{\odot }$ the radius of Schwarzschild
sphere in isotropic coordinates is%
\begin{equation}
R_{S2}=\frac{Gm}{2c^{2}}=25.7\text{ km}  \label{rs2}
\end{equation}%
which is much smaller then separation between objects at the onset of the
waveform ringdown stage $r_{\text{max}}\approx 300$ km. In the
post-Newtonian formalism the metric (\ref{sm2}) is expanded in the small
parameter 
\begin{equation}
\epsilon =\frac{V^{2}}{c^{2}}=\frac{Gm}{2c^{2}r}=\frac{R_{S2}}{r},
\label{ep}
\end{equation}%
where $V$ is the object velocity in the binary system which in the Newtonian
gravity is given by%
\begin{equation}
V^{2}=\frac{Gm}{2r}.  \label{Vng}
\end{equation}

For $r=r_{\max }$ Eq. (\ref{ep}) yields 
\begin{equation}
\epsilon =\frac{R_{S2}}{r_{\max }}=0.08\ll 1  \label{ep1}
\end{equation}%
and, therefore, decay of the radiation waveform in the LIGO signal actually
begins at a relatively weak gravity. One should note that estimate (\ref{ep1}%
) is independent of the value of mass $m$ which factors out from equations.

As a consequence, interpretation of the decaying part of the radiation
waveform in isotropic coordinates is qualitatively different. Namely, the
decay occurs at the stage of orbital inspiral when two objects are yet
considerably far from their merger (see top part of Fig. \ref{vgmerge}). The
LIGO signal becomes smaller than noise before the two objects actually start
to merge.

To compare vector gravity with general relativity we plot $g_{00}$ component
of the metric, given by Eqs. (\ref{vgm1}) and (\ref{sm2}), as a function of
separation between stars $r$ for both theories. The result is shown in Fig. %
\ref{vgmerge} (bottom). General relativity yields solid line, while $g_{00}$
for vector gravity is shown as dashed line. Vertical lines in the plot
separate three regions of the orbital inspiral, ringdown inspiral and merger
of two neutron stars of $60$ km radii and masses $35M_{\odot }$. The figure
demonstrates that $g_{00}$ in both theories is practically indistinguishable
upto the point of merger.

Next we calculate the radiation waveform in vector gravity and show that it
is compatible with the LIGO data. We assume that two compact stars with
masses $m$ move in the $x-y$ plane along circular orbits of diameter $r$
with tangential velocity $V=r\dot{\theta}/2$, where $\theta $ is the
azimuthal angle in the $x-y$ plane. Since the effects of gravity are
expected to be relatively weak even during the ringdown stage the loss of
the system's angular momentum $L=mrV$ can be accurately described by the
quadrupole formula \cite{Land95}%
\begin{equation*}
\frac{dL_{\alpha }}{dt}=-\frac{2G}{45c^{5}}e_{\alpha \beta \gamma }\dddot{D}%
_{\beta \delta }\ddot{D}_{\delta \gamma },
\end{equation*}%
where components of the quadrupole moment tensor are%
\begin{equation*}
D_{xx}=\frac{m}{2}r^{2}(3\cos ^{2}\theta -1),\quad D_{yy}=\frac{m}{2}%
r^{2}(3\sin ^{2}\theta -1),
\end{equation*}%
\begin{equation*}
D_{xy}=D_{yx}=\frac{3m}{4}r^{2}\sin (2\theta ),\quad D_{zz}=-\frac{m}{2}%
r^{2}.
\end{equation*}%
Keeping the leading order term we obtain%
\begin{equation}
m\frac{d}{dt}\left( rV\right) =-\frac{256Gm^{2}}{5c^{5}}\frac{V^{5}}{r}.
\label{aml}
\end{equation}%
For Newtonian gravity $V\propto 1/\sqrt{r}$ and the left hand side of Eq. (%
\ref{aml}) can be written as $\frac{d}{dt}\left( rV\right) =V\dot{r}/2$
which yields the following equation of the orbit decay 
\begin{equation}
\dot{r}=-\frac{512Gm}{5c^{5}}\frac{V^{4}}{r}.  \label{aml1}
\end{equation}

In the wave zone perturbation of the metric due to gravitational wave
propagating along the $x-$axis is given by \cite{Land95}%
\begin{equation*}
h_{0y}\propto \ddot{D}_{yx}\propto V^{2}\sin (2\theta ),\quad h_{0z}=0\text{.%
}
\end{equation*}%
The signal of the LIGO-like interferometer with perpendicular arms laying in
the $xy$ plane is proportional to 
\begin{equation}
h=h_{0y}\propto V^{2}\sin (2\theta ),  \label{aml2}
\end{equation}%
where for the orbital motion 
\begin{equation}
\dot{\theta}=\frac{2V}{r}.  \label{aml3}
\end{equation}

In Eqs. (\ref{aml1})-(\ref{aml3}) we need to specify how orbital velocity $V$
depends on the interstellar separation $r$. For Newtonian gravity the
relation is given by Eq. (\ref{Vng}). To go beyond Newtonian gravity we
replace $V(r)$ in Eqs. (\ref{aml1})-(\ref{aml3}) by the expression that
follows from the exact equation of motion of mass $m$ in the metric 
\begin{equation}
ds^{2}=g_{00}(r)c^{2}dt^{2}-F(r)(dx^{2}+dy^{2}+dz^{2})  \label{gs1}
\end{equation}%
produced by approximately static companion star at the distance $r$. Metric (%
\ref{gs1}) is given by Eq. (\ref{sm2}) in the case of general relativity and
by Eq. (\ref{vgm1}) for vector gravity. Equation of motion of a particle in
metric $g_{ik}$ \cite{Land95}%
\begin{equation*}
\frac{d^{2}x^{b}}{ds^{2}}=\frac{1}{2}g^{bl}\left[ \frac{\partial g_{ik}}{%
\partial x^{l}}-\frac{\partial g_{lk}}{\partial x^{i}}-\frac{\partial g_{il}%
}{\partial x^{k}}\right] \frac{dx^{i}}{ds}\frac{dx^{k}}{ds}
\end{equation*}%
yields the following relation 
\begin{equation*}
\frac{V^{2}}{c^{2}}=\frac{\partial g_{00}}{\partial r}\frac{1}{\frac{4F}{r}+%
\frac{\partial F}{\partial r}}.
\end{equation*}%
Introducing dimensionless velocity, distance and time%
\begin{equation*}
V\rightarrow Vc,\quad r\rightarrow R_{S2}r,\quad t\rightarrow \frac{R_{S2}}{c%
}t
\end{equation*}%
we find for the case of vector gravity%
\begin{equation}
V^{2}=\frac{e^{-8/r}}{r-1}  \label{vvg}
\end{equation}%
and 
\begin{equation}
V^{2}=\frac{r^{4}(r-1)}{(r+1)^{6}}  \label{vgr}
\end{equation}%
for general relativity. Substituting this into Eqs. (\ref{aml1})-(\ref{aml3}%
) we obtain equations for the orbit decay $r(t)$ and generated radiation
waveform $h(t)$. In the dimensionless coordinates the equations read%
\begin{equation}
\dot{r}=-\frac{1024}{5}\frac{e^{-16/r}}{r(r-1)^{2}},  \label{rt1}
\end{equation}%
\begin{equation}
\dot{\theta}=\frac{2e^{-4/r}}{r\sqrt{r-1}},
\end{equation}%
\begin{equation}
h=A\frac{e^{-8/r}}{r-1}\sin (2\theta +\varphi _{0})  \label{rt2}
\end{equation}%
for vector gravity, and%
\begin{equation}
\dot{r}=-\frac{1024}{5}\frac{r^{7}(r-1)^{2}}{(r+1)^{12}},  \label{rt3}
\end{equation}%
\begin{equation}
\dot{\theta}=\frac{2r\sqrt{r-1}}{(r+1)^{3}},
\end{equation}%
\begin{equation}
h=A\frac{r^{4}(r-1)}{(r+1)^{6}}\sin (2\theta +\varphi _{0})  \label{rt4}
\end{equation}%
for general relativity. In these equations $A$ and $\varphi _{0}$ are free
(fitting) parameters that depend, in particular, on the unknown distance to
the binary system and its initial phase of motion. Mass $m$ is another free
parameter that determines the scale of dimensional coordinates.

It turns out that Eqs. (\ref{rt1})-(\ref{rt2}) and (\ref{rt3})-(\ref{rt4})
are sufficiently accurate to describe the observed LIGO signal. In Fig. \ref%
{VG-GR-waveform} we plot radiation wave strain $h(t)$ (in arbitrary units)
as a function of time obtained by numerical solution of Eqs. (\ref{rt1})-(%
\ref{rt2}) in vector gravity (solid line) and Eqs. (\ref{rt3})-(\ref{rt4})
for the case of general relativity (dashed line). The free parameters are
chosen to get the best fit of the GW150914 event signal. Figure shows that
radiation waveforms obtained in both theories are practically
indistinguishable. One should note that the decaying part of the waveform
corresponds to the orbital inspiral rather than damped oscillations of the
merged system. Radiation waveform decay occurs because, according to Eqs. (%
\ref{vvg}) and (\ref{vgr}), deviation from the Newtonian gravity results in
slowing down the orbital motion which, according to Eq. (\ref{aml2}),
reduces the wave amplitude.

\begin{figure}[h]
\centering
\includegraphics[width=0.5\textwidth]{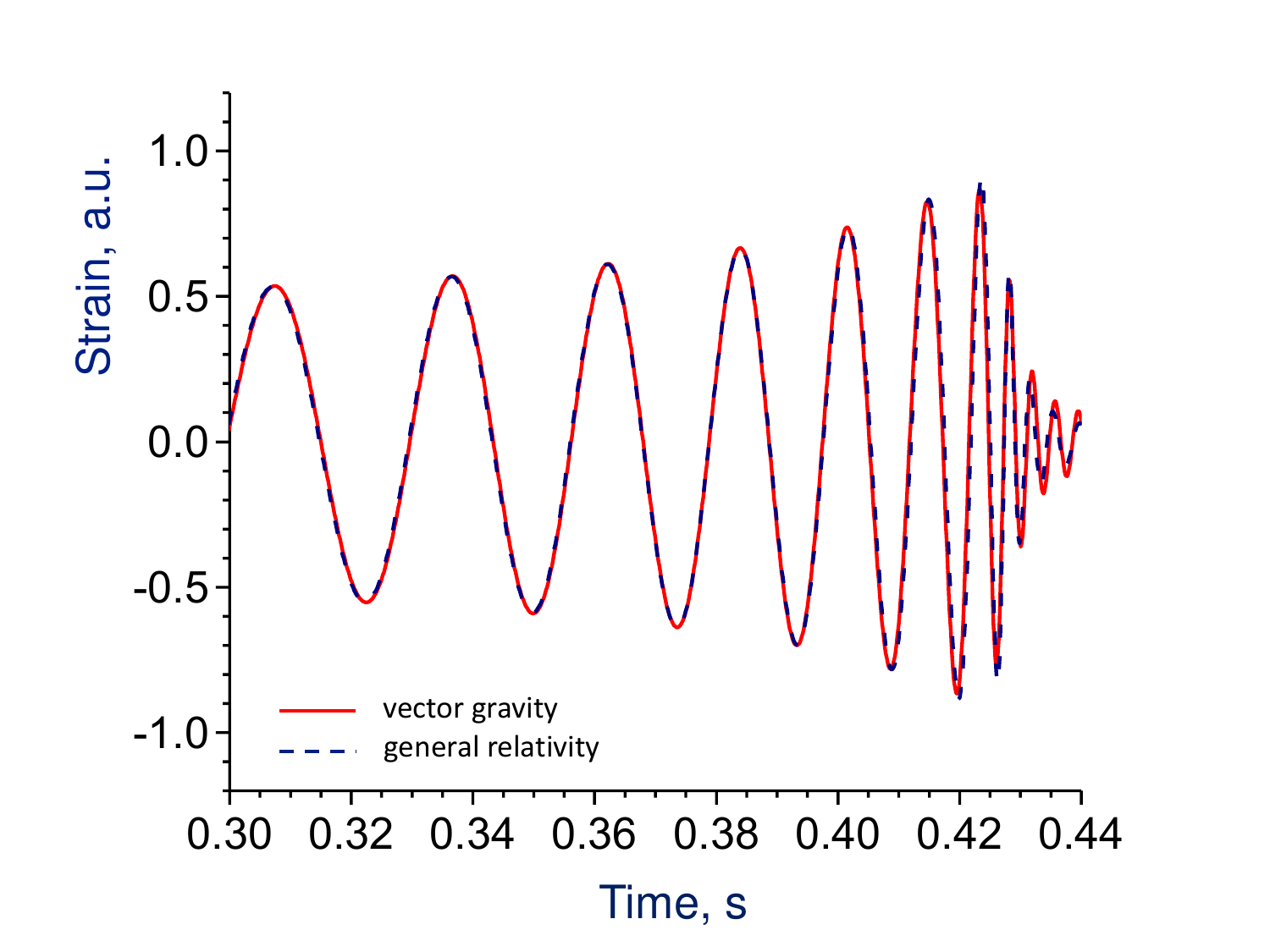}
\caption{Radiation wave strain $h(t)$ (in arbitrary units) as a function of
time obtained by numerical solution of Eqs. (\protect\ref{rt1})-(\protect\ref%
{rt2}) (solid line) and Eqs. (\protect\ref{rt3})-(\protect\ref{rt4}) (dashed
line). Free parameters $A$, $\protect\varphi _{0}$ and $m$ are chosen to
obtain the best fit of the LIGO GW150914 event signal and varied
independently for vector gravity and general relativity.}
\label{VG-GR-waveform}
\end{figure}

\begin{figure}[h]
\centering
\includegraphics[width=0.5\textwidth]{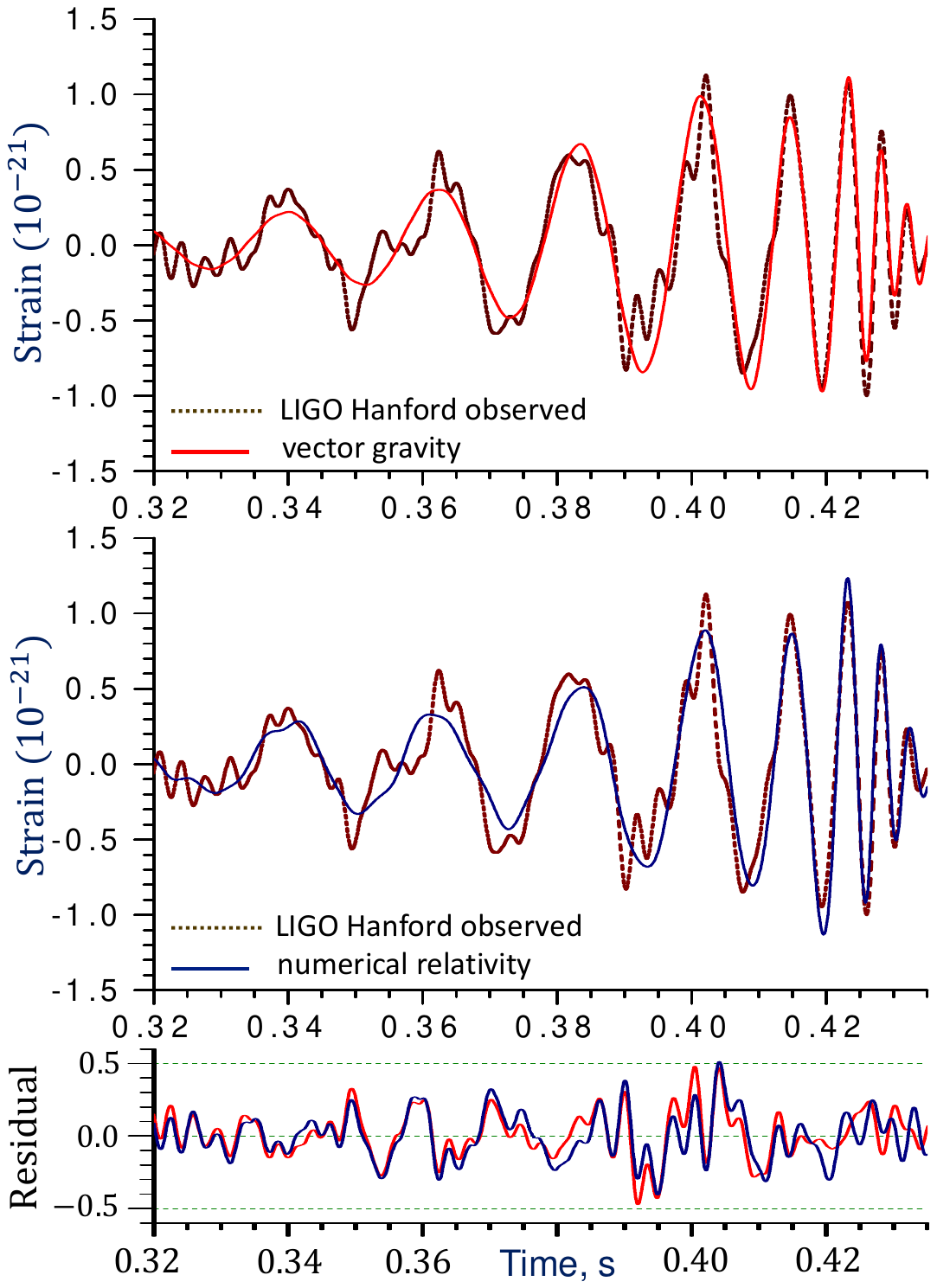}
\vspace{-0.5cm} \vspace{-0.3cm}
\caption{The gravitational-wave event GW150914 observed by the LIGO Hanford
detector. All time series are filtered with a bandpass and band-reject
filters in the same way as in Ref. \protect\cite{Abbo16}. \textit{Top row:}
Strain in Hanford detector as a function of time taken from Ref. 
\protect\cite{Abbo16} (dotted line) and filtered vector gravity radiation
waveform of Fig. \protect\ref{VG-GR-waveform} (red solid line). \textit{%
Second row}: Strain in Hanford detector (dotted line) and the best fit
numerical relativity waveform taken from Ref. \protect\cite{Abbo16} (blue
solid line). \textit{Bottom row:} Residuals after subtracting the filtered
vector gravity and numerical relativity waveforms from the filtered detector
time series. Red solid line is the residual obtained for vector gravity,
while the blue solid line is the residual for general relativity.}
\label{LIGO-Vector-GR}
\end{figure}

To show that our result is compatible with the LIGO data we process the
vector gravity radiation waveform of Fig. \ref{VG-GR-waveform} using
bandpass and spectral whitening algorithm described in the LIGO tutorial on
signal processing \cite{LIGOTutor} and compare the obtained waveform with
the respectively filtered LIGO\ signal reported in Ref. \cite{Abbo16}. The
results are summarized in Fig. \ref{LIGO-Vector-GR}. Top row shows strain $%
h(t)$ as a function of time for the gravitational-wave event GW150914
observed by the LIGO Hanford detector (dotted line) \cite{Abbo16} and the
best fit waveform obtained in vector gravity (red solid line). All time
series are filtered with a bandpass and band-reject filters in the same way
as in Ref. \cite{Abbo16}. Second row compares the LIGO Hanford signal with
the best fit waveform of numerical relativity taken from Ref. \cite{Abbo16}.
Bottom row shows residuals after subtracting the filtered vector gravity and
numerical relativity waveforms from the filtered detector time series. One
can see that within the limits of detector noise both theories yield
radiation waveforms which are compatible with the LIGO\ signal.

In Sec. \ref{NSML} we show that in vector gravity the masses of compact
objects in the merging binary systems obtained by the LIGO team based on the
gravitational wave detection reported so far ($29_{-4}^{+4}M_{\odot }$ and $%
36_{-4}^{+5}M_{\odot }$ \cite{Abbo16}; $14.2_{-3.7}^{+8.3}M_{\odot }$ and $%
7.5_{-2.3}^{+2.3}M_{\odot }$ \cite{Abbo16a}; $31.2_{-6.0}^{+8.4}M_{\odot }$
and $19.4_{-5.9}^{+5.3}M_{\odot }$ \cite{Abbo17}) fit in the mass intervals
for which neutron stars are stable. Thus, interpretation of the LIGO signals
as orbital inspiral and merger of two massive neutron stars, rather than
black holes, is plausible in vector gravity.

Polarization of gravitational waves in vector gravity differs from those in
general relativity (see Sec. \ref{gwave}). Since only two interferometers
were involved in the recent detection of gravitational waves by LIGO, the
wave polarization was not measured. Simultaneous detection of gravitational
waves by multiple instruments is able to distinguish between vector gravity
and general relativity. We discuss details of such experiment in Section \ref%
{GWtest}.

\section{Galactic centers and dark matter problem}

\label{dark}

In the present theory, static gravitational field is described by the
equivalent exponential metric (\ref{s10}). Metric (\ref{s10}) was also
obtained in Refs. \cite%
{Yilm58,Yilm71,Rose71,Rast75,Rast77,Chan80,Chan80b,Lind81,Svid09}.
Exponential metric (\ref{s10}) predicts no black holes, but rather compact
objects with no event horizon and finite gravitational redshift.

In recent years, the evidence for the existence of an ultra-compact
concentration of dark mass at centers of galaxies has become very strong.
However, a proof that such objects are black holes rather then compact
objects without event horizon is lacking. If the present theory of gravity
is correct then the compact supermassive objects at galactic centers are
unlikely composed of baryonic matter. Indeed, as we show in Sec. \ref{NSML},
mass of a compact (neutron star like) baryonic object with \textquotedblleft
causal\textquotedblright\ equation of state in vector gravity does not
exceed a few dozen solar masses, but the objects at galactic centers possess
masses upto a few $10^{9}$M$_{\odot }$. Even though there is no general
proof that state of matter must obey the \textquotedblleft
causality\textquotedblright\ constraint $dP/d\varepsilon \leq 1$ it is
unlikely that equation of state can be so stiff to make neutron star-like
objects of billion solar masses stable. Hence, likely those objects are made
of dark matter of non baryonic origin. This fact gives us an opportunity to
determine composition of dark matter based on observations of supermassive
objects at galactic centers.

In the previous paper \cite{Svid07} we found that properties of compact
objects at galactic centers can be explained quantitatively assuming they
are made of dark matter axions and the axion mass is about $0.6$ meV.
Analysis of Ref. \cite{Svid07} is based on the assumption that static
gravitational field is described by the exponential metric (\ref{s10})
rather then by general relativity. A full time-dependent theory of gravity
was unnecessary for calculations made in Ref. \cite{Svid07}. The present
paper provides such a theory and justifies our previous choice of the
exponential metric.

Axions are one of the leading particle candidates for the cold dark matter
in the Universe \cite{Brad03}. Interaction of axions with QCD instantons
generates the axion mass $m$ and periodic interaction potential \cite{Kim87} 
\begin{equation}
V(\varphi )=m^{2}F^{2}[1-\cos (\varphi /F)],  \label{p1}
\end{equation}%
where $\varphi $ is a real scalar axion field and $F$ is the Peccei-Quinn
symmetry breaking scale. The interaction potential (\ref{p1}) has degenerate
minima $V=0$ at $\varphi =2\pi nF$, where $n$ is an integer number. As a
consequence, axions can form bubbles. Bubble mass is concentrated in a thin
surface (interface between two degenerate vacuum states). In the exponential
metric the potential energy of a spherical bubble with radius $R$ is given
by \cite{Svid07}%
\begin{equation*}
U(R)=4\pi \sigma R^{2}\exp \left( \frac{GM}{c^{2}R}\right) ,
\end{equation*}%
where $\sigma $ is the surface energy density and $M$ is the fixed bubble
mass. $U(R)$ has a shape of a well. At $R\gg GM/c^{2}$ one can omit gravity
and $U(R)\simeq 4\pi \sigma R^{2}$ is just a surface energy (tension) which
tends to contract the bubble. At $R\ll GM/c^{2}$ gravity effectively
produces large repulsive potential which forces the bubble to expand. As a
result, the bubble radius $R(t)$ oscillates between two turning points.

In Ref. \cite{Svid07}, based on quantitative analysis of available data, we
argued that such oscillating axion bubbles, rather then supermassive black
holes, could be present at galactic centers. Recent observations of
near-infrared and X-ray flares from Sagittarius A$^{\ast }$, which is
believed to be a $4\times 10^{6}$M$_{\odot }$ black hole at the Galactic
center, show that the source exhibits about $20$-minute periodic variability 
\cite{Genz03,Gill06,Bela06}. An oscillating axion bubble can explain such
variability. Known value of the bubble mass at the center of our Galaxy and
its oscillation period yields the axion mass of about $0.6$ meV. Size of the
axion bubble at the center of the Milky Way oscillates between $R_{\min
}\approx 1R_{\odot }$ and $R_{\max }\approx 1$AU $\approx 210R_{\odot }$.

Further, as shown in Ref. \cite{Svid07}, the axion bubbles with no free
parameters (if we fix $m=0.6$ meV based on Sagittarius A* flare variability)
quantitatively explain the upper limit (a few $10^{9}$M$_{\odot }$) on the
supermassive \textquotedblleft black hole" mass found in analysis of the
measured mass distribution \cite{Grah07}. Also, with no free parameters the
bubble scenario explains observed lack of supermassive \textquotedblleft
black holes" with mass $M\lesssim 10^{6}$M$_{\odot }$. For such low-mass
bubbles the decay time $t\propto M^{9/2}$ becomes much shorter then the age
of the Universe and, as a result, such objects are very rare.

One should note that results of Ref. \cite{Svid07} describe bubbles which
are already formed and relatively isolated. Thus, Active Galactic Nuclei
whose bubbles are currently under formation or strongly interact with the
galactic environment should be excluded. A sample of predominantly inactive
galaxies for which direct supermassive \textquotedblleft black
hole\textquotedblright\ mass measurements have been catalogued shows lack of
such objects with $M\lesssim 10^{6}$M$_{\odot }$ (see Fig. 1 in \cite{Grah08}%
). On the other hand, both limits on the bubble mass in Active Galactic
Nuclei can be somewhat wider. For instance, in such galaxies, fast
vaporization of a low-mass bubble could be reduced by the back flow of
galactic axions into the bubble which extends its lifetime. Such a scenario
leads to observable consequences. Namely, it predicts that at the low-mass
end of the \textquotedblleft black hole\textquotedblright\ vs host galaxy
bulge mass diagram the \textquotedblleft black hole\textquotedblright\
masses must be lower than predicted by the relation established using
galaxies having predominantly higher-mass \textquotedblleft black
holes\textquotedblright . In addition, at the low-mass end it should not be
correlation between the \textquotedblleft black hole\textquotedblright\ mass
and bulge luminosity because bubbles loose mass fast on a time scale of the
bulge evolution. Thus, a wide range of bubble masses can exist at almost the
same bulge luminosity at the low-mass end. Observations support both these
predictions \cite{Jian11}. \textquotedblleft Black holes\textquotedblright\
with estimated mass in the range $10^{5}-10^{6}$M$_{\odot }$ have been found
in Active Galactic Nuclei and their masses lie substantially below the
scaling relation defined by the massive systems \cite{Jian11,Grah15,Grah15b}%
. In addition, at the lower end the measured \textquotedblleft black
hole\textquotedblright\ masses span a much wider range at fixed bulge
luminosity \cite{Jian11}.

Observation of the Galactic center with a Very Long Baseline Interferometry
(VLBI) within the next few years will be capable to test theories of
gravitation in the strong field limit. Such an observation will allow us to
distinguish between the black hole (predicted by general relativity) and the
oscillating axion bubble scenario. A defining characteristic of a black hole
is the event horizon. To a distant observer, the event horizon casts a
relatively large \textquotedblleft shadow\textquotedblright\ over the
background source with an apparent diameter of about $10GM/c^{2}\approx
80R_{\odot }$ due to bending of light. The predicted size of this shadow for
Sagittarius A* approaches the resolution of current radio-interferometers.
Hence, there exists a realistic expectation of imaging the shadow of a black
hole with the Event Horizon Telescope (EHT), a project to assemble VLBI
network of millimeter wavelength dishes that can resolve strong
gravitational field signatures near the supermassive object. As planned, the
EHT will include enough dishes to enable imaging of the black hole shadow
within the next few years \cite{Falc00,Falc01,Shen05,Huan07,Doel09,Akiy15}.
If the axion bubble, rather then a black hole, is present at the Galactic
center, the steady shadow will not be observed. Instead, the shadow will
appear and disappear periodically with a period of about $20$ $\min $.
Discovery of periodic appearance of the shadow from the Galactic center
object will also be a strong evidence for the axion nature of dark matter
and will lead to an accurate prediction of the axion mass.

One should mention that intrinsic size of Sagittarius A* at a wavelength of $%
1.3$ mm was determined using VLBA \cite{Doel08}. The intrinsic diameter of
Sagittarius A* was found to be $<0.3$AU$\approx 65R_{\odot }$ which is less
than the expected apparent size of the event horizon of the presumed black
hole. Such observation might indicate lack of black holes, in agreement with
the present theory.

Existence of dark matter axions with the predicted mass of about $0.6$ meV
can be experimentally tested in future ARIADNE \cite{Arva14} and Orpheus 
\cite{Rybk15} experiments.

\section{Testing vector gravity with gravitational wave interferometers}

\label{GWtest}

In spite of fundamental differences, vector gravity and general relativity
yield for the experimentally tested regimes quantitatively very close
predictions which allowed both theories to pass available tests. However,
prediction of the matter behavior at strong gravity and interpretation of
the universe evolution on large scales depends on the theory of gravity we
are using. Thus, there is a need for a feasible test which can distinguish
between vector gravity and general relativity and rule out one of the two
theories. Here we propose such an experiment that can be done in the nearest
years using gravitational wave interferometers.

Both in general relativity and vector gravity the polarization of
gravitational waves emitted by orbiting binary objects is transverse, that
is wave yields motion of test particles in the plane perpendicular to the
direction of wave propagation. However, as we show here, dependence of the
laser interferometer signal on the orientation of the interferometer arms
relative to the propagation direction of the gravitational wave is different
in the two theories.

Gravitational wave produces motion of the interferometer mirrors and changes
phase velocity of light. Both of these effects contribute to the relative
phase shift of light traveling in the perpendicular arms of the Michelson
interferometer. For certain propagation directions of the gravitational wave
relative to the arms the two contributions cancel each other yielding zero
net phase shift. Those are the directions of zero response of the
interferometer for which gravitational wave can not be detected for any
transverse polarization. As we show, directions of the zero response are
different for gravitational waves in vector gravity and general relativity.
Detection of a wave in the direction of the zero response predicted by a
theory of gravity will rule out such theory.

In vector gravity for a weak transverse plane gravitational wave propagating
along the $x-$axis the equivalent metric is given by Eq. (\ref{met1}) 
\begin{equation}
g_{ik}=\eta _{ik}+\left( 
\begin{array}{cccc}
0 & 0 & h_{0y}(t,x) & h_{0z}(t,x) \\ 
0 & 0 & 0 & 0 \\ 
h_{0y}(t,x) & 0 & 0 & 0 \\ 
h_{0z}(t,x) & 0 & 0 & 0%
\end{array}%
\right) ,  \label{met1A}
\end{equation}%
where $\eta _{ik}$ is Minkowski metric and $h_{0y}$, $h_{0z}$ are small
perturbations obeying the wave equation. A rest particle (or mirrors of an
interferometer) will move under the influence of the gravitational wave (\ref%
{met1A}) with a time-dependent velocity $V^{\alpha }=h_{0\alpha }c$ ($\alpha
=x$, $y$, $z$) perpendicular to the direction of the wave propagation. By
making a coordinate transformation into the co-moving frame of the test
particle 
\begin{equation*}
x^{\prime \alpha }=x^{\alpha }-\int^{t}V^{\alpha }dt,
\end{equation*}%
the metric (\ref{met1A}) reduces to%
\begin{equation}
g_{ik}=\eta _{ik}+\left( 
\begin{array}{cccc}
0 & 0 & 0 & 0 \\ 
0 & 0 & h_{xy}(t,x) & h_{xz}(t,x) \\ 
0 & h_{xy}(t,x) & 0 & 0 \\ 
0 & h_{xz}(t,x) & 0 & 0%
\end{array}%
\right) ,  \label{met1aA}
\end{equation}%
where $h_{xy}=h_{0y}$, $h_{xz}=h_{0z}$. Metric (\ref{met1aA}) is written in
the coordinate system in which test particles do not move under the
influence of the gravitational wave.

On the other hand, in general relativity for a weak gravitational wave
propagating along the $x-$axis the metric in the co-moving frame evolves as 
\cite{Land95}%
\begin{equation}
g_{ik}=\eta _{ik}+\left( 
\begin{array}{cccc}
0 & 0 & 0 & 0 \\ 
0 & 0 & 0 & 0 \\ 
0 & 0 & h_{yy}(t,x) & h_{yz}(t,x) \\ 
0 & 0 & h_{yz}(t,x) & -h_{yy}(t,x)%
\end{array}%
\right) .  \label{wa3A}
\end{equation}

Here we investigate a response of a laser interferometer with perpendicular
arms on a gravitational wave in the two theories of gravity. In
gravitational field with a metric $g_{ik}$ the Maxwell's equations for the
electromagnetic field vector $A^{k}$ in the absence of charges read%
\begin{equation}
\frac{\partial }{\partial x^{k}}\left[ \sqrt{-g}g^{kl}g^{im}\left( \frac{%
\partial A_{m}}{\partial x^{l}}-\frac{\partial A_{l}}{\partial x^{m}}\right) %
\right] =0,  \label{max1}
\end{equation}%
while the Lorenz gauge equation is%
\begin{equation}
\frac{\partial }{\partial x^{k}}\left( \sqrt{-g}A^{k}\right) =0,
\label{max2}
\end{equation}%
where $A^{k}=g^{km}A_{m}$ and $g=$det$(g_{ik})$. Gravitational wave causes
oscillation of $g_{ik}$ in space and time. However, since frequency of the
gravitational waves is much smaller then frequency of the electromagnetic
waves traveling in the interferometer one can disregard derivatives of the
metric in Eqs. (\ref{max1}) and (\ref{max2}). Then Eqs. (\ref{max1}) and (%
\ref{max2}) reduce to%
\begin{equation}
g^{kl}\frac{\partial ^{2}A^{i}}{\partial x^{k}\partial x^{l}}-g^{im}\frac{%
\partial ^{2}A^{k}}{\partial x^{m}\partial x^{k}}=0,  \label{max3}
\end{equation}%
\begin{equation}
\frac{\partial A^{k}}{\partial x^{k}}=0.  \label{max4}
\end{equation}%
Combining them together we obtain%
\begin{equation}
g^{kl}\frac{\partial ^{2}A^{i}}{\partial x^{k}\partial x^{l}}=0.
\label{max5}
\end{equation}

In the co-moving frame there is no motion of the interferometer mirrors and
the phase shift of light traveling along the two arms appears due to
difference in the light phase velocity. Calculation of the gravitational
wave signal can be done in any reference frame because the phase shift%
\begin{equation}
\Delta \varphi =\int k_{i}dx^{i}  \label{psh}
\end{equation}
is invariant under general coordinate transformations and, thus, the
interferometer signal is independent of the frame. In Eq. (\ref{psh}) $%
k_{i}=(\omega /c,\mathbf{k})$ is the photon four dimensional wave vector.

Substitute $g^{kl}=\eta ^{kl}+h^{kl}$ in Eq. (\ref{max5}), where $h^{kl}$ is
a small perturbation that has only spatial components $h^{\alpha \beta }$ ($%
\alpha $, $\beta =x$, $y$, $z$), yields the following propagation equation
for the electromagnetic wave%
\begin{equation}
\left( \frac{1}{c^{2}}\frac{\partial ^{2}}{\partial t^{2}}-\nabla
^{2}+h^{\alpha \beta }\frac{\partial ^{2}}{\partial x^{\alpha }\partial
x^{\beta }}\right) A^{i}=0.  \label{max7}
\end{equation}%
In this equation, $h^{\alpha \beta }$ can be approximately treated as
constants since they vary slowly as compared to the fast variation of $A^{i}$%
. Looking for solution of Eq. (\ref{max7}) in the form $A^{i}\propto
e^{-i\omega t+i\mathbf{k}\cdot \mathbf{r}}$, where $\mathbf{k}$ is the wave
vector of the electromagnetic wave, we obtain the following dispersion
relation for light%
\begin{equation*}
\frac{\omega ^{2}}{c^{2}}=k^{2}-h^{\alpha \beta }k_{\alpha }k_{\beta },
\end{equation*}%
and, hence, the phase velocity of the electromagnetic wave is (see also \cite%
{Gert63})%
\begin{equation}
V_{\text{ph}}=\frac{\omega }{k}\approx c\left( 1-\frac{1}{2}h^{\alpha \beta }%
\hat{k}_{\alpha }\hat{k}_{\beta }\right) ,  \label{max8}
\end{equation}%
where $\hat{k}=\mathbf{k}/k$.

Equation (\ref{max8}) shows that presence of the gravitational wave leads to
the change of the phase velocity of light which depends on the direction of
the light propagation $\hat{k}$. If arms of the interferometer are oriented
along unit vectors $\hat{a}$ and $\hat{b}$ then difference in the phase
velocities of the laser light propagating along the two arms is 
\begin{equation*}
\Delta V_{\text{ph}}=\frac{c}{2}h^{\alpha \beta }\left( \hat{a}_{\alpha }%
\hat{a}_{\beta }-\hat{b}_{\alpha }\hat{b}_{\beta }\right) .
\end{equation*}

Signal of the LIGO-like (Michelson) interferometer with arms of length $L$
oriented along the directions $\hat{a}$ and $\hat{b}$ is proportional to the
relative phase shift $\Delta \varphi $ of electromagnetic waves traveling a
roundtrip distance $2L$ along the two arms%
\begin{equation}
\Delta \varphi =k\frac{2L}{c}\Delta V_{\text{ph}}=\frac{\omega L}{c}%
h^{\alpha \beta }\left( \hat{a}_{\alpha }\hat{a}_{\beta }-\hat{b}_{\alpha }%
\hat{b}_{\beta }\right) ,  \label{wa4A}
\end{equation}%
where $\omega $ is the frequency of electromagnetic wave and $h^{\alpha
\beta }$ is the spatial perturbation of the metric in the reference frame in
which interferometer mirrors do not move (frame of Eqs. (\ref{met1aA}) and (%
\ref{wa3A})).

For the gravitational wave propagating along the $x-$axis, Eq. (\ref{wa4A})
yields for the gravitational wave (\ref{wa3A}) in general relativity%
\begin{equation}
\Delta \varphi =\frac{\omega L}{c}\left[ h^{yy}\left( \hat{a}_{y}^{2}-\hat{b}%
_{y}^{2}+\hat{b}_{z}^{2}-\hat{a}_{z}^{2}\right) +2h^{yz}\left( \hat{a}_{y}%
\hat{a}_{z}-\hat{b}_{y}\hat{b}_{z}\right) \right] ,  \label{wa5A}
\end{equation}%
while for the transverse wave (\ref{met1aA}) in vector gravity we obtain%
\begin{equation}
\Delta \varphi =\frac{2\omega L}{c}\left[ h^{xy}\left( \hat{a}_{x}\hat{a}%
_{y}-\hat{b}_{x}\hat{b}_{y}\right) +h^{xz}\left( \hat{a}_{x}\hat{a}_{z}-\hat{%
b}_{x}\hat{b}_{z}\right) \right] .  \label{wa6A}
\end{equation}

Equations (\ref{wa5A}) and (\ref{wa6A}) show that vector gravity and general
relativity predict qualitatively different effect of the gravitational wave
on the interferometer signal. Namely, general relativistic gravitational
wave of any polarization (arbitrary $h^{yy}$ and $h^{yz}$) produces no
signal when gravitational wave propagates parallel to the interferometer
plane at $45^{\circ }$ angle relative to one of its perpendicular arms (see
Fig. \ref{Arms}\textit{a}). E.g., this is the case for $\hat{a}=\left(
1,1,0\right) /\sqrt{2}$ and $\hat{b}=\pm \left( 1,-1,0\right) /\sqrt{2}$.
For these orientations the gravitational wave in vector gravity can produce
signal, namely, 
\begin{equation}
\Delta \varphi =\frac{2\omega L}{c}h^{xy}.
\end{equation}

On the other hand, gravitational wave in vector gravity (for arbitrary $%
h^{xy}$ and $h^{xz}$) yields no signal if gravitational wave propagates in
the direction perpendicular to the interferometer plane, e.g. $\hat{a}%
=\left( 0,1,0\right) $ and $\hat{b}=\left( 0,0,1\right) $ (see Fig. \ref%
{Arms}\textit{b}), or along one of the interferometer arms, e.g. $\hat{a}%
=\left( 1,0,0\right) $, $\hat{b}=\left( 0,0,1\right) $ or $\hat{a}=\left(
0,1,0\right) $, $\hat{b}=\left( 1,0,0\right) $. For these orientations the
gravitational wave in general relativity can produce signal.

\begin{figure}[t]
\centering
\includegraphics[width=0.55\textwidth]{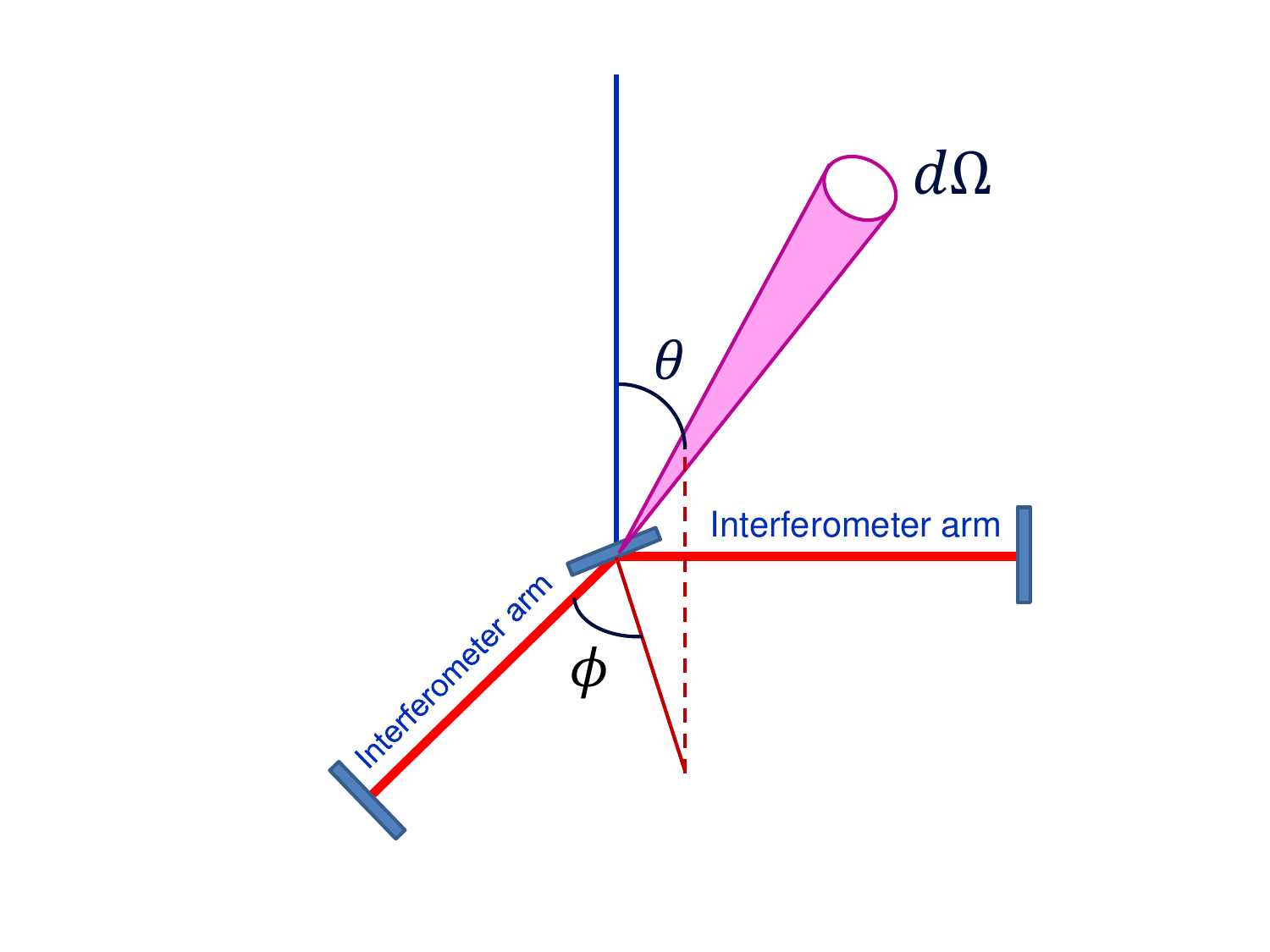}
\vspace{-1cm}
\caption{Spherical coordinate system in the frame of interferometer arms.}
\label{N}
\end{figure}

This difference can be used to test theories of gravity in polarization
experiments with several LIGO-like interferometers. The experiment can be
conducted with three interferometers. Simultaneous detection of the
gravitational wave by all three instruments allows us to determine the
direction of the wave propagation by measuring the wave arrival times at the
interferometer locations and using information about the waveform \cite%
{Gurs89,Cava06,Mark08}. In vector gravity the gravitational waveform
produced by inspiral of two objects can be specified to the same extent as
in general relativity and, thus, the wave source can be localized on the sky
with a similar accuracy in both theories.

The issue of localization of gravitational wave signals with a detector
network has been discussed previously in many publications (see, e.g., \cite%
{Gurs89,Cava06,Mark08,Sear08,Sear09,Fair09,Wen10,Klim11,Niss11,Croc12,Veit12,Grov14,Kyut14,Sing16,Abbo16bA}%
). The accuracy with which the source can be localized on the sky depends
upon the timing accuracy in each of the detectors, the network geometry and
the angle between the plane of the detectors and the signal location. A
detector network with widely separated detectors yields the best
localization ability. Depending on the source orientation the LIGO-Virgo
network can determine source location with accuracy $20\div 100$ deg$^{2}$ 
\cite{Fair09}. Prospects for observing and localizing gravitational-wave
transients to areas of $5$ deg$^{2}$ to $20$ deg$^{2}$ with Advanced LIGO
and Advanced Virgo gravitational-wave detectors over the next decade are
discussed in \cite{Abbo16cA}.

By detecting many events one can accumulate statistics and find a
distribution of the direction of the detected gravitational waves relative
to the interferometer arms. Namely, one can measure the distribution
function $N(\theta ,\phi )$ defined as $dN=$ $N(\theta ,\phi )d\Omega $,
where $dN$ is the number of events for which detected gravitational waves
propagate inside the solid angle $d\Omega =\sin (\theta )d\theta d\phi $.
Here $\theta $ and $\phi $ are the polar and azimuth angles in the spherical
coordinate system in the frame of the interferometer arms (see Fig. \ref{N}).

\begin{figure}[t]
\centering
\includegraphics[width=0.55\textwidth]{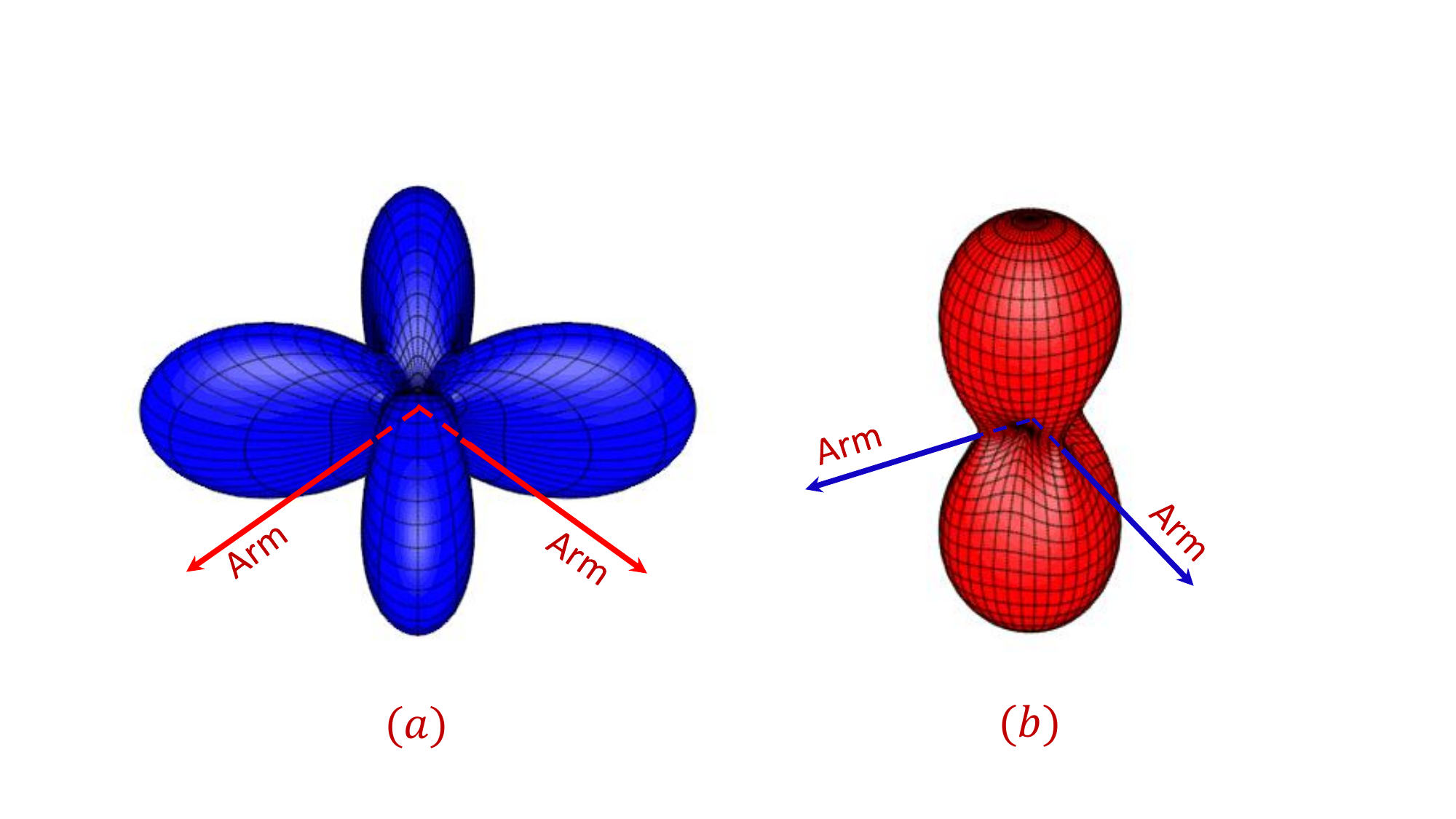}
\vspace{-0.5cm}
\caption{Distribution of the direction of the detected transverse
gravitational waves relative to the laser interferometer arms predicted by
vector gravity (a) and general relativity (b). }
\label{P}
\end{figure}

Taking square of Eqs. (\ref{wa5A}), (\ref{wa6A}) and averaging over
polarization of the gravitational waves we obtain that distribution function
in vector gravity is 
\begin{equation}
N(\theta ,\phi )=\sin ^{2}(\theta )\left[ 1-\sin ^{2}(\theta )\cos
^{2}(2\phi )\right] ,  \label{vg}
\end{equation}%
while for general relativity we find%
\begin{equation}
N(\theta ,\phi )=\cos ^{2}(\theta )+\frac{1}{4}\sin ^{4}(\theta )\cos
^{2}(2\phi ).  \label{gr}
\end{equation}%
Functions (\ref{vg}) and (\ref{gr}) are normalized such that $N_{\max }=1$.
We plot $N(\theta ,\phi )$ given by Eqs. (\ref{vg}) and (\ref{gr}) in Fig. %
\ref{P}. The two distributions look very different and, hence, the
experiment we are proposing should be able to distinguish between them with
available localization accuracy.

Vector gravity predicts that distribution $N(\theta ,\phi )$ will have dips
in the directions perpendicular to the interferometer plane - $\theta =0$, $%
\pi $ and along the interferometer arms - $\theta =\pi /2$, $\phi =0$, $\pi
/2$, $\pi $, $3\pi /2$ (see Fig. \ref{P}\textit{a}). The dips appear because
for these propagation directions the interferometer can not detect the
transverse gravitational wave. In the case of general relativity the dips
will appear in the directions for which wave propagates in the
interferometer plane at $45^{\circ }$ angle relative to one of the
interferometer arms - $\theta =\pi /2$ and $\phi =\pi /4$, $3\pi /4$, $5\pi
/4$, $7\pi /4$ (Fig. \ref{P}\textit{b}).

\begin{figure}[t]
\centering
\includegraphics[width=0.44\textwidth]{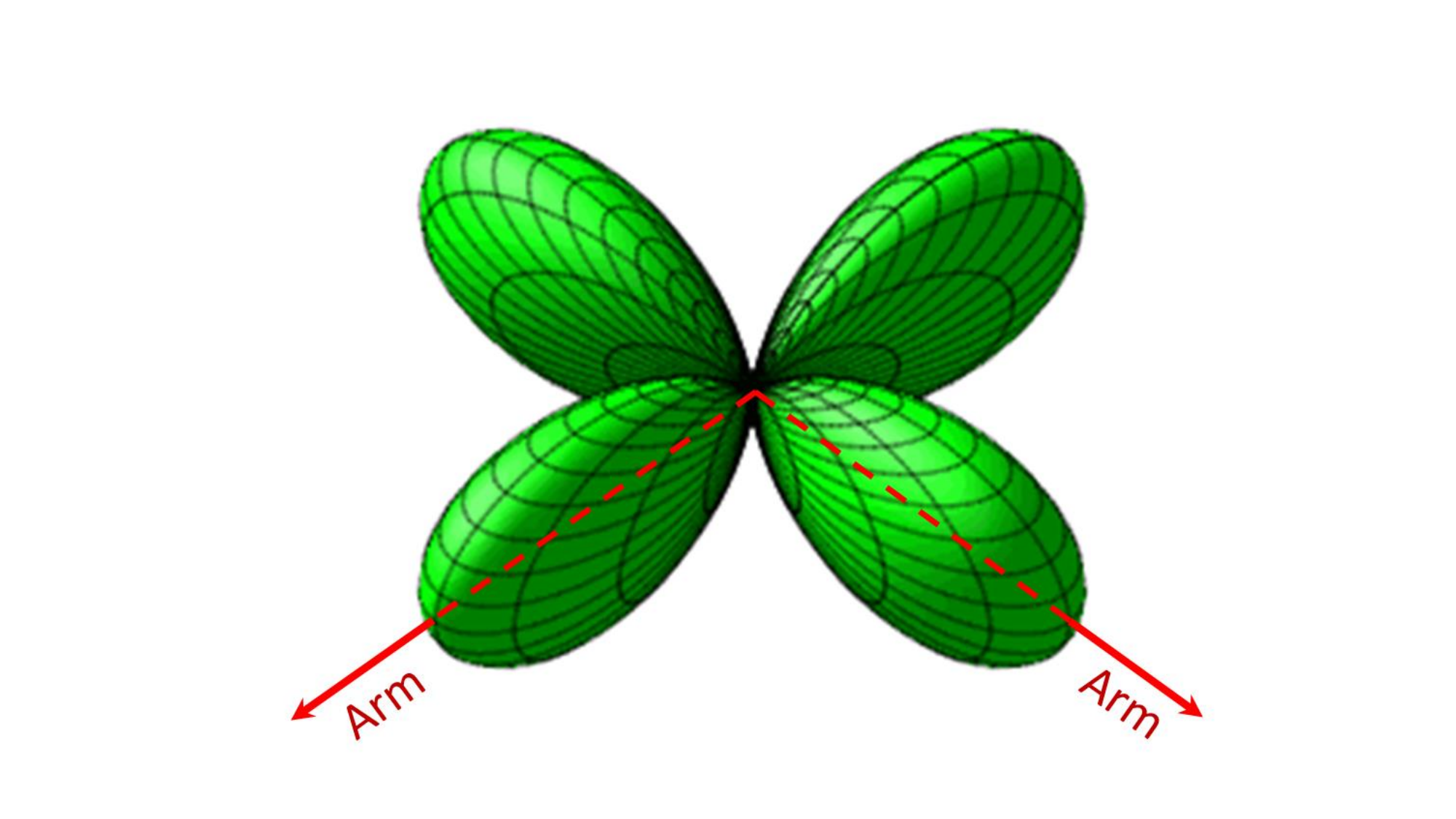}
\vspace{-0.5cm}
\caption{Probability to detect longitudinal gravitational wave propagating
in the direction $(\protect\theta ,\protect\phi )$ relative to the
perpendicular interferometer arms. $\protect\theta $ and $\protect\phi $ are
the polar and azimuth angles in the spherical coordinate system in the frame
of the interferometer arms shown in Fig. \protect\ref{N}.}
\label{PL}
\end{figure}

One should mention that vector gravity also predicts existence of
longitudinal gravitational waves which are not emitted by orbiting binary
starts. However, they might be generated during the star mergers or in early
universe. For such a wave propagating along the $x-$axis the equivalent
metric in the co-moving frame reads 
\begin{equation}
g_{ik}=\eta _{ik}+\left( 
\begin{array}{cccc}
0 & 0 & 0 & 0 \\ 
0 & -2h(t,x) & 0 & 0 \\ 
0 & 0 & h(t,x) & 0 \\ 
0 & 0 & 0 & h(t,x)%
\end{array}%
\right) .  \label{Lw}
\end{equation}%
For the longitudinal wave (\ref{Lw}) the interferometer signal is 
\begin{equation}
\Delta \varphi =\frac{3\omega Lh}{c}\left( \hat{b}_{x}^{2}-\hat{a}%
_{x}^{2}\right) ,
\end{equation}%
which yields the following distribution function $N$ 
\begin{equation}
N(\theta ,\phi )=\sin ^{4}(\theta )\cos ^{2}(2\phi ).  \label{vgL}
\end{equation}%
We plot this function in Fig. \ref{PL}.

Longitudinal gravitational waves produce no interferometer signal if they
propagate at equal angles relative to the interferometer arms (more exactly
when $\hat{k}\cdot \hat{a}=\pm \hat{k}\cdot \hat{b}$). Thus, propagation
direction perpendicular to the interferometer plane shown in Fig. \ref{Arms}%
\textit{b} is the direction of zero response for all kind of gravitational
waves in vector gravity. However, waves propagating along the arms can
produce signal if they are longitudinal.

The experiment we are proposing is crucial for our understanding of the
nature of gravity and can test whether gravity has a tensor or a vector
origin. Simultaneous detection of gravitational waves in at least three
instruments is necessary for the experiment. A joint scientific run of the
two LIGO interferometers in the US and the Virgo interferometer in Italy is
capable of distinguishing between tensor and vector origin of gravity. Such
joint runs can begin in 2017 when Virgo instrument will reach the required
sensitivity.

\section{Summary}

Einstein's general relativity is an elegant theory of gravity which is based
on the assumption that space-time geometry is a tensor gravitational field.
However, beauty of the theory does not guarantee that theory describes the
nature. So far general relativity has passed all available tests of gravity.
To the best of our knowledge, the vector theory of gravity we are proposing
in this paper also passes all available tests as we discuss in Sec. \ref%
{test}. General relativity, however, can not explain the nature of dark
energy. In contrast, vector gravity is free of such drawback.

Our alternative theory of gravity is based on the assumption that gravity is
a vector field in a fixed background four dimensional Euclidean space which
is coupled to matter universally and minimally through the equivalent metric 
$f_{ik}$ which is a functional of the vector field. We show that present
theory is the only possibility that can be obtained from this assumption.

There are several motivations for the vector theory of gravity. It provides
an appealing explanation of how the difference between space and time
appeared in the originally totally symmetric Euclidean Universe. Namely, the
vector gravitational field breaks the symmetry of the four dimensional
Euclidean space. Direction of the vector field gives the time coordinate,
while perpendicular directions are spatial coordinates.

Vector gravity also suggests a natural mechanism of matter generation at the
Big Bang. Namely, vector theory of gravity yields that at the moment of Big
Bang the energy of gravitational waves is negative and, thus, matter can be
created at the expense of generation of the negative energy gravitons. This
mechanism has an analogy with emission of electromagnetic waves by an
electric dipole (or a quadrupole) placed in a dispersionless medium with
negative refractive index. In a dispersionless medium with negative
dielectric constant $\varepsilon $ and negative magnetic permeability $\mu $
the energy density of the electromagnetic field%
\begin{equation*}
w=\frac{1}{8\pi }(\varepsilon E^{2}+\mu H^{2})
\end{equation*}%
is negative. As a result, photons emitted by an oscillating dipole placed in
such medium carry away negative energy yielding exponential growth of the
dipole oscillations \cite{Smit00}. Thus, system is unstable with respect to
generation of electromagnetic waves and acceleration of electric charges
placed in such medium.

For vector gravity the vacuum of empty fermion states acts as a
dispersionless medium with negative refraction (see Appendix \ref{AP4}).
Such vacuum is unstable with respect to generation of gravitational waves
and heating up the Universe. Vector gravity suggests that at the Big Bang
the Universe was heated up by this mechanism. The vacuum instability leads
to exponential growth of the gravitational field and matter generation. This
is the era of cosmological inflation. Thus, vector gravity predicts the
inflation stage. In vector gravity there is no need for an additional
cosmological field that would supply energy for matter generation. However,
such additional field, the inflaton, is a necessary ingredient of
cosmological models based on general relativity for which graviton energy is
always positive \cite{Lind90}.

At some point the heating of the Universe came to an end. Thus, it must be a
mechanism which stopped the heating process. A need for it motivated us to
postulate, by the analogy with the composite theory of photon, that in
vector gravity the graviton is a composite particle formed of
fermion-antifermion pairs and graviton emission corresponds to creation of
such pairs. The constituent fermion is an elementary spin $1/2$ massless
particle which has positive and negative energy states. This assumption
explains why heating of the Universe stopped. Since no more than one fermion
can occupy the same quantum state the matter generation at the Big Bang has
continued until fermion states were filled and the Universe became extremely
hot. Subsequent evolution of the Universe is described by the usual hot
Universe theory. The following expansion of the Universe practically did not
change the fermion occupation number and fermion states remain filled in the
present epoch.

Transition from the originally four dimensional Euclidean geometry to the
equivalent metric of Minkowski character occurred at the point of Big Bang.
Recall that in terms of the unit vector $u_{k}$ and the scalar $\phi $ the
equivalent metric in vector gravity reads%
\begin{equation}
f_{ik}=-e^{-2\phi }\delta _{ik}+2\cosh (2\phi )u_{i}u_{k}.  \label{metr1}
\end{equation}%
Before the Big Bang the Universe should be described quantum mechanically so
that $f_{ik}$ and $u_{k}$ are replaced with operators

\begin{equation}
\hat{f}_{ik}=-e^{-2\phi }\delta _{ik}+2\cosh (2\phi )\hat{u}_{i}\hat{u}_{k}.
\label{f1r}
\end{equation}%
Before the Big Bang the vector gravitational field had no preferred
direction and was undergoing quantum fluctuations. For this state of the
Universe the quantum mechanical average of the field operator is equal to
zero%
\begin{equation*}
\left\langle \hat{u}_{k}\right\rangle =0,
\end{equation*}%
while%
\begin{equation*}
\left\langle \hat{u}_{i}\hat{u}_{k}\right\rangle =\frac{1}{4}\delta _{ik}.
\end{equation*}%
For such state the expectation value of the metric operator (\ref{f1r}) is%
\begin{equation*}
\left\langle \hat{f}_{ik}\right\rangle =\frac{1}{4}\left( e^{2\phi
}-3e^{-2\phi }\right) \delta _{ik}.
\end{equation*}%
That is before the Big Bang the equivalent metric has Euclidean character.

Big Bang is the point of phase transition at which the gravitational field
vector acquires nonzero expectation value $\left\langle \hat{u}%
_{k}\right\rangle =u_{k}\neq 0$. This expectation value can serve as a
transition order parameter. Now the four dimensional space has a preferred
direction $u_{k}$. If fluctuations of the field around $u_{k}$ are small one
can use a mean-field description and replace operators with their mean
values: $\hat{u}_{k}\rightarrow u_{k}$, etc. Choosing direction of the
vector gravitational field as a time coordinate we have $u_{k}=(1,0,0,0)$
and the equivalent metric (\ref{metr1}) now reads 
\begin{equation*}
f_{ik}=\left( 
\begin{array}{cccc}
e^{2\phi } & 0 & 0 & 0 \\ 
0 & -e^{-2\phi } & 0 & 0 \\ 
0 & 0 & -e^{-2\phi } & 0 \\ 
0 & 0 & 0 & -e^{-2\phi }%
\end{array}%
\right) .
\end{equation*}%
That is geometry has the character of the Minkowski space-time in the
ordered phase of the Universe. In Minkowski geometry the initial vacuum of
empty fermion states is unstable towards generation of matter in the expense
of production of the negative energy gravitons. Universe enters the stage of
inflation which ends when fermion states become filled.

These filled states act as a new vacuum for the evolution of the Universe
after the heating stage. As we show in Sec. \ref{quantization}, for the
filled vacuum the graviton energy is positive and, thus, vacuum is stable.
For such vacuum, emission of a graviton corresponds to creation of
fermion-antifermion hole pairs out of the filled fermion states. Binary
stars orbiting each other are loosing their energy by emitting positive
energy gravitons. As we show in Sec. \ref{binary}, the rate of the energy
loss by the binary system is given by the same quadrupole formula as
obtained in general relativity. Thus, vector gravity also passes the binary
pulsars tests.

For the Universe expansion the present theory gives the same answer as
general relativity with cosmological constant and zero spatial curvature.
However, zero spatial curvature of the Universe is a solution of our
equations, while in general relativity the spatial curvature is a free
parameter. Thus, vector theory of gravity does not have the Euclidicity
problem, that is why space is almost perfectly Euclidean on large scales.

Moreover, the vector theory of gravity solves the dark energy problem.
Namely, the theory yields, with no free parameters, the value of the
cosmological constant $\Omega _{\Lambda }=2/3\approx 0.67$\footnote{%
This value of $\Omega _{\Lambda }$ was obtained assuming that matter in the
Universe is nonrelativistic. However, small fraction of matter (e.g.
neutrinos) is relativistic which can slightly modify $\Omega _{\Lambda }$.
Measuring deviation of $\Omega _{\Lambda }$ from $2/3$ in future more
accurate observations could give us information about the amount of
relativistic matter.} which agrees with the recent Planck result $\Omega
_{\Lambda }=0.686\pm 0.02$ \cite{Planck14}. General relativity failed to
predict the value of $\Omega _{\Lambda }$, but the present vector theory of
gravity passes this cosmological test. This result is crucial since it
points to the vector nature of gravity rather than a tensor field.

\begin{figure}[t]
\centering
\includegraphics[width=0.52\textwidth]{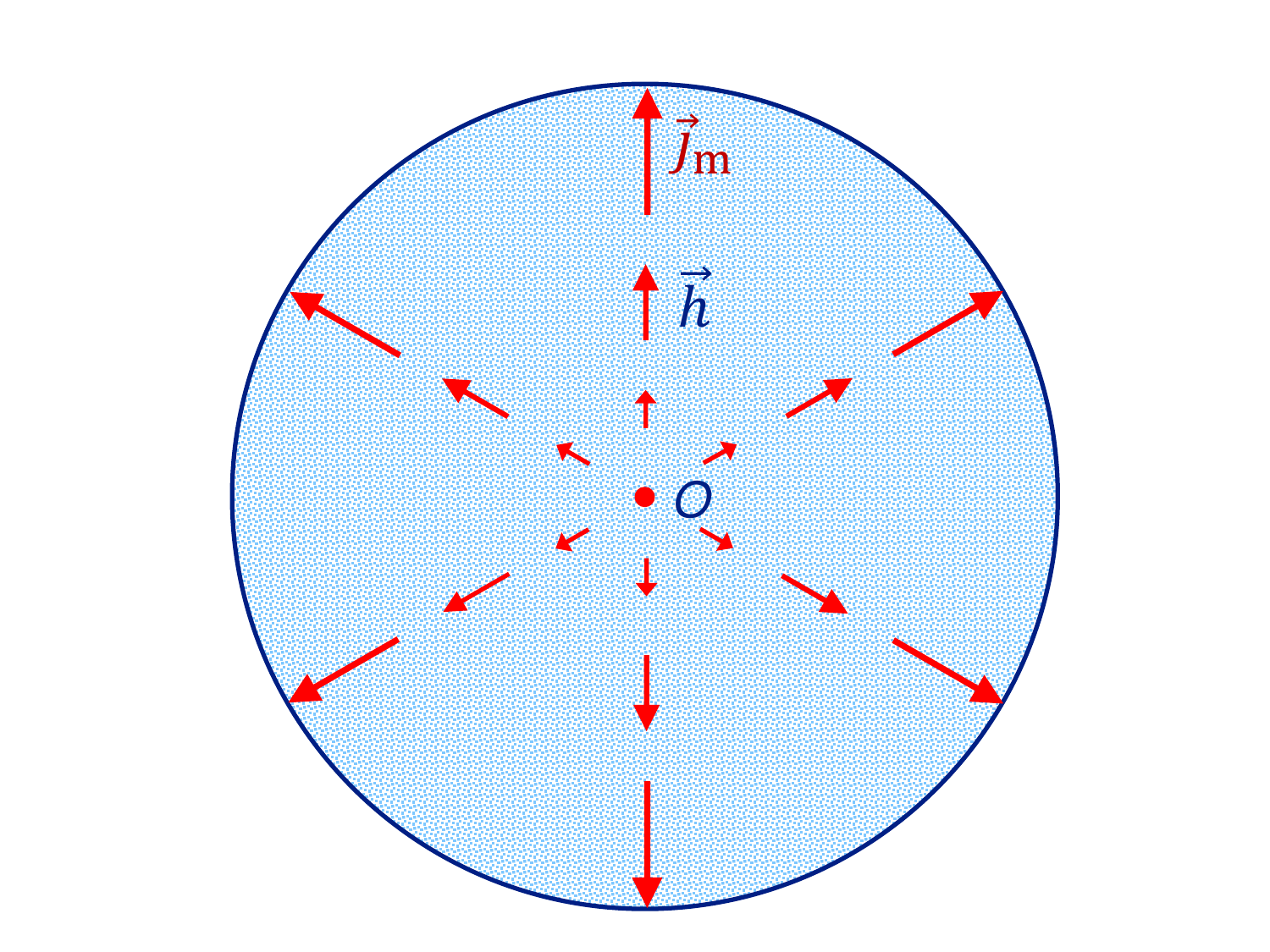}
\caption{Explanation of dark energy in vector gravity. Expansion of the
Universe generates matter current $\mathbf{j}_m$ directed away from an
observer $O$. Such current induces longitudinal gravitational field $\mathbf{%
h}=h^{0\protect\alpha }$ in a similar way as electric current creates vector
potential in classical electrodynamics. Spatial averaging of $\mathbf{h}$
over the local (shaded) region yields zero and, therefore, averaged metric
is spatially isotropic. However, average energy density associated with $%
\mathbf{h}$, $w_{h}=-c^{2}\mathbf{\dot{h}}^{2}/32\protect\pi G$, does not
vanish after spatial averaging. This energy is the mysterious dark energy.
Contrary to matter, it has negative energy density and accelerates expansion
of the Universe.}
\label{DarkEnergy}
\end{figure}

The present theory provides an explanation of the dark energy as the energy
of longitudinal (div $\mathbf{h}\neq 0$) gravitational field induced by the
Universe expansion (see Fig. \ref{DarkEnergy}). Namely, time variation of
the spatial scale caused by the Universe expansion produces matter current
directed away from an observer. Such current generates longitudinal part of
the vector gravitational field (similarly to generation of the vector
potential by the electric current in classical electrodynamics) which
possesses negative energy and accelerates expansion of the Universe\footnote{%
Universe expansion induces non radiative longitudinal gravitational field
which is not quantized. This is different from graviton which is a quantized
transverse field. Since graviton is a composite particle there are
constraints imposed by the Pauli exclusion principle on its generation.
However, for the classical longitudinal field there are no such constraints.}%
.

Since value of the current depends on a reference frame the value of the
cosmological constant $\Lambda $ depends on the time $t_{0}$ at which the
observer measures $\Lambda $. As we show in Sec. \ref{cosmology}, the value
of $\Lambda $ is given by $\Lambda =2\rho /a^{3}(t_{0})$ and the
contribution from the cosmological term to the expansion rate of the
Universe at time $t_{0}$ is twice larger than those of matter in any
reference frame. Therefore, according to our theory, Universe will expand
forever at a continually decelerating rate, with expansion asymptotically
approaching zero. This is what is expected for the flat Universe in absence
of exotic forms of energy.

Mathematically, the cosmological term appears in vector gravity as a result
of spatial averaging of the gravitational field equations. Namely, exact
spatially inhomogeneous equations yield that in the vicinity of time $t_{0}$
and $\mathbf{r}=0$ the solution for the equivalent metric is%
\begin{equation*}
f_{ik}=\left( 
\begin{array}{cccc}
1+h_{00} & h_{0x} & h_{0y} & h_{0z} \\ 
h_{0x} & -1+h_{00} & 0 & 0 \\ 
h_{0y} & 0 & -1+h_{00} & 0 \\ 
h_{0z} & 0 & 0 & -1+h_{00}%
\end{array}%
\right) ,
\end{equation*}%
where, according to Eq. (\ref{cos9}),%
\begin{equation}
h_{0\alpha }(t,\mathbf{r})=\frac{2}{c}\ddot{h}_{00}(t_{0})(t-t_{0})x^{\alpha
}.  \label{ev1}
\end{equation}%
That is $h_{0\alpha }$ is induced by the Universe expansion (more exactly by
the acceleration of expansion $\ddot{h}_{00}$). Spatial averaging of $%
h_{0\alpha }$ in the local region yields zero because $h_{0\alpha }$ is an
odd function of spatial coordinates $x^{\alpha }$. Therefore, the averaged
metric $f_{ik}$ is spatially isotropic. However, since $h_{0\alpha }$ enters
the evolution equation (\ref{cos5}) as divergence its contribution does not
vanish in the equation after spatial averaging and Eq. (\ref{cos5}) yields 
\begin{equation}
3\frac{\partial ^{2}}{\partial x^{0}\partial x^{0}}\left\langle
h_{00}\right\rangle -2\frac{\partial }{\partial x^{0}}\left\langle \frac{%
\partial h_{0\beta }}{\partial x^{\beta }}\right\rangle =\frac{8\pi G}{c^{4}}%
\left\langle T_{\text{now}}^{00}\right\rangle .  \label{ev2}
\end{equation}%
The second term in Eq. (\ref{ev2}) is the cosmological (dark energy) term
which, according to Eq. (\ref{ev1}), is equal to $-4\ddot{h}%
_{00}(t_{0})/c^{2}$. The dark energy term appears because Universe expansion
induces $h_{0\alpha }$ which itself affects Universe evolution. The value of
the cosmological constant $\Lambda $ in the effective nonlinear evolution
equation (\ref{qq1}) is determined by matching this equation with the local
evolution of the Universe in the vicinity of the observer's time $t_{0}$. As
a consequence, the value of $\Lambda $ depends on the average matter density
at time $t_{0}$, that is it depends on the observer's reference frame.

According to the vector gravity, the contents of the Universe are somewhat
different from those predicted by general relativity. The total energy
density $w$ of the Universe in the effective cosmological model is given by
Eq. (\ref{hd})%
\begin{equation}
w=-\frac{3c^{2}}{8\pi G}a^{2}\dot{a}^{2}+c^{2}\Lambda a^{2}+\frac{\rho c^{2}%
}{a},  \label{en1}
\end{equation}%
which is the energy density in the fixed background Euclidean space. The net
energy of the Universe is equal to zero ($w=0$), and positive energy of
matter is compensated by the negative energy of the gravitational field (see
Fig. \ref{Ucont}).

Introducing $X=a^{2}$ one can rewrite Eq. (\ref{en1}) as an equation of
energy conservation for a particle in an external potential $U(X)$%
\begin{equation*}
\frac{3}{32\pi G}\dot{X}^{2}+U(X)=\text{const},
\end{equation*}%
where 
\begin{equation}
U(X)=-\Lambda X-\frac{\rho }{\sqrt{X}}.  \label{en2}
\end{equation}%
The term with $\Lambda $ in $U(X)$ decreases with increasing $X$, while the
matter term increases. Thus, the cosmological $\Lambda -$term accelerates
expansion of the Universe while matter causes deceleration.

In the post-Newtonian limit the vector gravity gives the same answer as
general relativity. As we explain in Sec. \ref{PPN}, the reason for such a
coincidence is symmetry of the action $S_{\text{matter}}$. Namely, in the
post-Newtonian limit the symmetries of $S_{\text{matter}}$ coincide in both
theories. By construction of both theories, the symmetries of $S_{\text{%
matter}}$ uniquely determine the whole classical theory of gravity. Thus in
the post-Newtonian limit when symmetries of $S_{\text{matter}}$ coincide the
two theories become equivalent.

For strong field, vector gravity gives a very different result and yields no
singularities such as black holes. A defining characteristic of a black hole
is the event horizon. So far there were no observations of the event horizon
and, thus, a proof of black holes existence is lacking.

The current vector theory is not equivalent to general relativity even in
the weak field limit. For example, it predicts different polarization of
weak gravitational waves. As we show in Sec. \ref{quantization},
quantization of the vector gravitational field can be performed in a way
similar to the quantization of electromagnetic field in the composite photon
theory. Namely, gravitational field is decomposed into free field
(corresponding to radiation) which is quantized assuming that graviton is
composed of fermion-antifermion pairs and the residual non radiative part of
the field which remains classical. In particular, the post-Newtonian limit
as well as cosmological evolution of the Universe are described by the part
of the gravitational field which is not quantized. As a result, classical
field equations (\ref{sss1}) are applicable for these problems.

We show in Sec. \ref{quantization} that quantization of the free transverse
gravitational field for the filled vacuum yields quantum theory which is
equivalent to QED. At the moment of Big Bang the vacuum fermion states are
empty. This is the classical limit of quantum vector gravity which yields
classical evolution equations for the free gravitational field with negative
energy of gravitational waves. As we show in Appendix \ref{AP4}, classical
equations for the weak gravitational field are analogous to Maxwell's
equations in a medium with $\varepsilon =\mu =-1$. For the filled vacuum
(quantum limit) the classical equations (\ref{sss1}) no longer describe
radiation field. In this case the quantum mechanical treatment must be used
to describe evolution of the radiation (quantized) part of the gravitational
field. As we show in Sec. \ref{quantization}, quantum mechanical analysis
yields that for the filled vacuum the equations for the radiation field are
analogous to Maxwell's equations with $\varepsilon =\mu =1$ and the graviton
energy is positive.

The present theory, if confirmed, can also lead to a break through in the
problem of dark matter. Namely, the theory predicts that likely the
supermassive compact objects at galactic centers have non baryonic origin
and, thus, yet undiscovered dark matter particle is a likely ingredient of
their composition. As a result, observations of such objects can allow us to
predict the nature of dark matter. In the previous paper \cite{Svid07} we
showed that properties of compact objects at galactic centers can be
explained quantitatively assuming they are made of dark matter axions and
the axion mass is about $0.6$ meV. Analysis of Ref. \cite{Svid07} was based
on the exponential metric (\ref{s10}) for the static gravitational field
rather than general relativity. The present theory of gravity justifies our
previous use of the exponential metric.

The vector theory of gravity can be tested in several ways. For example, one
can examine gravity beyond the post-Newtonian limit in the solar system by
improving the accuracy of Shapiro time delay experiment (time delay of a
radar signal traveling near the Sun), or improving precision of the light
deflection measurements by placing an optical interferometer with
microarcsecond resolution into Earth orbit \cite{Reas88}. Vector gravity
differs from general relativity in the post-post-Newtonian regime which has
not been accurately tested to date.

Future detection of gravitational waves from binary mergers with improved
sensitivity or detection of merger events with louder signals might be able
to constrain the higher-order post-Newtonian parameters with a reasonable
accuracy \cite{Abbo16c}. One can also test the vector theory of gravity by
measuring propagation direction of gravitational waves relative to the
interferometer arms (see Sec. \ref{gwave}). Such measurement can be
performed by detecting a signal from the same event by several LIGO-like
interferometers \cite{Abbo16b} and is able to distinguish between vector
gravity and general relativity (see Fig. \ref{Arms} and Section \ref{GWtest}%
).

Another possibility is to resolve the supermassive object at the center of
our Galaxy with the Event Horizon Telescope. If general relativity is
correct we must see a steady shadow from a black hole. If the present theory
is right then shadow might appear and disappear periodically with a period
of about $20$ min as we predicted in \cite{Svid07}. Observation of such
oscillations will also provide evidence for dark matter axion with mass in
meV range.

Finally, we want to emphasize that vector gravity and general relativity are
constructed in a unique way and have no adjustable parameters. Thus, both
theories make fixed predictions and every new test of the theory is
potentially a deadly test. A verified discrepancy between observation and
prediction would kill the theory.

Despite fundamental differences, vector gravity and general relativity yield
for the experimentally tested regimes quantitatively very close predictions
which allowed both theories to hold up under extensive experimental
scrutiny. In order to determine whether gravitational field has a vector or
a tensor origin additional tests are required. Such tests are crucial for
understanding of our Universe.

In Section \ref{GWtest} we propose an experiment which can rule out one of
the two theories in the next few years. The test is based on measurement of
the gravitational wave propagation direction relative to the interferometer
arms using several instruments and can be performed in a joint run of the
LIGO-Virgo interferometer network. Such joint runs can start in 2017 when
Virgo instrument will reach the required sensitivity.

Although it is remarkable that general relativity (GR), born 100 years ago,
has managed to pass many unambiguous observational and experimental tests,
it actually has unwanted issues. For example,

\begin{itemize}
\item GR is not compatible with quantum mechanics.

\item GR can not explain why Universe is spatially flat. Models involving
cosmic inflation are needed to fix the problem. In contrast, in vector
gravity the spatially flat Universe comes out as a solution of equations.

\item GR does not provide a mechanism of matter generation at the Big Bang.
An additional field with negative energy, the inflaton, is required to
resolve the issue. In vector gravity the mechanism of matter generation is
part of the theory. No extra fields are necessary.

\item GR can not explain the value of the cosmological term. In contrast,
vector gravity predicts, with no free parameters, the value of the
cosmological constant that agrees with observations.

\item Existence of space-time singularities for which geometry is
ill-defined is a generic feature of GR. Schwarzschild solution describing a
static black hole is an example of a curvature singularity, where
geometrical quantities characterizing space-time curvature, such as the
Ricci scalar, take on infinite values. In GR such a singularity is
unavoidable once the gravitational collapse of an object with realistic
matter properties has proceeded beyond a certain stage. In contrast, stars
in vector gravity do not collapse into a singularity. In vector gravity
black holes do not exist and the end point of the gravitational collapse is
a stable star with a reduced mass.

\item Energy and momentum of the gravitational field in GR do not form a
tensor quantity under an arbitrary coordinate transformation. Recall that
conservation laws reflect a symmetry of the background space-time, namely
its homogeneity and isotropy. In GR, which identifies gravitational field
with the metric tensor $g_{\mu \nu }$, the real space is a space with
Riemannian geometry, and this does not admit symmetries corresponding to
displacements and rotations \cite{Logu87}. As a consequence, conservation
laws do not hold in GR. If we want to make a theory compatible with the
conservation laws we must postulate existence of a fixed symmetric
background geometry \cite{Logu87}. We do so in the present vector theory of
gravity.
\end{itemize}

The mentioned arguments point on the vector, rather than a tensor, nature of
the gravitational field.

I am very grateful to the Institute for Quantum Science and Engineering of
Texas A\&M University for providing opportunity and resources for conducting
research. This work was supported by the Office of Naval Research (Award No.
N00014-16-1-3054) and the Robert A. Welch Foundation (Award A-1261).


\appendix

\section{Derivation of the equivalent metric}

\label{AP1}

Metric tensor determines the line element for infinitesimal coordinate
displacement $dx^{k}$%
\begin{equation*}
ds^{2}=f_{ik}dx^{i}dx^{k}.
\end{equation*}

Let us consider vector field $A_{k}$. The most general form of the
equivalent metric $f_{ik}$ which can be constructed from the background
Euclidean metric $\delta _{ik}=$diag$(1,1,1,1)$ and $A_{k}$ is%
\begin{equation}
f_{ik}=-F\delta _{ik}+\left( F+G\right) \frac{A_{i}A_{k}}{A^{2}},  \label{m1}
\end{equation}%
where $F$ and $G$ are scalar functions of%
\begin{equation*}
A=\sqrt{A_{i}A_{k}\delta ^{ik}}.
\end{equation*}
If we chose $x^{0}$ axis along the direction of $A_{k}$ then $%
A_{k}=(A,0,0,0) $ and the equivalent metric is diagonal%
\begin{equation}
f_{ik}=\text{diag}(G,-F,-F,-F).  \label{m2}
\end{equation}

To find a relation between $F$ and $G$ one can consider a particular case of
gravitational field which has two nonzero components $A_{0}$ and $A_{x}$ and
apply Einstein equivalence principle. Let us assume that a test particle
moves along the $x-$axis under the influence of such field. Particle
velocity is a function of time $V=V(t)$ and the line element for the
particle in the gravitational field reads%
\begin{equation}
ds^{2}=f_{00}c^{2}dt^{2}+2f_{0x}cdtdx+f_{xx}dx^{2}.  \label{b0}
\end{equation}%
The same motion is obtained if the particle is at rest in Minkowski
space-time, but the reference frame moves with velocity $V(t)$. Making a
change of coordinate $x\rightarrow x+\int^{t}V(t^{\prime })dt^{\prime }$ in
the Minkowski line element%
\begin{equation*}
ds^{2}=c^{2}dt^{2}-dx^{2}
\end{equation*}%
we obtain that the interval in the moving frame is

\begin{equation}
ds^{2}=dt^{2}\left( c^{2}-V^{2}\right) -2Vdtdx-dx^{2}.  \label{b1}
\end{equation}%
According to the Einstein equivalence principal, the interval (\ref{b1})
must be equal to (\ref{b0}) which yields three equations%
\begin{equation}
f_{00}=1-V^{2}/c^{2},\quad f_{0x}=-V/c,\quad f_{xx}=-1.  \label{b2}
\end{equation}%
Taking into account Eq. (\ref{m1}) and $A_{0}^{2}+A_{x}^{2}=A^{2}$, Eqs. (%
\ref{b2}) give%
\begin{equation*}
F=\sqrt{\frac{V^{4}}{4c^{4}}+1}+\frac{V^{2}}{2c^{2}},\quad G=\sqrt{\frac{%
V^{4}}{4c^{4}}+1}-\frac{V^{2}}{2c^{2}},
\end{equation*}%
that is

\begin{equation}
GF=1.  \label{b3}
\end{equation}%
Equation (\ref{b3}) fixes the relation between $G$ and $F$ for arbitrary $A$.

One should note that normalization of $A_{k}$ is not unique. Namely, $A_{k}$
can be multiplied by an arbitrary scalar function of $A$ which yields
another vector field. Our theory is independent of the field normalization
and Eq. (\ref{b3}) is the only constraint on the equivalent metric we have.
For example, one can choose norm of $A_{k}$ such that 
\begin{equation}
G=A.  \label{norm}
\end{equation}%
Then equivalent metric $f_{ik}$ is given by%
\begin{equation}
f_{ik}=-\frac{\delta _{ik}}{A}+\left( A+\frac{1}{A}\right) \frac{A_{i}A_{k}}{%
A^{2}}.  \label{m4}
\end{equation}%
Metric $\tilde{f}^{ik}$ inverse to $f_{ik}$, defined as $\tilde{f}%
^{ik}f_{im}=\delta _{m}^{k}$, reads 
\begin{equation}
\tilde{f}^{ik}=-A\delta ^{ik}+\left( A+\frac{1}{A}\right) \frac{A^{i}A^{k}}{%
A^{2}},  \label{m5}
\end{equation}%
where $A^{i}=\delta ^{ik}A_{k}$ and the following relationships are
satisfied 
\begin{equation}
A_{i}A_{m}\tilde{f}^{im}=A,\quad A_{k}\tilde{f}^{ik}=\frac{A^{i}}{A},\quad 
\sqrt{-f}=\frac{1}{A},  \label{m6}
\end{equation}%
where $f=$det$(f_{ik})$. In Cartesian coordinate system if we chose $x^{0}$
axis along the direction of $A_{k}$ the equivalent metric reads%
\begin{equation}
f_{ik}=\text{diag}\left( A,-\frac{1}{A},-\frac{1}{A},-\frac{1}{A}\right) .
\label{eqm1}
\end{equation}

\section{Derivation of gravitational field action}

\label{AP2}

\subsection{Weak field limit}

First we find gravitational field action for small deviation of the
gravitational field from a constant value $\phi =\phi _{0}$ and $%
u_{k}=(1,0,0,0)$. For small deviation the equivalent metric is%
\begin{equation}
f_{ik}=\eta _{ik}+\left( 
\begin{array}{cccc}
h_{00} & h_{01} & h_{02} & h_{03} \\ 
h_{01} & h_{00} & 0 & 0 \\ 
h_{02} & 0 & h_{00} & 0 \\ 
h_{03} & 0 & 0 & h_{00}%
\end{array}%
\right) ,  \label{wm}
\end{equation}%
where $|h_{0k}|\ll 1$. One can obtain the weak field action for the
gravitational field from the requirement that the action must be invariant
under the gauge transformation (\ref{gg1})%
\begin{equation*}
h_{00}\rightarrow h_{00}+2\frac{\partial \psi }{\partial x^{0}},\quad
h_{0\alpha }\rightarrow h_{0\alpha }+\frac{\partial \psi }{\partial
x^{\alpha }}
\end{equation*}%
upto the $V^{2}/c^{2}$ order. One can look for the weak-field action as a
combination of gauge invariant terms. Introducing%
\begin{equation*}
B_{0}=\frac{h_{00}}{2},\quad B_{\alpha }=h_{0\alpha }
\end{equation*}%
and a guage-invariant combination%
\begin{equation*}
F_{ik}=\frac{\partial B_{k}}{\partial x^{i}}-\frac{\partial B_{i}}{\partial
x^{k}}
\end{equation*}%
the general form of the action which is gauge invariant upto the $%
V^{2}/c^{2} $ order is%
\begin{equation*}
S_{\text{gravity}}=C\int d^{4}x\left[ F^{ik}F_{ik}+C_{1}\frac{\partial h_{00}%
}{\partial x^{0}}\frac{\partial h_{00}}{\partial x^{0}}\right.
\end{equation*}%
\begin{equation}
+\left. C_{2}\left( \frac{\partial h_{0\alpha }}{\partial x^{0}}\frac{%
\partial h_{00}}{\partial x^{\alpha }}-\frac{1}{4}\frac{\partial h_{00}}{%
\partial x^{\alpha }}\frac{\partial h_{00}}{\partial x^{\alpha }}\right) %
\right] ,  \label{ppp1}
\end{equation}%
where terms with $C_{1}$ and $C_{2}$ break gauge invariance in the $%
V^{3}/c^{3}$ order. One can find coefficients $C_{1}$ and $C_{2}$ from the
requirement of the low-velocity Lorentz invariance. Straightforward
calculations yield that action (\ref{ppp1}) 
\begin{equation*}
S_{\text{gravity}}=C\int d^{4}x\left[ -\frac{1}{2}\left( 1+\frac{C_{2}}{2}%
\right) \frac{\partial h_{00}}{\partial x^{\alpha }}\frac{\partial h_{00}}{%
\partial x^{\alpha }}\right.
\end{equation*}%
\begin{equation*}
-2\frac{\partial h_{0\alpha }}{\partial x^{0}}\frac{\partial h_{0\alpha }}{%
\partial x^{0}}+2\frac{\partial h_{0\alpha }}{\partial x^{\beta }}\frac{%
\partial h_{0\alpha }}{\partial x^{\beta }}-2\frac{\partial h_{0\alpha }}{%
\partial x^{\beta }}\frac{\partial h_{0\beta }}{\partial x^{\alpha }}
\end{equation*}%
\begin{equation}
\left. +\left( 2+C_{2}\right) \frac{\partial h_{0\alpha }}{\partial x^{0}}%
\frac{\partial h_{00}}{\partial x^{\alpha }}+C_{1}\frac{\partial h_{00}}{%
\partial x^{0}}\frac{\partial h_{00}}{\partial x^{0}}\right]  \label{p1a}
\end{equation}%
is invariant under transformations (\ref{gg2}) and (\ref{gg3}) provided%
\begin{equation*}
C_{1}=-6,\quad C_{2}=6.
\end{equation*}%
Indeed, under transformation%
\begin{equation}
\frac{\partial }{\partial x^{0}}\rightarrow \frac{\partial }{\partial x^{0}}-%
\frac{\mathbf{V}}{c}\nabla ,\quad \frac{\partial }{\partial \mathbf{r}}%
\rightarrow \frac{\partial }{\partial \mathbf{r}}-\frac{\mathbf{V}}{c}\frac{%
\partial }{\partial x^{0}}+\frac{\mathbf{V}}{2c^{2}}\left( \mathbf{V}\frac{%
\partial }{\partial \mathbf{r}}\right)
\end{equation}%
\begin{equation}
h_{00}\rightarrow h_{00}\left( 1+\frac{2V^{2}}{c^{2}}\right) -2\frac{%
V^{\alpha }}{c}h_{0\alpha },\quad h_{0\alpha }\rightarrow h_{0\alpha }-2%
\frac{V^{\alpha }}{c}h_{00},
\end{equation}%
the action (\ref{p1a}) transforms as (we keep terms upto $V^{3}/c^{3}$ order)%
\begin{equation*}
S_{\text{gravity}}\rightarrow S_{\text{gravity}}+C\int d^{4}x\left[ -\left(
C_{2}-6\right) \frac{V^{2}}{c^{2}}\left( \frac{\partial h_{00}}{\partial
x^{\alpha }}\right) ^{2}\right.
\end{equation*}%
\begin{equation*}
-\left( 2C_{1}+\frac{3}{2}C_{2}+3\right) \frac{V^{\alpha }}{c}\frac{\partial
h_{00}}{\partial x^{\alpha }}\frac{\partial h_{00}}{\partial x^{0}}+\left(
C_{2}-6\right) \frac{V^{\beta }}{c}\frac{\partial h_{00}}{\partial x^{\alpha
}}\frac{\partial h_{0\beta }}{\partial x^{\alpha }}
\end{equation*}%
\begin{equation*}
\left. +\left( 6-C_{2}\right) \frac{V^{\beta }}{c}\frac{\partial h_{00}}{%
\partial x^{\alpha }}\frac{\partial h_{0\alpha }}{\partial x^{\beta }}+\left[
C_{1}+\frac{7}{4}C_{2}-\frac{9}{2}\right] \left( \frac{\mathbf{V}}{c}\nabla
h_{00}\right) ^{2}\right] .
\end{equation*}%
Thus if $C_{1}=-6$ and $C_{2}=6$ the action is invariant.

The overall constant factor $C$ in Eq. (\ref{p1a}) is obtained by matching
the action (\ref{p1a}) with the Newtonian limit. The factor $C$ must be
independent of the background field $\phi _{0}$ because such independence is
one of the symmetries of $S_{\text{matter}}$. The final expression for the
action of weak gravitational field reads%
\begin{equation*}
S_{\text{gravity}}=\frac{c^{3}}{32\pi G}\int d^{4}x\left( -\frac{\partial
h_{00}}{\partial x^{\alpha }}\frac{\partial h_{00}}{\partial x^{\alpha }}-%
\frac{\partial h_{0\alpha }}{\partial x^{0}}\frac{\partial h_{0\alpha }}{%
\partial x^{0}}+\right.
\end{equation*}%
\begin{equation}
\left. \frac{\partial h_{0\alpha }}{\partial x^{\beta }}\frac{\partial
h_{0\alpha }}{\partial x^{\beta }}-\frac{\partial h_{0\alpha }}{\partial
x^{\beta }}\frac{\partial h_{0\beta }}{\partial x^{\alpha }}+4\frac{\partial
h_{0\alpha }}{\partial x^{0}}\frac{\partial h_{00}}{\partial x^{\alpha }}-3%
\frac{\partial h_{00}}{\partial x^{0}}\frac{\partial h_{00}}{\partial x^{0}}%
\right) ,  \label{p2}
\end{equation}%
where $G$ is the gravitational constant.

\subsection{Arbitrary gravitational field}

The most general form of the gravitational field action that can be
constructed from the scalar $\phi $ and the unit vector $u_{k}$ in four
dimensional Euclidean space $\delta _{ik}$ is%
\begin{equation*}
S_{\text{gravity}}=\frac{c^{3}}{64\pi G}\int d^{4}x\left[ \frac{\partial
\phi }{\partial x^{i}}\frac{\partial \phi }{\partial x^{k}}\left( F_{1}(\phi
)\delta ^{ik}+F_{2}(\phi )u^{i}u^{k}\right) \right.
\end{equation*}%
\begin{equation*}
+\frac{\partial u_{i}}{\partial x^{k}}\frac{\partial u_{m}}{\partial x^{l}}%
\left( F_{3}(\phi )\delta ^{im}u^{k}u^{l}+F_{4}(\phi )\delta ^{im}\delta
^{kl}+F_{5}(\phi )\delta ^{il}\delta ^{km}\right)
\end{equation*}%
\begin{equation}
\left. +F_{6}(\phi )\frac{\partial \phi }{\partial x^{i}}\frac{\partial u_{m}%
}{\partial x^{k}}\delta ^{im}u^{k}+F_{7}(\phi )\frac{\partial \phi }{%
\partial x^{k}}\frac{\partial u_{m}}{\partial x^{i}}\delta ^{im}u^{k}\right]
.  \label{GA}
\end{equation}%
The unknown functions $F_{1}$, \ldots , $F_{5}$ can be obtained by matching
the action with the weak field limit. For small deviations of $\phi $ from a
constant $\phi _{0}$ and $u_{k}$ from $(1,0,0,0)$ the equivalent metric reads%
\begin{equation*}
f_{ik}=\left( 
\begin{array}{cccc}
e^{2\phi _{0}} & 0 & 0 & 0 \\ 
0 & -e^{-2\phi _{0}} & 0 & 0 \\ 
0 & 0 & -e^{-2\phi _{0}} & 0 \\ 
0 & 0 & 0 & -e^{-2\phi _{0}}%
\end{array}%
\right) +2\cosh (2\phi _{0})\times
\end{equation*}%
\begin{equation*}
\left( 
\begin{array}{cccc}
0 & u_{1} & u_{2} & u_{3} \\ 
u_{1} & 0 & 0 & 0 \\ 
u_{2} & 0 & 0 & 0 \\ 
u_{3} & 0 & 0 & 0%
\end{array}%
\right) +2\delta \phi \left( 
\begin{array}{cccc}
e^{2\phi _{0}} & 0 & 0 & 0 \\ 
0 & e^{-2\phi _{0}} & 0 & 0 \\ 
0 & 0 & e^{-2\phi _{0}} & 0 \\ 
0 & 0 & 0 & e^{-2\phi _{0}}%
\end{array}%
\right) ,
\end{equation*}%
while action (\ref{GA}) reduces to

\begin{equation*}
S_{\text{gravity}}=\frac{c^{3}}{64\pi G}\int d^{4}x\left[ \left(
F_{1}+F_{2}\right) \frac{\partial \phi }{\partial x^{0}}\frac{\partial \phi 
}{\partial x^{0}}\right.
\end{equation*}%
\begin{equation*}
+F_{1}\frac{\partial \phi }{\partial x^{\alpha }}\frac{\partial \phi }{%
\partial x^{\alpha }}+\left( F_{3}+F_{4}\right) \frac{\partial u_{\alpha }}{%
\partial x^{0}}\frac{\partial u_{\alpha }}{\partial x^{0}}
\end{equation*}%
\begin{equation}
\left. +F_{4}\frac{\partial u_{\alpha }}{\partial x^{\beta }}\frac{\partial
u_{\alpha }}{\partial x^{\beta }}+F_{5}\frac{\partial u_{\alpha }}{\partial
x^{\beta }}\frac{\partial u_{\beta }}{\partial x^{\alpha }}+(F_{6}+F_{7})%
\frac{\partial \phi }{\partial x^{\alpha }}\frac{\partial u_{\alpha }}{%
\partial x^{0}}\right] ,  \label{GA1}
\end{equation}%
where $F_{1}$, \ldots , $F_{7}$ are taken at $\phi =\phi _{0}$. In the
rescaled coordinates%
\begin{equation*}
x^{0}\rightarrow e^{-\phi _{0}}x^{0},\qquad x^{\alpha }\rightarrow e^{\phi
_{0}}x^{\alpha }
\end{equation*}%
the action (\ref{GA1}) reads%
\begin{equation*}
S_{\text{gravity}}=\frac{c^{3}}{64\pi G}\int d^{4}x\left[ \left(
F_{1}+F_{2}\right) e^{4\phi _{0}}\frac{\partial \phi }{\partial x^{0}}\frac{%
\partial \phi }{\partial x^{0}}\right.
\end{equation*}%
\begin{equation*}
+F_{1}\frac{\partial \phi }{\partial x^{\alpha }}\frac{\partial \phi }{%
\partial x^{\alpha }}+\left( F_{3}+F_{4}\right) e^{4\phi _{0}}\frac{\partial
u_{\alpha }}{\partial x^{0}}\frac{\partial u_{\alpha }}{\partial x^{0}}
\end{equation*}%
\begin{equation}
\left. +F_{4}\frac{\partial u_{\alpha }}{\partial x^{\beta }}\frac{\partial
u_{\alpha }}{\partial x^{\beta }}+F_{5}\frac{\partial u_{\alpha }}{\partial
x^{\beta }}\frac{\partial u_{\beta }}{\partial x^{\alpha }}%
+(F_{6}+F_{7})e^{2\phi _{0}}\frac{\partial \phi }{\partial x^{\alpha }}\frac{%
\partial u_{\alpha }}{\partial x^{0}}\right]  \label{a1}
\end{equation}%
and the equivalent metric is%
\begin{equation*}
f_{ik}=\eta _{ik}+2\delta \phi \delta _{ik}+2\cosh (2\phi _{0})\left( 
\begin{array}{cccc}
0 & u_{1} & u_{2} & u_{3} \\ 
u_{1} & 0 & 0 & 0 \\ 
u_{2} & 0 & 0 & 0 \\ 
u_{3} & 0 & 0 & 0%
\end{array}%
\right) .
\end{equation*}%
Thus in Eq. (\ref{wm})%
\begin{equation*}
h_{00}=2\delta \phi ,\quad h_{0\alpha }=2\cosh (2\phi _{0})u_{\alpha }\text{.%
}
\end{equation*}%
In terms of $h_{00}$ and $h_{0\alpha }$ the action (\ref{a1}) reads%
\begin{equation*}
S_{\text{gravity}}=\frac{c^{3}}{64\pi G}\int d^{4}x\left[ \frac{F_{1}}{4}%
\frac{\partial h_{00}}{\partial x^{\alpha }}\frac{\partial h_{00}}{\partial
x^{\alpha }}+\right.
\end{equation*}%
\begin{equation*}
\frac{\left( F_{3}+F_{4}\right) e^{4\phi _{0}}}{4\cosh ^{2}(2\phi _{0})}%
\frac{\partial h_{0\alpha }}{\partial x^{0}}\frac{\partial h_{0\alpha }}{%
\partial x^{0}}+\frac{F_{4}}{4\cosh ^{2}(2\phi _{0})}\frac{\partial
h_{0\alpha }}{\partial x^{\beta }}\frac{\partial h_{0\alpha }}{\partial
x^{\beta }}
\end{equation*}%
\begin{equation*}
+\frac{F_{5}}{4\cosh ^{2}(2\phi _{0})}\frac{\partial h_{0\alpha }}{\partial
x^{\beta }}\frac{\partial h_{0\beta }}{\partial x^{\alpha }}+\frac{%
(F_{6}+F_{7})e^{2\phi _{0}}}{4\cosh (2\phi _{0})}\frac{\partial h_{0\alpha }%
}{\partial x^{0}}\frac{\partial h_{00}}{\partial x^{\alpha }}
\end{equation*}%
\begin{equation*}
\left. +\frac{1}{4}\left( F_{1}+F_{2}\right) e^{4\phi _{0}}\frac{\partial
h_{00}}{\partial x^{0}}\frac{\partial h_{00}}{\partial x^{0}}\right] .
\end{equation*}%
Matching this with the weak field limit action (\ref{p2}) we obtain%
\begin{equation}
F_{1}=-8,\quad F_{2}=8-24e^{-4\phi },\quad F_{3}=-16e^{-2\phi }\cosh
^{3}(2\phi ),  \label{F1}
\end{equation}%
\begin{equation}
F_{4}=8\cosh ^{2}(2\phi ),\quad F_{5}=-8\cosh ^{2}(2\phi ),  \label{F2}
\end{equation}%
\begin{equation}
F_{6}+F_{7}=32e^{-2\phi }\cosh (2\phi ).  \label{F3}
\end{equation}

The functions $F_{6}(\phi )$ and $F_{7}(\phi )$ yet remain undetermined,
only their sum. In order to find these functions we need to investigate
symmetries of the action in the higher order in the post-Newtonian expansion
parameter $\epsilon $. The \textquotedblleft order of
smallness\textquotedblright\ is determined according to the rules that
matter velocity is of order $V\sim \epsilon ^{1/2}$ and gravitational
constant $G\sim \epsilon $. Making change of functions 
\begin{equation}
h_{0\alpha }=2\cosh (2\phi )u_{\alpha },  \label{nf1}
\end{equation}%
\begin{equation*}
e^{2\Phi }=e^{2\phi }-2\cosh (2\phi )u_{\alpha }^{2},
\end{equation*}%
or 
\begin{equation}
\Phi \approx \phi -\frac{h_{0\alpha }^{2}}{2\left( 1+e^{4\phi }\right) }
\label{nf2}
\end{equation}%
and taking into account that $h_{0\alpha }\sim \epsilon ^{3/2}$ and $%
dx^{\alpha }/dx^{0}\sim \epsilon ^{1/2}$ we obtain the following expression
for the square of the interval upto the terms of the $\epsilon ^{3}$ order%
\begin{equation}
ds^{2}=e^{2\Phi }(dx^{0})^{2}+2h_{0\alpha }dx^{0}dx^{\alpha }-e^{-2\Phi }(d%
\mathbf{r})^{2}.  \label{int0}
\end{equation}%
Interval (\ref{int0}), and hence $S_{\text{matter}}$, is invariant under
transformation 
\begin{equation}
x^{0}\rightarrow e^{-a}x^{0},\qquad x^{\alpha }\rightarrow e^{a}x^{\alpha },
\label{tr1}
\end{equation}%
\begin{equation}
\Phi \rightarrow \Phi +a,\qquad h_{0\alpha }\rightarrow h_{0\alpha },
\label{tr2}
\end{equation}%
where $a$ is an arbitrary constant. Therefore, action for the gravitational
field $S_{\text{gravity}}$ must also possess such symmetry in the $\epsilon
^{3}$ order. Taking into account Eqs. (\ref{F1})-(\ref{F3}) and making
change of functions (\ref{nf1}) and (\ref{nf2}) we obtain that $S_{\text{%
gravity}}$ upto the terms of the $\epsilon ^{3}$ order reads 
\begin{equation*}
S_{\text{gravity}}=\frac{c^{3}}{8\pi G}\int d^{4}x\left[ -\frac{\partial
\Phi }{\partial x^{\alpha }}\frac{\partial \Phi }{\partial x^{\alpha }}%
-3e^{-4\Phi }\frac{\partial \Phi }{\partial x^{0}}\frac{\partial \Phi }{%
\partial x^{0}}+\right.
\end{equation*}%
\begin{equation*}
\frac{1}{4}\left( \frac{\partial h_{0\alpha }}{\partial x^{\beta }}\right)
^{2}-\frac{1}{4}\frac{\partial h_{0\alpha }}{\partial x^{\beta }}\frac{%
\partial h_{0\beta }}{\partial x^{\alpha }}+2e^{-2\Phi }\frac{\partial \Phi 
}{\partial x^{\alpha }}\frac{\partial h_{0\alpha }}{\partial x^{0}}+
\end{equation*}%
\begin{equation*}
h_{0\beta }\frac{\partial \Phi }{\partial x^{\alpha }}\frac{\partial
h_{0\alpha }}{\partial x^{\beta }}-h_{0\alpha }\frac{\partial h_{0\alpha }}{%
\partial x^{\beta }}\frac{\partial \Phi }{\partial x^{\beta }}-2e^{-2\Phi
}h_{0\alpha }\frac{\partial \Phi }{\partial x^{\alpha }}\frac{\partial \Phi 
}{\partial x^{0}}-
\end{equation*}%
\begin{equation}
\left. \frac{e^{-4\Phi }}{4}\frac{\partial h_{0\alpha }}{\partial x^{0}}%
\frac{\partial h_{0\alpha }}{\partial x^{0}}+\frac{F_{7}(\Phi )h_{0\beta }}{%
2\cosh ^{2}(2\Phi )}\left( \frac{\partial \Phi }{\partial x^{\beta }}\frac{%
\partial h_{0\alpha }}{\partial x^{\alpha }}-\frac{\partial \Phi }{\partial
x^{\alpha }}\frac{\partial h_{0\alpha }}{\partial x^{\beta }}\right) \right]
.  \label{ac1}
\end{equation}%
Please note that there is a factor $1/G$ in front of the integral and,
hence, expression under the integral must be calculated upto the $\epsilon
^{4}$ order. Action (\ref{ac1}) is invariant under transformation (\ref{tr1}%
) and (\ref{tr2}) provided 
\begin{equation*}
F_{7}(\Phi )=2F\cosh ^{2}(2\Phi ),
\end{equation*}%
where $F$ is a constant independent of $\Phi $. To find $F$ we ought to dig
symmetries deeper.

Let us consider stationary gravitational field for which equivalent metric
is independent of time and make the following gauge transformation%
\begin{equation}
h_{0\alpha }\rightarrow h_{0\alpha }+e^{2\Phi }\frac{\partial \psi }{%
\partial x^{\alpha }},  \label{gt100}
\end{equation}%
where $\psi \sim \epsilon ^{3/2}$ is a function of spatial coordinates.
Taking into account that

\begin{equation*}
\delta S_{\text{matter}}=-\frac{1}{2c}\int d^{4}x\sqrt{-f}T^{ik}\delta
f_{ik},
\end{equation*}%
where $T^{ik}$ is the energy-momentum tensor of matter, we obtain that under
the gauge transformation (\ref{gt100}) the action $S_{\text{matter}}$
transforms as 
\begin{equation*}
S_{\text{matter}}\rightarrow S_{\text{matter}}-\frac{1}{c}\int d^{4}x\sqrt{-f%
}T^{0\alpha }e^{2\Phi }\frac{\partial \psi }{\partial x^{\alpha }}.
\end{equation*}%
Using $\sqrt{-f}=e^{-2\Phi }$ and integrating by parts we find%
\begin{equation}
S_{\text{matter}}\rightarrow S_{\text{matter}}+\frac{1}{c}\int d^{4}x\frac{%
\partial T^{0\alpha }}{\partial x^{\alpha }}\psi .  \label{tr3}
\end{equation}%
Conservation equation $T_{i;k}^{k}=0$ yields 
\begin{equation*}
\frac{\partial }{\partial x^{k}}\left( \sqrt{-f}T_{i}^{k}\right) =\frac{%
\sqrt{-f}}{2}T^{kl}\frac{\partial f_{kl}}{\partial x^{i}}
\end{equation*}%
which for stationary field reduces to 
\begin{equation*}
\frac{\partial }{\partial x^{\alpha }}\left( \sqrt{-f}T_{0}^{\alpha }\right)
=0.
\end{equation*}%
Keeping in mind that $T_{0}^{\alpha }=f_{00}T^{0\alpha }+f_{0\beta }T^{\beta
\alpha }$ and $f_{00}=e^{2\Phi }$ we find%
\begin{equation}
\frac{\partial T^{0\alpha }}{\partial x^{\alpha }}=-\frac{\partial }{%
\partial x^{\alpha }}\left[ e^{-2\Phi }h_{0\beta }T^{\beta \alpha }\right]
\sim \epsilon ^{5/2}.  \label{vvv1}
\end{equation}%
Therefore, the last term in Eq. (\ref{tr3}) is of the order of $\epsilon
^{4} $ and, thus, action $S_{\text{matter}}$ is invariant under gauge
transformation (\ref{gt100}) in the $\epsilon ^{3}$ order for stationary
field. Now we apply gauge transformation (\ref{gt100}) to the action (\ref%
{ac1}). Keeping terms upto the $\epsilon ^{3}$ order we obtain that for
stationary field $S_{\text{gravity}}$ transforms as 
\begin{equation*}
S_{\text{gravity}}\rightarrow S_{\text{gravity}}+F\frac{c^{3}}{8\pi G}\int
d^{4}xe^{2\Phi }\frac{\partial \Phi }{\partial x^{\beta }}\left[ h_{0\beta }%
\frac{\partial ^{2}\psi }{\partial x^{\alpha }\partial x^{\alpha }}\right.
\end{equation*}%
\begin{equation*}
-h_{0\alpha }\frac{\partial ^{2}\psi }{\partial x^{\beta }\partial x^{\alpha
}}+\frac{\partial \psi }{\partial x^{\beta }}\frac{\partial h_{0\alpha }}{%
\partial x^{\alpha }}-\frac{\partial \psi }{\partial x^{\alpha }}\frac{%
\partial h_{0\beta }}{\partial x^{\alpha }}
\end{equation*}%
\begin{equation*}
\left. +e^{2\Phi }\left( \frac{\partial \psi }{\partial x^{\beta }}\frac{%
\partial ^{2}\psi }{\partial x^{\alpha }\partial x^{\alpha }}-\frac{\partial
\psi }{\partial x^{\alpha }}\frac{\partial ^{2}\psi }{\partial x^{\beta
}\partial x^{\alpha }}\right) \right] .
\end{equation*}%
Hence, $S_{\text{gravity}}$ is gauge invariant provided $F=0$, that is $%
F_{7}(\phi )=0$ and, according to Eq. (\ref{F3}), 
\begin{equation}
F_{6}=32e^{-2\phi }\cosh (2\phi ).
\end{equation}
Now all functions in the action (\ref{GA}) are uniquely determined.

The final expression for the gravitational field action in Euclidean space is%
\begin{equation*}
S_{\text{gravity}}=\frac{c^{3}}{8\pi G}\int d^{4}x\left[ \frac{\partial \phi 
}{\partial x^{i}}\frac{\partial \phi }{\partial x^{k}}\left( -\delta
^{ik}+\left( 1-3e^{-4\phi }\right) u^{i}u^{k}\right) \right.
\end{equation*}%
\begin{equation*}
+\cosh ^{2}(2\phi )\frac{\partial u_{i}}{\partial x^{k}}\frac{\partial u_{m}%
}{\partial x^{l}}\Big(\delta ^{im}\delta ^{kl}-\delta ^{il}\delta ^{km}-
\end{equation*}%
\begin{equation*}
\left. \left( 1+e^{-4\phi }\right) \delta ^{im}u^{k}u^{l}\Big)+2\left(
1+e^{-4\phi }\right) \frac{\partial \phi }{\partial x^{i}}\frac{\partial
u_{m}}{\partial x^{k}}\delta ^{im}u^{k}\right] .
\end{equation*}

\section{Equations for gravitational field}

\label{AP6}

Here we sketch how to derive equations for the gravitational field. The
Lagrangian density has two parts%
\begin{equation*}
L=L_{\text{g}}+L_{\text{matter}},
\end{equation*}%
where $L_{\text{matter}}$ depends on the gravitational field via the
equivalent metric $f_{ik}$, while $L_{\text{g}}$ depends on the field
explicitly. We treat $\phi $ and $u_{\alpha }$ ($\alpha =1,$ $2,$ $3$) as
independent functions. Then $u_{0}^{2}=1-u_{1}^{2}-u_{2}^{2}-u_{3}^{2}$.
Equations for the gravitational field are obtained by taking variation of
the action with respect to $\phi $ and $u_{\alpha }$%
\begin{equation}
\frac{\partial L_{\text{g}}}{\partial \phi }+W^{ik}\frac{\partial f_{ik}}{%
\partial \phi }=0,  \label{z1}
\end{equation}%
\begin{equation}
\frac{\partial L_{\text{g}}}{\partial u_{\alpha }}-\frac{\partial L_{\text{g}%
}}{\partial u_{0}}\frac{u^{\alpha }}{u_{0}}+W^{ik}\left( \frac{\partial
f_{ik}}{\partial u_{\alpha }}-\frac{\partial f_{ik}}{\partial u_{0}}\frac{%
u^{\alpha }}{u_{0}}\right) =0,  \label{z2}
\end{equation}%
where%
\begin{equation*}
W^{ik}=\frac{\partial L_{\text{matter}}}{\partial f_{ik}}
\end{equation*}%
and we used%
\begin{equation*}
\frac{\partial u_{0}}{\partial u_{\alpha }}=-\frac{u^{\alpha }}{u_{0}}.
\end{equation*}
Variational derivatives of $L_{\text{g}}$ in Eqs. (\ref{z1}) and (\ref{z2})
deal with derivatives of functions in $L_{\text{g}}$ in a usual way. Taking
into account that \cite{Land95}%
\begin{equation*}
\delta S_{\text{matter}}=-\frac{1}{2c}\int d^{4}x\sqrt{-f}T^{ik}\delta f_{ik}
\end{equation*}
we obtain%
\begin{equation*}
W^{ik}=-\frac{1}{2}\sqrt{-f}T^{ik}.
\end{equation*}

Using 
\begin{equation*}
f_{ik}=-e^{-2\phi }\delta _{ik}+2\cosh (2\phi )u_{i}u_{k},
\end{equation*}%
we find%
\begin{equation*}
\frac{\partial f_{ik}}{\partial \phi }=2e^{-2\phi }\delta _{ik}+4\sinh
(2\phi )u_{i}u_{k}=-2f_{ik}+4e^{2\phi }u_{i}u_{k},
\end{equation*}%
\begin{equation*}
\frac{\partial f_{ik}}{\partial u_{m}}=2\cosh (2\phi )\left( \delta
_{i}^{m}u_{k}+\delta _{k}^{m}u_{i}\right) .
\end{equation*}%
Plugging this in Eqs. (\ref{z1}) and (\ref{z2}) yields%
\begin{equation}
\frac{\partial L_{\text{g}}}{\partial \phi }-2W+4e^{2\phi
}W^{ik}u_{i}u_{k}=0,  \label{z3}
\end{equation}%
\begin{equation}
\frac{\partial L_{\text{g}}}{\partial u_{\alpha }}-\frac{\partial L_{\text{g}%
}}{\partial u_{0}}\frac{u^{\alpha }}{u_{0}}+4\cosh (2\phi )\left( W^{\alpha
k}u_{k}-W^{0k}u_{k}\frac{u^{\alpha }}{u_{0}}\right) =0,  \label{z4}
\end{equation}%
where 
\begin{equation*}
W=W^{ik}f_{ik}.
\end{equation*}

Equations (\ref{z3}) and (\ref{z4}) can be written in the form%
\begin{equation}
\left[ W^{ik}-F\tilde{f}^{ik}\right] u_{k}-B^{i}=0,  \label{z5}
\end{equation}%
where $F$ is a scalar and $B^{i}$ is a vector which we find next. Equation (%
\ref{z5}) gives%
\begin{equation}
W^{ik}u_{k}=Fe^{-2\phi }u^{i}+B^{i},  \label{z6}
\end{equation}%
\begin{equation}
W^{ik}u_{i}u_{k}=Fe^{-2\phi }+B^{i}u_{i},  \label{z7}
\end{equation}%
where we used%
\begin{equation*}
\tilde{f}^{ik}u_{k}=e^{-2\phi }u^{i},
\end{equation*}%
\begin{equation*}
\tilde{f}^{ik}u_{i}u_{k}=e^{-2\phi }.
\end{equation*}%
Substitution of Eqs. (\ref{z6}) and (\ref{z7}) in Eqs. (\ref{z3}) and (\ref%
{z4}) yields%
\begin{equation}
\frac{\partial L_{\text{g}}}{\partial \phi }-2W+4e^{2\phi }\left( Fe^{-2\phi
}+B^{i}u_{i}\right) =0,  \label{z8}
\end{equation}%
\begin{equation}
\frac{\partial L_{\text{g}}}{\partial u_{\alpha }}+4\cosh (2\phi )B^{\alpha
}-\left( \frac{\partial L_{\text{g}}}{\partial u_{0}}+4\cosh (2\phi
)B^{0}\right) \frac{u^{\alpha }}{u_{0}}=0.  \label{z9}
\end{equation}%
Equation (\ref{z9}) gives%
\begin{equation}
B^{i}=-\frac{1}{4\cosh (2\phi )}\frac{\partial L_{\text{g}}}{\partial u_{i}}.
\label{z11}
\end{equation}%
Substituting this into Eq. (\ref{z8}) we obtain%
\begin{equation}
F=\frac{W}{2}+\frac{e^{2\phi }}{4\cosh (2\phi )}\frac{\partial L_{\text{g}}}{%
\partial u_{m}}u_{m}-\frac{1}{4}\frac{\partial L_{\text{g}}}{\partial \phi }.
\label{z12}
\end{equation}

Equations (\ref{z11}) and (\ref{z12}) determine $B^{i}$ and $F$ in Eq. (\ref%
{z5}). Substituting them in Eq. (\ref{z5}) we can write equations for the
gravitational field as%
\begin{equation*}
2e^{2\phi }\left( W^{ik}-\frac{W}{2}\tilde{f}^{ik}\right) u_{k}+\frac{1}{%
1+e^{-4\phi }}\left[ \frac{\partial L_{\text{g}}}{\partial u_{i}}-\frac{%
\partial L_{\text{g}}}{\partial u_{m}}u_{m}u^{i}\right]
\end{equation*}%
\begin{equation*}
+\frac{1}{2}\frac{\partial L_{\text{g}}}{\partial \phi }u^{i}=0,
\end{equation*}%
or%
\begin{equation*}
\frac{1}{1+e^{-4\phi }}\left[ \frac{\partial L_{\text{g}}}{\partial u_{m}}%
u_{m}u^{i}-\frac{\partial L_{\text{g}}}{\partial u_{i}}\right] -\frac{1}{2}%
\frac{\partial L_{\text{g}}}{\partial \phi }u^{i}
\end{equation*}%
\begin{equation}
=-\left( T^{ik}-\frac{T}{2}\tilde{f}^{ik}\right) u_{k},  \label{z13}
\end{equation}%
where $T^{ik}$ is the energy-momentum tensor of matter and $T=T^{ik}f_{ik}$.

What is left is to calculate variational derivatives of $L_{\text{g}}$ and
substitute them into Eqs. (\ref{z13}). Straightforward but lengthy algebra
yields Eqs. (\ref{sss1}).

\section{Motion of particles in gravitational field}

\label{motion}

Here we obtain how a test particle with rest mass $m$ moves in an external
gravitational field $f_{ik}$. Interaction of the particle with the field is
described by the action 
\begin{equation}
S_{\text{matter}}=-mc\int \sqrt{f_{ik}dx^{i}dx^{k}},  \label{r8}
\end{equation}%
where the integral is taken along the particle trajectory. One can find
equation of particle motion varying the action (\ref{r8}) at fixed $f_{ik}$ 
\cite{Land95}%
\begin{equation*}
\delta S_{\text{matter}}=
\end{equation*}%
\begin{equation*}
-\frac{mc}{2}\int \left[ \frac{dx^{i}}{ds}dx^{k}\delta f_{ik}+f_{ik}\left( 
\frac{dx^{k}}{ds}d\delta x^{i}+\frac{dx^{i}}{ds}d\delta x^{k}\right) \right]
,
\end{equation*}%
where 
\begin{equation}
ds=\sqrt{f_{ik}dx^{i}dx^{k}}.  \label{r11}
\end{equation}%
Next we take into account $\delta f_{ik}=\left( \partial f_{ik}/\partial
x^{l}\right) \delta x^{l}$ and integrate the second term by parts%
\begin{equation*}
\delta S_{\text{matter}}=
\end{equation*}%
\begin{equation*}
-\frac{mc}{2}\int \left[ \left( \frac{\partial f_{ik}}{\partial x^{l}}-\frac{%
\partial f_{lk}}{\partial x^{i}}-\frac{\partial f_{il}}{\partial x^{k}}%
\right) \frac{dx^{i}}{ds}\frac{dx^{k}}{ds}-2f_{lk}\frac{d^{2}x^{k}}{ds^{2}}%
\right] ds\delta x^{l}.
\end{equation*}%
Principle of least action $\delta S_{\text{matter}}=0$ yields the following
equation%
\begin{equation}
f_{lk}\frac{d^{2}x^{k}}{ds^{2}}=\frac{1}{2}\left[ \frac{\partial f_{ik}}{%
\partial x^{l}}-\frac{\partial f_{lk}}{\partial x^{i}}-\frac{\partial f_{il}%
}{\partial x^{k}}\right] \frac{dx^{i}}{ds}\frac{dx^{k}}{ds}.  \label{r9}
\end{equation}%
Multiplying both sides of Eq. (\ref{r9}) by tensor inverse to $f_{lk}$ we
find%
\begin{equation}
\frac{d^{2}x^{b}}{ds^{2}}=\frac{1}{2}\tilde{f}^{bl}\left[ \frac{\partial
f_{ik}}{\partial x^{l}}-\frac{\partial f_{lk}}{\partial x^{i}}-\frac{%
\partial f_{il}}{\partial x^{k}}\right] \frac{dx^{i}}{ds}\frac{dx^{k}}{ds}.
\label{r10}
\end{equation}%
This is equation of motion of a particle in gravitational field $f_{ik}$.

From Eq. (\ref{r8}) we obtain the following Lagrangian of the particle 
\begin{equation}
L=-mc\sqrt{f_{ik}\frac{dx^{i}}{dt}\frac{dx^{k}}{dt}}.  \label{r12}
\end{equation}

Action (\ref{r8}) and Eq. (\ref{r10}) are invalid for massless particles.
Let us consider a massless scalar field $\chi $. In the gravitational field $%
f_{ik}$ the action for $\chi $ reads%
\begin{equation}
S=\frac{1}{8\pi }\int d^{4}x\sqrt{-f}\tilde{f}^{\mu \nu }\frac{\partial \chi
^{\ast }}{\partial x^{\mu }}\frac{\partial \chi }{\partial x^{\nu }}.
\label{r13}
\end{equation}%
Variation of Eq. (\ref{r13}) yields the following equation of motion for the
field $\chi $%
\begin{equation}
\frac{\partial }{\partial x^{\mu }}\left( \sqrt{-f}\tilde{f}^{\mu \nu }\frac{%
\partial \chi }{\partial x^{\nu }}\right) =0.  \label{r14}
\end{equation}%
For geometrical optics one can write $\chi $ as $\chi =|\chi |e^{i\psi }$,
where $\psi $ (eikonal) has a large value. Substituting this into Eq. (\ref%
{r14}) and keeping only the leading term we obtain eikonal equation in
gravitational field 
\begin{equation}
\tilde{f}^{\mu \nu }\frac{\partial \psi }{\partial x^{\mu }}\frac{\partial
\psi }{\partial x^{\nu }}=0.  \label{r15}
\end{equation}

\section{Motion of particles in static gravitational field}

\label{StaticMotion}

Here we consider motion of a particle with rest mass $m$ in static
gravitational field $\phi (\mathbf{r})$. Equation of particle motion in
general case is obtained in Appendix \ref{motion}. Equation (\ref{r10}) for
static field (\ref{s10}) reduces to%
\begin{equation}
\frac{d(e^{2\phi }\gamma )}{dt}=0,  \label{u12}
\end{equation}%
\begin{equation}
\frac{d\left( \gamma e^{-2\phi }\mathbf{V}\right) }{dt}=-\gamma c^{2}\left[
e^{2\phi }+\frac{V^{2}}{c^{2}}e^{-2\phi }\right] \nabla \phi ,  \label{u11}
\end{equation}%
where $\nabla \phi =\partial \phi /\partial \mathbf{r}$, $\mathbf{r}%
=x^{\alpha }$, $\mathbf{V}=\partial \mathbf{r}/\partial t$ is the particle
velocity and 
\begin{equation}
\gamma =\frac{e^{-\phi }}{\sqrt{1-\frac{V^{2}}{c^{2}}e^{-4\phi }}}.
\label{u14}
\end{equation}%
One can also find equation of particle motion (\ref{u11}) directly from
Lagrange's equation $\frac{d}{dt}\frac{\partial L}{\partial \mathbf{V}}=%
\frac{\partial L}{\partial \mathbf{r}},$ where the Lagrangian (\ref{r12})
for static gravitational field reads 
\begin{equation}
L=-mc^{2}\sqrt{e^{2\phi }-\frac{V^{2}}{c^{2}}e^{-2\phi }}.  \label{u15}
\end{equation}%
Equation (\ref{u12}) follows from Eq. (\ref{u11}) if we multiply both sides
of Eq. (\ref{u11}) by $\gamma e^{-2\phi }\mathbf{V}$ and make simple
algebraic transformations.

Lagrangian (\ref{u15}) gives the following expression for the particle
generalized momentum $\mathbf{p}=\frac{\partial L}{\partial \mathbf{V}}$%
\begin{equation}
\mathbf{p}=\gamma e^{-2\phi }m\mathbf{V},  \label{u16}
\end{equation}%
and particle Hamiltonian $H=\mathbf{V}\frac{\partial L}{\partial \mathbf{V}}%
\mathbf{-}L$%
\begin{equation}
H=e^{2\phi }\gamma mc^{2}=\sqrt{m^{2}c^{4}e^{2\phi }+p^{2}c^{2}e^{4\phi }}.
\label{u17}
\end{equation}%
Thus, Eq. (\ref{u12}) is the equation of energy conservation $W=$const,
where 
\begin{equation}
W=e^{2\phi }\gamma mc^{2}=\frac{e^{\phi }mc^{2}}{\sqrt{1-\frac{V^{2}}{c^{2}}%
e^{-4\phi }}}  \label{u19}
\end{equation}%
is the particle energy and Eq. (\ref{u11}) is the equation for momentum.

For a massless particle one should use Eq. (\ref{r14}) which for a static
field reads%
\begin{equation}
e^{-4\phi }\frac{\partial ^{2}\chi }{\partial t^{2}}-c^{2}\Delta \chi =0.
\label{u20}
\end{equation}%
Equation (\ref{u20}) describes propagation of a massless particle with speed 
\begin{equation}
v=ce^{2\phi }.  \label{u21}
\end{equation}

One can see that speed of light depends on the gravitational field $\phi $
and $v\leq c$ if $\phi $ is given by Eq. (\ref{r7}) with positive masses. By
proper rescaling of coordinates in Eq. (\ref{u20}) one can remove the factor 
$e^{-4\phi }$ at any given point. Let us fix $\phi =0$ at infinite distance
from masses. If an observer at infinity sends a light signal towards the Sun
then near the solar surface $\phi <0$ and light will propagate with a
smaller speed. This is the explanation of Shapiro time delay in the present
theory of gravity. Light signal traveling the same distance arrives with a
delay if the light trajectory passes near the Sun. The delay occurs because
the speed of light is smaller near the solar surface.

Since Eq. (\ref{u20}) does not contain $t$ explicitly the photon frequency $%
\omega _{0}$ (measured in time $t$) remains the same during light
propagation. However, physical processes occur with different rates at
different $\phi $. Gravitational field (\ref{s10}) can be removed at a given
point by rescaling time in the factor $\sqrt{f_{00}}=e^{\phi }$ ($t=\tau
/e^{\phi }$) and spatial coordinates by $\sqrt{-f_{\alpha \alpha }}=e^{-\phi
}$. In such rescaled coordinates identical atoms emit light with equal
frequencies $\omega \propto $ $\partial \chi /\partial \tau =e^{-\phi
}\partial \chi /\partial t$. Thus we obtain 
\begin{equation}
\omega =\omega _{0}e^{-\phi }\text{,}  \label{p10}
\end{equation}%
where $\omega _{0}$ is the photon frequency measured in time $t$.

Equation (\ref{p10}) shows that if light emitted by an atom propagates into
a region with larger gravitational potential then the detected light
frequency is smaller then those an identical atom emits at the detection
point. This phenomenon is known as gravitational redshift of light. Equation
(\ref{p10}) also shows that in our theory there are no black holes. Indeed
for the gravitational field created by a point mass $M$: $\phi =-GM/c^{2}r$.
Therefore if a photon is emitted at a distance $r$ from the mass $M$ with
frequency $\omega $ then an observer at infinity will detect the photon with
the energy 
\begin{equation}
\hbar \omega _{0}=\hbar \omega e^{-GM/c^{2}r}\text{.}  \label{p11}
\end{equation}%
According to Eq. (\ref{p11}) no matter how close the photon is emitted to
the mass $M$ the photon's energy at infinity never becomes zero. This means
that photon can escape from the mass $M$ from any distance. Such a
conclusion is dramatically different from prediction of general relativity.
In Einstein's theory photons become trapped by the mass $M$ if they are
emitted from a distance smaller then the event horizon (that is point mass $%
M $ behaves as a black hole).

\section{Equations for metric in post-Newtonian limit}

\label{AP7}

Here we show that Einstein equations%
\begin{equation}
R_{ik}=\frac{8\pi G}{c^{4}}\left( T_{ik}-\frac{1}{2}g_{ik}T\right)
\label{pp1}
\end{equation}%
and equations of the vector theory of gravity (\ref{sss1}) are the same in
the post-Newtonian limit. In such limit, components of the Ricci tensor are%
\begin{equation*}
R_{00}=\frac{\partial }{\partial x^{0}}\left( \frac{\partial h_{0}^{\alpha }%
}{\partial x^{\alpha }}-\frac{1}{2}\frac{\partial h_{\alpha }^{\alpha }}{%
\partial x^{0}}\right) +\frac{1}{2}\Delta h_{00}+\frac{1}{2}h^{\alpha \beta }%
\frac{\partial ^{2}h_{00}}{\partial x^{\alpha }\partial x^{\beta }}
\end{equation*}%
\begin{equation}
-\frac{1}{4}\left( \nabla h_{00}\right) ^{2}-\frac{1}{4}\frac{\partial h_{00}%
}{\partial x^{\beta }}\left( 2\frac{\partial h_{\beta }^{\alpha }}{\partial
x^{\alpha }}-\frac{\partial h_{\alpha }^{\alpha }}{\partial x^{\beta }}%
\right) ,  \label{pp2}
\end{equation}%
\begin{equation}
R_{0\alpha }=\frac{1}{2}\frac{\partial ^{2}h_{\alpha }^{\beta }}{\partial
x^{0}\partial x^{\beta }}+\frac{1}{2}\frac{\partial ^{2}h_{0}^{\beta }}{%
\partial x^{\alpha }\partial x^{\beta }}-\frac{1}{2}\frac{\partial
^{2}h_{\beta }^{\beta }}{\partial x^{0}\partial x^{\alpha }}+\frac{1}{2}%
\Delta h_{0\alpha },  \label{pp3}
\end{equation}%
where $h_{\alpha }^{\beta }=\eta ^{\beta \gamma }h_{\gamma \alpha }$. Taking
into account Eq. (\ref{pp4}) we obtain%
\begin{equation*}
R_{00}=\frac{1}{2}\Delta h_{00}+\frac{3}{2}\frac{\partial ^{2}h_{00}}{%
\partial x^{0}\partial x^{0}}-\frac{\partial ^{2}h_{0\beta }}{\partial
x^{0}\partial x^{\beta }}+\frac{1}{2}h_{00}\Delta h_{00}-\frac{1}{2}\left(
\nabla h_{00}\right) ^{2},
\end{equation*}%
\begin{equation*}
R_{0\alpha }=\frac{1}{2}\Delta h_{0\alpha }+\frac{1}{2}\frac{\partial
^{2}h_{0\beta }}{\partial x^{\alpha }\partial x^{\beta }}+\frac{\partial
^{2}h_{00}}{\partial x^{0}\partial x^{\alpha }}.
\end{equation*}%
As a result, Einstein equations in the post-Newtonian limit read%
\begin{equation*}
\frac{1}{2}\Delta h_{00}+\frac{3}{2}\frac{\partial ^{2}h_{00}}{\partial
x^{0}\partial x^{0}}-\frac{\partial ^{2}h_{0\beta }}{\partial x^{0}\partial
x^{\beta }}+\frac{1}{2}h_{00}\Delta h_{00}-\frac{1}{2}\left( \nabla
h_{00}\right) ^{2}
\end{equation*}%
\begin{equation}
=\frac{8\pi G}{c^{4}}\left( T_{00}-\frac{1}{2}g_{00}T\right) ,  \label{e1}
\end{equation}%
\begin{equation}
\frac{1}{2}\Delta h_{0\alpha }-\frac{1}{2}\frac{\partial ^{2}h_{0\beta }}{%
\partial x^{\alpha }\partial x^{\beta }}+\frac{\partial ^{2}h_{00}}{\partial
x^{0}\partial x^{\alpha }}=\frac{8\pi G}{c^{4}}T_{0\alpha }.  \label{e2}
\end{equation}

On the other hand in the cosmological reference frame for small deviations
of $\phi $ from a constant value $\phi _{0}$ ($\delta \phi =\phi -\phi _{0}$%
) and $|u_{\alpha }|\ll 1$, keeping post-Newtonian terms, and taking into
account that%
\begin{equation*}
T^{00}=\tilde{f}^{00}\tilde{f}^{00}T_{00}=e^{-4\phi }T_{00},\quad \tilde{f}%
^{00}=e^{-4\phi }f_{00},
\end{equation*}%
\begin{equation*}
T^{\alpha 0}=-T_{\alpha 0},
\end{equation*}%
Eqs. (\ref{sss1}) of the vector theory of gravity yield%
\begin{equation*}
\Delta \phi +3e^{-4\phi _{0}}\frac{\partial ^{2}\phi }{\partial
x^{0}\partial x^{0}}-2e^{-2\phi _{0}}\cosh (2\phi _{0})\frac{\partial
^{2}u^{\beta }}{\partial x^{\beta }\partial x^{0}}
\end{equation*}

\begin{equation}
=\frac{8\pi G}{c^{4}}e^{-4\phi }\left( T_{00}-\frac{T}{2}f_{00}\right) ,
\label{eee1}
\end{equation}%
\begin{equation}
e^{2\phi _{0}}\cosh (2\phi _{0})\left( \frac{\partial ^{2}u^{\beta }}{%
\partial x_{\alpha }\partial x^{\beta }}-\Delta u^{\alpha }\right) -2\frac{%
\partial ^{2}\phi }{\partial x^{\alpha }\partial x^{0}}=-\frac{8\pi G}{c^{4}}%
T_{\alpha 0}.  \label{eee2}
\end{equation}%
Next we rescale coordinates as%
\begin{equation*}
x^{0}\rightarrow e^{-\phi _{0}}x^{0},\quad x^{\alpha }\rightarrow e^{\phi
_{0}}x^{\alpha }.
\end{equation*}%
In new coordinates the equivalent metric $f_{ik}$ has the form of Eq. (\ref%
{eee3}) with%
\begin{equation*}
h_{00}=2\delta \phi +2(\delta \phi )^{2},\quad h_{0\alpha }=2\cosh (2\phi
_{0})u_{\alpha },
\end{equation*}%
\begin{equation*}
\delta \phi =\frac{h_{00}}{2}-\frac{h_{00}^{2}}{4}.
\end{equation*}%
In the rescaled coordinates Eqs. (\ref{eee1}) and (\ref{eee2}) reduce to%
\begin{equation*}
\frac{1}{2}\Delta h_{00}-\frac{1}{4}\Delta h_{00}^{2}+\frac{3}{2}\frac{%
\partial ^{2}h_{00}}{\partial x^{0}\partial x^{0}}-\frac{\partial
^{2}h_{0\beta }}{\partial x^{\beta }\partial x^{0}}
\end{equation*}%
\begin{equation}
=\frac{8\pi G}{c^{4}}e^{4(\phi _{0}-\phi )}\left( T_{00}-\frac{T}{2}%
f_{00}\right) ,  \label{e7a}
\end{equation}%
\begin{equation}
\frac{1}{2}\Delta h_{0\alpha }-\frac{1}{2}\frac{\partial ^{2}h_{0\beta }}{%
\partial x^{\alpha }\partial x^{\beta }}+\frac{\partial ^{2}h_{00}}{\partial
x^{\alpha }\partial x^{0}}=\frac{8\pi G}{c^{4}}T_{\alpha 0}.  \label{e7}
\end{equation}%
Multiplying both sides of Eq. (\ref{e7a}) by $e^{-4(\phi _{0}-\phi
)}=e^{2h_{00}}$ and expanding the exponential factor\ we obtain%
\begin{equation*}
\frac{1}{2}\Delta h_{00}+h_{00}\Delta h_{00}-\frac{1}{4}\Delta h_{00}^{2}+%
\frac{3}{2}\frac{\partial ^{2}h_{00}}{\partial x^{0}\partial x^{0}}-\frac{%
\partial ^{2}h_{0\beta }}{\partial x^{\beta }\partial x^{0}}
\end{equation*}%
\begin{equation*}
=\frac{8\pi G}{c^{4}}\left( T_{00}-\frac{T}{2}f_{00}\right) .
\end{equation*}%
Using 
\begin{equation*}
\Delta h_{00}^{2}=2(\nabla h_{00})^{2}+2h_{00}\Delta h_{00}
\end{equation*}%
we finally find%
\begin{equation*}
\frac{1}{2}\Delta h_{00}+\frac{3}{2}\frac{\partial ^{2}h_{00}}{\partial
x^{0}\partial x^{0}}-\frac{\partial ^{2}h_{0\beta }}{\partial x^{\beta
}\partial x^{0}}+\frac{1}{2}h_{00}\Delta h_{00}-\frac{1}{2}(\nabla
h_{00})^{2}
\end{equation*}%
\begin{equation}
=\frac{8\pi G}{c^{4}}\left( T_{00}-\frac{T}{2}f_{00}\right) .  \label{e6}
\end{equation}

Equations (\ref{e6}) and (\ref{e7}) of the vector theory of gravity are
identical to the Einstein equations (\ref{e1}) and (\ref{e2}). Boundary
conditions are also the same. Thus, in the post-Newtonian limit both
theories are equivalent.

\section{Analogy of weak gravity in the classical limit with
electromagnetism in medium with negative refractive index}

\label{AP4}

In a medium with dielectric constant $\varepsilon $ and magnetic
permeability $\mu $ Maxwell equations describing electromagnetic field read 
\begin{equation}
\text{curl}\mathbf{E}=-\frac{1}{c}\frac{\partial \mathbf{B}}{\partial t},
\label{mm1}
\end{equation}%
\begin{equation}
\text{div}(\varepsilon \mathbf{E})=4\pi \rho _{e},
\end{equation}%
\begin{equation}
\text{curl}\left( \frac{\mathbf{B}}{\mu }\right) =\frac{1}{c}\frac{\partial
(\varepsilon \mathbf{E})}{\partial t}+\frac{4\pi }{c}\rho _{e}\mathbf{V},
\label{mm2}
\end{equation}%
where $\rho _{e}$ is the electric charge density. In terms of the vector $%
\mathbf{A}$ and scalar $\varphi $ potentials%
\begin{equation*}
\mathbf{E}=-\nabla \varphi -\frac{\partial \mathbf{A}}{\partial x^{0}},\quad 
\mathbf{B=}\text{curl}\mathbf{(A})\text{,}
\end{equation*}%
the Maxwell equations (\ref{mm1})-(\ref{mm2}) are%
\begin{equation*}
\Delta \varphi +\frac{\partial }{\partial x^{0}}\text{div}\mathbf{A}=-\frac{%
4\pi }{\varepsilon }\rho _{e},
\end{equation*}%
\begin{equation*}
\left( \Delta -\varepsilon \mu \frac{\partial ^{2}}{\partial x^{0}\partial
x^{0}}\right) \mathbf{A}-\nabla \left( \varepsilon \mu \frac{\partial
\varphi }{\partial x^{0}}+\text{div}\mathbf{A}\right) =-\frac{4\pi \mu }{c}%
\rho _{e}\mathbf{V}.
\end{equation*}%
In the Lorenz gauge%
\begin{equation}
\frac{\partial \varphi }{\partial x^{0}}+\text{div}\mathbf{A}=0  \label{Lg}
\end{equation}%
and for $\varepsilon =\mu =-1$ equations reduce to%
\begin{equation}
\left( \Delta -\frac{\partial ^{2}}{\partial x^{0}\partial x^{0}}\right)
\varphi =4\pi \rho _{e},  \label{mm3}
\end{equation}%
\begin{equation}
\left( \Delta \mathbf{-}\frac{\partial ^{2}}{\partial x^{0}\partial x^{0}}%
\right) \mathbf{A}=\frac{4\pi }{c}\rho _{e}\mathbf{V}.  \label{mm4}
\end{equation}

On the other hand, equations for the weak classical gravitational field and
non relativistic motion of matter with density $\rho _{\text{m}}$ and
velocity $\mathbf{V}$ are [see Eqs. (\ref{wf1}) and (\ref{wf2})]%
\begin{equation}
\left( \Delta +3\frac{\partial ^{2}}{\partial x^{0}\partial x^{0}}\right)
h_{00}-2\frac{\partial ^{2}h_{0\beta }}{\partial x^{0}\partial x^{\beta }}=%
\frac{8\pi G}{c^{2}}\rho _{\text{m}},  \label{w3aa}
\end{equation}%
\begin{equation}
\left( \frac{\partial ^{2}}{\partial x^{0}\partial x^{0}}-\Delta \right) 
\mathbf{h}+2\nabla \left( \frac{\partial h_{00}}{\partial x^{0}}-\frac{1}{2}%
\frac{\partial h_{0\beta }}{\partial x^{\beta }}\right) =-\frac{16\pi G}{%
c^{3}}\rho _{\text{m}}\mathbf{V},  \label{w4aa}
\end{equation}%
where $\mathbf{h}=h^{0\alpha }$, $\mathbf{V}=V^{\alpha }$ and $\nabla
=\partial /\partial x^{\alpha }$. Introducing 
\begin{equation*}
\tilde{\varphi}=\frac{c^{2}}{2}h_{00},\quad \mathbf{\tilde{A}}=\frac{c^{2}}{4%
}\mathbf{h}
\end{equation*}%
we obtain%
\begin{equation}
\left( \Delta +3\frac{\partial ^{2}}{\partial x^{0}\partial x^{0}}\right) 
\tilde{\varphi}+4\frac{\partial }{\partial x^{0}}\text{div}\mathbf{\tilde{A}}%
=4\pi G\rho _{\text{m}},  \label{w3a}
\end{equation}%
\begin{equation}
\left( \Delta -\frac{\partial ^{2}}{\partial x^{0}\partial x^{0}}\right) 
\mathbf{\tilde{A}}-\nabla \left( \frac{\partial \tilde{\varphi}}{\partial
x^{0}}+\text{div}\mathbf{\tilde{A}}\right) =\frac{4\pi G}{c}\rho _{\text{m}}%
\mathbf{V}.  \label{w4a}
\end{equation}%
Taking $\partial /\partial x^{0}$ from Eq. (\ref{w3a}) and $(1/2)$div from
Eq. (\ref{w4a}), adding these equations together and using the continuity
equation%
\begin{equation*}
\frac{\partial \rho }{\partial t}+\text{div}\left( \rho _{\text{m}}\mathbf{V}%
\right) =0
\end{equation*}%
we find%
\begin{equation*}
\frac{\partial ^{2}}{\partial x^{0}\partial x^{0}}\left( \frac{\partial 
\tilde{\varphi}}{\partial x^{0}}+\text{div}\mathbf{\tilde{A}}\right) =0.
\end{equation*}%
Therefore one can take 
\begin{equation*}
\frac{\partial \tilde{\varphi}}{\partial x^{0}}+\text{div}\mathbf{\tilde{A}}%
=0
\end{equation*}%
which is the same equation as the Lorenz gauge condition (\ref{Lg}) in
electromagnetism. Then Eqs. (\ref{w3a}) and (\ref{w4a}) reduce to 
\begin{equation}
\left( \Delta -\frac{\partial ^{2}}{\partial x^{0}\partial x^{0}}\right) 
\tilde{\varphi}=4\pi G\rho _{\text{m}},  \label{gm1}
\end{equation}%
\begin{equation}
\left( \Delta -\frac{\partial ^{2}}{\partial x^{0}\partial x^{0}}\right) 
\mathbf{\tilde{A}}=\frac{4\pi G}{c}\rho _{\text{m}}\mathbf{V},  \label{gm2}
\end{equation}%
which have the same form as Maxwell equations (\ref{mm3}) and (\ref{mm4}) in
the left handed medium with $\varepsilon =\mu =-1$.

The analogy becomes more transparent if we compare expressions for the
energy density and energy flux. According to Eqs. (\ref{ff4}) and (\ref{ff5}%
), the energy density and the energy density flux of the transverse
gravitational wave in terms of $\tilde{\varphi}$ and $\mathbf{\tilde{A}}$ are%
\begin{equation*}
w_{\text{tr}}=-\frac{1}{2\pi G}\left[ \left( \frac{\partial \mathbf{\tilde{A}%
}}{\partial x^{0}}\right) ^{2}+\text{curl}^{2}\mathbf{\tilde{A}}\right] ,
\end{equation*}
\begin{equation*}
\mathbf{S}_{\text{tr}}\mathbf{=}\frac{c}{\pi G}\frac{\partial \mathbf{\tilde{%
A}}}{\partial x^{0}}\times \text{curl}\mathbf{(\tilde{A}}),
\end{equation*}%
which are similar to those for a transverse (div$\mathbf{A}=0$)
electromagnetic wave in the left handed medium with $\varepsilon =\mu =-1$%
\begin{equation*}
w_{\text{em}}=-\frac{1}{8\pi }\left[ \left( \frac{\partial \mathbf{A}}{%
\partial x^{0}}\right) ^{2}+\text{curl}^{2}\mathbf{A}\right] ,
\end{equation*}%
\begin{equation*}
\mathbf{S}_{\text{em}}\mathbf{=}\frac{c}{4\pi }\frac{\partial \mathbf{A}}{%
\partial x^{0}}\times \text{curl}\mathbf{(A}).
\end{equation*}%
In such left handed dispersionless medium the energy density of the
electromagnetic field is also negative. However, equation of mass motion in
weak gravitational field (\ref{em}) is somewhat different from the equation
of motion of a charge in electromagnetic field.

\section{Energy density and energy flux for weak classical gravitational
field}

\label{AP5}

One can obtain the energy density and energy flux for classical
gravitational field using general formula for the energy-momentum tensor.
Namely, if action of the system has the form 
\begin{equation*}
S=\frac{1}{c}\int d^{4}xL\left( A_{l},\frac{\partial A_{l}}{\partial x^{k}}%
\right) ,
\end{equation*}%
where the Lagrangian density $L$ is some function of the quantities $A_{l}$,
describing the state of the system, and of their first derivatives, then the
energy-momentum tensor $T^{ik}$ of the system can be calculated using
equation \cite{Land95}

\begin{equation}
T_{i}^{k}=\sum_{l}\frac{\partial A_{l}}{\partial x^{i}}\frac{\partial L}{%
\partial \frac{\partial A_{l}}{\partial x^{k}}}-\delta _{i}^{k}L.
\label{emt}
\end{equation}%
$T_{i}^{k}$ obeys the conservation law%
\begin{equation*}
\frac{\partial T_{i}^{k}}{\partial x^{k}}=0
\end{equation*}%
and, therefore, $T^{00}$ can be interpreted as the energy density of the
system, while vector $S^{\alpha }=cT^{0\alpha }$ (the Poynting vector) is
the flux density (the amount of energy passing through unit surface per unit
time).

For weak gravitational field and nonrelativistic motion of masses the
Lagrangian density reads 
\begin{equation*}
L=\frac{c^{4}}{32\pi G}\left( -3\frac{\partial h_{00}}{\partial x^{0}}\frac{%
\partial h_{00}}{\partial x^{0}}-\frac{\partial h_{00}}{\partial x^{\alpha }}%
\frac{\partial h_{00}}{\partial x^{\alpha }}-\frac{\partial h_{0\alpha }}{%
\partial x^{0}}\frac{\partial h_{0\alpha }}{\partial x^{0}}\right.
\end{equation*}%
\begin{equation*}
+\left. \frac{\partial h_{0\alpha }}{\partial x^{\beta }}\frac{\partial
h_{0\alpha }}{\partial x^{\beta }}-\frac{\partial h_{0\alpha }}{\partial
x^{\beta }}\frac{\partial h_{0\beta }}{\partial x^{\alpha }}+2\left[ \frac{%
\partial h_{0\alpha }}{\partial x^{0}}\frac{\partial h_{00}}{\partial
x^{\alpha }}+\frac{\partial h_{0\alpha }}{\partial x^{\alpha }}\frac{%
\partial h_{00}}{\partial x^{0}}\right] \right)
\end{equation*}%
\begin{equation}
-\rho c^{2}-\frac{1}{2}\rho c^{2}h_{00}-\rho cV^{\alpha }h_{0\alpha }+\frac{1%
}{2}\rho V^{2}
\end{equation}%
and components of the equivalent metric $h_{0k}$ can be treated as function
describing the state of the gravitational field. Applying Eq. (\ref{emt}) we
find%
\begin{equation*}
T^{00}=\frac{c^{4}}{32\pi G}\left( -3\frac{\partial h_{00}}{\partial x^{0}}%
\frac{\partial h_{00}}{\partial x^{0}}-\frac{\partial h_{0\alpha }}{\partial
x^{0}}\frac{\partial h_{0\alpha }}{\partial x^{0}}+\frac{\partial h_{00}}{%
\partial x^{\alpha }}\frac{\partial h_{00}}{\partial x^{\alpha }}\right.
\end{equation*}%
\begin{equation}
-\left. \frac{\partial h_{0\alpha }}{\partial x^{\beta }}\frac{\partial
h_{0\alpha }}{\partial x^{\beta }}+\frac{\partial h_{0\alpha }}{\partial
x^{\beta }}\frac{\partial h_{0\beta }}{\partial x^{\alpha }}\right) +\rho
c^{2}+\frac{1}{2}\rho c^{2}h_{00}+\frac{1}{2}\rho V^{2},
\end{equation}%
\begin{equation*}
T^{0\alpha }=\frac{c^{4}}{16\pi G}\left( -\frac{\partial h_{00}}{\partial
x^{\alpha }}\frac{\partial h_{00}}{\partial x^{0}}+\frac{\partial h_{0\beta }%
}{\partial x^{\alpha }}\frac{\partial h_{0\beta }}{\partial x^{0}}-\frac{%
\partial h_{0\alpha }}{\partial x^{\beta }}\frac{\partial h_{0\beta }}{%
\partial x^{0}}\right.
\end{equation*}%
\begin{equation}
\left. +2\frac{\partial h_{0\alpha }}{\partial x^{0}}\frac{\partial h_{00}}{%
\partial x^{0}}\right) +\rho cV^{\alpha },
\end{equation}%
where $\rho $ is the mass density, $\mathbf{V}$ is the velocity of
nonrelativistic motion of matter and $\alpha =1$, $2$, $3$. Introducing
vector%
\begin{equation*}
\mathbf{h}=h^{_{0\alpha }},
\end{equation*}%
we obtain the following expression for the energy density of the weak
classical gravitational field%
\begin{equation*}
T^{00}=-\frac{c^{4}}{32\pi G}\left[ 3\left( \frac{\partial h_{00}}{\partial
x^{0}}\right) ^{2}-\left( \nabla h_{00}\right) ^{2}+\left( \frac{\partial 
\mathbf{h}}{\partial x^{0}}\right) ^{2}+\text{curl}^{2}\mathbf{h}\right]
\end{equation*}%
\begin{equation}
+\rho c^{2}+\frac{1}{2}\rho c^{2}h_{00}+\frac{1}{2}\rho V^{2}.
\end{equation}%
The energy density flux is given by%
\begin{equation}
\mathbf{S=}\frac{c^{5}}{16\pi G}\left[ -\left( 2\frac{\partial \mathbf{h}}{%
\partial x^{0}}+\nabla h_{00}\right) \frac{\partial h_{00}}{\partial x^{0}}+%
\frac{\partial \mathbf{h}}{\partial x^{0}}\times \text{curl }\mathbf{h}%
\right] +\rho c\mathbf{V}.
\end{equation}

In the Newtonian limit the energy density of field and matter reads%
\begin{equation}
w=\rho c^{2}+\frac{\rho V^{2}}{2}+\rho c^{2}\phi +\frac{c^{4}}{8\pi G}%
(\nabla \phi )^{2},
\end{equation}%
where $c^{2}\phi =c^{2}h_{00}/2$ is the Newtonian gravitational potential.

\section{Cosmological suppression of preferred frame and preferred location
effects}

\label{cossup}

Here we show lack of the preferred frame and preferred location effects for
neutron starts orbiting each other. Since gravitational field of a neutron
star is not weak we must find symmetries of the action valid when spatial
change of $\phi $ is of the order of unity. Spatial variation of $\phi $
produced by a neutron star yet substantially smaller than cosmological value 
$\phi _{\text{cosm}}$. Indeed, due to expansion of the Universe the spatial
scale has been magnified in a factor $e^{-\phi _{\text{cosm}}}\sim 10^{40}$.
Thus, at the present epoch $e^{-\phi }\ggg 1$, and, therefore, we can
disregard exponentially small number $e^{\phi }$ compared to the
exponentially large value of $e^{-\phi }$.

In terms of components the equivalent metric (\ref{met}) reads%
\begin{equation*}
f_{00}=e^{2\phi }-2\cosh (2\phi )u_{\alpha }^{2},
\end{equation*}%
\begin{equation*}
f_{0\alpha }=2\cosh (2\phi )u_{0}u_{\alpha },
\end{equation*}%
\begin{equation*}
f_{\alpha \beta }=-e^{-2\phi }\delta _{\alpha \beta }+2\cosh (2\phi
)u_{\alpha }u_{\beta }.
\end{equation*}%
Taking into account that $e^{-\phi }\ggg 1$ and introducing new function%
\begin{equation*}
h_{0\alpha }=e^{-2\phi }u_{\alpha }
\end{equation*}
one can write the equivalent metric as 
\begin{equation}
f_{00}=e^{2\phi }\left( 1-h_{0\alpha }^{2}\right) ,\quad f_{0\alpha
}=h_{0\alpha },\quad f_{\alpha \beta }=-e^{-2\phi }\delta _{\alpha \beta }.
\label{www1}
\end{equation}%
Thus, the square of the interval is%
\begin{equation}
ds^{2}=e^{2\phi }\left( 1-h_{0\alpha }^{2}\right) (dx^{0})^{2}+2h_{0\alpha
}dx^{0}dx^{\alpha }-e^{-2\phi }d\mathbf{r}^{2}.  \label{aaaa1}
\end{equation}

Motion of stars orbiting each other is not relativistic and, therefore, $V/c$
is another small parameter in our problem. \ Keeping terms upto $V^{3}/c^{3}$
and taking into account that $e^{-\phi }\ggg 1$ the gravitational field
action (\ref{fa2}) reduces to%
\begin{equation*}
S_{\text{gravity}}=\frac{c^{3}}{8\pi G}\int d^{4}x\left[ -\frac{\partial
\phi }{\partial x^{\alpha }}\frac{\partial \phi }{\partial x^{\alpha }}%
-3e^{-4\phi }\frac{\partial \phi }{\partial x^{0}}\frac{\partial \phi }{%
\partial x^{0}}\right.
\end{equation*}%
\begin{equation*}
+2e^{-2\phi }\frac{\partial \phi }{\partial x^{\alpha }}\frac{\partial
h_{0\alpha }}{\partial x^{0}}+\frac{1}{4}\frac{\partial h_{0\alpha }}{%
\partial x^{\beta }}\frac{\partial h_{0\alpha }}{\partial x^{\beta }}-\frac{1%
}{4}\frac{\partial h_{0\alpha }}{\partial x^{\beta }}\frac{\partial
h_{0\beta }}{\partial x^{\alpha }}
\end{equation*}%
\begin{equation*}
+h_{0\beta }\frac{\partial \phi }{\partial x^{\alpha }}\frac{\partial
h_{0\alpha }}{\partial x^{\beta }}+h_{0\alpha }\frac{\partial h_{0\alpha }}{%
\partial x^{\beta }}\frac{\partial \phi }{\partial x^{\beta }}
\end{equation*}%
\begin{equation}
-2e^{-2\phi }h_{0\alpha }\frac{\partial \phi }{\partial x^{\alpha }}\frac{%
\partial \phi }{\partial x^{0}}+\left. h_{0\alpha }h_{0\alpha }\frac{%
\partial \phi }{\partial x^{\beta }}\frac{\partial \phi }{\partial x^{\beta }%
}\right] .  \label{aaa1}
\end{equation}

The interval (\ref{aaaa1}) and the gravitational field action (\ref{aaa1})
are invariant under scaling transformation 
\begin{equation}
x^{0}\rightarrow e^{-a}x^{0},\qquad x^{\alpha }\rightarrow e^{a}x^{\alpha },
\label{sc1}
\end{equation}%
\begin{equation}
\phi \rightarrow \phi +a,\qquad h_{0\alpha }\rightarrow h_{0\alpha },
\label{sc2}
\end{equation}%
where $a$ is an arbitrary constant parameter, not necessarily small.

There is also an additional Lorentz-like symmetry of the action valid for
the strong gravitational field of a neutron star. Straightforward but
lengthy calculation yields that upto the terms of the order of $V^{3}/c^{3}$
the total action $S_{\text{gravity}}+S_{\text{matter}}$ is invariant under a
coordinate transformation for which derivatives transform as 
\begin{equation*}
\frac{\partial }{\partial x^{0}}\rightarrow \left( 1+\frac{V^{2}}{2c^{2}}%
\right) \frac{\partial }{\partial x^{0}}-\frac{\mathbf{V}}{c}\nabla ,\quad 
\frac{\partial }{\partial \mathbf{r}}\rightarrow \frac{\partial }{\partial 
\mathbf{r}}-\frac{\mathbf{V}}{c}\frac{\partial }{\partial x^{0}}+
\end{equation*}%
\begin{equation*}
\left( \left[ 1+8e^{-4\phi }-9e^{4\phi }\right] \frac{V^{2}}{2c^{2}}-\frac{%
e^{2\phi }}{c}\mathbf{V\cdot h}\right) \frac{\partial }{\partial \mathbf{r}}+
\end{equation*}%
\begin{equation}
e^{4\phi }\frac{\mathbf{V}}{2c^{2}}\left( \mathbf{V}\frac{\partial }{%
\partial \mathbf{r}}\right) +e^{2\phi }\frac{\mathbf{V}}{c}\left( \mathbf{h}%
\frac{\partial }{\partial \mathbf{r}}\right) ,  \label{tr7}
\end{equation}%
where $\mathbf{V}=V^{\alpha }$ is a constant (velocity) vector and $\mathbf{h%
}=h^{0\alpha }$. Under this transformation the equivalent metric $f_{ik}$,
given by Eq. (\ref{www1}), transforms as a covariant tensor. Namely, $f_{00}$
transforms as $\frac{\partial }{\partial x^{0}}\frac{\partial }{\partial
x^{0}}$, $f_{0\alpha }$ transforms as $\frac{\partial }{\partial x^{0}}\frac{%
\partial }{\partial x^{\alpha }}$ and so on. Since $f_{ik}$ transforms as a
tensor the interval $ds$, and hence $S_{\text{matter}}$, are invariant.

Keeping terms of the proper order the metric transformation reads 
\begin{equation}
h_{0\alpha }\rightarrow h_{0\alpha }-2\frac{V^{\alpha }}{c}\text{sinh}(2\phi
),  \label{tr4}
\end{equation}%
\begin{equation*}
e^{2\phi }\rightarrow e^{2\phi }+2\frac{V^{2}}{c^{2}}\text{sinh}(2\phi
)-2e^{4\phi }\frac{V^{\alpha }}{c}h_{0\alpha },
\end{equation*}%
or%
\begin{equation}
\phi \rightarrow \phi +\frac{V^{2}}{c^{2}}e^{-2\phi }\text{sinh}(2\phi )-%
\frac{V^{\alpha }}{c}e^{2\phi }h_{0\alpha }.  \label{tr5}
\end{equation}%
Under transformation (\ref{tr7}) the volume element $d^{4}x$ in the action $%
S_{\text{gravity}}$ transforms as $d^{4}x\rightarrow d^{4}x/J$, where $J$ is
the Jacobian of the transformation%
\begin{equation*}
J=1+\left( 1+12e^{-4\phi }-13e^{4\phi }\right) \frac{V^{2}}{c^{2}}+2e^{2\phi
}\frac{V^{\alpha }}{c}h_{0\alpha }.
\end{equation*}

In the post-Newtonian limit (far away from the neutron star) the
transformation (\ref{tr7})-(\ref{tr5}) reduces to the low-velocity Lorentz
transformation%
\begin{equation}
x^{0}\rightarrow \left( 1+\frac{V^{2}}{2c^{2}}\right) x^{0}+\frac{1}{c}%
\mathbf{V\cdot r},\quad \mathbf{r}\rightarrow \mathbf{r}+\frac{\mathbf{V}}{c}%
x^{0},  \label{lor0}
\end{equation}%
\begin{equation}
\phi \rightarrow \left( 1+\frac{2V^{2}}{c^{2}}\right) \phi -\frac{V^{\alpha }%
}{c}h_{0\alpha },  \label{lor1}
\end{equation}%
\begin{equation}
h_{0\alpha }\rightarrow h_{0\alpha }-4\frac{V^{\alpha }}{c}\phi .
\label{lor2}
\end{equation}

Scaling transformation (\ref{sc1}) and (\ref{sc2}) combined with the
Lorentz-like transformation (\ref{tr7})-(\ref{tr5}) allow us to eliminate
the preferred frame and preferred location from the equations describing
motion and gravitational field of a neutron star. Indeed, let us consider a
reference frame in which background gravitational field is $\phi ^{\text{back%
}}=$const and $h_{0\alpha }^{\text{back}}=$const. The background field is
produced by the companion star and the cosmological part. Since the total
action is invariant under rotations in the four dimensional Euclidean space $%
\delta _{ik}$ one can eliminate $h_{0\alpha }^{\text{back}}$ by making such
a rotation. After this transformation the background field becomes $\phi ^{%
\text{back}}=\phi _{0}=$const and $h_{0\alpha }^{\text{back}}=0$. Next we
perform scaling transformation (\ref{sc1}) and (\ref{sc2}) with $a=-\phi
_{0} $ which makes $\phi ^{\text{back}}=0$, that is now the background
metric is Minkowski metric $\eta _{ik}$. In the new frame, however, the
neutron star moves with some velocity $\mathbf{V}$. Finally, the
Lorentz-like transformation (\ref{tr7})-(\ref{tr5}) eliminates $\mathbf{V}$,
that is in the new reference frame the neutron star is at rest. At the same
time, the Lorentz-like transformation does not change the background
Minkowski metric $\eta _{ik}$. Indeed, far from the star the field
transformation reduces to Eqs. (\ref{lor1}) and (\ref{lor2}) for which the
asymptotic values $\phi =0$ and $h_{0\alpha }=0$ remain invariant.

We found that transformations which keep the total action invariant
eliminate the background field from the boundary conditions. Thus, field
equations and the equation of the star motion do not yield preferred frame
and preferred location effects for they are obtained by taking variation of
the total action. As a consequence, one can choose a reference frame in
which star is static and metric is asymptotically Minkowski. In this frame,
in the outer region of a nonrotating static star the gravitational field is
described by the equation $\Delta \phi =0$ with asymptotic boundary
conditions $\phi ^{\text{back}}=0$ and $h_{0\alpha }^{\text{back}}=0$ which
yields%
\begin{equation}
\phi (\mathbf{r})=-\frac{GM}{c^{2}r},  \label{stat}
\end{equation}%
where $M$ is a Kepler-measured mass of the star. Thus, solution for the
relativistic structure and gravitational field of the star is independent of
the background gravitational field.

To obtain gravitational field in a frame in which star moves with velocity $%
V\ll c$ one can make Lorentz-like transformation (\ref{tr7})-(\ref{tr5})
which yields analytical solution for arbitrary values of $\phi $. Using Eqs.
(\ref{www1}) we find the following expression for the equivalent metric
produced by the moving star 
\begin{equation}
f_{00}=e^{2\phi }+2\frac{V^{2}}{c^{2}}\text{sinh}(2\phi )\left( 2-e^{4\phi
}\right) ,  \label{fm1}
\end{equation}%
\begin{equation}
f_{0\alpha }=-2\frac{V^{\alpha }}{c}\text{sinh}(2\phi ),  \label{fm2}
\end{equation}%
\begin{equation}
f_{\alpha \beta }=\left( -e^{-2\phi }+2e^{-4\phi }\frac{V^{2}}{c^{2}}\text{%
sinh}(2\phi )\right) \delta _{\alpha \beta },  \label{fm3}
\end{equation}%
where $\phi $ is the field of the static star written in terms of
coordinates in the moving frame. In the far field in the post-Newtonian
limit Eqs (\ref{fm1})-(\ref{fm3}) reduce to%
\begin{equation}
h_{00}=-\frac{2GM}{c^{2}|\mathbf{r}-\mathbf{V}t|}\left( 1+\frac{2V^{2}}{c^{2}%
}\right) +\frac{2G^{2}M^{2}}{c^{4}|\mathbf{r}-\mathbf{V}t|^{2}},
\label{ppnf}
\end{equation}%
\begin{equation}
h_{0\alpha }=\frac{4G}{c^{3}}\frac{MV^{\alpha }}{|\mathbf{r}-\mathbf{V}t|},
\label{ppnh}
\end{equation}%
\begin{equation}
h_{\alpha \beta }=-\frac{2GM}{c^{2}|\mathbf{r}-\mathbf{V}t|}\delta _{\alpha
\beta }.  \label{fm4}
\end{equation}%
Our result coincides with those obtained in general relativity (in the gauge 
$2\partial h_{00}/\partial x^{0}-\partial h_{0\beta }/\partial x^{\beta }=0$%
) in the post-Newtonian limit far from the star. It is independent of the
original background metric as well as motion of the reference frame relative
to the background.

\section{Post-Newtonian limit of vector gravity in the parametrized
post-Newtonian formalism}

\label{PPNF}

Clifford Will in his book on \textit{\textquotedblleft Theory and experiment
in gravitational physics\textquotedblright\ }\cite{Will93} presented a
method of calculation of the post-Newtonian (PN) limits of any metric theory
of gravity. Vector gravity is a metric theory and, thus, approach of Ref. 
\cite{Will93} can be applied here as well.

The method outlined in the Will's book consists of $9$ steps. Namely, one
should start from the basic field equations of a metric theory of gravity,
solve them in the PN limit for the equivalent metric and compare the answer
with the parametrized post-Newtonian (PPN) expansion. In Appendix \ref{AP7}
we carried out the first four steps and found that equations of vector
gravity in the PN limit are identical to those of general relativity, so do
the boundary conditions. If this is the case then equivalence of vector
gravity and general relativity in the PN limit is proven. Calculations of
steps $5-9$ dealing with the actual solution of the equations are simply
identical to those in general relativity.

However, one of the referees of our paper believes that calculations of
Appendix \ref{AP7} are not sufficient and all $9$ steps must be included. We
present them here closely following prescription of Ref. \cite{Will93}. For
completeness of the presentation we repeat the first four steps as well.

In the PN formalism the metric is expanded in a small parameter $\epsilon $.
The \textquotedblleft order of smallness\textquotedblright\ is determined
according to the rules that matter velocity is of order $V\sim \epsilon
^{1/2}$ and gravitational constant $G\sim \epsilon $. A consistent PN limit
requires determination of $g_{00}$ correction through $O(\epsilon ^{2})$, $%
g_{0\alpha }$ through $O(\epsilon ^{3/2})$, and $g_{\alpha \beta }$ through $%
O(\epsilon )$.

Recall, that we use the following convention. Lower case Latin indices ($i$, 
$k$, $m$, ...) label four dimensional coordinates (range $0$, $1$, $2$, $3$%
), while lower case Greek letters $\alpha $, $\beta $, $\gamma $ denote
spatial coordinates (range $1$, $2$, $3$).

\textbf{Step 1.} \textit{Identify the variables.}

In vector gravity the scalar $\phi $ and the unit vector $u_{k}$ are
dynamical gravitational variables and flat background metric $\delta _{ik}$
is the prior-geometrical variable.

\textbf{Step 2. }\textit{Set the cosmological boundary conditions. Assume a
homogeneous isotropic cosmology, and at a chosen moment of time define the
values of the variables far from the PN system. Rest frame of the universe
is a convenient choice of the coordinate system. }

Please note that PN expansion of a metric theory of gravity and comparison
with general relativity\ can be made in any convenient reference frame. This
is true for any metric theory of gravity, including vector gravity. This
question is explained well in \cite{Will93}.

For vector gravity, the cosmological boundary conditions are:%
\begin{equation*}
\phi \rightarrow \phi _{0},\quad u_{k}\rightarrow (1,0,0,0)
\end{equation*}%
far from the PN system.

\textbf{Step 3.} \textit{Expand in a post-Newtonian series about the
asymptotic values.}

Expansion of the dynamical gravitational variables in vector gravity is%
\begin{equation*}
\phi =\phi _{0}+\tilde{\phi},\quad u_{k}=(1,0,0,0)+\tilde{u}_{k},
\end{equation*}%
where $|\tilde{\phi}|,|\tilde{u}_{k}|\ll 1$.

\textbf{Step 4.} \textit{Substitute these forms into the field equations,
keeping only such terms as are necessary to obtain a final, consistent PN
solution for }$h_{ik}$\textit{.}

Keeping the post-Newtonian terms, and taking into account that%
\begin{equation*}
T^{00}=\tilde{f}^{00}\tilde{f}^{00}T_{00}=e^{-4\phi }T_{00},\quad \tilde{f}%
^{00}=e^{-4\phi }f_{00},
\end{equation*}%
\begin{equation*}
T^{\alpha 0}=-T_{\alpha 0},
\end{equation*}%
the basic field equations of vector gravity (\ref{sss1}) reduce to the
following equations for $\tilde{\phi}$ and $\tilde{u}_{k}$%
\begin{equation*}
\Delta \tilde{\phi}+3e^{-4\phi _{0}}\frac{\partial ^{2}\tilde{\phi}}{%
\partial x^{0}\partial x^{0}}-2e^{-2\phi _{0}}\cosh (2\phi _{0})\frac{%
\partial ^{2}\tilde{u}^{\beta }}{\partial x^{0}\partial x^{\beta }}
\end{equation*}

\begin{equation}
=\frac{8\pi G}{c^{4}}e^{-4(\phi _{0}+\tilde{\phi})}\left( T_{00}-\frac{T}{2}%
f_{00}\right) ,  \label{eee1A}
\end{equation}%
\begin{equation}
e^{2\phi _{0}}\cosh (2\phi _{0})\left( \frac{\partial ^{2}\tilde{u}^{\beta }%
}{\partial x_{\alpha }\partial x^{\beta }}-\Delta \tilde{u}^{\alpha }\right)
-2\frac{\partial ^{2}\tilde{\phi}}{\partial x^{\alpha }\partial x^{0}}=-%
\frac{8\pi G}{c^{4}}T_{0\alpha }.  \label{eee2A}
\end{equation}%
Next we rescale coordinates as%
\begin{equation*}
x^{0}\rightarrow e^{-\phi _{0}}x^{0},\quad x^{\alpha }\rightarrow e^{\phi
_{0}}x^{\alpha }.
\end{equation*}%
In the new coordinates the equivalent metric $f_{ik}$ has the form 
\begin{equation}
f_{ik}=\eta _{ik}+\left( 
\begin{array}{cccc}
h_{00} & h_{01} & h_{02} & h_{03} \\ 
h_{01} & h_{00} & 0 & 0 \\ 
h_{02} & 0 & h_{00} & 0 \\ 
h_{03} & 0 & 0 & h_{00}%
\end{array}%
\right) ,  \label{eee3A}
\end{equation}%
where $\eta _{ik}=$diag$(1,-1,-1,-1)$ is Minkowski metric and%
\begin{equation*}
h_{00}=2\tilde{\phi}+2\tilde{\phi}^{2},\quad h_{0\alpha }=2\cosh (2\phi _{0})%
\tilde{u}_{\alpha },
\end{equation*}%
\begin{equation*}
\tilde{\phi}=\frac{h_{00}}{2}-\frac{h_{00}^{2}}{4}.
\end{equation*}%
In the rescaled coordinates in terms of $h_{00}$ and $h_{0\alpha }$ Eqs. (%
\ref{eee1A}) and (\ref{eee2A}) read%
\begin{equation*}
\frac{1}{2}\Delta h_{00}-\frac{1}{4}\Delta h_{00}^{2}+\frac{3}{2}\frac{%
\partial ^{2}h_{00}}{\partial x^{0}\partial x^{0}}-\frac{\partial
^{2}h_{0\beta }}{\partial x^{0}\partial x^{\beta }}
\end{equation*}%
\begin{equation}
=\frac{8\pi G}{c^{4}}e^{-4\tilde{\phi}}\left( T_{00}-\frac{T}{2}%
f_{00}\right) ,  \label{e7aA}
\end{equation}%
\begin{equation}
\frac{1}{2}\Delta h_{0\alpha }-\frac{1}{2}\frac{\partial ^{2}h_{0\beta }}{%
\partial x^{\alpha }\partial x^{\beta }}+\frac{\partial ^{2}h_{00}}{\partial
x^{\alpha }\partial x^{0}}=\frac{8\pi G}{c^{4}}T_{0\alpha }.  \label{e7A}
\end{equation}%
Multiplying both sides of Eq. (\ref{e7aA}) by $e^{4\tilde{\phi}}\approx
e^{2h_{00}}$ and expanding the exponential factor\ we obtain%
\begin{equation*}
\frac{1}{2}\Delta h_{00}+h_{00}\Delta h_{00}-\frac{1}{4}\Delta h_{00}^{2}+%
\frac{3}{2}\frac{\partial ^{2}h_{00}}{\partial x^{0}\partial x^{0}}-\frac{%
\partial ^{2}h_{0\beta }}{\partial x^{\beta }\partial x^{0}}
\end{equation*}%
\begin{equation*}
=\frac{8\pi G}{c^{4}}\left( T_{00}-\frac{T}{2}f_{00}\right) .
\end{equation*}%
Using 
\begin{equation*}
\Delta h_{00}^{2}=2(\nabla h_{00})^{2}+2h_{00}\Delta h_{00}
\end{equation*}%
we find%
\begin{equation*}
\frac{1}{2}\Delta h_{00}+\frac{3}{2}\frac{\partial ^{2}h_{00}}{\partial
x^{0}\partial x^{0}}-\frac{\partial ^{2}h_{0\beta }}{\partial x^{0}\partial
x^{\beta }}+\frac{1}{2}h_{00}\Delta h_{00}-\frac{1}{2}(\nabla h_{00})^{2}
\end{equation*}%
\begin{equation}
=\frac{8\pi G}{c^{4}}\left( T_{00}-\frac{T}{2}f_{00}\right) .  \label{e6A}
\end{equation}

Equations (\ref{e7A}), (\ref{e6A}) of vector gravity are identical to the
Einstein equations 
\begin{equation}
R_{ik}=\frac{8\pi G}{c^{4}}\left( T_{ik}-\frac{1}{2}g_{ik}T\right)
\label{pp1A}
\end{equation}%
in the PN limit. Indeed, let us consider small deviations $h_{ik}$ of the
tensor gravitational field $g_{ik}$ from the Minkowski metric $\eta _{ik}$%
\begin{equation*}
g_{ik}=\eta _{ik}+h_{ik}.
\end{equation*}%
In the PN limit, components of the Ricci tensor are%
\begin{equation*}
R_{00}=\frac{\partial }{\partial x^{0}}\left( \frac{\partial h_{0}^{\alpha }%
}{\partial x^{\alpha }}-\frac{1}{2}\frac{\partial h_{\alpha }^{\alpha }}{%
\partial x^{0}}\right) +\frac{1}{2}\Delta h_{00}+\frac{1}{2}h^{\alpha \beta }%
\frac{\partial ^{2}h_{00}}{\partial x^{\alpha }\partial x^{\beta }}
\end{equation*}%
\begin{equation}
-\frac{1}{4}\left( \nabla h_{00}\right) ^{2}-\frac{1}{4}\frac{\partial h_{00}%
}{\partial x^{\beta }}\left( 2\frac{\partial h_{\beta }^{\alpha }}{\partial
x^{\alpha }}-\frac{\partial h_{\alpha }^{\alpha }}{\partial x^{\beta }}%
\right) ,  \label{pp2A}
\end{equation}%
\begin{equation}
R_{0\alpha }=\frac{1}{2}\frac{\partial ^{2}h_{\alpha }^{\beta }}{\partial
x^{0}\partial x^{\beta }}+\frac{1}{2}\frac{\partial ^{2}h_{0}^{\beta }}{%
\partial x^{\alpha }\partial x^{\beta }}-\frac{1}{2}\frac{\partial
^{2}h_{\beta }^{\beta }}{\partial x^{\alpha }\partial x^{0}}+\frac{1}{2}%
\Delta h_{0\alpha },  \label{pp3A}
\end{equation}%
\begin{equation}
R_{\alpha \beta }=\frac{1}{2}\frac{\partial ^{2}h_{\alpha }^{m}}{\partial
x^{\beta }\partial x^{m}}+\frac{1}{2}\frac{\partial ^{2}h_{\beta }^{m}}{%
\partial x^{\alpha }\partial x^{m}}-\frac{1}{2}\frac{\partial ^{2}h_{m}^{m}}{%
\partial x^{\alpha }\partial x^{\beta }}+\frac{1}{2}\Delta h_{\alpha \beta },
\label{pp3a}
\end{equation}%
where $h_{\alpha }^{\beta }=\eta ^{\beta \gamma }h_{\gamma \alpha }$. If we
impose the three gauge conditions ($\gamma =1,2,3$)%
\begin{equation}
\frac{\partial h_{\gamma }^{m}}{\partial x^{m}}-\frac{1}{2}\frac{\partial
h_{m}^{m}}{\partial x^{\gamma }}=0
\end{equation}%
equation (\ref{pp3a}) becomes%
\begin{equation*}
R_{\alpha \beta }=\frac{1}{2}\Delta h_{\alpha \beta }
\end{equation*}%
and, in the order $O(\epsilon )$, Einstein equations (\ref{pp1A}) with $%
ik=\alpha \beta $ reduce to%
\begin{equation}
\Delta h_{\alpha \beta }=-\frac{8\pi G}{c^{4}}T\eta _{\alpha \beta }.
\label{sos1}
\end{equation}%
On the other hand, in this order, Einstein equations with $ik=00$ yield%
\begin{equation}
\Delta h_{00}=\frac{8\pi G}{c^{4}}\left( 2T_{00}-T\right) =\frac{8\pi G}{%
c^{4}}T.  \label{sos2}
\end{equation}%
Comparing Eqs. (\ref{sos1}) and (\ref{sos2}) we find that in the PN limit of
general relativity 
\begin{equation}
h_{\beta }^{\alpha }=-h_{00}\delta _{\beta }^{\alpha }  \label{pp4A}
\end{equation}%
and, hence, the metric is given by 
\begin{equation}
g_{ik}=\eta _{ik}+\left( 
\begin{array}{cccc}
h_{00} & h_{01} & h_{02} & h_{03} \\ 
h_{01} & h_{00} & 0 & 0 \\ 
h_{02} & 0 & h_{00} & 0 \\ 
h_{03} & 0 & 0 & h_{00}%
\end{array}%
\right) .  \label{sos3}
\end{equation}%
Metric (\ref{sos3}) has the same form as the equivalent metric (\ref{eee3A})
in vector gravity. Plugging Eq. (\ref{pp4A}) into Eqs. (\ref{pp2A}) and (\ref%
{pp3A}) we obtain%
\begin{equation*}
R_{00}=\frac{1}{2}\Delta h_{00}+\frac{3}{2}\frac{\partial ^{2}h_{00}}{%
\partial x^{0}\partial x^{0}}-\frac{\partial ^{2}h_{0\beta }}{\partial
x^{0}\partial x^{\beta }}+\frac{1}{2}h_{00}\Delta h_{00}-\frac{1}{2}\left(
\nabla h_{00}\right) ^{2},
\end{equation*}%
\begin{equation*}
R_{0\alpha }=\frac{1}{2}\Delta h_{0\alpha }+\frac{1}{2}\frac{\partial
^{2}h_{0\beta }}{\partial x^{\alpha }\partial x^{\beta }}+\frac{\partial
^{2}h_{00}}{\partial x^{\alpha }\partial x^{0}}.
\end{equation*}%
As a result, Einstein equations (\ref{pp1A}) with $i=0$ and $k=0,1,2,3$ in
the PN limit read%
\begin{equation*}
\frac{1}{2}\Delta h_{00}+\frac{3}{2}\frac{\partial ^{2}h_{00}}{\partial
x^{0}\partial x^{0}}-\frac{\partial ^{2}h_{0\beta }}{\partial x^{0}\partial
x^{\beta }}+\frac{1}{2}h_{00}\Delta h_{00}-\frac{1}{2}\left( \nabla
h_{00}\right) ^{2}
\end{equation*}%
\begin{equation}
=\frac{8\pi G}{c^{4}}\left( T_{00}-\frac{T}{2}g_{00}\right) ,  \label{e1A}
\end{equation}%
\begin{equation}
\frac{1}{2}\Delta h_{0\alpha }-\frac{1}{2}\frac{\partial ^{2}h_{0\beta }}{%
\partial x^{\alpha }\partial x^{\beta }}+\frac{\partial ^{2}h_{00}}{\partial
x^{\alpha }\partial x^{0}}=\frac{8\pi G}{c^{4}}T_{0\alpha }.  \label{e2A}
\end{equation}%
Equations (\ref{e1A}) and (\ref{e2A}) for the four unknown functions $h_{00}$%
, $h_{0\alpha }$ coincide with Eqs. (\ref{e6A}) and (\ref{e7A}) of vector
gravity. Boundary conditions are also the same in both theories, namely, far
from the PN system 
\begin{equation*}
h_{00}\rightarrow 0,\quad h_{0\alpha }\rightarrow 0.
\end{equation*}

This is sufficient to conclude that vector gravity and general relativity
are equivalent in the PN limit.

\textbf{Step 5. }\textit{Solve for }$h_{00}$\textit{\ to }$O(\epsilon )$%
\textit{. }

Only the lowest PN order equation is needed. In this order Eq. (\ref{e6A})
of vector gravity reduces to 
\begin{equation}
\Delta h_{00}=\frac{16\pi G}{c^{4}}\left( T_{00}-\frac{T}{2}\right) =\frac{%
8\pi G}{c^{2}}\rho ,  \label{h00e}
\end{equation}%
where $\rho (t,\mathbf{r})$ is the matter density (measured in a frame
momentarily comoving with the matter). Using 
\begin{equation*}
\Delta \frac{1}{|\mathbf{r}-\mathbf{r}^{\prime }|}=-4\pi \delta (\mathbf{r}-%
\mathbf{r}^{\prime })
\end{equation*}%
we obtain that solution of Eq. (\ref{h00e}) is%
\begin{equation}
h_{00}=-\frac{2}{c^{2}}U,  \label{h00}
\end{equation}%
where%
\begin{equation*}
U(t,\mathbf{r})=G\int \frac{\rho (t,\mathbf{r}^{\prime })}{|\mathbf{r}-%
\mathbf{r}^{\prime }|}d\mathbf{r}^{\prime }
\end{equation*}%
is the Newtonian gravitational potential with minus sign.

\textbf{Step 6. }\textit{Solve for }$h_{\alpha \beta }$\textit{\ to }$%
O(\epsilon )$\textit{\ and }$h_{0\alpha }$ to $O(\epsilon ^{3/2}).$

According to Eqs. (\ref{eee3A}) and (\ref{h00}), in vector gravity%
\begin{equation*}
h_{\alpha \beta }=h_{00}\delta _{\alpha \beta }=-\frac{2}{c^{2}}U\delta
_{\alpha \beta }\text{.}
\end{equation*}

Next we note that with the PN accuracy Eqs. (\ref{e7A}), (\ref{e6A}) are
invariant under the gauge transformation%
\begin{equation*}
h_{00}\rightarrow h_{00}+2\frac{\partial \psi }{\partial x^{0}},\quad
h_{0\alpha }\rightarrow h_{0\alpha }+\frac{\partial \psi }{\partial
x^{\alpha }},
\end{equation*}%
where $\psi $ is an arbitrary function of the order of $O(\epsilon )$. Thus,
we can impose one gauge fixing condition which we choose as in Refs. \cite%
{Will93,Land95}%
\begin{equation*}
\frac{\partial h_{0}^{\alpha }}{\partial x^{\alpha }}-\frac{1}{2}\frac{%
\partial h_{\alpha }^{\alpha }}{\partial x^{0}}=0
\end{equation*}%
or 
\begin{equation}
\frac{\partial h_{0\alpha }}{\partial x^{\alpha }}=\frac{3}{2}\frac{\partial
h_{00}}{\partial x^{0}}.  \label{hg}
\end{equation}%
Then, using Eq. (\ref{h00}), Eq. (\ref{e7A}) reduces to%
\begin{equation}
\Delta h_{0\alpha }-\frac{1}{c^{3}}\frac{\partial ^{2}U}{\partial x^{\alpha
}\partial t}=\frac{16\pi G}{c^{4}}T_{0\alpha },  \label{h03}
\end{equation}%
where with the required accuracy%
\begin{equation*}
T_{0\alpha }=\rho cV_{\alpha }
\end{equation*}%
and $V_{\alpha }=dx_{\alpha }/dt$ is the velocity of matter.

Solution of Eq. (\ref{h03}) satisfying the proper boundary condition is%
\begin{equation}
h_{0\alpha }=-\frac{4G}{c^{3}}\int \frac{\rho (t,\mathbf{r}^{\prime
})V_{\alpha }(t,\mathbf{r}^{\prime })}{|\mathbf{r}-\mathbf{r}^{\prime }|}d%
\mathbf{r}^{\prime }+\frac{1}{c^{3}}\frac{\partial ^{2}F}{\partial x^{\alpha
}\partial t},  \label{h0a}
\end{equation}%
where $F$ is the solution of the auxiliary equation 
\begin{equation*}
\Delta F=U=G\int \frac{\rho (t,\mathbf{r}^{\prime })}{|\mathbf{r}-\mathbf{r}%
^{\prime }|}d\mathbf{r}^{\prime }.
\end{equation*}%
Using the relation $\Delta r=2/r$, we find 
\begin{equation*}
F(t,\mathbf{r})=\frac{G}{2}\int \rho (t,\mathbf{r}^{\prime })|\mathbf{r}-%
\mathbf{r}^{\prime }|d\mathbf{r}^{\prime }.
\end{equation*}

Applying the continuity equation%
\begin{equation*}
\frac{\partial }{\partial t}\rho (t,\mathbf{r})+\frac{\partial }{\partial 
\mathbf{r}}(\rho \mathbf{V})=0,
\end{equation*}%
where $\mathbf{V}\equiv V^{\alpha }$, we obtain%
\begin{equation*}
\frac{\partial }{\partial t}F(t,\mathbf{r})=\frac{G}{2}\int d\mathbf{r}%
^{\prime }|\mathbf{r}-\mathbf{r}^{\prime }|\frac{\partial }{\partial t}\rho
(t,\mathbf{r}^{\prime })=
\end{equation*}%
\begin{equation*}
=-\frac{G}{2}\int d\mathbf{r}^{\prime }|\mathbf{r}-\mathbf{r}^{\prime }|%
\frac{\partial }{\partial \mathbf{r}^{\prime }}\left[ \rho (t,\mathbf{r}%
^{\prime })\mathbf{V}(t,\mathbf{r}^{\prime })\right] =
\end{equation*}%
\begin{equation*}
=\frac{G}{2}\int d\mathbf{r}^{\prime }\rho (t,\mathbf{r}^{\prime })\mathbf{V}%
(t,\mathbf{r}^{\prime })\frac{\partial }{\partial \mathbf{r}^{\prime }}|%
\mathbf{r}-\mathbf{r}^{\prime }|=
\end{equation*}%
\begin{equation*}
=-\frac{G}{2}\int d\mathbf{r}^{\prime }\rho (t,\mathbf{r}^{\prime })\frac{%
\mathbf{V}(t,\mathbf{r}^{\prime })\cdot (\mathbf{r}-\mathbf{r}^{\prime })}{|%
\mathbf{r}-\mathbf{r}^{\prime }|}.
\end{equation*}%
Taking derivative with respect to $x^{\alpha }$ we have%
\begin{equation*}
\frac{\partial ^{2}F(t,\mathbf{r})}{\partial x^{\alpha }\partial t}=-\frac{G%
}{2}\int d\mathbf{r}^{\prime }\frac{\rho (t,\mathbf{r}^{\prime })V^{\alpha
}(t,\mathbf{r}^{\prime })}{|\mathbf{r}-\mathbf{r}^{\prime }|}+
\end{equation*}%
\begin{equation*}
+\frac{G}{2}\int d\mathbf{r}^{\prime }\rho (t,\mathbf{r}^{\prime })\frac{%
\left[ \mathbf{V}(t,\mathbf{r}^{\prime })\cdot (\mathbf{r}-\mathbf{r}%
^{\prime })\right] (x^{\alpha }-x^{\prime \alpha })}{|\mathbf{r}-\mathbf{r}%
^{\prime }|^{3}}.
\end{equation*}%
Substituting this into Eq. (\ref{h0a}) and taking into account that $%
V^{\alpha }=-V_{\alpha }$, $x^{\alpha }=-x_{\alpha }$ we finally find%
\begin{equation*}
h_{0\alpha }=-\frac{7G}{2c^{3}}\int \frac{\rho (t,\mathbf{r}^{\prime
})V_{\alpha }(t,\mathbf{r}^{\prime })}{|\mathbf{r}-\mathbf{r}^{\prime }|}d%
\mathbf{r}^{\prime }-
\end{equation*}%
\begin{equation*}
-\frac{G}{2c^{3}}\int \rho (t,\mathbf{r}^{\prime })\frac{\left[ \mathbf{V}(t,%
\mathbf{r}^{\prime })\cdot (\mathbf{r}-\mathbf{r}^{\prime })\right]
(x_{\alpha }-x_{\alpha }^{\prime })}{|\mathbf{r}-\mathbf{r}^{\prime }|^{3}}d%
\mathbf{r}^{\prime }.
\end{equation*}

\textbf{Step 7. }\textit{Solve for }$h_{00}$\textit{\ to }$O(\epsilon ^{2})$%
\textit{. }

In the gauge (\ref{hg}), keeping terms of the $O(\epsilon ^{2})$ order, Eq. (%
\ref{e6A}) reduces to%
\begin{equation*}
\Delta h_{00}+h_{00}\Delta h_{00}-(\nabla h_{00})^{2}=\frac{16\pi G}{c^{4}}%
\left( T_{00}-\frac{T}{2}f_{00}\right) .
\end{equation*}%
Using 
\begin{equation*}
(\nabla h_{00})^{2}=\frac{1}{2}\Delta h_{00}^{2}-h_{00}\Delta h_{00}
\end{equation*}%
we obtain%
\begin{equation*}
\Delta \left( h_{00}-\frac{1}{2}h_{00}^{2}\right) +2h_{00}\Delta h_{00}=%
\frac{16\pi G}{c^{4}}\left( T_{00}-\frac{T}{2}f_{00}\right) .
\end{equation*}%
Substituting in the higher-order terms the known lower-order solution $%
h_{00}=-2U/c^{2}$, $\Delta h_{00}=\frac{8\pi G}{c^{2}}\rho $ and taking into
account that $f_{00}\approx 1+h_{00}$, we get%
\begin{equation*}
\Delta \left( h_{00}-\frac{2}{c^{4}}U^{2}\right) =
\end{equation*}%
\begin{equation}
\frac{16\pi G}{c^{4}}\left[ T_{00}-\frac{T}{2}\left( 1-\frac{2U}{c^{2}}%
\right) +2\rho U\right] .  \label{h2d}
\end{equation}

We will use a perfect fluid as a model of matter. Then in curved space-time
with the equivalent metric $f_{ik}$ the energy-momentum tensor of matter
reads%
\begin{equation*}
T_{ik}=\left( \varepsilon +P\right) v_{i}v_{k}-Pf_{ik},
\end{equation*}%
where $\varepsilon =\rho c^{2}(1+\Pi )$ is the rest energy density of the
fluid, $\Pi $ is the specific density of thermal energy, $P$ is the
isotropic pressure and $v^{k}=dx^{k}/ds$ is the four-velocity of the fluid
element. Here $ds=\sqrt{f_{ik}dx^{i}dx^{k}}$ and $v_{i}=f_{ik}v^{k}$. Trace
of $T_{ik}$ is $T=\varepsilon -3P$.

With the required accuracy $ds\approx cdt\sqrt{1+h_{00}-V^{2}/c^{2}}$ and,
therefore,%
\begin{equation*}
v_{0}^{2}=f_{00}^{\text{ }2}\frac{c^{2}dt^{2}}{ds^{2}}\approx \frac{\left(
1+h_{00}\right) ^{2}}{1+h_{00}-V^{2}/c^{2}}\approx 1-\frac{2U}{c^{2}}+\frac{%
V^{2}}{c^{2}},
\end{equation*}%
which yields%
\begin{equation*}
T_{00}\approx \varepsilon \left( 1-\frac{2U}{c^{2}}+\frac{V^{2}}{c^{2}}%
\right) .
\end{equation*}%
Substituting this into Eq. (\ref{h2d}) we obtain to the required accuracy%
\begin{equation*}
\Delta \left( h_{00}-\frac{2}{c^{4}}U^{2}\right) =
\end{equation*}%
\begin{equation}
\frac{8\pi G}{c^{4}}\left[ \rho c^{2}\left( 1+\Pi \right) +2\rho \left(
V^{2}+U\right) +3P\right] .  \label{h2de}
\end{equation}%
Solution of Eq. (\ref{h2de}) is%
\begin{equation*}
h_{00}=-\frac{2}{c^{2}}U+\frac{2}{c^{4}}U^{2}-\frac{G}{c^{4}}\int \left[
4\rho (t,\mathbf{r}^{\prime })\left( V^{2}(t,\mathbf{r}^{\prime })+U(t,%
\mathbf{r}^{\prime })\right) \right.
\end{equation*}%
\begin{equation*}
+\left. 2c^{2}\rho (t,\mathbf{r}^{\prime })\Pi (t,\mathbf{r}^{\prime })+6P(t,%
\mathbf{r}^{\prime })\right] \frac{d\mathbf{r}^{\prime }}{|\mathbf{r}-%
\mathbf{r}^{\prime }|}.
\end{equation*}%
\textbf{Steps 8 and 9. }\textit{Equivalent metric and PPN parameters. }

The final form for the equivalent metric is 
\begin{equation}
f_{00}=1-\frac{2}{c^{2}}U+\frac{2}{c^{4}}U^{2}-4\Phi _{1}-4\Phi _{2}-2\Phi
_{3}-6\Phi _{4},  \label{f00}
\end{equation}%
\begin{equation}
f_{0\alpha }=-\frac{7}{2}\mathcal{V}_{\alpha }-\frac{1}{2}W_{\alpha },
\label{f01}
\end{equation}%
\begin{equation}
f_{\alpha \beta }=-\left( 1+\frac{2U}{c^{2}}\right) \delta _{\alpha \beta },
\label{f02}
\end{equation}%
where 
\begin{equation*}
U=G\int \frac{\rho (t,\mathbf{r}^{\prime })}{|\mathbf{r}-\mathbf{r}^{\prime
}|}d\mathbf{r}^{\prime },\quad \mathcal{V}_{\alpha }=\frac{G}{c^{3}}\int 
\frac{\rho (t,\mathbf{r}^{\prime })V_{\alpha }(t,\mathbf{r}^{\prime })}{|%
\mathbf{r}-\mathbf{r}^{\prime }|}d\mathbf{r}^{\prime },
\end{equation*}%
\begin{equation*}
W_{\alpha }=\frac{G}{c^{3}}\int \rho (t,\mathbf{r}^{\prime })\frac{\left[ 
\mathbf{V}(t,\mathbf{r}^{\prime })\cdot (\mathbf{r}-\mathbf{r}^{\prime })%
\right] (x_{\alpha }-x_{\alpha }^{\prime })}{|\mathbf{r}-\mathbf{r}^{\prime
}|^{3}}d\mathbf{r}^{\prime },
\end{equation*}%
\begin{equation*}
\Phi _{1}=\frac{G}{c^{4}}\int \frac{\rho (t,\mathbf{r}^{\prime })V^{2}(t,%
\mathbf{r}^{\prime })}{|\mathbf{r}-\mathbf{r}^{\prime }|}d\mathbf{r}^{\prime
},
\end{equation*}%
\begin{equation*}
\Phi _{2}=\frac{G}{c^{4}}\int \frac{\rho (t,\mathbf{r}^{\prime })U(t,\mathbf{%
r}^{\prime })}{|\mathbf{r}-\mathbf{r}^{\prime }|}d\mathbf{r}^{\prime },
\end{equation*}%
\begin{equation*}
\Phi _{3}=\frac{G}{c^{2}}\int \frac{\rho (t,\mathbf{r}^{\prime })\Pi (t,%
\mathbf{r}^{\prime })}{|\mathbf{r}-\mathbf{r}^{\prime }|}d\mathbf{r}^{\prime
},\quad \Phi _{4}=\frac{G}{c^{4}}\int \frac{P(t,\mathbf{r}^{\prime })}{|%
\mathbf{r}-\mathbf{r}^{\prime }|}d\mathbf{r}^{\prime }
\end{equation*}%
are metric potentials defined in the same way as in Ref. \cite{Will93}. One
should note that in \cite{Will93} the Minkowski metric is chosen as $\eta
_{ik}=$diag$(-1,1,1,1)$ and, as a result, formulas for $f_{00}$ and $%
f_{\alpha \beta }$ have the opposite sign.

The metric (\ref{f00})-(\ref{f02}) is written in the standard PPN gauge and,
hence, the PPN parameters can be read off immediately%
\begin{equation*}
\gamma =\beta =1,\quad \zeta =0,
\end{equation*}%
\begin{equation*}
\alpha _{1}=\alpha _{2}=\alpha _{3}=\zeta _{1}=\zeta _{2}=\zeta _{3}=\zeta
_{4}=0.
\end{equation*}%
They are the same as in general relativity. Thus, vector gravity is a fully
conservative theory of gravity and predicts no preferred-frame effects in
the PN limit. Moreover, in Appendix \ref{cossup} we show that in vector
gravity there are no preferred-frame effects in the $V^{2}/c^{2}$ order in
the matter velocity for arbitrary large values of the gravitational
potential.

\end{document}